\documentclass[aps,prb,twocolumn,superscriptaddress,nofootinbib,floatfix]{revtex4-2}
\usepackage[T1]{fontenc} 
\usepackage{float}
\usepackage{graphicx}
\usepackage{dcolumn}
\usepackage{bm}
\usepackage{color}
\usepackage{amsmath}
\usepackage{amssymb}
\usepackage{xcolor}
\usepackage{glossaries}
\usepackage{MnSymbol}
\usepackage{braket}
\usepackage{hyperref}
\hypersetup{colorlinks=true,linkcolor=blue,urlcolor=red,citecolor=blue,anchorcolor=blue}

\usepackage{mathdots}

\newacronym{QMC}{QMC}{quantum Monte Carlo}
\newacronym{CDW}{CDW}{charge-density-wave}
\newacronym{DQMC}{DQMC}{determinant quantum Monte Carlo}
\newacronym{DMRG}{DMRG}{density matrix renormalization group}
\newacronym{DMFT}{DMFT}{dynamical mean-field theory}
\newacronym{DCA}{DCA}{dynamical cluster approximation}
\newacronym{ML}{ML}{machine learning}
\newacronym{AI}{AI}{artificial intelligence}
\newacronym{PCA}{PCA}{Principal component analysis}
\newacronym{HS}{HS}{Hubbard-Stratonovich}

\begin{document}
\title{A perspective on machine learning and data science for strongly correlated electron problems}
\author{Steven Johnston}
\email{sjohn145@utk.edu}
\affiliation{Department of Physics and Astronomy, The University of Tennessee, Knoxville, Tennessee 37996, USA}
\affiliation{Institute for Advanced Materials and Manufacturing, University of Tennessee, Knoxville, Tennessee 37996, USA\looseness=-1}
\author{Ehsan Khatami}
\affiliation{Department of Physics and Astronomy, San Jose State University, San Jose, California 95192, USA\looseness=-1}

\author{Richard Scalettar}
\affiliation{Department of Physics and Astronomy, University of California, Davis, California 95616, USA\looseness=-1}

\date{\today}

\begin{abstract}
Numerical approaches to the correlated electron problem have achieved considerable success, yet are still constrained by several bottlenecks, including high order polynomial or exponential scaling in system size, long autocorrelation times, challenges in recognizing novel phases, and the Fermion sign problem. Methods in machine learning (ML), artificial intelligence, and data science promise to help address these limitations and open up a new frontier in strongly correlated quantum system simulations. In this paper, we review some of the progress in this area. We begin by examining these approaches in the context of classical models, where their underpinnings and application can be easily illustrated and benchmarked. We then discuss cases where ML methods have enabled scientific discovery. Finally, we will examine their applications in accelerating model solutions in state-of-the-art quantum many-body methods like quantum Monte Carlo and discuss potential future research directions. 
\end{abstract}

\maketitle

\tableofcontents
\section{Introduction}\label{sec:intro}

The development of nonperturbative numerical methods and steady growth in available computational power achieved over the last half-century have provided us with the tools necessary to obtain reliable solutions to simplified correlated electron models. Through the combined use of methods like \gls*{QMC}, \gls*{DMRG}, \gls*{DMFT} and its cluster extensions, and others, we now have a great deal of insight into the physics of the single-band Hubbard \cite{WhitePRB1989, PhysRevLett.101.186403, Jarrell_2001, IdoPRB2018, WhitePRL1989a, PhysRevB.78.241101, WhitePRL1989b, HuangQuantMat2018, MaierPRL2005, LiCommPhys2021, MaiPNAS2022, HuangPreprint, Qin2020, JiangPNAS2021, Zheng1155, WhitePRL2003, XiaoPreprint, Wietek2021}, Holstein \cite{ScalettarPRB1989, PhysRevB.74.045106, Hague_2005,DeeCommPhys2021, NosarzewskiPRB2021, BradleyPRB2021, CohenSteadPRB2021}, Su-Schrieffer-Heeger~\cite{marchand10, weber15, johnsousmonaberciu2018, LiQM2020, cai_2021,GoetzAssaad2021, XingBatrouni2021, Zhang_Sous2022}, and periodic Anderson models~\cite{jarrell1993periodic,vekic1995competition,held2000similarities,sun2003extended,luitz2010weak,wu2015d}. Substantial progress is also being made towards understanding models where several different, competing, interactions are present~\cite{PhysRevLett.99.146404, PhysRevLett.99.146404, fehske08, PhysRevB.90.195134, greitemann15, hohenadler16,WangDemler2020, costa2020, JohnstonPRB2013, KarakuzuPreprint2022, LiPreprint2022}, as well as those that extend beyond  the ``standard''  interactions~\cite{PhysRevB.65.214510, LiPRB2015, PhysRevB.87.125149, PakiPRB2019, dee2020relative,  ChenScience2021, LiPreprint2022}. For example, while not everything is settled, it is now clear that the single-band Hubbard model in the intermediate coupling regime,\footnote{$U/W\approx 1$ where $U$ is the strength of the Hubbard interaction and $W$ is the electronic bandwidth, see Sec.~\ref{sec:models}.} is a minimal model for the low-temperature electronic properties of cuprates, and many aspects of their phase diagram \cite{WhitePRL1989a, WhitePRL1989b, WhitePRL2003, Jarrell_2001, Zheng1155, HuangScience2017, XiaoPreprint, IdoPRB2018, HuangQuantMat2018, Wietek2021, MaiPNAS2022, MaierPRL2005, HuangPreprint, KarakuzuPreprint2022} (with the important possible exception of superconductivity \cite{MaierPRL2005, Qin2020, JiangPNAS2021}). This knowledge, however, was only won after decades of method development and research studies by groups from around the world. 

While impressive progress has been made toward understanding multi-orbital extensions of these models, achieving the same level of success for other strongly correlated materials will require new approaches. For example, the suitable minimal effective models remain unknown in many cases. And even when they are known, they often cannot be treated using existing numerical methods owning to issues like the Fermion ``sign problem''~\cite{loh90, wu05, troyer05, chandrasekharan10, li16, Iazzi2016, Hangleiter2020, MondainiScience2022}. 

Many researchers hope that methods in data science, \gls*{ML}, and \gls*{AI} can be exploited to reach the next stage of quantum simulations. Research in this area has developed along with several particularly opportune areas, including (but not limited to) using 
\gls*{ML} to 1) implement general schemes to perform ``global moves'' -- analogs of loop and Wolff-Swendsen-Wang methods~\cite{swendsen87,wolff88,edwards88} -- in classical and QMC simulations that update many degrees of freedom simultaneously with high acceptance rates; 
(2) recognize phase transitions in models, especially in cases when the relevant order parameter(s) are unknown; (3) determine optimal constrained wave functions in variational approaches to deal with the sign problem with the minimal introduction of bias; and
(4) extract suitable low-energy effective models from experimental data or high-energy models.

In response to this rapidly emerging field, Oak Ridge National Laboratory held a workshop on \emph{Artificial Intelligence in Multi-Fidelity, Multi-Scale and Multi-Physics Simulations of Materials} in August of 2021 as part of the Joint Nanoscience and Neutron Scattering User Meeting. This special issue 
on {\it Digital Twins in Materials and Chemical Sciences} 
is the result of that workshop. 
At that meeting, we gave consecutive talks on applications of \gls*{ML} for the many-body problem, drawing from our research areas. In that spirit, this article synthesizes our talks in a mini-review, focusing on where these methods intersect with our work and interests in magnetic, charge, and pairing order in classical and quantum many-body systems. The fields of correlated electrons and \gls*{ML} applications are broad and rapidly evolving. We have not attempted to be exhaustive in our references or discussion, especially given the rapid pace at which this field is developing. For the interested reader, we note that several comprehensive reviews and books of \gls*{ML} applications in physics have recently been published \cite{CarleoRMP2019, Carrasquilla2020, FeickertPreprint, Karniadakis2021, ChenReview2021, PDLT-2022, DawidPreprint}. We apologize in advance for any work that we have left omitted. 

The organization of this paper is as follows. Section \ref{sec:models} will provide a brief overview of the models discussed throughout this paper. Since the majority of our research involving \gls*{ML} methods focuses on Markov Chain Monte Carlo methods, Sec.~\ref{sec:MCMC} provides a brief description, including the path integral mapping of the quantum partition function in $d$ spatial dimensions to an equivalent classical problem in $d+1$ dimensions. Sec.~\ref{sec:ML_Methods} then provides an overview of the \gls*{ML} methods discussed throughout this work while Sec.~\ref{sec:benchmarks} discusses several proof-of-concept studies applying these methods to problems in statistical and correlated electron physics. 
We then highlight a few new scientific discoveries that have been enabled by 
\gls*{ML} methods in Sec.~\ref{sec:discovery}. Sec.~\ref{sec:SLMC} focuses on self-learning Monte Carlo methods, emphasizing recent applications to the Holstein model as an example. Finally, Sec.~\ref{sec:outlook} provides our perspective on future work in this area.

\section{Models}\label{sec:models}
We will discuss several ML and data science applications to understanding correlated electron systems, often tested on canonical model Hamiltonians. This section provides an overview of the relevant models to acquaint the reader and establish our notation, beginning with classical systems. Due to the nature of this review, our discussion will be brief. The reader familiar with this material can safely skip to the next section. 

\subsection{The Ising model}
The Ising Hamiltonian
\begin{equation}\label{eq:ising}
    H = - \sum_{ij} J_{ij} s_i s_j - \sum_{i} Bs_i, 
\end{equation}
describes a collection of localized classical degrees of freedom, which are often understood to represent the magnetic moments (or spins) of individual atoms on a lattice of sites $i=1,2,\dots, N$. 
These spins are constrained to point in one of two directions such that $s_i = \pm 1$. Each spin individually couples to an external magnetic field $B$ so that the orientation
$s_i=+1$ is energetically favored. Pairs of spins interact via a
magnetic coupling $J_{ij}$ between 
different lattice sites. For $J_{ij} > 0$ ($<0$), the interactions are ferromagnetic (antiferromagnetic) in nature. 

The Ising model is perhaps the simplest model exhibiting spontaneous symmetry breaking, which occurs at a phase transition. For example, in the {\it absence} of a field $B=0$, the total energy is invariant under a global change of variables $s_i \leftrightarrow -s_i$.  Nevertheless in $d \geq 2$, the system falls into one of these two equivalent states as $T$ is lowered, and for the ferromagnetic case, the magnetization $m = 1/N \sum_i \langle s_i \rangle$ becomes non-zero.
This behavior occurs only in the thermodynamic limit, $N\rightarrow\infty$, so one of the challenges of Monte Carlo and in the use of \gls*{ML} is in the finite-size scaling needed to extrapolate finite $N$ simulations.

The Ising model has known analytical solutions in certain cases~\cite{Onsager}. For example, on a square lattice for nearest-neighbor ferromagnetic coupling and in zero field, it undergoes a second-order transition between paramagnetic ($m=0$) and ferromagnetic ($m \neq 0$) phases at $T_c=2.269 \, J$. 
The correlation length $\xi$, which measures the decay in the magnetic correlations 
$\langle s_i s_{i+r} \rangle \sim e^{-r/\xi}$, diverges at this transition as $\xi \sim 1/(T-T_c)^{\nu}$,
as does the magnetic susceptibility $\chi = [ \langle m^2 \rangle - \langle m \rangle^2 ]/T \sim
1/(T-T_c)^\gamma$.
The exponents $\nu=1$ and $\gamma = 7/4$ are known analytically, as is the exponent $\beta =1/8$ describing
the onset of nonzero magnetization $m \sim (T_c-T)^\beta$ below $T_c$.
There is no known analytic solution for the Ising model in $d>2$.

The Ising model combines simplicity with deep conceptual physics. It also provides a textbook application for Markov chain Monte Carlo methods, through which researchers can readily produce training and validation data in large quantities. The fact that the model has an exact solution in $d\le 2$ while being unsolved in general means that researchers can use it as a high-precision benchmark for new numerical approaches while also offering the opportunity to sharpen our knowledge of unknown critical points and exponents.  
These aspects have positioned the Ising model as an essential test for many new ML-based approaches \cite{SLMC, wetzel2017machine, kim2018smallest, Morningstar2018, alexandrou2020critical, yevick2021variational, Agrawal2022}. 

Because the (ferromagnetic) Ising model on other 2D lattices falls into the same universality class (i.e.~identical critical exponents), it is probably sufficient to consider the square lattice in testing \gls*{ML} approaches to this problem. It is worth noting, however, that realizations of the Ising model on different lattices exist, which offer rich opportunities to compare traditional simulation methods with \gls*{ML}-driven methodologies. These include Kagome and triangular lattices~\cite{stephenson1964ising,landau1983critical,moessner2001ising}, which exhibit frustration in the antiferromagnetic (AF) case of $J_{ij} < 0$, and spin-glass behavior when the $J_{ij}$ are chosen with random ferromagnetic and antiferromagnetic signs~\cite{sherrington1975solvable,singh1986critical,mcmillan1984scaling}.

\subsection{The Blume-Capel model}
The Blume-Capel model \cite{Blume, Capel} is a generalization of the Ising model, where the magnetic moments can align parallel, antiparallel, or orthogonal to an external magnetic field. Its Hamiltonian is 
\begin{equation}\label{eq:blume-capel}
    H = -\Delta \sum_{i}\left(1-s_i^2\right) - 
    \sum_{i,j} J_{ij} s_is_j - B \sum_i s_i.
\end{equation}
where $s_i = 0,\pm 1$ and $\Delta$ is the zero-field splitting, which measures the energy difference between the singlet $s_i = 0$ and doublet $s_i = \pm 1$ moments. 

The parameter $\Delta$ controls the density of $s_i=0$ sites.  When $\Delta \rightarrow -\infty$ such sites are energetically highly unfavorable. The Blume-Capel model approaches the Ising model in this limit, exhibiting the same second-order magnetic phase transition.  As $\Delta$ increases,
more $s_i=0$ sites are introduced and $T_c$ decreases.  Ultimately, at $\Delta=J/2$ there is no long-range order at any $T>0$, i.e.~the critical
temperature $T_c=0$.
A fascinating feature of this model is that along the phase boundary in the $T-\Delta$ plane, the nature of the transition
changes from second to first order at a ``tricritical point.''
Thus the Blume-Capel model offers several new vistas for
\gls*{ML}: first, in providing a simple model to explore methods that can locate a one-dimensional locus of critical temperatures, as opposed to an isolated critical point, and second, as means to examine whether these methods can distinguish different types of phase transitions.

\subsection{The Hubbard model}
The Hubbard Hamiltonian~\cite{Hubbard} is a simple model for describing the effect
of electron-electron interactions on itinerant lattice electrons. 
In the case of a single band,
\begin{equation}
    \hat H = \hat H_\mathrm{0} + \hat H_U, 
\end{equation}
where 
\begin{equation}\label{eq:Hkin}
    \hat H_\mathrm{0} = \sum_{i,j,\sigma} t^{\phantom{\dagger}}_{i,j} c^\dagger_{i,\sigma} c^{\phantom\dagger}_{j,\sigma} -\mu \sum_{i,\sigma} \hat{n}_{i,\sigma} 
\end{equation}
describes electrons propagating through a one-orbital lattice 
and
\begin{equation}\label{eq:Hint}
    \hat H_U = U\sum_{i}\hat{n}_{i,\uparrow}\hat{n}_{i,\downarrow}
\end{equation}
is a local Coulomb interaction between electrons on the same site. 
Here, $c^\dagger_{i,\sigma}$ ($c^{\phantom\dagger}_{i,\sigma}$) creates (destroys) an electron with spin $\sigma$ on lattice site $i$, $\hat{n}_{i,\sigma}$ is the Fermion number operator at site $i$, $\mu$ is the chemical potential, $t_{i,j}$ is hopping integral between sites $i$ and $j$, and $U$ is the strength of the Hubbard interaction, which can either be repulsive ($U > 0$) or attractive ($U < 0$). 
The range of values of the energy levels of $H_0$ is referred to 
as the bandwidth $W$.

Despite its simplicity, analytical solutions to the Hubbard model have only been obtained in one dimension \cite{LiebPRL1968}. In higher dimensions, the most reliable results have been obtained with nonperturbative numerical methods, particularly in the intermediate coupling regime $U/W \approx 1$. Nevertheless, significant progress has been made toward understanding the physics of the Hubbard model in this regime. For example, the doped two-dimensional Hubbard model contains robust antiferromagnetic Mott correlations \cite{WhitePRB1989, SordiPRL2012, Wietek2021}, spin- and charge-stripes \cite{WhitePRL2003, Robert2014, ZhengScience2017, HuangQuantMat2018, MaiPNAS2022, HuangPreprint, Qin2020, Wietek2021}, strange metal behavior \cite{HuangScience2019, MondainiScience2022}, a  pseudogap \cite{SordiPRL2012, GullPRL2013, Wietek2021}, and unconventional superconductivity  \cite{MaierPRL2005, GullPRL2013, ChungPRB2020, Mai2021}. 

The Hubbard model can also be easily generalized to include longer-range interactions, 
\begin{equation}\label{eq:ext_hubbard}
    \hat H_V = \sum_{i\ne j,\sigma,\sigma^\prime} V_{i,j}
    \hat{n}_{i,\sigma}\hat{n}_{j,\sigma^\prime},
\end{equation}
or additional orbital degrees of freedom by modifying the underlying tight-binding model. 
From a computational perspective, the inclusion of further orbitals is `trivial'
in the determinant QMC formalism on which we focus - 
the orbital index plays an identical role as a site label.
However, longer-range interactions have a dramatic effect.
They necessitate a significant restructuring of the Hubbard-Stratonovich transformation,
described below, and make the fermion sign problem much worse.

\subsection{The Holstein model}
The Holstein Hamiltonian is a simplified model for describing electrons coupled to the lattice ~\cite{Holstein}. 
Like the Hubbard model, it approximates the noninteracting electronic degrees of freedom using a single orbital tight-binding model. The noninteracting lattice degrees of freedom are described using simple harmonic oscillators at each site, while the electron-lattice interaction is introduced through a linear coupling between 
the on-site electron density and the lattice displacement. The corresponding Hamiltonian is 
\begin{equation}
    \hat H = \hat H_{0}+\hat H_\mathrm{lat}+\hat H_{e-\mathrm{lat}}, 
\end{equation} 
where $H_0$ describes the electronic degrees of freedom as in Eq.~\eqref{eq:Hkin}, 
\begin{equation}
    \hat H_\mathrm{lat} = \sum_i\left[\frac{\hat{P}_i^2}{2M}
    +\frac{1}{2}M\omega_0^2 \hat{X}_i^2\right]
\end{equation}
describes the noninteracting lattice degrees of freedom and 
\begin{equation}
    \hat H_{e-\mathrm{lat}} = \sum_{i,\sigma} \alpha \, \hat{n}_{i,\sigma}\hat{X}_i 
\end{equation}
describes the electron-phonon interaction. 
Here, $\hat{P}_i$ and $\hat{X}_i$ are the momentum and position operators for the oscillator at site $i$, $\omega_0$ is the energy of the oscillator ($\hbar = 1$), and $\alpha$ is the electron-phonon coupling strength. 

Two dimensionless ratios are frequently quoted when simulating the Holstein model. The first is the so-called adiabatic ratio $\omega_0/E_\mathrm{F}$, which measures the relative energies of the lattice and electron degrees of freedom. The second is the dimensionless $e$-ph coupling constant $\lambda = 2\alpha^2 \, / \,(MW\omega_0^2)$, 
where $W$ is again the noninteracting bandwidth. 

Like the Hubbard model, the Holstein model exhibits a rich collection of phases. The model has a metal to a \gls*{CDW} insulator transition near half-filling \cite{ScalettarPRB1991, BradleyPRB2021}, conventional superconductivity away from half-filling \cite{DeePRB2019}, and polaron and bipolaron formation~\cite{Esterlis2018, BradleyPRB2021, NosarzewskiPRB2021}. In the ML context, this model provides an excellent platform for testing new algorithms where electrons are coupled to continuous fields (the lattice displacements) and where there can be significant differences in the time scales associated with the electron and lattice dynamics. It is also worth noting that a fast and scalable hybrid quantum Monte Carlo algorithm has recently been made available for studying Holstein models and other electron-phonon coupled systems~\cite{CohenSteadPRE2022}. This approach can treat these models on large system sizes, thus providing additional opportunities for testing proposed ML-accelerated algorithms.  

\subsection{The periodic Anderson model}\label{sec:PAM_model}
The periodic Anderson model (PAM) is a variant of the Hubbard Hamiltonian
containing two orbitals per site. In this case, one orbital ($c_{i,\sigma}$) is `metallic'  and has no on-site interaction, while the other orbital ($f_{i,\sigma}$) is `localized' and has a large on-site Hubbard repulsion $U_f \neq 0$. 
Its Hamiltonian is 
\begin{eqnarray}\nonumber
    \hat H&=&-t\sum_{\langle i,j\rangle}c^\dagger_{i,\sigma}c^{\phantom\dagger}_{j,\sigma} - \mu \sum_{i,\sigma} \left[n^c_{i,\sigma} + n^f_{i,\sigma}\right] \\
    &+&V\sum_{i,\sigma}\left[c^\dagger_{i,\sigma}f^{\phantom\dagger}_{i,\sigma} + h.c.\right]
    +U_f\sum_i n^f_{i,\uparrow}n^f_{i,\downarrow},  
\end{eqnarray}
where $V$ is the hybridization between the localized and metallic orbitals and 
$n^c_{i,\sigma} = c^\dagger_{i,\sigma}c^{\phantom\dagger}_{i,\sigma}$ and 
$n^f_{i,\sigma} = f^\dagger_{i,\sigma}f^{\phantom\dagger}_{i,\sigma}$ are the number operators for each type of orbital. 

In its ground state and at half-filling, the PAM undergoes a quantum phase transition as a function of $V$. Here, the system transitions from a state with antiferromagnetic order on the $f$-orbitals at small $V$ to a singlet phase at large $V$. In the latter, the conduction and localized $f$ electrons form local spin-0 singlets on a site, which disrupts the long-range antiferromagnetic order. Previous determinant QMC studies have placed the quantum critical point (QCP) for this transition at $V_c \sim t$ for $U_f \sim 4t$~\cite{vekic1995competition}.

\section{Monte Carlo methods}\label{sec:MCMC}
\subsection{Overview}
Markov chain Monte Carlo (MCMC) methods are a powerful
class of algorithms for 
simulating physical systems and have found widespread use
 in condensed matter physics
 \cite{gubernatis_kawashima_werner_2016, becca_sorella_2017}.
 These techniques perform a random walk through some configuration space  while statistically sampling the relevant observables in a way that guarantees the correct probability distribution is generated. 
 
\subsection{Classical Monte Carlo}
In a classical Monte Carlo simulation, one aims to evaluate the thermodynamic expectation value of an observable $O$ with respect to a set of microstates $|m\rangle$ that follow a Boltzmann probability distribution,
\begin{equation}\label{eq:classicalMC}
   \langle O \rangle  = \frac{1}{Z}\sum_m \langle m |O_m e^{-\beta E_m}|m\rangle. 
\end{equation}
Here, $E_m$ denotes the energy of the microstate, $O_m$ denotes the value of the observable for the microstate, and 
$Z = \sum_m \exp(-\beta E_m)$ is the partition function. 
 
The sum in Eq.~\eqref{eq:classicalMC} must be taken over all accessible microstates, which is intractable for most systems of interest. Instead, one uses MCMC methods to evaluate the sum stochastically. 
In a classical simulation of the Ising model, for example, a
 standard procedure is to start with a random lattice of up or
 down spins and then select individual spins to flip.
 These flips may be done either by visiting each site in a random order
 or by going through the lattice in some specified pattern. 
 The change in energy $\Delta E$ that is induced by flipping
 the spin is then evaluated, and the proposed move is accepted 
 with probability $p={\rm min} (1,e^{-\beta \Delta E})$, where
 $\beta=1/T$ is the inverse temperature in units where $k_{\rm B}=1$. 
This prescription is the `Metropolis-Hastings'
 algorithm ~\cite{Hastings}, and it ensures that the statistical
 distribution of the spins
 follows a Boltzmann distribution. Alternatively,
 one can use a `heat-bath' probability
 $p=e^{-\beta \Delta E}/(1+ e^{-\beta \Delta E})$.
 
 A crucial feature of classical Monte Carlo is that updating the
 entire lattice scales {\it linearly} with system size $N$, making simulations of large lattices practical\footnote{There can, however, be `hidden' factors of $N$. 
 Most commonly, 
 near a critical point, the autocorrelation time $\tau$ 
 of the simulation diverges 
 as $L^z$, where $L$ is the linear system size and
 $z$ is the dynamical critical exponent of the algorithm
 being used. $z$ often takes the value $z\sim 2$, leading to
 very long $\tau$. However, in many cases, special larger-scale
 ``global'' or ``block'' moves have been developed to address this
 \cite{swendsen87,wolff88,edwards88}, leading to $z \sim 0$.
 }.  
 This linear scaling follows from the locality of the energy, 
 which implies that evaluating $\Delta E$ is independent of $N$. 
 In contrast, QMC generally scales as $N^3$ for interacting fermions.
 As we shall discuss later, this scaling results from the fact that the action determining 
 the probability is {\it non-local}, involving the 
 determinant of a matrix of dimension $N$.
 
\begin{figure}[t]
    \centering
    \includegraphics[width=0.8\columnwidth]{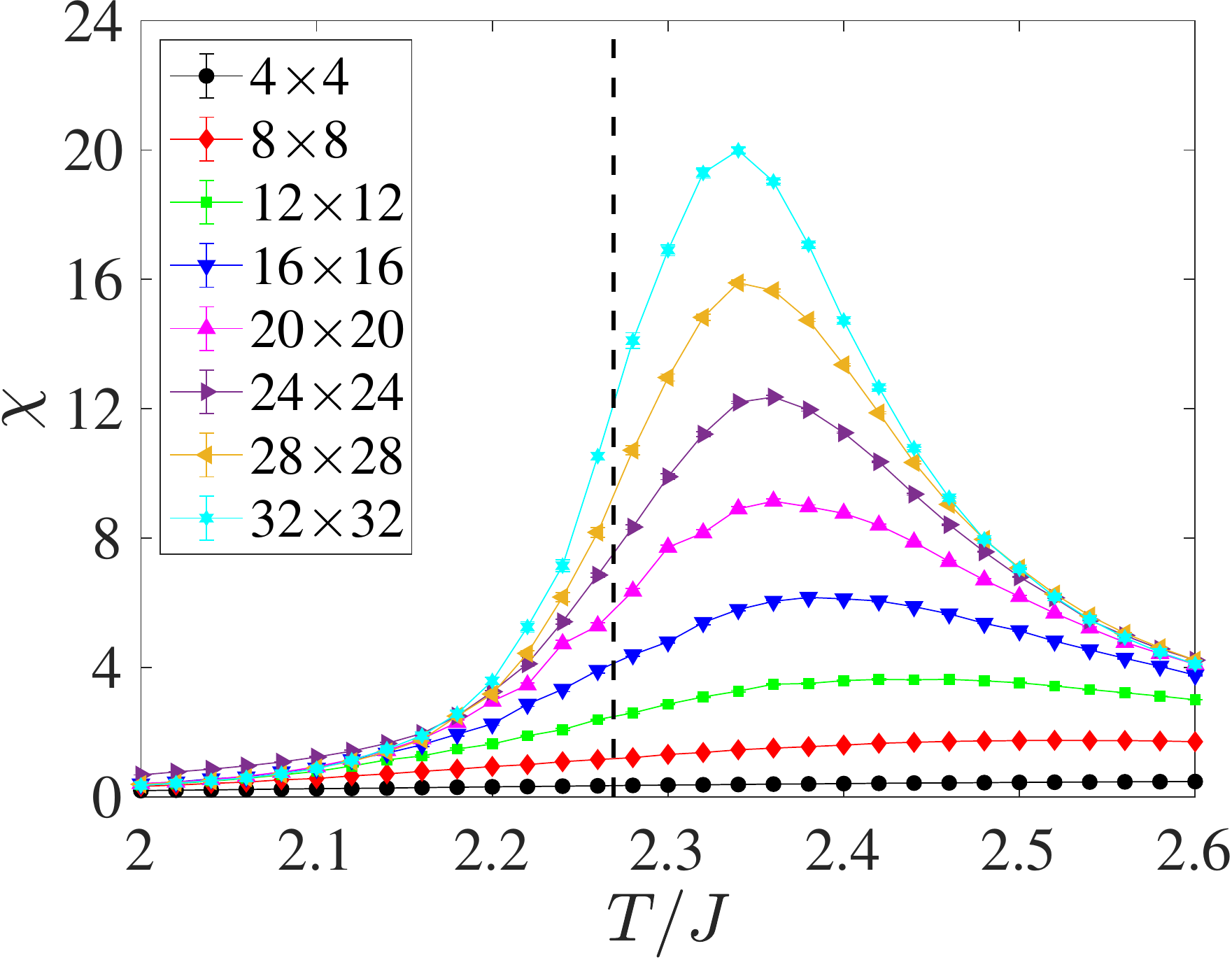}
    \caption{
    The magnetic susceptibility $\chi = d\langle M \rangle/dB
    = \beta (\langle M^2 \rangle - \langle M \rangle^2 )$
    of the 2D Ising model using a primitive single spin flip
    Metropolis Monte Carlo algorithm. $\chi$ is maximal in the vicinity
    of the known $T_c=2.269$, and the peak shifts towards $T_c$ as 
    the lattice size $N$ grows.
    }
    \label{fig:Isingchi}
\end{figure}

\begin{figure}[t]
    \centering
    \includegraphics[width=0.8\columnwidth]{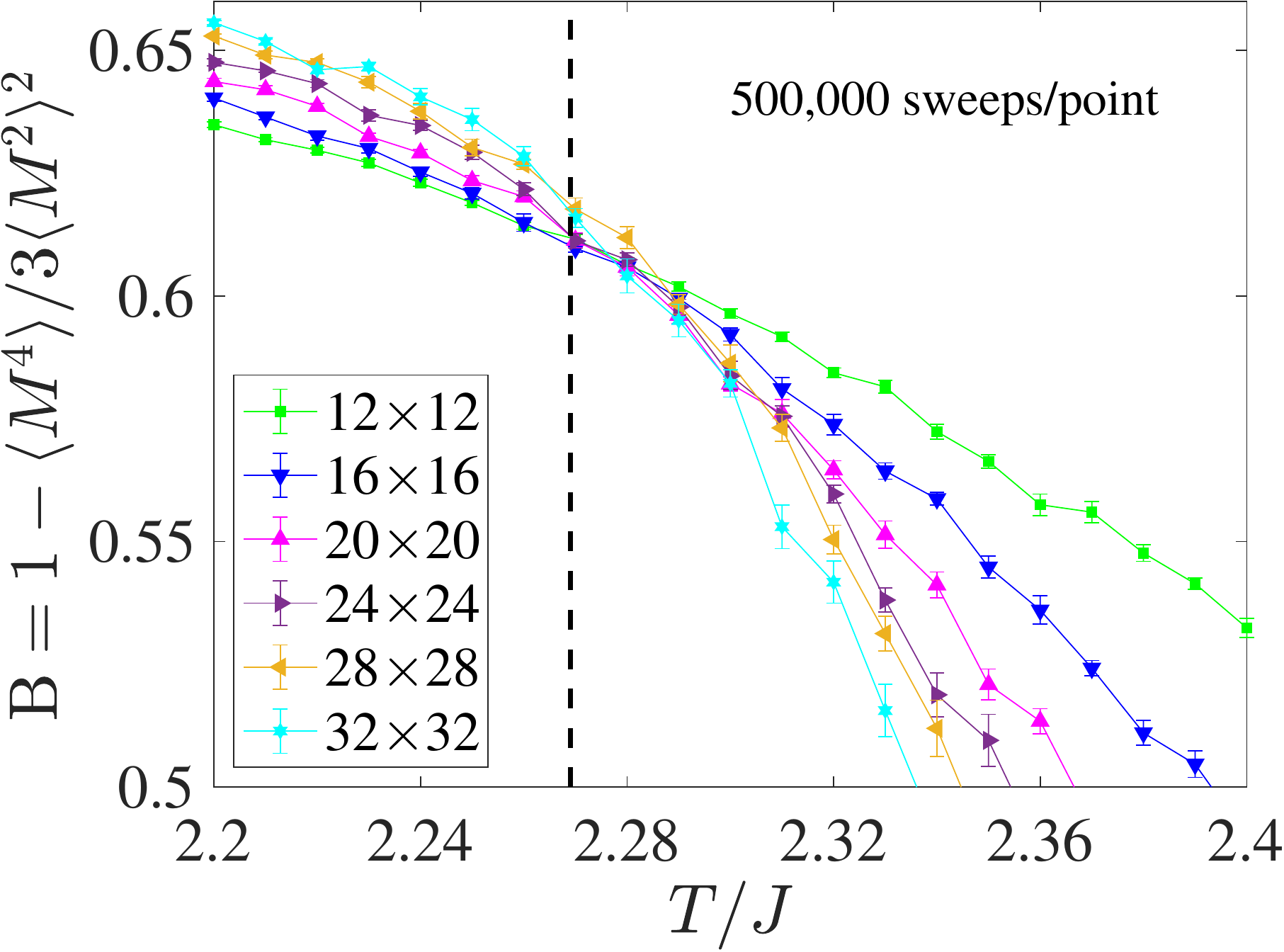}
    \caption{
    Analysis of the crossing of the Binder ratio 
    $B=1-\langle M^4 \rangle / 3 \langle M^2 \rangle^2$ gives
    a more accurate estimate of $T_c$. The data here could be vastly
    improved by using a cluster method to avoid critical slowing down; however, even with this straightforward implementation, $T_c$ is given to 
    accuracy of one percent or so, with simulation run times on
    the order of minutes on a desktop computer.
    }
    \label{fig:Isingbinder}
\end{figure}

Figure \ref{fig:Isingchi} shows a typical result for a classical simulation obtained using the traditional classical Monte Carlo methodology. In this case, the magnetic
 susceptibility $\chi$ of the 2D Ising model is plotted as a function of temperature. Scale-invariant measurements such as
 the Binder ratio provides improved ways to locate the critical point
 on finite lattices~\cite{binder1993monte}, as indicated in Fig.~\ref{fig:Isingbinder}.
 The key takeaway is that a determination of the critical temperature to
 one percent accuracy is easily attained using these simplest classical models, even with unsophisticated
 single spin flip moves, and typical desktop computers.
 These models, therefore, form interesting testbeds for
 \gls*{ML} methods and provided the first evidence of their potential.
 However, \gls*{ML}, while adding interesting insight, are unlikely
 to constitute breakthrough applications in this context, given the efficacy
 of existing tools.
 
\subsection{Quantum Monte Carlo: the DQMC method}
This section presents a brief overview of the 
standard determinant quantum Monte Carlo (DQMC) algorithm while highlighting the critical aspects for understanding the ML applications discussed later. 
For more complete discussions of the DQMC algorithm, we refer the reader to Refs.~\cite{WhitePRB1989,assaad2002quantum,dossantos03,assaad2008world,JohnstonPRB2013,gubernatis_kawashima_werner_2016,CohenSteadPRE2022}. 

The goal of  DQMC is to evaluate the expectation values of thermodynamic observables of quantum many-body Hamiltonians such as
the Hubbard, periodic Anderson, and Holstein models. That is,
it allows the computation, within the grand canonical ensemble, of 
\begin{equation*}
    O \equiv \langle \hat{O} \rangle =\frac{1}{\mathcal{Z}}\mathrm{Tr} \, \hat{O}
    \, e^{-\beta  \hat H}, 
\end{equation*}
where $\mathcal{Z} = \mathrm{Tr}\, e^{-\beta \hat H}$ is the grand partition function.
Observables of interest include the Hamiltonian itself, giving the
energy and specific heat, as well as pairing, charge, and spin correlation
function and their susceptibilities, which signal transitions
into low-temperature ordered phases.
We will first focus on evaluating $\mathcal{Z}$. Once the recipe is established, it is straightforward to generalize it to evaluate $O$. 

It is convenient first to partition the Hamiltonian into two parts, 
$\hat H = \hat H_0 + \hat H_\mathrm{int}$, where $\hat H_0$ contains the noninteracting terms (i.e.~those which are quadratic in the
fermion creation and destruction operators) 
and $\hat H_\mathrm{int}$ contains the interaction terms. 
(Note that if lattice terms are present in the model - e.g., 
as in the Holstein model -- then the Hamiltonian is further partitioned as 
$\hat H = \hat H_0 + \hat H_\mathrm{lat} + \hat H_\mathrm{int}$.)
Next, we divide the imaginary time interval into $L$ discrete imaginary time steps such that $\tau = l\Delta\tau$, with $l = 0,1,\dots,L-1$ and $\Delta \tau =\beta/L$. We can then approximate $\mathcal{Z}$ using the Trotter 
formula~\cite{trotter1959product,suzuki1976relationship,fye1986new},
\begin{equation}
    \mathcal{Z}=\mathrm{Tr}\, e^{-\beta \hat H} 
    = \mathrm{Tr}\, (e^{-\Delta\tau \hat H_0}e^{-\Delta\tau 
    \hat H_\mathrm{int}})^L + O(\Delta\tau^2)\label{Eq:Z}
\end{equation}
Here, $O(\Delta\tau^2)$ is a controllable Suzuki-Trotter error introduced by the neglected commutation of the Hamiltonian terms.  

We first consider the case where the model only has local Hubbard interactions. In this case, one needs to recast the quartic terms in the Hamiltonian into quadratic ones by introducing an auxiliary field $s_{i,l}$ at every spatial $i$ ($=1,2,\dots, N)$ and imaginary time $l$ point. While many auxiliary fields are possible, in DQMC, one usually adopts the discrete \gls*{HS} transformation for an on-site Hubbard interaction~\cite{HirschPRB1983}. For $U > 0$, $s_{i,l}$ couples
to the $z$-component of spin,
\begin{equation*}
    e^{-U\Delta\tau\left(n_{i,\uparrow}-\tfrac{1}{2}\right)\left(n_{i,\downarrow}-\tfrac{1}{2}\right)} = 
    \frac{1}{2}e^{-\tfrac{\Delta\tau U}{4}}\sum_{s_{i,l}} 
    e^{\lambda s_{i,l}(n_{i,\uparrow}-n_{i,\downarrow})}
\end{equation*}
and for $U = -|U| < 0$, $s_{i,l}$ couples to the charge,
\begin{equation*}
    e^{|U|\Delta\tau\left(n_{i,\uparrow}-\tfrac{1}{2}\right)\left(n_{i,\downarrow}-\tfrac{1}{2}\right)} = 
    \frac{1}{2}e^{-\tfrac{\Delta\tau |U|}{4}}\sum_{s_{i,l}} 
    e^{\lambda s_{i,l}(n_{i,\uparrow}+n_{i,\downarrow}-1)}. 
\end{equation*}
Regardless of the sign of $U$, $s_{i,l} = \pm 1$ and $\lambda$ is a constant satisfying $\cosh(\lambda) = \exp(\Delta\tau |U|/2)$. From this point forward, we will focus on the repulsive case with $U > 0$. The generalization to negative values of $U$ is straightforward. 

Once the \gls*{HS} transformation is applied, the partition function only involves quadratic terms in the fermion operators, which comes at the expense of having to sum over the auxiliary fields. For a fixed configuration $\{ s_{i,l}\}$, the trace over the fermion degrees of freedom can  be evaluated analytically~\cite{BlankenbeclerPRD1981} to yield a product of determinants (hence the name). 
The traces over up and down fermions yield separate determinants
as long as $\hat H_0$ contains no terms that hybridize the two spin species.
The partition function thus reduces to 
\begin{equation}\label{eq:ZHubbard}
    \mathcal{Z} = \sum_{\{s_{i,l}\}} \mathrm{det}(M_\uparrow) \, \mathrm{det}(M_\downarrow) \equiv 
    \sum_{\{s_{i,l}\}} W(\{s_{i,l}\}). 
\end{equation}
Here, the matrices $M_\sigma$ are defined as 
\begin{equation*}
    M_\sigma = \mathbb{I} + B_\sigma(L-1)_\sigma B_\sigma(L-2)\dots B_\sigma(0), 
\end{equation*}
where $B_{\sigma}(l) = e^{\mp \Delta\tau \lambda v(l)}e^{-\Delta\tau K}$, $K$ is the matrix representation of $H_0$, $\mathbb{I}$ is the $N\times N$ identity matrix, and 
$v(l)$ is a diagonal matrix whose elements are given by $[v(l)]_{i,j} = \delta_{i,j}s_{i,l}$.  The $\mp$ in the exponential refers to spin $\sigma=\uparrow,\downarrow$.

All that remains is to evaluate the summation over all \gls*{HS} configurations appearing in Eq.~\eqref{eq:ZHubbard}, which is accomplished using standard MCMC methods. 
A central quantity in this sampling procedure is the equal time Green's function $G^\sigma_{ij}(\tau =  l\Delta\tau) = \langle c_{j,\sigma}^{\phantom\dagger}(\tau)c_{i,\sigma}^{\dagger}(\tau)\rangle$. For a given auxiliary field configuration, it can be expressed as   
\begin{eqnarray}\nonumber
    G^\sigma_{i,j}(l) = [\mathbb{I}&+&B_\sigma(l)_\sigma B(l-1)\dots B_\sigma(0)\\  &&\times B_\sigma(L-1)\dots B_\sigma(l+1)]^{-1}_{i,j}. \label{eq:G}
\end{eqnarray}
This quantity can be interpreted as describing the propagation of a free electron through the potential established by the given HS field configuration. Note that Eq.~\eqref{eq:G} implies that the Green's function on successive time slices satisfies the equation
\begin{equation}\label{eq:recursion}
    G^\sigma(l+1) = B_\sigma(l+1) G^\sigma(l) B^{-1}_\sigma(l+1). 
\end{equation}
It also establishes that the weight of the Monte Carlo configuration can be equated to the product of the determinant of the inverse Green's functions 
\begin{equation*}
W(\{s_{i,l}\}) = \mathrm{det}(G^{-1}_\uparrow) \,
\mathrm{det}(G^{-1}_\downarrow).
\end{equation*}

These relationships form the basis for the MCMC sampling procedure, where one performs a random walk through the configurations $\{s_{i,l}\}$ following the probability distribution $W(\{s_{i,l}\})$. First, the time index is fixed to a particular slice $l$, and the corresponding equal time Green's functions $G_\sigma(l\Delta\tau)$ are computed. Next, one visits every spatial site $i$ in the lattice, proposing changes in the local auxiliary fields $s_{i,l}\rightarrow s^\prime_{i,l}$. (For the case of the HS fields, this is a local spin flip move $s_{i,l}\rightarrow -s_{i,l}$.) These moves can be accepted with the Metropolis probability
$p = \mathrm{min}[1,W(\{ s^\prime_{i,l}\})/W(\{s_{i,l}\})]$, or with the heat-bath prescription,
$p = W(\{ s^\prime_{i,l}\})/
\big(\, W(\{ s^\prime_{i,l}\}+  W(\{s_{i,l}\})\,\big)$, both of which obey detailed
balance. 
Similarities with the Monte Carlo procedure for a classical
model (like Ising) are evident. Two key differences are (i) that simulation
of the original quantum model in $d$ dimensions requires sampling 
a field $s_{i,l}$ with an `imaginary time' index $l$ in addition
to the spatial site label $i$; and (ii) the weight involves
a non-local quantity - the fermion determinants. 

In principle, updating a single $s_{i,l}$ requires the $O(N^3)$
operations required in evaluating a determinant.
However,
the computation of $p$ involves only
the ratio of the determinants, for which there is a simple
expression in terms of the equal time Green's functions. 
Once updates have been proposed at every site, the algorithm advances to the next time slice using Eq.~\eqref{eq:recursion}, and the process is repeated.  
Because
of the locality of the change in the matrix, the determinant ratios can be evaluated
in $O(N^2)$ operations, so that updating all $NL$ components 
$s_{i,l}$ of the HS field takes $O(N^3 L)$ steps.  This is the
fundamental system-size scaling of the DQMC algorithm.

The Holstein model does not have $e$-$e$ interaction terms and thus does not necessitate the introduction of the Ising-like \gls*{HS} fields. Instead, one must deal with the $e$-ph terms, which couple the local (quadratic) fermion density operator to the lattice displacement. We again begin with Eq.~\eqref{Eq:Z}, but this time we insert a complete set of phonon position and momentum states at each spacetime point $(i,l)$. At this stage, the trace over the phonon momenta and quadratic electron degrees of freedom can be performed analytically, reducing the partition function to a familiar form 
\begin{equation}\label{Eq:ZHolstein}
    \mathcal{Z} = \int \mathrm{d}X e^{S_\mathrm{lat}}\mathrm{det}(M_\uparrow)\,\mathrm{det}(M_\downarrow) = \int \mathrm{d}X W(\{X_{i,l}\}).
\end{equation}
In this case, the matrices $B_\sigma(l) = e^{\Delta\tau g X(l)}e^{-\Delta\tau K}$, where $X(l)$ denotes a diagonal matrix whose diagonal elements are $[X(l)]_{i,j} = X_{i,l}\delta_{i,j}$, $\{X_{i,l}\}$ denotes a given configuration of the lattice sites, and $\mathrm{d}X$ is short hand for an $N \times L$ multi-dimensional integral over all lattice displacements $\{ X_{i,l}\}$.
The structure is very similar to that of the Hubbard Hamiltonian.
Note, the additional term $\exp({S_\mathrm{lat}})$ where 
\begin{equation*}
    S_\mathrm{lat} = \Delta \tau \left[,\frac{\omega_0^2}{2}\sum_{i,l}X_{i,l} + \sum_{i,l}\left(\frac{X_{i,l+1}-X_{i,l}}{2\Delta\tau}\right)^2 \,\right]
\end{equation*} 
in the configuration weight resulting from noninteracting lattice terms $\hat H_{\rm lat}$. Apart from these changes, the remainder of the DQMC algorithm is unchanged.

The evaluation of $S_\mathrm{lat}$ is very rapid compared to dealing
with the fermion determinant, so it contributes very little
to the computational workload. Indeed, its simplicity also makes adaptation
of the Holstein model to other forms of
$\hat H_{\rm lat}$, e.g.~including anharmonic terms in the phonon
potential energy, almost trivial.  
$S_\mathrm{lat}$ does play a profound role in controlling
the imaginary time fluctuations of the phonon field
in comparison to the HS field, for which an analog of
$S_\mathrm{lat}$ 
is absent. As a consequence, the fermion matrices are much better
conditioned, opening the door for efficient Langevin updates of electron-phonon Hamiltonians~\cite{CohenSteadPRB2021, GoetzAssaad2021, CohenSteadPRE2022}, which
are much more challenging in the Hubbard model.

It is important to stress that we have skipped over many technical details in this brief overview, which must be addressed when implementing the DQMC algorithm. For example, the $B_\sigma(l)$ matrices are stiff, especially as the product $U\beta$ becomes large. To evaluate the long products of these matrices, one must use $QR$ factorizations and other numerical tricks to stabilize the numerics. Additional details on these technical details can be found in Refs.~\cite{WhitePRB1989,assaad2002quantum,dossantos03,assaad2008world,JohnstonPRB2013,gubernatis_kawashima_werner_2016,Tomas,CohenSteadPRE2022}. 

We will briefly mention one issue since it is the primary
limitation to the application of DQMC:  the signs of the configuration weights  $\mathrm{det}(M_\uparrow)\, \mathrm{det}(M_\downarrow)$ are not always positive-definite and, therefore, cannot be directly interpreted as a probability. When this occurs, the auxiliary fields are sampled according to a new probability distribution given by the absolute value of the original distribution $|W(\{s_{i,l}\})|$. The two distributions are related by $W = \mathrm{sign}W\, |W|$. This change requires us to re-weight any observable as 
\begin{equation}
    \langle O \rangle_W = \frac{\langle O\, \mathrm{sign}W\rangle_{|W|}}{\langle \mathrm{sign}W\rangle_{|W|}},
\end{equation}
where  the subscript $|W|$ in the expectation value emphasizes the
configurations are now generated with probability $\propto |W|$.
When the system size increases or the temperature decreases, the sign of $W$ is positive and negative with nearly equal probability, causing $\langle \mathrm{sign}W\rangle\rightarrow 0$. The numerator must also
vanish exponentially since $\langle O \rangle$ is well defined.
The process of taking the ratio of two very small quantities,
each with finite error bars ultimately yields values
that are meaningless. 
This behavior is a reflection of the Fermion sign 
problem~\cite{loh90,troyer05}. 
(The sign problem is not restricted to fermionic models,
but also occurs for frustrated quantum spins~\cite{henelius00}.)

It is worth noting that the determinants $\mathrm{det}(M_\uparrow) = \mathrm{det}(M_\downarrow)$ for the Holstein model because the phonon field couples in the same way to the spin up and spin down fermions. 
This is also true of the attractive Hubbard model at any filling, and of the half-filled Hubbard with nearest-neighbor hopping only. Other `sign-problem free' models and the symmetries and choices of bases
from which they originate have been 
discovered~\cite{chandrasekharan99,wu03,wu05,berg12,chandrasekharan12,cai13,huffman14,wang15,kaul15,ZiXiang2015,Iazzi2016,li16,Wei2016,li19,kim20,Hangleiter2020,Levy2021,zhang2021sign}.
In such cases, $W(\{X_{i,l}\})$ is always greater than zero and 
DQMC can access the physics down to arbitrarily low temperatures. In these cases, one can obtain essentially 
exact solutions of the corresponding quantum many-body 
problem.

\subsection{Challenges and limitations}
Despite their power and versatility, MCMC methods are limited in several notable ways. First, they frequently employ finite-size clusters when studying correlated electron models like those discussed in Sec.~\ref{sec:models}. Because of this, a finite-size analysis is required to extrapolate results to the thermodynamic limit. The thermodynamic limit can be approached more directly by embedding the cluster in a self-consistent dynamical mean field that approximates correlations beyond the cluster \cite{MaierRMP2005, GeorgesRMP1996}. However, such embedding schemes can still exhibit considerable finite-size effects \cite{MaierPRL2005}. Because of this, addressing the thermodynamic limit requires large system sizes, which can be prohibitively expensive for many classes of MC algorithms.  

Another fundamental limitation of any MC method is its decorrelation time. 
A MC algorithm must draw measurements from statistically independent configurations to obtain unbiased estimators for an observable and its statistical error. Because of this, the autocorrelation time $\tau$ - defined as the number of updates needed to generate such configurations - is a crucial measure of a MC simulation's efficiency. Many MC applications suffer from prohibitively long autocorrelation times (e.g., $e$-ph or frustrated spin models, continuum limit lattice gauge theory simulations, and confined quantum liquids). 

Autocorrelation times can also depend strongly on the parameter regime of a particular model and the sampling method. The Holstein model in the adiabatic limit ($\omega_0/t \ll 1$) is quite challenging for traditional Metropolis-Hastings sampling but more amenable to hybrid Monte Carlo methods~\cite{CohenSteadPRE2022}. Autocorrelation times also tend to grow in the vicinity of a phase transition, where correlation lengths extend beyond the size of the cluster and single-site updates are no longer capable of efficiently moving MC configurations out of meta-stable minima. 
This latter problem is known as ``critical slowing down.'' 
In some cases, ``global'' MC moves can be performed to mitigate the autocorrelation time. In such schemes, an extended region of configuration space is updated simultaneously to generate independent configurations quickly. However, the form of these updates is only known in some special cases. Examples include the Wolff update \cite{Wolff} for the classical Ising model as well as $\omega_n = 0$ shifts of auxiliary fields in \gls*{QMC} simulations of the Hubbard \cite{ScalettarPRB1991} and Holstein models \cite{JohnstonPRB2013}, or ``swap'' updates where lattice configurations on neighboring sites are interchanged ~\cite{CohenSteadPRE2022}. Importantly, these global moves can fail to reduce autocorrelation time if they only achieve low acceptance rates. It is usually not obvious how global MC moves should be proposed for a given model.  

Finally, QMC simulations must also contend with the aforementioned Fermion sign problem. 

These factors have prevented the widespread deployment of QMC algorithms for many effective models relevant to current materials of interest. However, it is hoped that \gls*{ML} methods can provide new routes forward. For example, the problem of constructing generalized global updates can be addressed using so-called self-learning Monte Carlo methods~\cite{SLMC}, as discussed in Sec.~\ref{sec:SLMC}.

\section{Machine learning methods}\label{sec:ML_Methods}
\subsection{Artificial neural networks}
Artificial neural networks (ANNs) are data structures capable of encoding highly-nonlinear functions of their input features. Originally motivated by models for the brain, ANNs usually consist of several interconnected layers of perceptrons. Like a neuron, a perceptron is an element of decision-making, providing an output ($y$) based on the weighted average of a set of input values ({$\bf x$})
\begin{eqnarray}
y=f({\bf x}\cdot{\bf W}+b),
\label{eq:yfx}
\end{eqnarray}
where {$\bf W$} are weights associated with the input values, $b$ is a bias variable, and $f$ is called the activation function, usually a nonlinear function such as $\tanh$, sigmoid, or the rectified linear unit (ReLU). The idea is that more complex decisions can be made by having a large number of these perceptrons in deeper (in terms of the number of layers) and more complicated ANNs. 

ANNs are usually designed for specific tasks or to make particular decisions, e.g., categorizing a large number of inputs. Training ANNs to make correct decisions takes place through observation. In supervised learning, a supervisor provides many inputs and their ``labels'' (correct categories) to the ANN. During this process, through backpropagation, the network gradually adjusts its many fitting parameters (weights {$\bf W$'s} and biases $b$'s) to match its output with the desired output (labels). For more details, see Refs.~\cite{ML, Youtube}.

\subsection{Convolutional neural networks}

Convolutional neural networks (CNNs) are a group of ANNs that use one or more convolutional layers in their architecture. In a convolutional layer, a kernel (also known as a filter), usually with the same dimensionality as the input data, sweeps across each input and convolves with portions of it. After going through an activation function, the results of these convulsions are combined in a ``feature map'', which is passed to the next layer of the neural network. Pixel values of the kernel are considered weights that can be adjusted during the training process. But importantly, the same weights are used for every convolution for a given kernel. A CNN can have more than one convolutional layer, with subsequent layers acting on feature maps of previous layers, or more than one kernel in each convolutional layer, where each kernel has a unique set of weights and creates its feature map.

The idea behind convolutional layers is that one can work with the input data in their original shape and take advantage of spatial correlations or patterns that may exist in them. Kernels in the first convolutional layers usually pick up the most basic features in the data that can be used for categorization, and subsequent layers use those features to create more complicated patterns. Having convolutional layers generally improves the training accuracy of neural networks, especially if the input data contain important spatial features (e.g., translational invariance in physical systems). For this reason, CNNs are widely used in image recognition. However, in Sec.~\ref{sec:discovery}, we will see examples of how the information encoded in convolutional kernels can be used to infer nonlocal correlations from snapshots of fermions on a lattice.

\subsection{Principal component analysis}
\gls*{PCA} is perhaps conceptually the simplest unsupervised \gls*{ML} approach.  
Within the context of classical statistical mechanics, it begins with a set of $M$ `snapshots' of the lattice, e.g.~a collection of $N$-component vectors 
${\bf S}_\gamma$, 
with $\gamma = 1,2,3,\cdots, M$, each representing a given configuration generated during a Monte Carlo simulation. (For example, the components ${\bf S}_\gamma$ could represent the $N=L\times L$ spin orientations at a given instance of a simulation of the 2D Ising model.)  
These $M$ vectors are assigned to rows of a matrix ${\cal X}$, with dimension $M \times N$. The $M$ vectors are typically chosen from simulations at $n_T$ different temperatures 
$\{ T_i \}$ that transit $T_\mathrm{c}$.  At each $T_i$, $R$ configurations
are chosen, so that $M = n_{T} \, R$.  To implement the PCA, the eigenvalues $\lambda_\alpha$
and eigenvectors ${\bf w}_\alpha$ of the $N \times N$ covariance matrix
${\cal X}^T {\cal X}$ are determined 
\begin{eqnarray}
{\cal X}^T {\cal X} \, {\bf w}_\alpha = \lambda_\alpha {\bf w}_\alpha. 
\end{eqnarray}
The overlaps of each configuration ${\bf S}_\gamma$ with a given eigenvector 
${\bf w}_\alpha$
are then computed to define weights, or principal components, $p_{\gamma\alpha} = {\bf S}_\gamma \cdot {\bf w}_{\alpha}$.

As we shall see below, the topology of scatter plots of $p_{\gamma\alpha}$ for the first few largest eigenvalues $\lambda_\alpha$
changes decisively through $T_c$.
The method works best when these eigenvalues are much larger
than the remaining ones, a condition that holds in many interesting cases.
In such a scenario, the original $N$-dimensional data contained in ${\cal X}$, the $M$ vectors ${\bf S}_\gamma$
of length $N$,
have been projected to a (much) smaller dimensional space of
$p_{\gamma\alpha}$ with, for example, $\gamma=1,2$.
%% of a few vectors, e.g.~ ${\bf w}_1, {\bf w}_2$  of length $N$.
PCA is an unsupervised \gls*{ML} method. No labeling of ${\bf S}_\gamma$ as 
being below or above $T_c$ is required, and one only utilizes the raw spin configurations. The critical point emerges spontaneously as a change
in the nature of the principal components in passing through the phase transition.

\subsection{Autoencoders}
\begin{figure}[t]
    \centering
    \includegraphics[width=0.8\columnwidth]{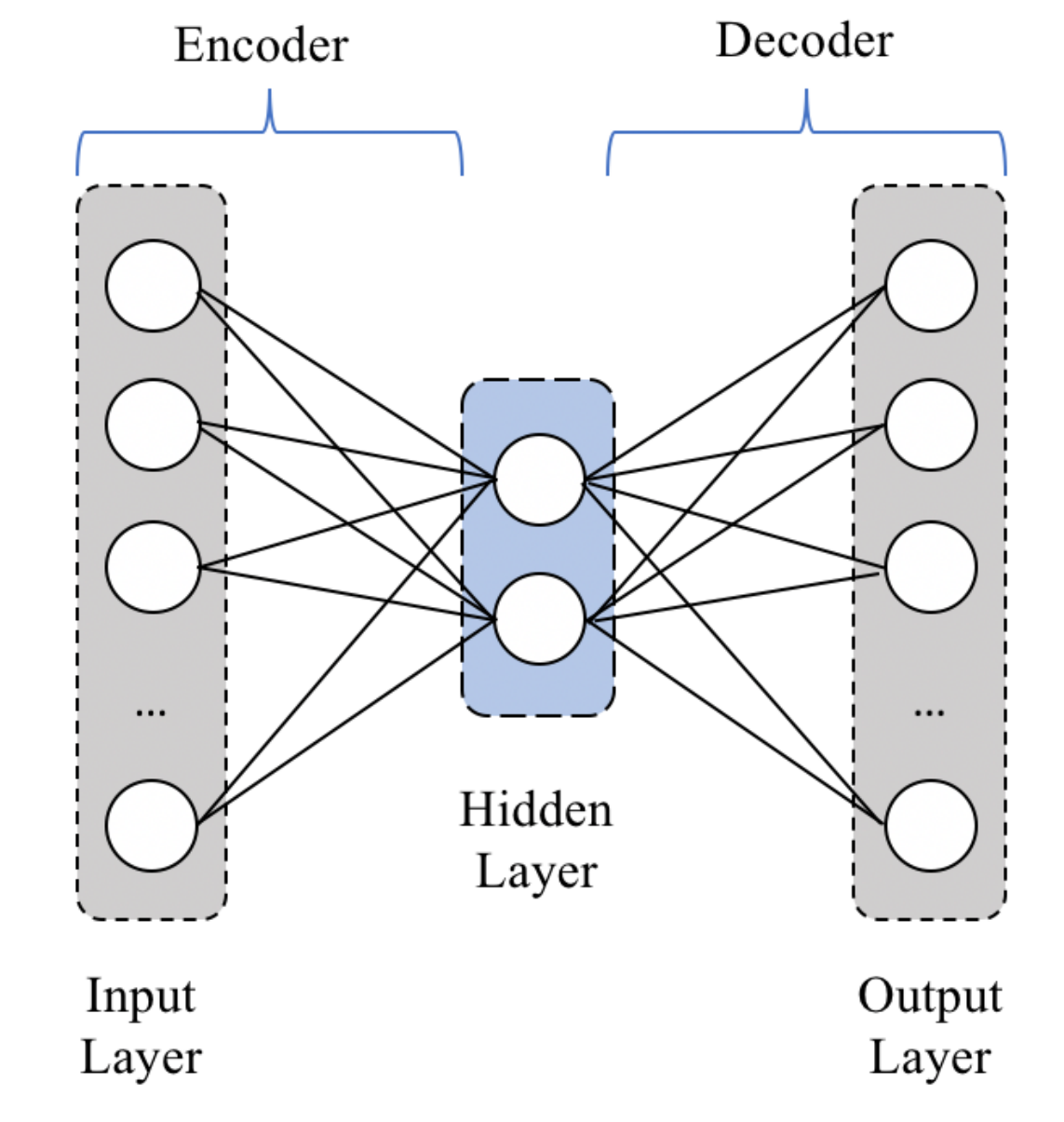}
    \caption{
    The topology of an autoencoder. High dimensional input data are compressed through
    a hidden layer with (far) fewer degrees of freedom, followed by a demand that
    the high dimensional output reproduces the input when fed back into the decoder layers.  
    }
    \label{fig:autoencoder}
\end{figure}

Like PCA, the autoencoder (AE) method is an unsupervised method for performing dimensional reduction. However, it is generally more powerful because it employs nonlinear transformations. In its simplest form, an AE consists of two  ANNs, the ``encoder'' and the ``decoder'', connected through a hidden layer in the middle, as shown in Fig.~\ref{fig:autoencoder}.
An AE analysis begins, like the PCA, with a collection ${\bf S}_\gamma$ of snapshots
of the system.
These snapshots are used as the `inputs' to the encoder from which the values ${\bf U}$ of a subsequent
(`hidden') layer are computed. Denoting 
the components of a given ${\bf S}_\gamma$ as $S_i$ (suppressing the
label $\gamma$), and the associated
hidden layer values by $U_j$, one computes for each $\gamma$
\begin{eqnarray}
U_j = f\left(\sum_{i} W^{(1)}_{ji} S^{\phantom\dagger}_i + b^{(1)}_j \right).
\end{eqnarray}
as in Eq.~\eqref{eq:yfx}.
Similarly, the values $V_k$ on an `output' layer out of the decoder are obtained from those on
the hidden layer using 
\begin{eqnarray}
V_k = f\left( \sum_{j} W^{(2)}_{kj} U^{\phantom\dagger}_j + b^{(2)}_k \right).
\end{eqnarray}
The weights $W^{(1)}$ between the input and hidden layers, and $W^{(2)}$ between 
hidden and output layers, as well as the biases $b^{(1)}$ and $b^{(2)}$
are adjusted, e.g.~through
backpropagation so that the output $V$ matches the input $S$.
(Hence the name `autoencoder.')  

In the AE approach, separate networks are trained for
distinct parameter choices (e.g.~temperature). The values characterizing the small number
of hidden layer nodes, e.g.~their activations, can be analyzed as a 
function of the control parameter. Typically these values change distinctively through a phase transition.

\subsection{t-distributed stochastic neighbor embedding}
Like PCA, t-distributed stochastic neighbor embedding (tSNE) is an unsupervised method used to reduce the dimensionality of data and represent them using a few projected values. But unlike PCA, tSNE does this nonlinearly by minimizing the difference between pairwise conditional probability distributions representing the similarity of points in the high- and low-dimensional spaces. The distributions are based on Student's $t$-distribution functions centered at each point and have widths adjusted to keep the number of effective neighbors of each point fixed throughout the configuration space. The user sets the latter as the ``perplexity'' number. A typical tSNE analysis starts with an initial PCA to reduce the dimensionality of data from the original value to around 50 before applying the tSNE algorithm, as it can be costly to work directly with an input dimension of hundreds or thousands. More details about the tSNE method can be found in Refs.~\cite{tSNE,tSNEHowTo}.

\subsection{Random trees embedding}

Random trees embedding is another unsupervised learning method that uses the notion of a tree to extract features from data. A tree refers to a graph with nodes repeatedly branching out in one direction. The parent node (the root of the tree) has all the data, while the data is divided into subsets at subsequent nodes representing tree branches. Each node corresponds to a feature in the data, so branches farther from the parent node containing smaller subsets of data correspond to finer features. 

In random trees embedding, data are projected onto an ensemble of random trees whose total number and maximum branching depth are parameters that can be tuned. Each tree makes an independent observation regarding the features, and the ensemble of trees ``votes'' for dominant features, judged by the amount of overlap between nodes at a certain depth from different trees. For more details see Refs.~\cite{f_moosmann_06,RandomTrees}.

\section{Proofs of concept and benchmarks}\label{sec:benchmarks}
\subsection{Classical models}\label{sec:benchmarks-classical}
In May and June of 2016, a series of groundbreaking papers~\cite{j_carrasquilla_16,Wang2016,Carleo2016,g_torlai_16} 
came out that demonstrated the power and potential of machine learning techniques 
in encoding information about the statistics of classical and quantum many-body systems 
and how they may be used for physics discovery. These works showed for the first time that 
one could think of the degrees of freedom in many-body systems -- e.g., individual 
spin orientations in the Ising model or auxiliary field configurations in QMC simulations -- as ``features'' in ML  
algorithms. This realization paved the way for utilizing ML methods developed and refined for industry applications to learn new physics. 

Carrasquilla and Melko~\cite{j_carrasquilla_16} employed fully-connected and convolutional neural networks to study phase transitions in models for magnetic systems. By coloring their 2D spin configurations obtained from 
a Monte Carlo simulation as hot or cold (referring to whether they were obtained at a temperature below the critical temperature of the model or not), they were able to
train the networks to classify never-before-seen configurations and pinpoint 
$T_\mathrm{c}$ with a high degree of accuracy. They further showed that predicted values of $T_c$ approached the analytical value in the thermodynamic limit as their system sizes increased and explored applications of training with models
exhibiting topological orders. They also demonstrated that a fully-connected neural network simply learns to compute the magnetization of the Ising model and uses it as a 
metric for classification. This observation helped explain its success in transferring the knowledge learned on a square lattice geometry to a triangular lattice geometry.

\begin{figure}[t]
    \centering
    \includegraphics[width=\columnwidth]{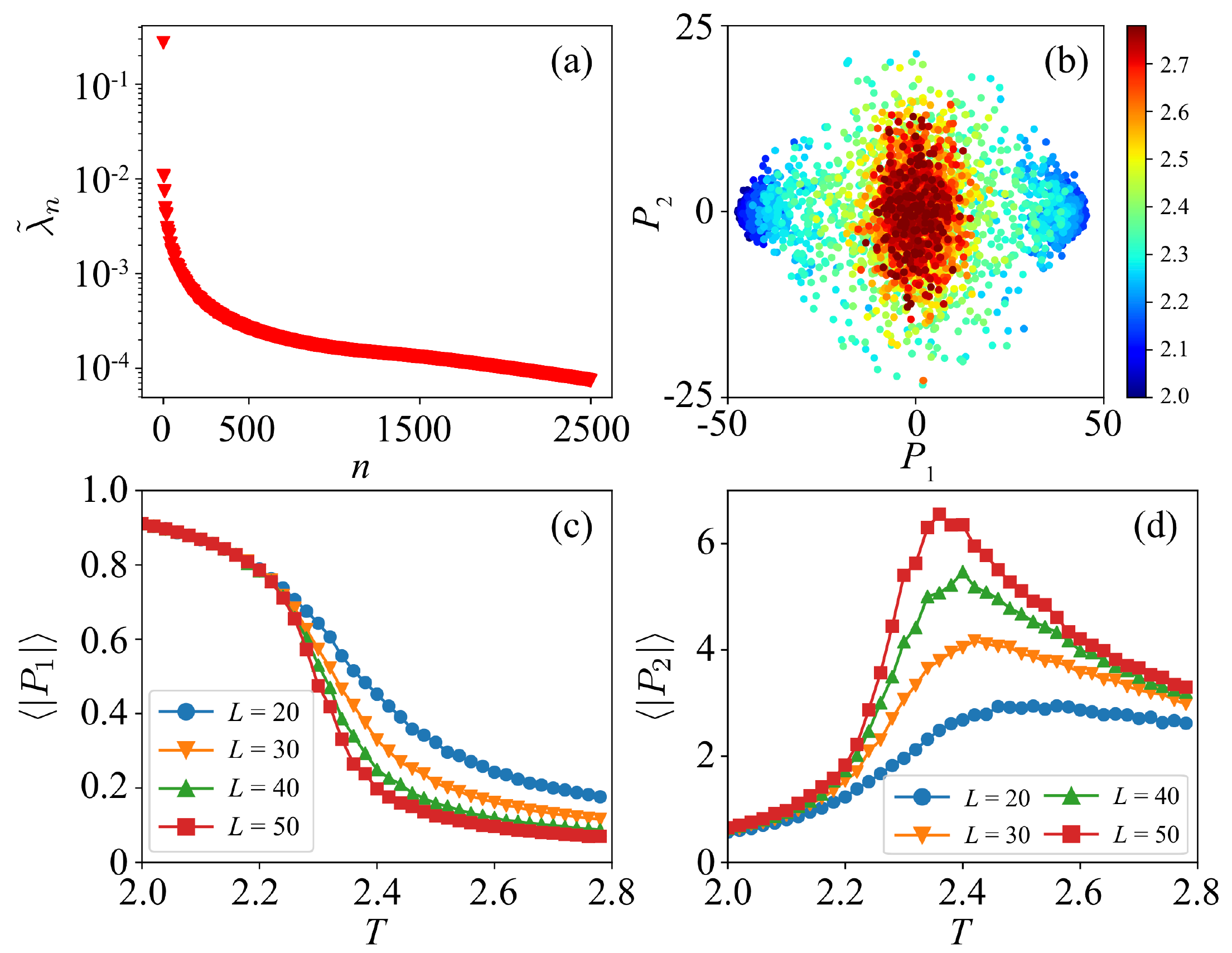}
    \caption{
    PCA results for the square lattice Ising model, which has an analytically known $T_c/J = 2.269$. 
    (a) Distribution of the lowest 2500 eigenvalues of the matrix ${\cal X}^T{\cal X}$.
    (b) Scatter plot of pairs $(p_{1\gamma},p_{2\gamma})$ of overlaps of Ising
    configurations ${\bf S}_\gamma$ with the first two eigenvectors ${\bf w}_1, {\bf w}_2$
    of ${\cal X}^T {\cal X}$.
    Results are shown for 100 configurations at each temperature.
    (c) Average overlap $p_{1\gamma}$ and (d)  $p_{2\gamma}$ as a function of temperature for four 
    linear lattice sizes $L$. Adapted from Ref.~\cite{HuPRE2017} with minor modifications.
    }
    \label{fig:IsingMLresults1}
\end{figure}

Lei Wang's application of the PCA technique to the Ising model~\cite{Wang2016} led to a similar conclusion: physical properties, such as magnetization or magnetic susceptibility, emerge in the first two principal components. What was remarkable about Wang's findings was that these properties could be inferred without providing knowledge about the problem's physics to the machine. That is, the spin configurations were not labeled in the PCA study. Later these ideas were applied to study frustrated classical magnetic models, such as in Ref.~\cite{c_wang_17}.

\begin{figure}[t]
    \centering
    \includegraphics[width=\columnwidth]{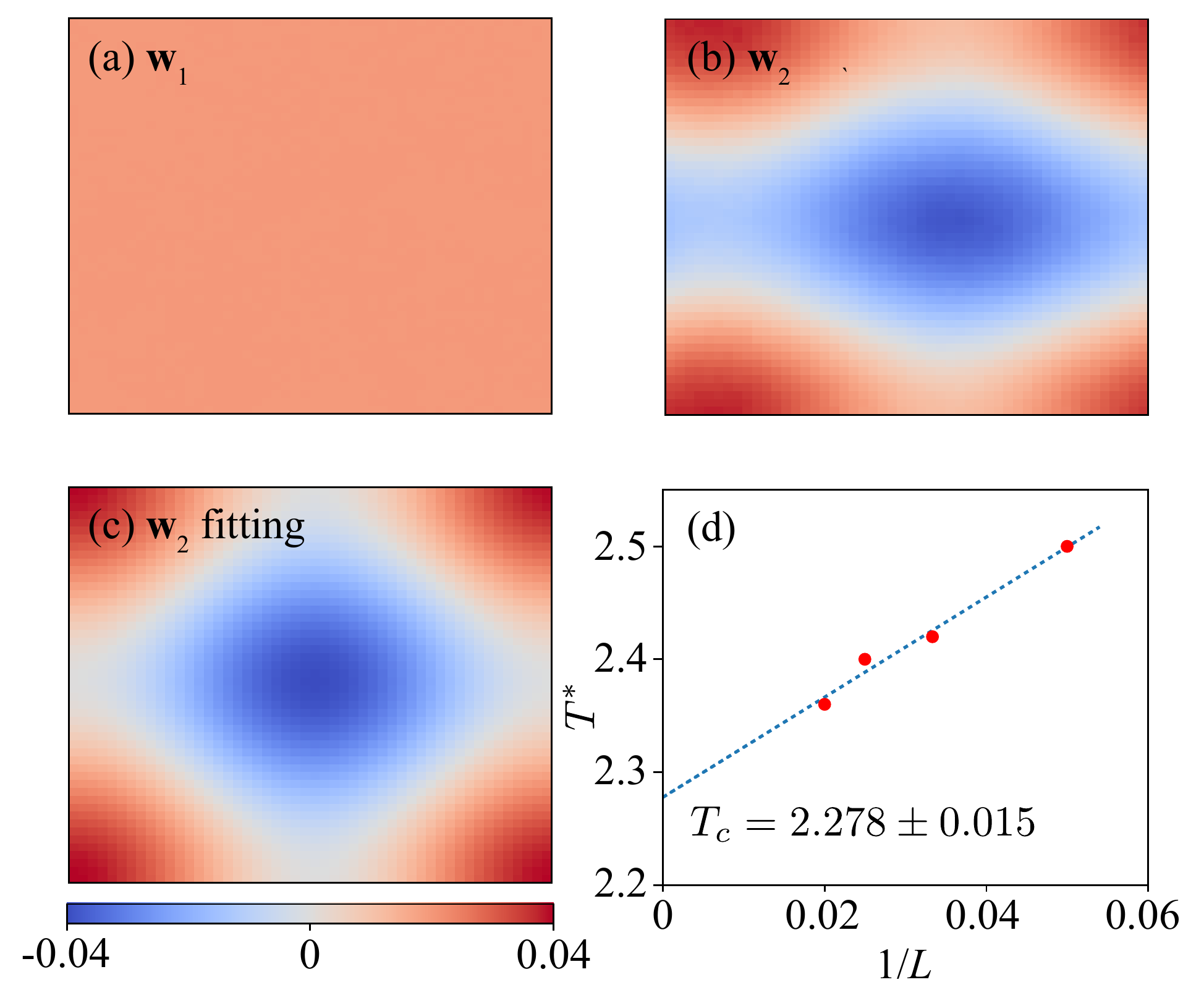}
    \caption{
    (a) The eigenvector ${\bf w}_1$ of largest eigenvalue in a PCA 
    is nearly uniform (${\bf k}=(0,0)$), so that its dot products $p_{1\gamma}$ with spin configurations 
    ${\bf S}_\gamma$ have the physical interpretation of the magnetization. 
    (b) The eigenvector ${\bf w}_2$ of next-largest eigenvalue in a PCA
    exhibits domain walls between spin-up and spin-down regions. It can
    be well fit by a vector consisting of a sum of two plane waves with
    $k_1 = (0,2 \pi/L)$ and $k_2=(2 \pi/L,0)$, i.e.~the $k$ values closest to
    the origin, as seen in panel (c). (d) shows the evolution of the
    peak positions of the average of $p_{2\gamma}$ 
    (Fig.~\ref{fig:IsingMLresults1}(d)) with inverse linear lattice size. Adapted from Ref.~\cite{HuPRE2017} with minor modifications.
    }
    \label{fig:IsingMLresults2}
\end{figure}

To illustrate these approaches more concretely, we now highlight a specific example, the 2D ferromagnetic Ising model~\cite{HuPRE2017,wetzel2017unsupervised,tanaka2017detection}. 
Fig.~\ref{fig:IsingMLresults1} shows results obtained for the phase transition using a PCA \cite{HuPRE2017}. 
Fig.~\ref{fig:IsingMLresults1}(a) plots the eigenvalues of
${\cal X}^T {\cal X}$, which fall off rapidly, a condition for the data compression inherent
in a PCA to be effective. A scatter plot of
pairs $(p_{1\gamma},p_{2\gamma})$ of projections of the spin
configurations ${\bf S}_\gamma$ on the PCA eigenvectors ${\bf w}_1,{\bf w}_2$ with the largest and next-largest eigenvalues
[Fig.~\ref{fig:IsingMLresults1}(b)] 
shows an evolution from a single clump centered at the origin for $T>T_c$ to
a bimodal distribution for $T<T_c$.
The average of $|p_{1\gamma}|$, shown in Fig.~\ref{fig:IsingMLresults1}(c), behaves like an order parameter (the magnetization),
evolving from zero at high $T$ to a non-zero value at low $T$. The transition
becomes increasingly sharp with $L$ (the total number of sites $N=L\times L$) and occurs near the known $T_c=2.269$ \cite{Onsager}. 
The average of $|p_{2\gamma}|$ [see Fig.~\ref{fig:IsingMLresults1}(d)] behaves like the susceptibility, peaking near $T_c$
    (compare with Fig.~\ref{fig:Isingchi}). 

\begin{figure}[t]
    \centering
    \includegraphics[width=\columnwidth]{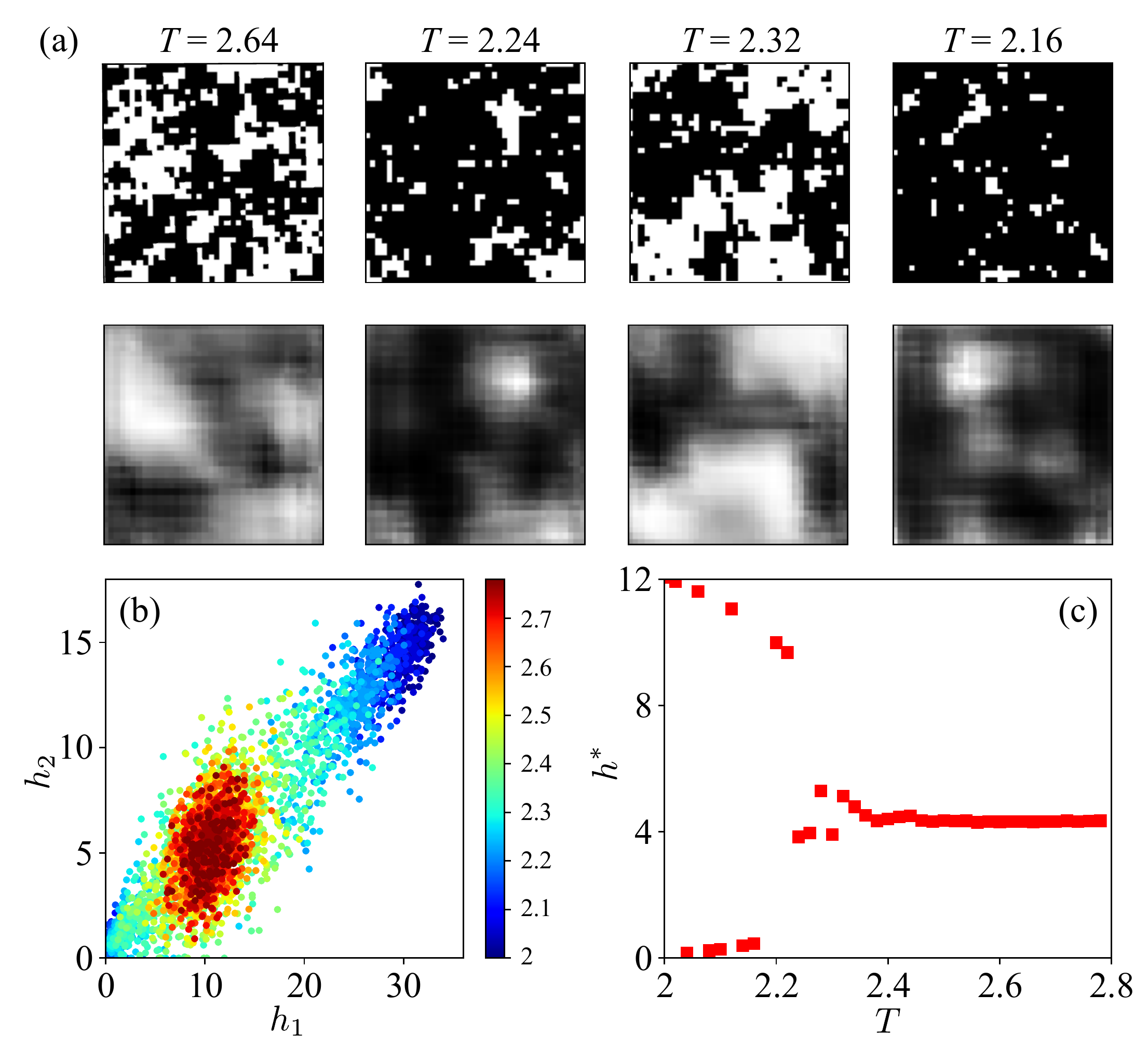}    
    \caption{Autoencoder results for the 2D square lattice Ising model.
    (a) After compression of the spin configuration of an $N=40 \times 40$ 
    lattice through 200 hidden layer neurons, the basic features of the
    input can still be replicated in the output.
    (b) Scatter plots of the activations $(h_1,h_2)$ of 
    a hidden layer with two neurons.  The distribution bifurcates at $T_c$.
    (c) The activation $h^*$ of a single hidden layer neuron
    follows a single trajectory for $T>T_c$ but then separates
    into two branches for $T<T_c$. Adapted from Ref.~\cite{HuPRE2017}  with minor modifications.
    }
    \label{fig:Isingautoencoder}
\end{figure}

    Figure \ref{fig:IsingMLresults2}(a) shows the first eigenvector ${\bf w}_1$.  It is nearly uniform,
    so that $p_{1\gamma} = {\bf w}_1 \cdot S_\gamma$ is essentially the magnetization, as already implied by
Fig.~\ref{fig:IsingMLresults1}(c).
One of the most promising possibilities of \gls*{ML} approaches to statistical mechanics
is the possibility that examining the learning mechanism might lend insight into the
nature of phase transitions, especially in cases where the order parameter is unknown.
    Figure~\ref{fig:IsingMLresults2}(b) shows, similarly, the second eigenvector, which can be compared with
    a linear combination of vectors with domain walls in the vertical and horizontal directions, 
    i.e.~${\bf k}_1=(0,2\pi/L)$ and ${\bf k}_2=(2\pi/L,0)$, as shown in  
    Fig.~\ref{fig:IsingMLresults2}(c). 
    Together with the ${\bf k}=(0,0)$ structure of the first eigenvector shown in Fig.~\ref{fig:IsingMLresults2}(a), one concludes PCA is constructing the low energy (small ${\bf k}$) Fourier
    components of the spin configurations. 
    Finally, Fig.~\ref{fig:IsingMLresults2}(d) shows an extrapolation of the peaks in 
     Fig.~\ref{fig:IsingMLresults1}(d) in inverse linear lattice size.
     The temperature intercept in the thermodynamic limit is within  error bars (of about 1\%)
     of the exact $T_c=2.269$.  This result illustrates that 
     \gls*{ML} can be combined with finite size scaling to reach the thermodynamic limit (as with older methods), but also that quite accurate results can be obtained without too much effort (i.e., using the simple PCA).

    As mentioned, the AE method can be viewed as a nonlinear generalization of PCA. Therefore, it should be no surprise that the AE method is also an effective means to study the Ising phase transition, as shown in Fig.~\ref{fig:Isingautoencoder}. Panel (a) illustrates the data compression of the AE by showing the actual input features (top row), in this case, $N=40\times 40$ spin values for configurations at four different temperatures, together with their replication 
    at the output stage (bottom row). Here, the AE uses two hundred hidden neurons, almost an order of magnitude reduction over the number of sites in the model. Figs.~\ref{fig:Isingautoencoder}(b) \& (c) demonstrate that the AE retains the ability to detect the phase transition even when the data is highly compressed. For example, in Fig.~\ref{fig:Isingautoencoder}(b), the hidden layer has only {\it two} neurons, yet a scatter plot in the plane of their activation ($h_1$, $h_2$) bifurcates in a manner similar to a PCA of Fig.~\ref{fig:IsingMLresults1}. The temperature at which this bifurcation occurs yields an estimate of $T_c$; however, even the activation $h^*$ of a single
    hidden layer neuron, panel Fig.~\ref{fig:Isingautoencoder}(c), shows a clear signal of the phase transition as $T$ is reduced.
    
\begin{figure}[t]
    \centering
    \includegraphics[width=0.85\columnwidth]{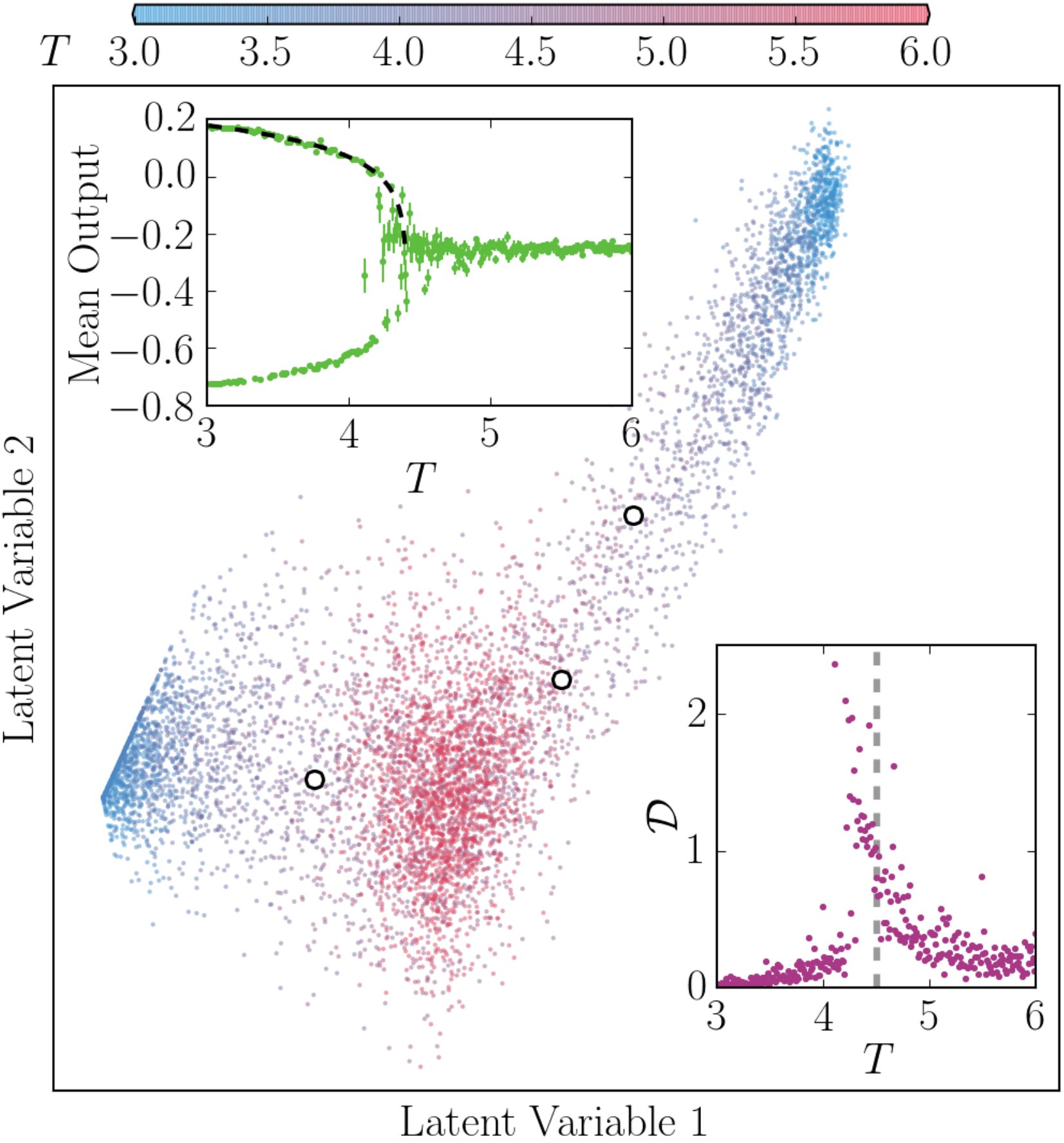}
    \caption{
    Projection of 3D Ising data for an $N=8\times 8\times 8$ lattice on the space of two latent variables (hidden neurons) of an AE. CNN's are used for the encoder/decoder of the AE. Different colors in the scatter plot correspond to different temperatures. Top inset: The output of a different AE, in which  the hidden layer consists of one single neuron, as a function of temperature. The dashed line is a fit to $A(B-T)^\beta +C$  with $A= 0.38$, $B=4.55$, $\beta =0.34$, and $C$ kept fixed at -0.25, which is the average output overall $T$. Bottom inset: Temperature dependence of a measure for the spread of data in the main panel. The vertical dashed lines mark the location of $T_c$.
    Taken from Ref.~\cite{HuPRE2017}.
    }
    \label{fig:3DIsing}
\end{figure}
    
Similar results have also been obtained for the 3D version of the antiferromagnetic Ising model~\cite{k_chng_18} (see Fig.~\ref{fig:3DIsing}). The top inset in Fig.~\ref{fig:3DIsing} shows that the single latent variable of a convolutional autoencoder can act as the order parameter; however, other indicators can also be defined, based on the distribution of more than one latent variable, that point to the transition temperature and even correlate with physical properties. For example, in the bottom inset of Fig.~\ref{fig:3DIsing}, the spread of the data in the space of two latent variables peaks around the critical temperature and closely follows the susceptibility curve.  
The mean output can be fit to give a critical temperature of $T_c=4.55$ and critical exponent of 
$\beta = 0.34$. These values should be compared with values $T_c=4.5115$ and $\beta=0.326$ obtained by Monte Carlo simulation~ \cite{binder2001monte}.
    
Agrawal \emph{et al}. have also used autoencoders to examine the related problem of detecting and identifying which symmetry is broken spontaneously across a phase transition~\cite{Agrawal2022}. To this end, they introduced an architecture called the group-equivariant autoencoder (GE-autoencoder). 
In this application, one first deduces a set of symmetries that will remain intact in all phases at all temperatures using group theory. This information is then used to constrain the hyperparameters of the GE-autoencoder such that the encoder learns an order parameter invariant to these symmetries. Benchmarking their method for the ferromagnetic and antiferromagnetic Ising model in 2D, they could construct GE-autoencoders whose size remained independent of the system size. By including additional symmetry regularization terms in the loss function, they found that the GE-autoencoder learns an order parameter that satisfied the remaining symmetries of the system. The authors extracted information about the associated spontaneous symmetry breaking by examining the group representation by which the learned order parameter transforms. 
The GE-autoencoder was also able to produce estimates for T$_c$ in the thermodynamic limit with greater accuracy, robustness, and time efficiency than a symmetry-agnostic autoencoder discussed above. 

\begin{figure}[t]
    \centering
     \includegraphics[width=\columnwidth]{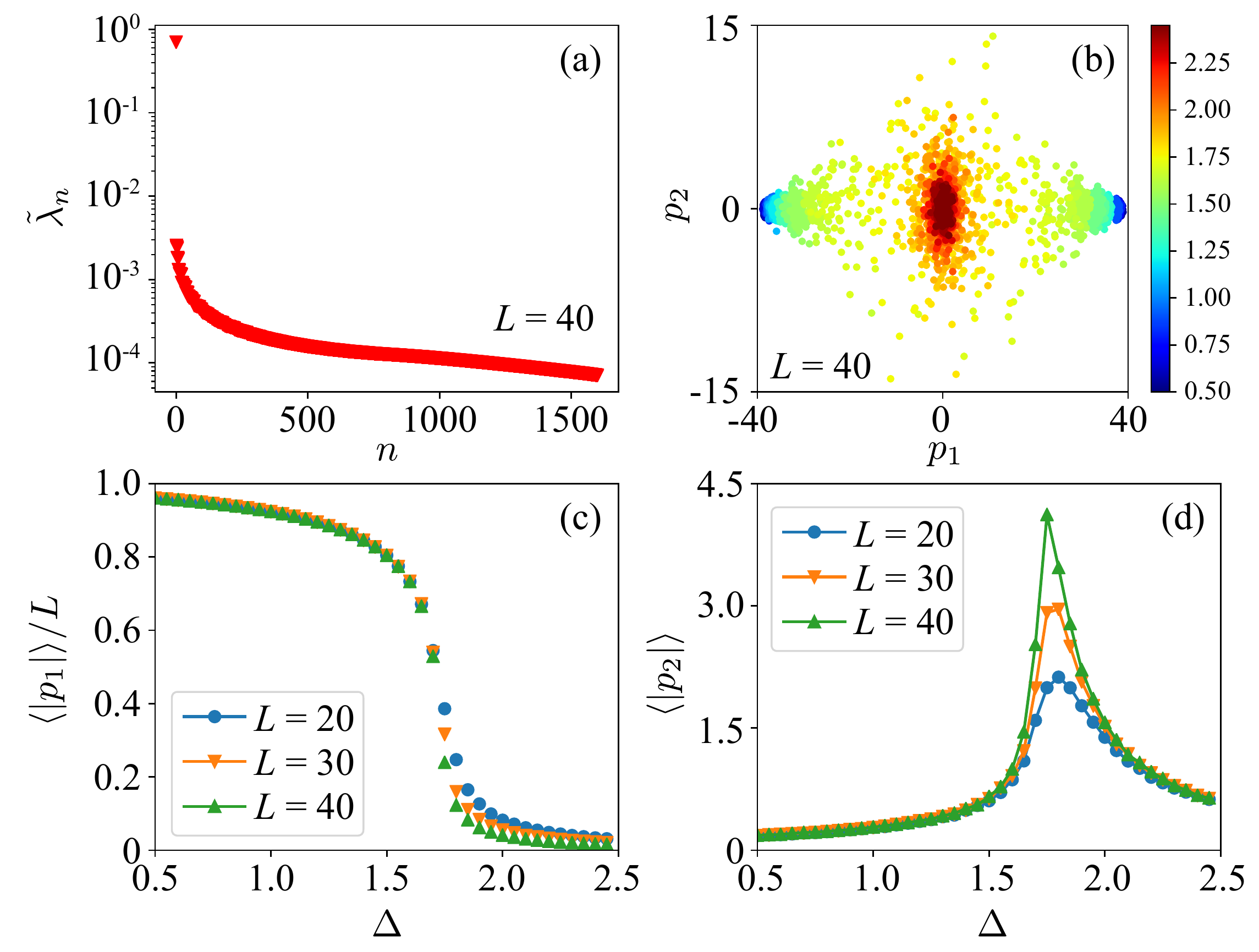}
    \caption{
    Same as Fig.~\ref{fig:IsingMLresults1}
    except for the Blume-Capel model and sweeping $\Delta$ at fixed $J=T=1$.
    As in the Ising model, there is a crisp separation of eigenvalue 
    scales (panel a) and
    a bifurcation of the scatter plot at $\Delta_c$ (panel b).
    The average overlaps once again serve as proxies for
    the magnetization ($\langle |p_1| \rangle$, panel c) and
    susceptibility ($\langle |p_2| \rangle$, panel d).
    The transition is second order for this $T=1$ sweep.  
    Adapted from Ref.~\cite{HuPRE2017} with minor modifications.
    }
    \label{fig:BlumeCapelresults1}
\end{figure}

\begin{figure}[t]
    \centering
     \includegraphics[width=\columnwidth]{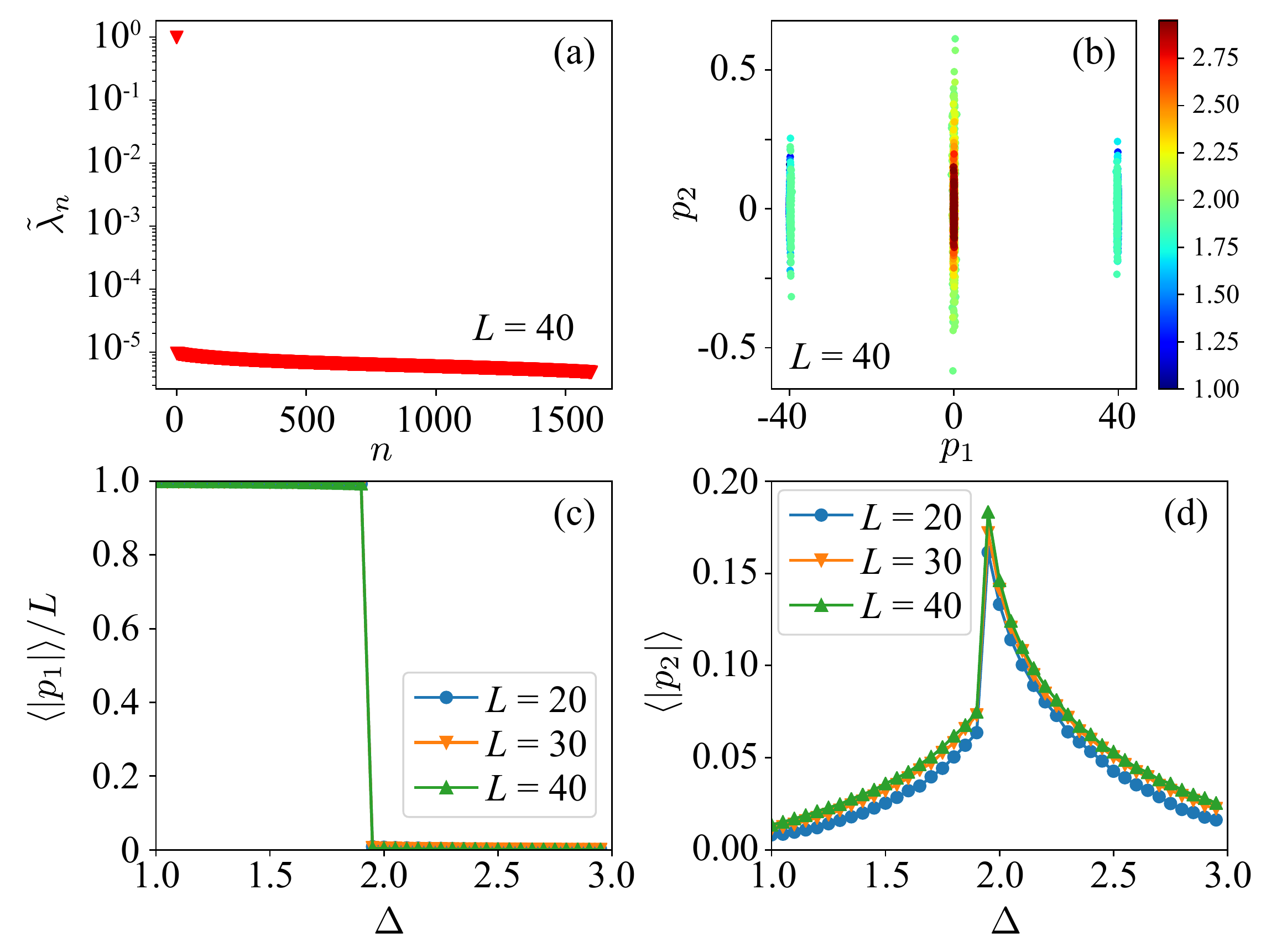}
    \caption{
    Similar to Fig.~\ref{fig:BlumeCapelresults1},
    except the transition is first order for this $T=0.4$ sweep. 
    Adapted from Ref.~\cite{HuPRE2017} with minor modifications. 
    }
    \label{fig:BlumeCapelresults2}
\end{figure}

The studies of the Ising model 
    (Figs.~\ref{fig:IsingMLresults1}, \ref{fig:IsingMLresults2}, and
    \ref{fig:3DIsing})
focused on using configurations at
different {\it temperatures} to determine $T_c$.  
In the Blume-Capel model, however, the phase boundary can be crossed by varying the zero-field splitting $\Delta$. PCA is also effective in such `parameter-tuned' 
transitions, as shown in Figs.~\ref{fig:BlumeCapelresults1} and   \ref{fig:BlumeCapelresults2}. 
The Blume-Capel model has a tricritical point at 
$(T/J,\Delta/J)=(0.609,1.965)$ along its phase boundary.
Fig.~\ref{fig:BlumeCapelresults1} cuts across at 
a second-order transition $(T/J=1.0)$, while 
Fig.~\ref{fig:BlumeCapelresults2} cuts across at 
a first-order transition $(T/J=0.4)$. 
Not only are the critical points easily identified, but their
orders are readily apparent.

For pedagogical reasons, we have focused here on the use of
\gls*{ML} for the well-known and characterized Ising
and Blume-Capel models.  It is worth noting, however, that these methods have also been used to elucidate the behavior 
of other, more challenging, classical models. 
These include the biquadratic spin exchange spin-$1$ Ising  
model of He$^3$-He$^4$ mixtures~\cite{HuPRE2017, Blume71}, 
frustrated magnetism~\cite{c_wang_17}, the Kosterlitz-Thouless transition of the 2D $XY$ Hamiltonian~\cite{HuPRE2017, wang2018machine, beach2018machine},
and topological order in Ising gauge theories~\cite{rodriguez2019identifying}. Together, the studies discussed here illustrate the rapid pace at which powerful variants of \gls*{ML} methods are being developed, in analogy with the (much longer) history of the evolution of Monte Carlo methods for studying phase transitions.

\subsection{Quantum models}\label{sec:benchmarks-quantum}
The use of restricted Boltzmann machines (RBM) as an ansatz for representing ground 
state wavefunctions of quantum many-body systems have been another exciting and fruitful approach. In a novel study, Carleo and Troyer~\cite{Carleo2016} used reinforcement learning to train their RBMs by minimizing the ground state energy of the transverse-field 
Ising and quantum Heisenberg models. In doing so, they showed that they could surpass the performance of then state-of-the-art conventional variational techniques by systematically increasing the density of the hidden layer. 

These works laid the groundwork for and inspired many other studies that 
followed shortly after~\cite{AnnasList}. For example, ideas of Ref.~\cite{j_carrasquilla_16}
were extended and applied to quantum many-body systems like the Fermi 
Hubbard model on the honeycomb and cubic lattices~\cite{p_broecker_16,k_chng_17} 
to learn quantum or thermal phase transitions. Broecker \emph{et al}.~\cite{p_broecker_16}  showed that the knowledge of the physics of the problem at the extremes, in that
case, deep in the semi-metal or AFM phases of the honeycomb lattice Hubbard model, 
can lead to an accurate estimate of the transition temperature by the artificial 
neural networks. They also found that input data engineering to guide the neural networks toward physical properties of interest can significantly affect 
the training.

\begin{figure}[t]
\centerline {\includegraphics*[width=3.3in]{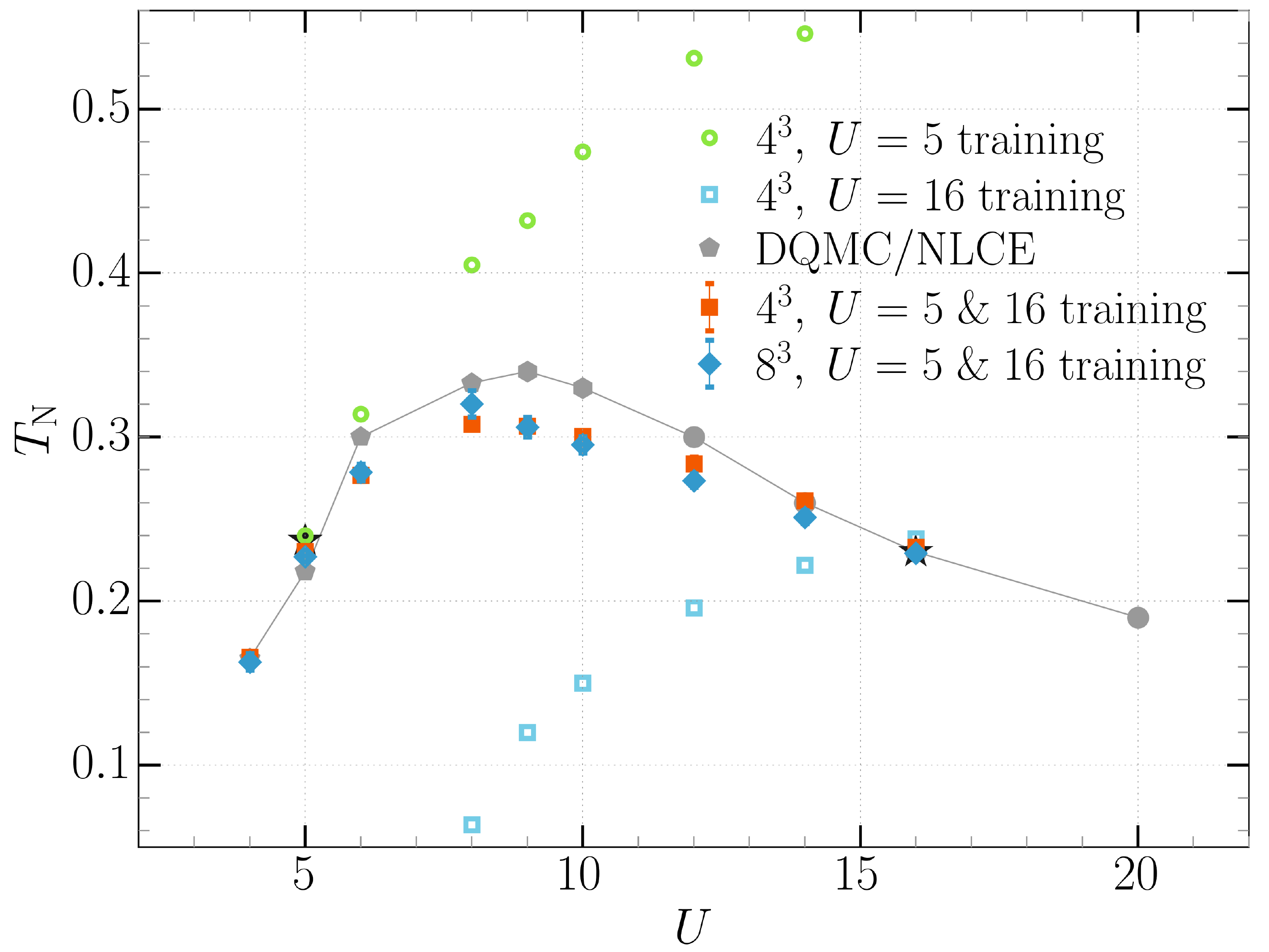}}
\caption{Prediction of the AF transition temperature in the 3D Hubbard model by a CNN. 
The CNN is trained using DQMC's auxiliary field data at different temperatures for fixed values of $U$. The prediction is made after three pieces of training: One with data from $U=5$, one from $U=16$, and one with mixed data from $U=5$ and $16$. The latter yields the nontrivial shape of the phase boundary when providing the CNN with data from other $U$ values it has not seen before. Estimates 
for $T_N$ in the thermodynamic limit are from past DQMC and numerical linked-cluster expansion calculations. Taken from Ref.~\cite{k_chng_17}
\label{fig:3DHubb}}
\end{figure}

Ch'ng \emph{et al}.~\cite{k_chng_17} used raw auxiliary field configurations
of the cubic lattice Hubbard model in 3+1 dimensions and by treating the time
slices along the quantum dimension as ``color channels'' in their CNNs, allowing the method 
to learn the finite-temperature N\'{e}el transition at half-filling. They then 
demonstrated the power of transfer learning by training a CNN using a mix of
configurations from two different $U$ values in the weak- and 
strong-coupling regimes, which were then used to estimate the nontrivial shape of 
the AFM phase boundary in the temperature-interaction strength phase diagram 
of the model (see Fig.~\ref{fig:3DHubb}).

\begin{figure}[t]
\centerline {\includegraphics*[width=3.3in]{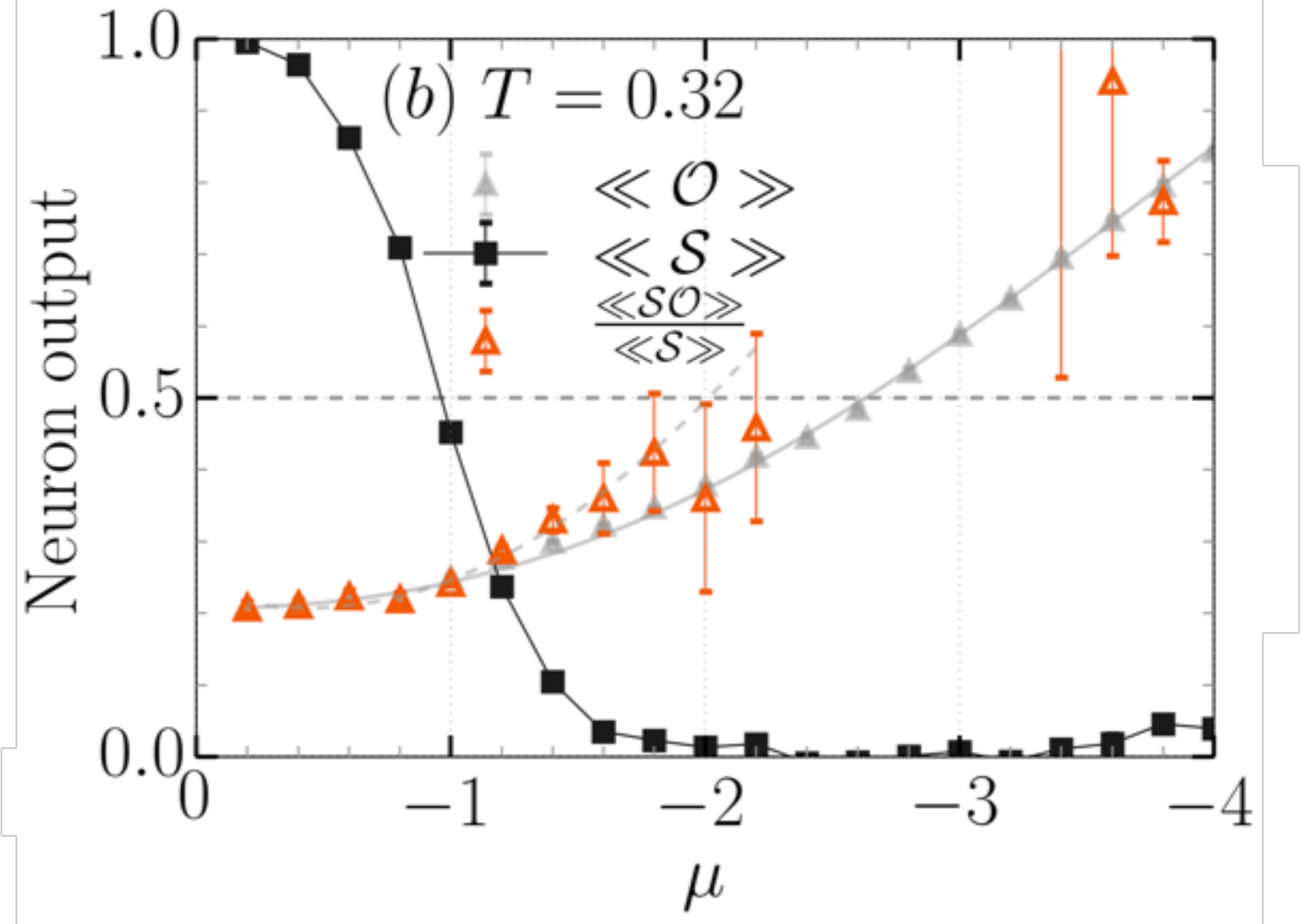}}
\caption{Average neuron output $\mathcal{O}$ calculated with and without taking the fermion sign into account
at $T=0.32$ (${\llangle \mathcal{S O} \rrangle}/{\llangle\mathcal{S} \rrangle}$ and $\llangle \mathcal{O} \rrangle$, respectively) when using a CNN trained at half-filling to detect the N\'{e}el transition away from half-filling. 0.5 crossing indicates the location of the transition. The black line shows the average fermion sign. Taken from Ref.~\cite{k_chng_17}
\label{fig:3DHubbAway}}
\end{figure}

Both of these works touched on the fermion sign problem in Fermi-Hubbard models away from symmetry points and their implications for machine learning. Ref.~\cite{p_broecker_16} demonstrated that, at least for some models and
properties of interest, the sign could essentially be ignored in the training
and classification. In Ref.~\cite{k_chng_17}, Ch'ng \emph{et al}.~avoided
training their machines in the sign-problematic parameter region away from 
half-filling. Instead, in using those CNNs trained at half-filling to track 
the magnetic transition away from half-filling, they treated their network output 
as another physical observable, arguing that the sign of the auxiliary field 
configurations should be incorporated into their averages. As shown in 
Fig.~\ref{fig:3DHubbAway}, ignoring the sign can lead to small but significant differences in the results in this case.

Early on, it was shown that topological states of matter could also be studied using artificial neural networks. RMBs were first used to represent topological states in one, two, and three dimensions~\cite{d_deng_17}, and it was found that the number of hidden parameters needed scales only linearly with the system size. It was further shown that RBMs could find the topological ground states of generic nonintegrable Hamiltonians through reinforcement learning and identify their topological phase transitions. RBMs were also used as a decoder of topological codes~\cite{g_torlai_17}. The challenge of capturing nonlocal properties of topological phases with neural networks led Zhang and Kim to introduce quantum loop topography, a procedure to construct a multidimensional image of the wavefunction based on two-point correlation functions that form loops, which was then used as input to neural networks to distinguish Chern insulators from trivial insulators~\cite{y_zhang_17}.

Algorithms for the unsupervised learning of phases and phase transitions (with no specific knowledge of the nature of phases or the whereabouts of the transition) that were based on supervised machine learning methods attracted much interest. In the ``confusion'' method~\cite{e_vanNieuwenburg_17}, neural networks are trained with data that have been deliberately mislabeled. Monitoring how the training accuracy varies as different locations for the phase transition are proposed allows one to identify the correct labels for the data and pinpoint the transition. Another method also used the training accuracy as a function of the tuning parameter, but for training performed on data from consecutive tuning parameters~\cite{p_broecker_17}. In this approach, a peak in the accuracy would indicate a sudden change in the character at the location of the true phase transition.

Traditional unsupervised learning methods, such as PCA, autoencoders, tSNE, and random trees embedding, showed a remarkable ability to reveal phase changes in the presence of quantum fluctuations. While an early application of the PCA to QMC data for the Heisenberg model led to no discernible features in the reduced dimensional space~\cite{e_vanNieuwenburg_17}, a thorough analysis of the raw auxiliary field DQMC data for the 2D and 3D Hubbard model demonstrated outstanding potential for nonlinear methods to shed light on the phases and phase transitions of quantum lattice models. It also showed that indicators that correlate with conventional properties could be defined using the data projected onto the reduced dimensional space~\cite{k_chng_18}.

\begin{figure}[t]
\centerline {\includegraphics*[width=3in]{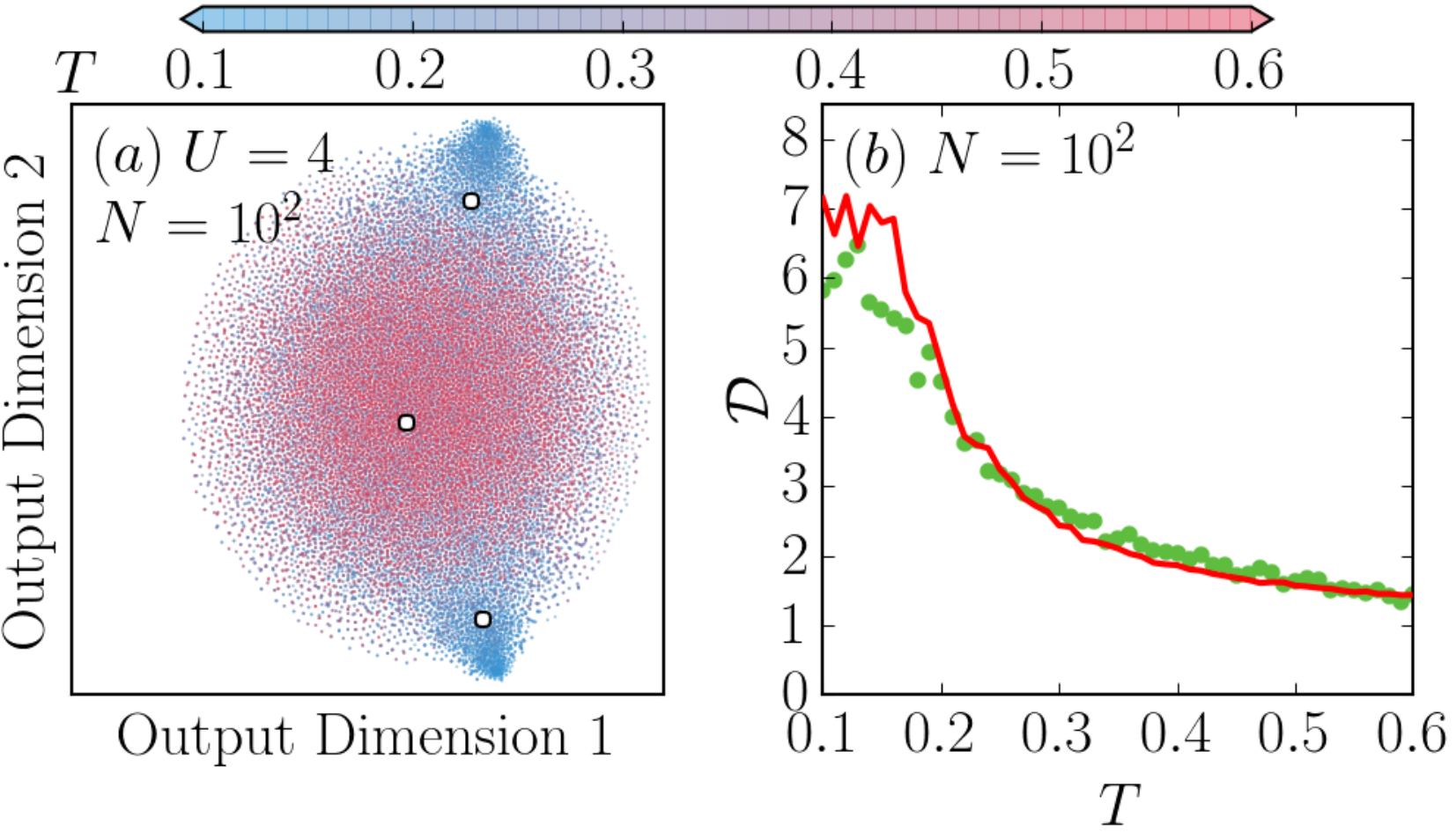}}
\caption{Projection of the auxiliary field data for the half-filled 2D Hubbard model to a two-dimensional latent space using the tSNE algorithm. The system size is $N=10^{2}$ and $U=4t$. 
Eight hundred configurations per temperature in a uniform grid of $T/t$ between 0.1 and 0.60 are used. 
(b) The indicator (green symbols -- see text) and the AFM structure factor (red line) as a function of temperature. Image reproduced from Ref.~\cite{k_chng_18}.}
\label{fig:tsne2DHubb}
\end{figure}

\begin{figure}[t]
\centerline {\includegraphics*[width=3.3in]{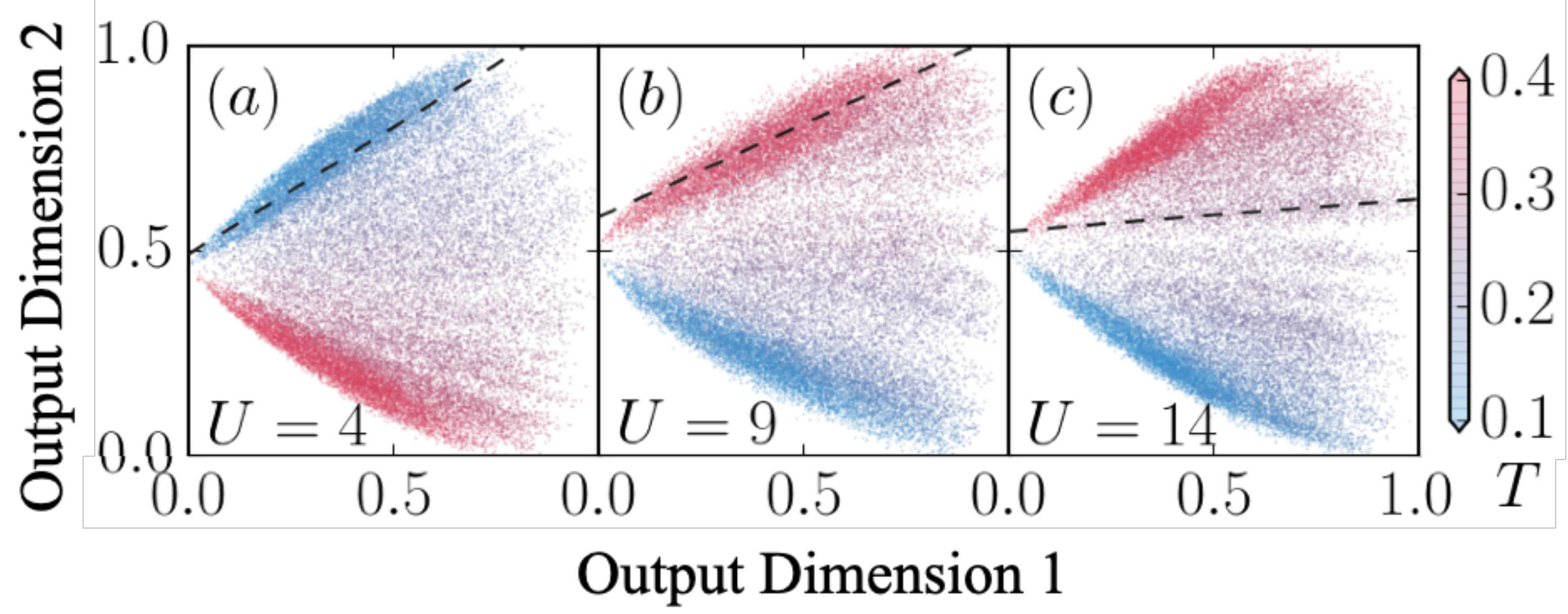}}
\caption{Projection of the four latent variables of a fully-connected AE  to two dimensions using the random trees embedding algorithm. The original data used to train the AE are auxiliary fields for the 3D 
Hubbard model with (a) $U=4$, (b) $U=9$, and (c) $U=14$ at half-filling. The dashed lines are line fits to data at the estimated N\'{e}el temperature for each $U$. Image reproduced from Ref.~\cite{k_chng_18}
\label{fig:RandomTrees}}
\end{figure}

Figures~\ref{fig:tsne2DHubb} and \ref{fig:RandomTrees} highlight some of those early findings. Fig.~\ref{fig:tsne2DHubb}(a) shows the results of tNSE applied to raw auxiliary field data for the half-filled 2D Hubbard model with $U=4t$ on a $10\times 10$ square lattice. The color indicates temperature and makes it clear that the data projected to the 2D space evolve from a large symmetric cluster at relatively high temperatures to two smaller ones on either side of the hot cluster at lower temperatures. They correspond to the two possible sublattice orientations of the N\'{e}el ordered phase. When applied to the data, the k-means algorithm identifies three clusters at each temperature. Fig.~\ref{fig:tsne2DHubb}(b) shows how the average distance of a cluster's center from the mean location of data closely follows the magnetic structure factor as a function of temperature. Note that no knowledge about the physics of the problem has been provided to the machine. 

Figure~\ref{fig:RandomTrees} shows temperature gradients in the 3D Hubbard data from an AE that have been further analyzed by the random trees embedding method. Here, the AE reduces the original auxiliary field data to four latent variables, and then the random trees embedding projects the latent variables to a 2D space. In this case, the output separates data points from different temperatures. 

\gls*{ML} methods have been applied to many other quantum models. For example, PCA has also been widely used to locate phase transitions in various quantum Hamiltonians, with some notable successes and failures, which we now discuss. 
We begin with the finite temperature CDW phase transition in the Holstein model and then examine magnetic quantum phase transitions tuned by changing model parameters in several different contexts, including: 
1)  the inter-orbital hybridization in the periodic Anderson model,
2)  the on-site interaction in the Hubbard model on a honeycomb lattice, 
and
3) the density in the Hubbard model on a Lieb lattice. 
`Topological data analysis' is a related \gls*{ML} method
recently applied to (1) and (2)~\cite{tirelli2021learning}. 
In all these cases, the input features are provided to the PCA 
are the Hubbard-Stratonovich field variables $\{s_{i,l}\}$ obtained from DQMC 
simulations of these models. 
With these success stories established, we then discuss the challenges
encountered in studying the Kosterlitz-Thouless transition
to a superconducting phase 
in the doped two-dimensional attractive Hubbard model~\cite{Fontenele2022}.
We note that for quantum models, DQMC simulations work in a path integral
representation of the partition function so that they sample a $d+1$ dimensional
space-imaginary time lattice, where $d$ is the spatial dimension. Unless otherwise indicated, the 
configurational vectors used in the PCA discussion that follow contain the entire lattice. 
In principal, one might also study the performance of the PCA for different
discretizations $\beta = L \Delta \tau$ of imaginary time; 
however, the results do not appear to be very sensitive to the size of the Suzuki-Trotter errors, provided $\Delta\tau$ is reasonably small. 

The Holstein Hamiltonian poses special difficulties to QMC simulations
owing to its long autocorrelation times. \gls*{ML} 
methods play an especially useful role in the acceleration of the simulations,
as discussed in Sec.~\ref{sec:SLMC}. In this section, we will confine ourselves to PCA's use in analyzing configurations generated by the conventional DQMC method.

\begin{figure}[t]
\centerline {\includegraphics*[width=\columnwidth]{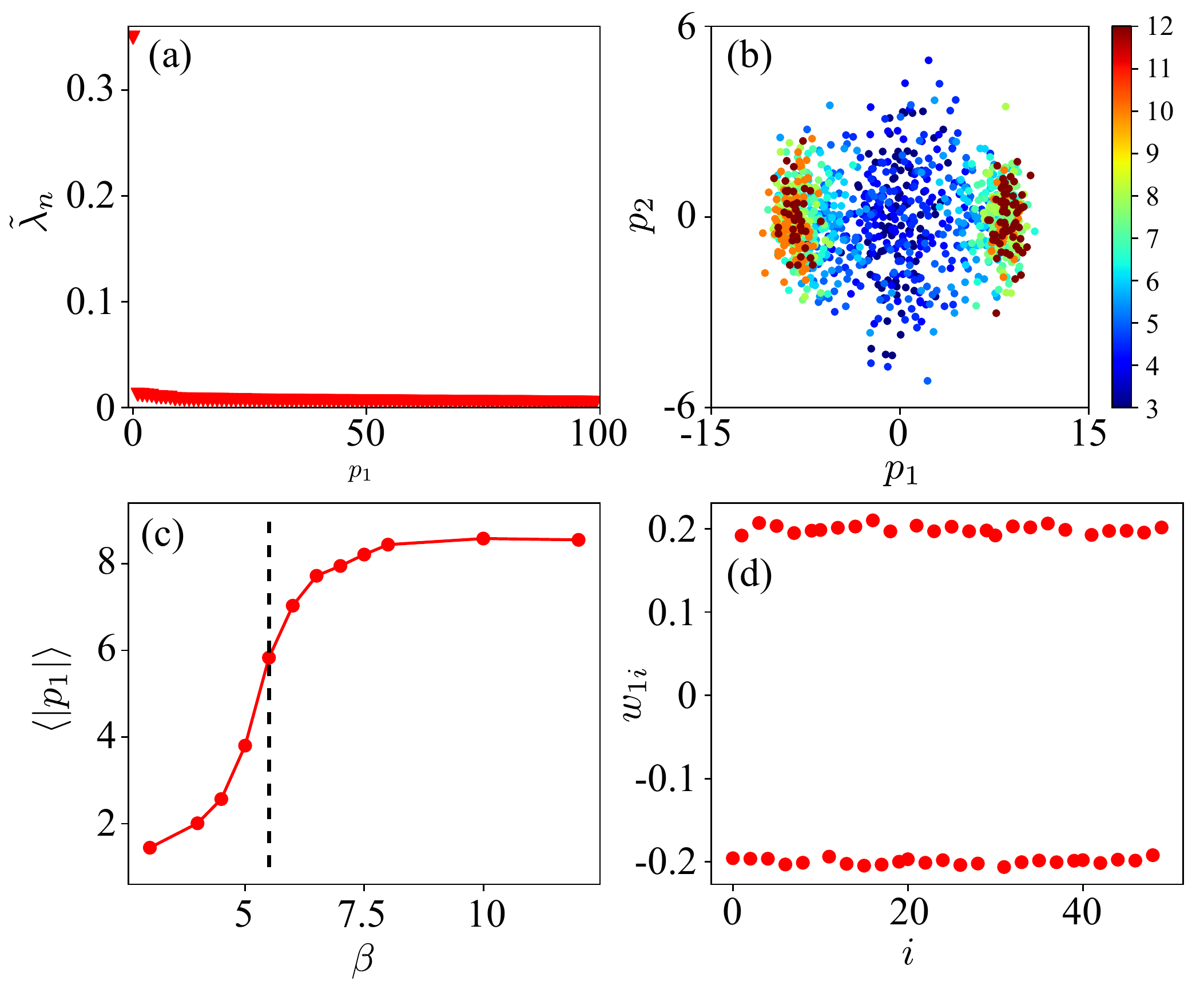}}
\caption{PCA for the Holstein Hamiltonian on a $10 \times 10$ lattice at half-filling, with $\omega_0=1$ and
$\lambda=1/2$.  The principal eigenvalues of panel (a) exhibit a rapid fall-off, 
suggesting data compression should be effective.
The vertical dashed line gives the critical temperature for the CDQ phase transition 
obtained from a traditional finite-size scaling analysis of the order parameter.
The \gls*{ML} calculation shows a sharp rise in the first principal component
(panel c)
and the development of a two-peak structure in a $(p_1,p_2)$ scatter plot
(panel b)
at the same $T_c$.  The overlaps of the spin configurations
with the first principal eigenvector having the staggered structure 
of CDW order (panel d).
Adapted from Ref.~\cite{costa17} with minor modifications. 
}
\label{fig:PCAHolstein}
\end{figure}

Figure \ref{fig:PCAHolstein} shows PCA analysis of the CDW transition in the Holstein model in a close analogy to that presented in Fig.~\ref{fig:IsingMLresults1}.
As with the previous examples, the sharp drop-off in the eigenvalues shown in Fig. \ref{fig:PCAHolstein}(a) (even more dramatic than that
in the Ising case) suggests that the PCA method will be able to compress the configuration space efficiently. The projection onto the plane of the two principal components is shown in panel (b). Here, the bifurcation of the distribution is observed as $T$ is lowered, reflecting the transition to the CDW phase. In this case, the average of the first
principal components as a function of (inverse) temperature, shown in panel (c), serves as an order parameter. At the same time, the spatial structure
of the components of the first eigenvector $w_{1i}$ at low $T$ oscillates in sign, reflecting the two sublattice structure of the CDW phase whose ordering wave-vector is ${\bf Q}_\mathrm{CDW} = (\pi,\pi)$.

\begin{figure}[t]
\centerline {\includegraphics*[width=\columnwidth]{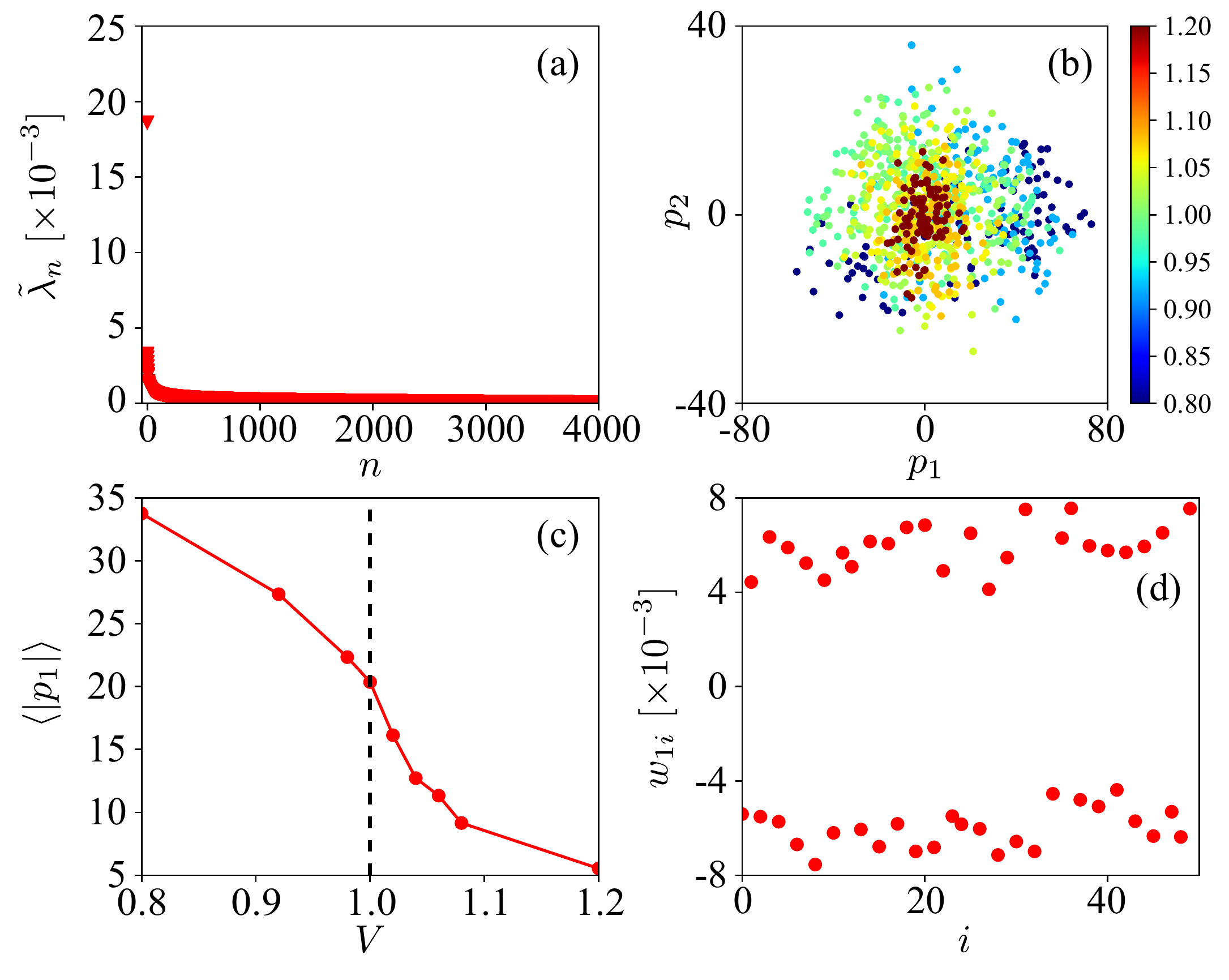}}
\caption{PCA for the periodic Anderson Model (PAM) on a $12 \times 12$ lattice with $U_f=4t$ and 
$\beta t =24$.  The inter-orbital hybridization $V$ is the tuning parameter. The PCA can isolate the QPT, which occurs between AF order at small $V$ and the singlet phase at large $V$. Adapted from Ref.~\cite{costa17} with minor modifications. 
}
\label{fig:PCAPAM}
\end{figure}

Figure \ref{fig:PCAPAM} examines PCA's ability to discern the quantum phase transition of the PAM (Sec.~\ref{sec:PAM_model})~\cite{costa17}. In this case, DQMC simulations at low temperature ($T=t/24)$ are used to generate data 
for varying hybridization values $V$. Applying the PCA to this data set at fixed $T$ produces results very similar to the Holstein model;
the principal eigenvalues fall off rapidly [Fig.~\ref{fig:PCAPAM}(a)], demonstrating that PCA can indeed achieve a large degree of data compression. Similarly, the scatter
plot of $(p_1,p_2)$ bifurcates as a function of the model parameter  $V$. At the same time, the first principal component $p_1$
[Fig.~\ref{fig:PCAPAM}(c)] 
behaves like an order parameter for the antiferromagnetic phase in that it goes to zero as $V$ increases across the known $V_c \sim 1$~\cite{vekic1995competition}. 
At low $V$, the first eigenvector exhibits a clear oscillatory pattern, reflecting strong AFM correlations.

The Hubbard model on a square lattice has AFM order at half-filling for all values of the on-site repulsion $U$ \cite{hirsch1985two,WhitePRB1989}. For weak coupling, this ordering is a consequence of the `perfect-nesting' of the Fermi surface (FS), where a large number of points ${\bf k}$ and ${\bf k} + (\pi,\pi)$ lie on the FS leading to an enhanced instability to AFM order. The logarithmic van-Hove singularity of the density of states $N(\omega=0)$ contributes to this process. No such nesting occurs on a honeycomb lattice, and $N(\omega=0)$ vanishes linearly with $\omega$ at half-filling (the so-called `Dirac spectrum')~\cite{Sorella_1992}. This electronic structure leads to a finite $U_c$ for the AF order.  The physics of the semi-metal to AF transition has engendered a 
great deal of investigation with numerical methods~\cite{Sorella_1992,paiva05,Otsuka2016,raczkowski2020hubbard,costa2021magnetism}, 
including early studies of a possible intervening 
spin-liquid phase~\cite{meng10} that does not appear to occur~\cite{Sorella2012}. Ref.~\cite{costa17} revisited this issue and studied the semi-metal/AF transition of Dirac fermions on a honeycomb lattice using PCA, with clear indications of a transition to an AFM state at a $U_c$ in agreement with the most accurate value $U_c = 3.87 t$ found in the literature~\cite{Sorella2012}.  

\begin{figure}[t]
\centerline {\includegraphics*[width=\columnwidth]{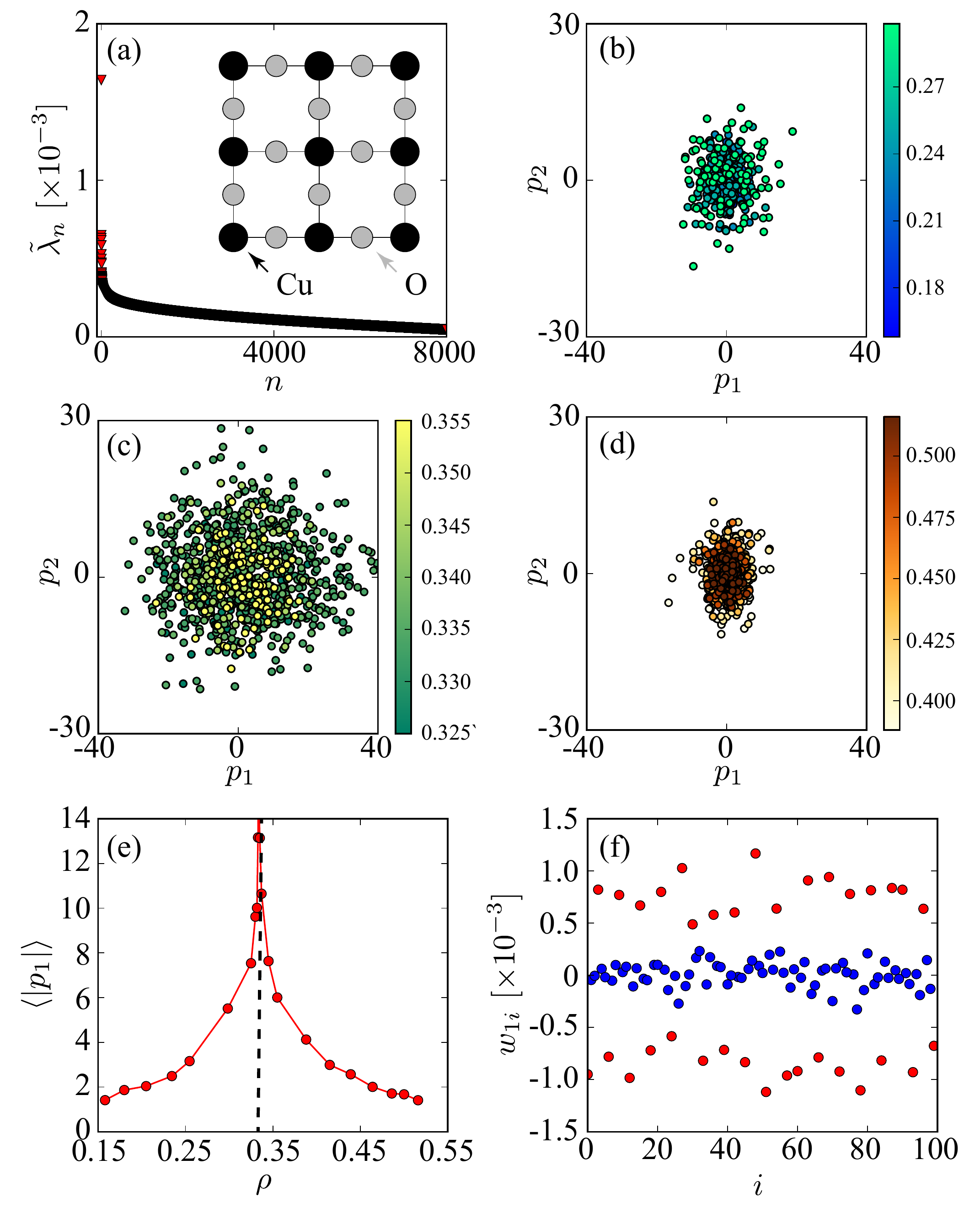}}
\caption{PCA for the antiferromagnetic transition on a Lieb lattice.  The geometry consisted of
$10 \times 10$ unit cells, each with three orbitals, as shown in the inset of panel (a). The inverse temperature $\beta t = 20$. The on-site $U_{d}=4t$ for the `copper sites'
and $U_{p}=0$ for the `oxygen sites'.  The charge-transfer energy 
$\epsilon_{pd} = 2t$.
Here the first eigenvector, shown in panel f, contains precise information
about the nature of the ordered phase. The copper atoms (red circles) 
hold the AF order, whereas the oxygen sites (blue circles) do not participate.
The convention for the density is such that
$\rho=1/3$ corresponds to one hole per (three sites) unit cell.
Adapted from Ref.~\cite{costa17} with minor modifications. 
}
\label{fig:PCALieb}
\end{figure}

The results above demonstrate that a PCA can resolve temperature- and model-parameter-driven  (quantum) phase transitions. 
However, many quantum materials can be doped, and the fermionic carrier 
density often functions as another tuning parameter. 
With this in mind, we now examine simulations of the Hubbard model on a Lieb lattice, the geometry of the CuO$_2$ planes in cuprate superconductors, as a function of carrier concentration.
The Lieb lattice has a square array of $d$ orbital `Cu' sites that are bridged by intervening $p$ orbital `O' sites, as shown in the inset of Fig.~\ref{fig:PCALieb}(a).
Each unit cell thus  has three orbitals and the filling corresponding to the AF
parent compounds of the cuprates is $\rho=1/3$ hole per unit cell.
The model is most commonly studied with $U_d  > U_p$, and on-site energy for the oxygen sites is higher (for holes) than for the copper sites, as is the case for cuprates~\cite{DopfPRB1990, KungRB2016, HuangScience2017}.

Figure~\ref{fig:PCALieb} shows the density-tuned transition through AF order in the Lieb lattice.
A somewhat different perspective here is obtained by showing the distribution of principal components both in the SDW phase $\rho=1/3$ (panel c), as well as below $\rho<1/3$ (panel b) and
panel $\rho>1/3$ (panel d).  
The tightness of the cluster at $\rho=1/3$, in contrast
to the bracketing densities, indicates the nature of the transition is disorder-order-disorder.
This is a distinction from the previous cases where the critical point separates disordered from ordered
regions. Another difference is the relative closeness of the first principal component to the succeeding ones -  it is only larger by a factor of two.

Application of PCA to Hubbard models has a sign problem in cases where the problem is not particle-hole symmetric~\cite{loh90, troyer05}, as is the case for the Lieb lattice~\cite{DopfPRB1990}. 
Thus, the results of Fig.~\ref{fig:PCALieb} carry the additional implication that
\gls*{ML} methods have some potential to address phase transitions in a quantum model, evading the sign problem. This aspect is an especially intriguing possibility since the \gls*{ML} 
analysis does not involve the measurement of the noisy ratio of two
quantities that are both exponentially small. Other approaches investigating novel observables that avoid the sign problem have recently
been proposed~\cite{MondainiScience2022, YiPreprint}. 

\begin{figure}[t]
\centerline {\includegraphics*[width=\columnwidth]{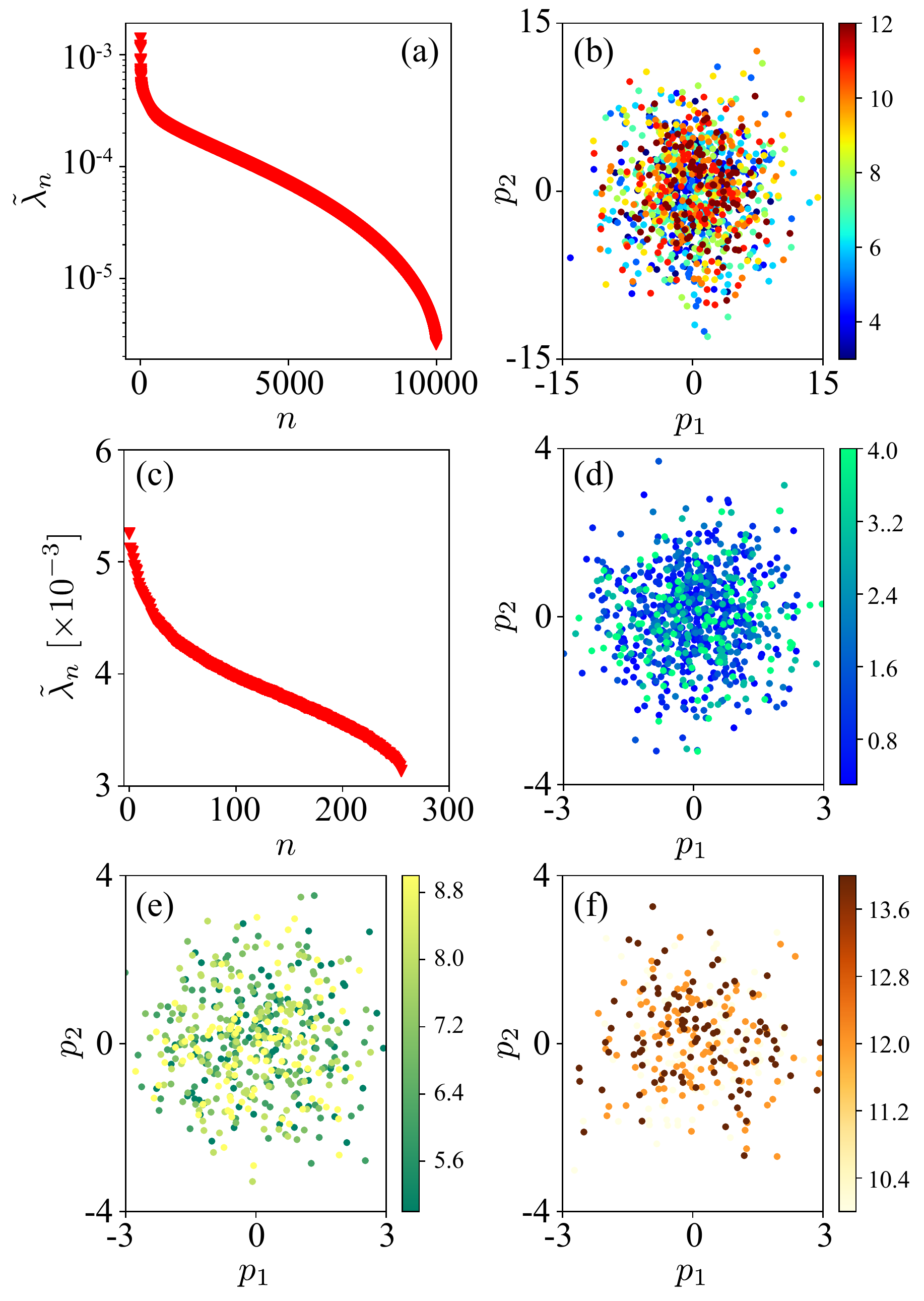}}
\caption{PCA for the attractive Hubbard model on a square lattice with $U=-4t$ and $\rho=0.80$.
Panels a,b are for a $12 \times 12$ lattice and panels c,d,e,f for a $16 \times 16$ lattice.
In the latter case, the distribution has been separated into three plots.
The critical value of the inverse temperature is $\beta_c \sim 7.5$, but the analysis provides
no signal there.
Adapted from Ref.~\cite{costa17} with minor modifications. 
}
\label{fig:PCAHubbardHSF}
\end{figure}

Our final example of using \gls*{ML} 
to examine fermionic quantum phase transitions concerns the transition into a superconducting phase in the attractive Hubbard model when doped away from half-filling. 
This transition has long been a challenging
problem in the field because the transition is in the Kosterlitz-Thouless universality 
class; the nature of the ordered phase, where the correlation functions
decay as power laws, is more delicate than with a ``true'' long-range order.  
Larger lattices are required to treat such phases 
accurately, and it is fair to say they are much less understood than the phase transitions discussed above. 
Indeed, quantitative values of superconducting $T_c$ in the attractive Hubbard model have varied by $20$-$30\%$
in various QMC studies~\cite{moreo1991two, scalapino1993insulator, singer1996bcs, kyung2001pairing, paiva2004critical, karakuzu2018study, Fontenele2022}. 

In Fig.~\ref{fig:PCAHubbardHSF}, we observe that the same difficulty is encountered
in a PCA. For one, the eigenvalue distribution [Fig.~\ref{fig:PCAHubbardHSF}(a)] does not exhibit a gap but instead decays slowly. Similarly, the principal
component distribution [Fig.~\ref{fig:PCAHubbardHSF}(b)] shows no clear signal as the inverse temperature $\beta$ is tuned through $\beta_c \sim 7.5$, where the superconductivity is believed to onset. 
In an attempt to discern the transition, analyses using input feature vectors containing the entire space-time lattice (panels a \& b) or just a single time slice (panels c-f) have been attempted with no success. 

\begin{figure}[t]
\includegraphics*[width=\columnwidth]{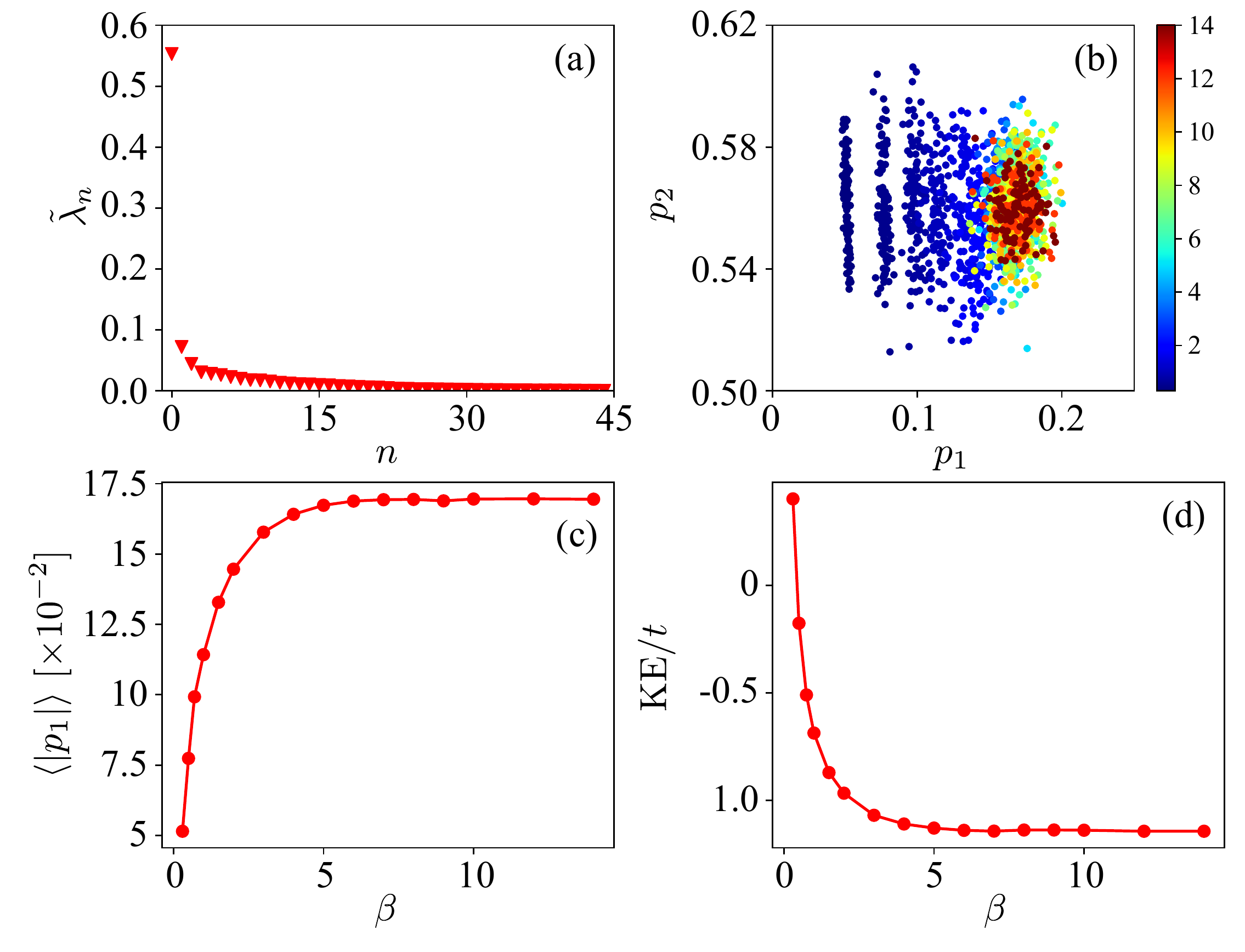}
\caption{PCA for  the attractive Hubbard model using vectors comprised of the equal time Greens 
function.
Although there is a gap in the eigenspectrum and an evolution of the scatter plot of
the first two principal components, there is no evidence of the transition. As might 
be expected, the average of the first principal component tracks the kinetic energy
(panel d) rather than the pairing order parameter.
Adapted from Ref.~\cite{costa17} with minor modifications. 
}
\label{fig:PCAHubbardGreen}
\end{figure}

\begin{figure}[t]
\centering
\includegraphics*[width=\columnwidth]{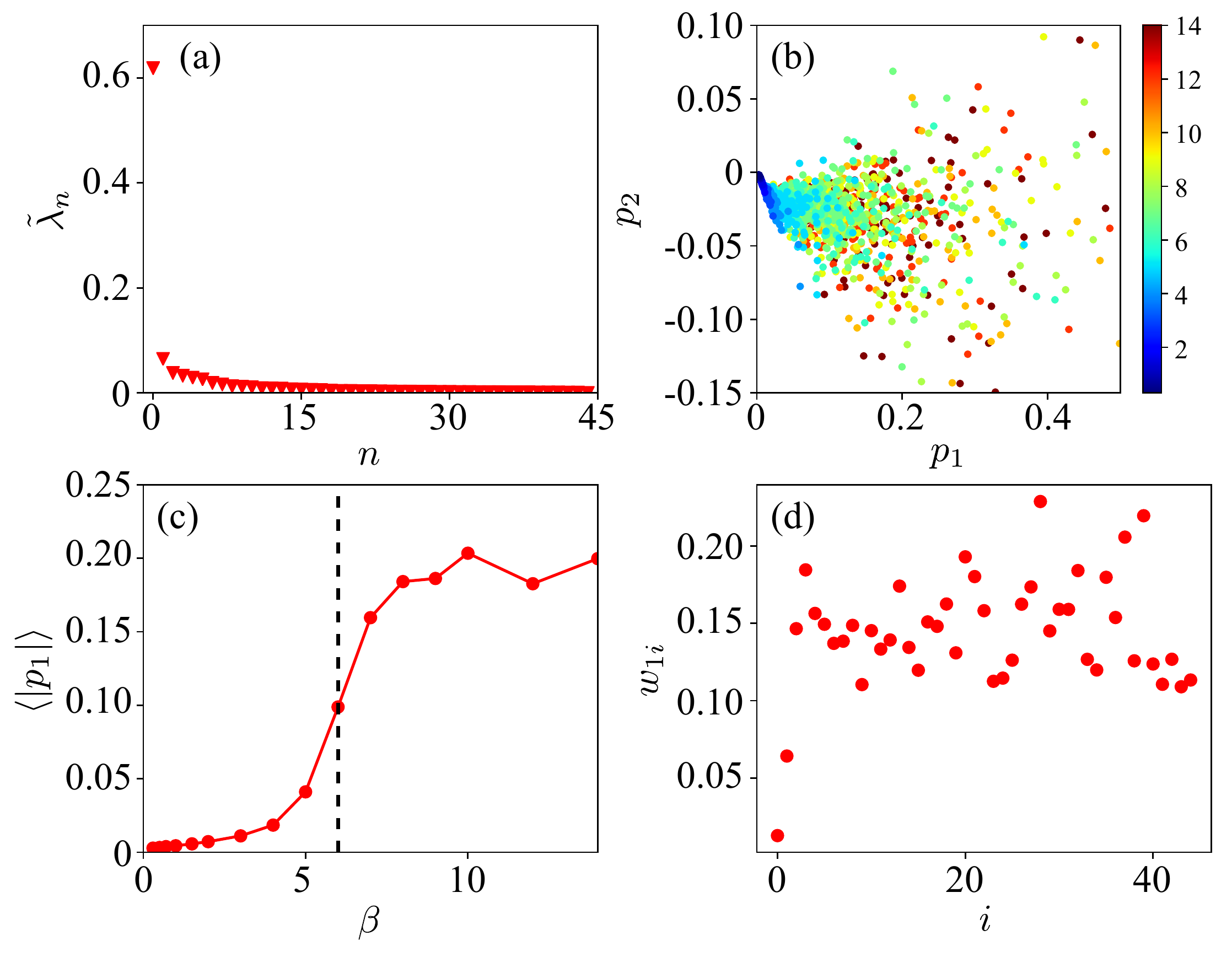}
\caption{PCA for  the attractive Hubbard model using vectors comprised of the equal time pair 
correlation function.
Here, finally, there is some indication of the superconducting transition.
Adapted from Ref.~\cite{costa17} with minor modifications. 
}
\label{fig:PCAHubbardPairing}
\end{figure}

Figures~\ref{fig:PCAHubbardGreen}
and \ref{fig:PCAHubbardPairing}
present further efforts to discern the superconducting transition in the attractive Hubbard model. In the former, rather than providing the HS field configurations as the vectors ${\bf S}_\gamma$, 
the authors instead use the equal time Greens function 
$G_{ij}=\langle c_{i\sigma}^{\phantom{\dagger}}c_{j\sigma}^{\dagger}\rangle$.
In the latter, they used the equal time pair correlation function,  
${\cal P}_{ij}=\langle \Delta_{i}^{\phantom{\dagger}}\Delta_{j}^{\dagger}\rangle$
with
$\Delta_{i}=c_{i\uparrow}^{\phantom{\dagger}}c_{i\uparrow}^{\phantom{\dagger}}$.\footnote{Both quantities require measuring physical observables, and would be impacted by the fermion sign if this approach is applied to models with a sign problem.} 
Only the second approach seems capable of capturing the transition. For example, 
the average of the first principal component in 
Fig.~\ref{fig:PCAHubbardPairing}(c) has a maximum slope near the known value of $\beta_c$. 

\section{AI-assisted phase discovery in strongly correlated systems}\label{sec:discovery}
\begin{figure}[t]
\centerline {\includegraphics*[width=3.3in]{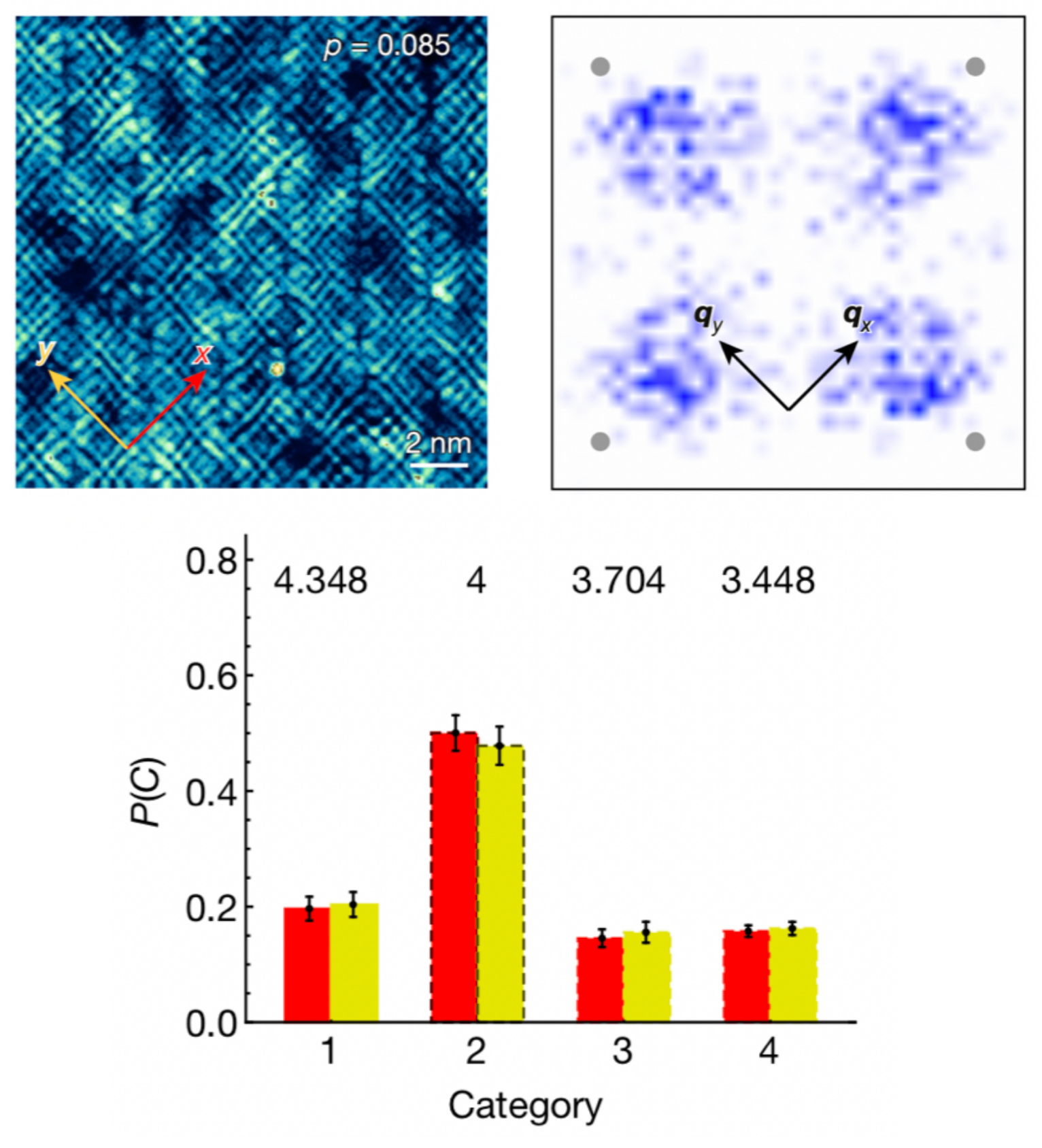}}
\caption{Top Left: Measured 440 pixel $\times$ 440 pixel STM image of Bi$_2$Sr$_2$CaCu$_2$O$_8$ at 8.5\% hole doping. Top Right: The d-symmetry Fourier transform of the image on the left. Bottom: Average output category of 81 ANNs. The numbers on top show the wavelength of each electronic ordering category in units of the lattice spacing. Red and yellow indicate the outputs for two different orientations of the input image, 90$^\circ$ rotated relative to each other. Image reproduced from Ref.~\cite{f_zhang_18}. 
\label{fig:STM}}
\end{figure}

The establishment of ML methods as viable tools for categorizing quantum data led scientists to utilize them for phase detection and physics discovery. Here, we highlight two studies in which experimental data were analyzed by AI to yield insight into the physics of strongly correlated systems. In the first study, Zhang {\it et al.}~\cite{f_zhang_18} used an ensemble of fully-connected neural networks to identify the dominant pattern in an archive of real scanning tunneling microscopy (STM) images of lightly-doped cuprates. These networks were trained on synthetic STM images created using theoretical models to represent four different categories of electronic ordering patterns. The authors showed that the ANNs could discover a lattice-commensurate, four-unit-cell periodic, translational-symmetry-breaking phase in the noisy experimental data. Moreover, they established the unidirectionality of the ordering pattern and how its dominance depends on the electron energy. Figure~\ref{fig:STM} shows a sample STM image analyzed by the ensemble of ANNs (top left panel). The linear Fourier transform (top right panel) reflects the level of noise and complexity that exists in the image and does not point to any particularly dominant ${\bf q}$ vector. On the other hand, the output of the ANNs (lower panel) demonstrates that the second category with a wavelength four times the lattice spacing is clearly dominant.

Linear Fourier transforms have been traditionally used for decades to analyze such images. The key to the success of ANNs in this study was the existence of non-linearities, allowing them to look beyond what a Fourier transform can provide. 

\begin{figure}[t]
\centerline {\includegraphics*[width=3.3in]{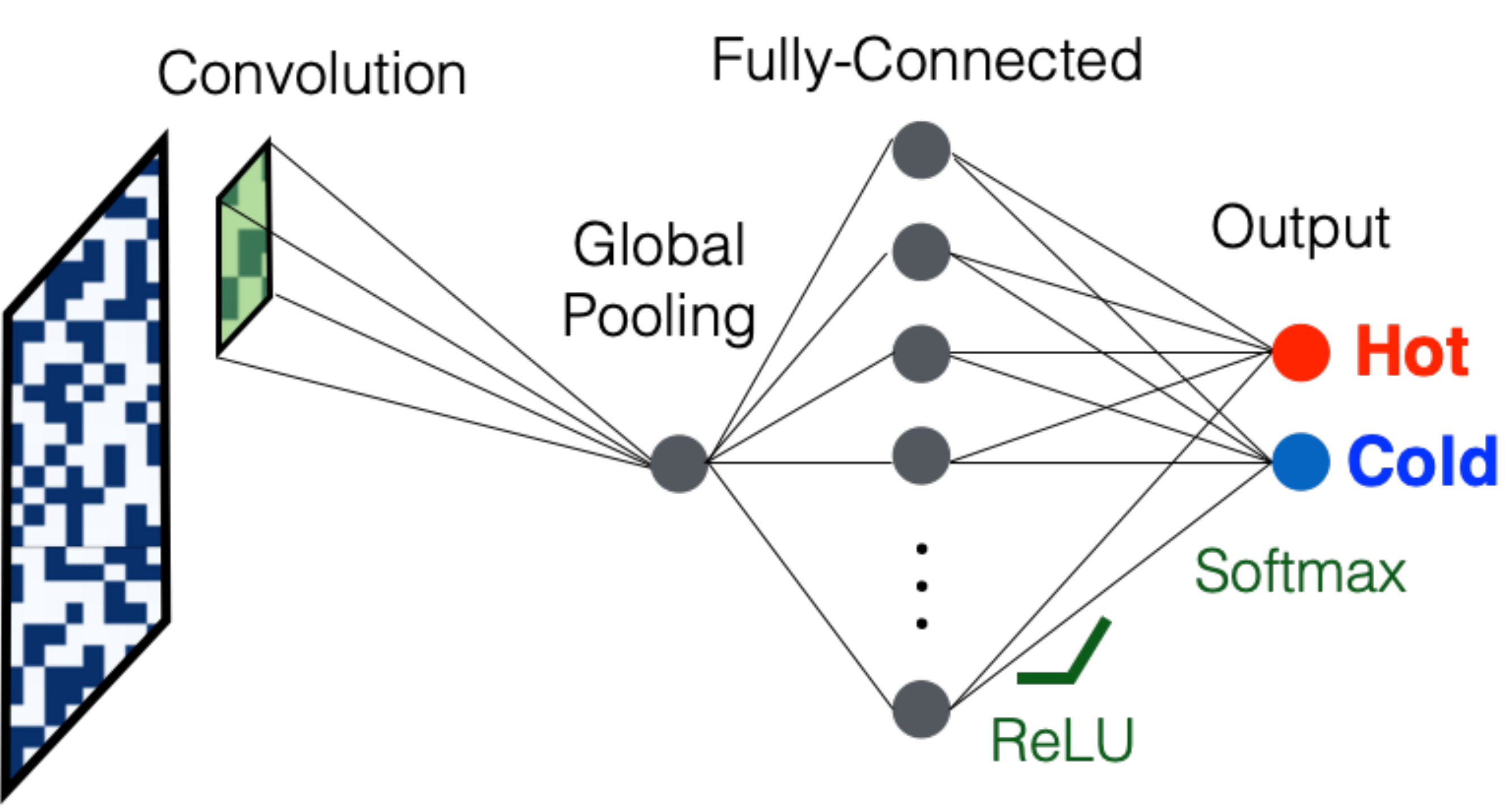}}
\caption{The main CNN used in Ref.~\cite{e_khatami_20}. It has one convolutional layer with one filter, followed by other pooling and fully-connected layers. It is trained to distinguish input snapshots of fermions at two extreme temperatures. After the training, patterns developed in the filter can correspond to physical correlations in the system at low temperatures. Image reproduced from Ref.~\cite{e_khatami_20}. 
\label{fig:CNN1}}
\end{figure}

The second study~\cite{e_khatami_20} aimed for the CNNs to have an unbiased take (not guided by any theory) on the ordering patterns and possible correlations of strongly correlated fermions in optical lattices. Following an early application of CNNs to help decide which one of two theories better describes patterns in snapshots taken in the pseudogap regime of the Hubbard model using quantum gas microscopy~\cite{a_bohrdt_18}, Khatami {\it et al.}~\cite{e_khatami_20} designed a simple CNN architecture and showed that patterns formed in filters of a CNN, trained to distinguish snapshots taken at low temperatures from those taken at high temperatures, can reflect the correlations favored by the systems as the temperature is lowered. Fig.~\ref{fig:CNN1} shows an example of the CNN used in their study. Having one/few filter(s) in the usually only convolutional layer directly connected to the input (physical snapshots of fermions) allows the scheme to work. The idea is that since there are no correlations at high temperatures when the system is entirely unordered, the filter will likely pick up patterns formed in the snapshots of the system at low temperatures to carry out the categorization accurately. By studying the trained filters, one can then infer relevant electronic correlations. 

\begin{figure}[t]
\centerline {\includegraphics*[width=3.3in]{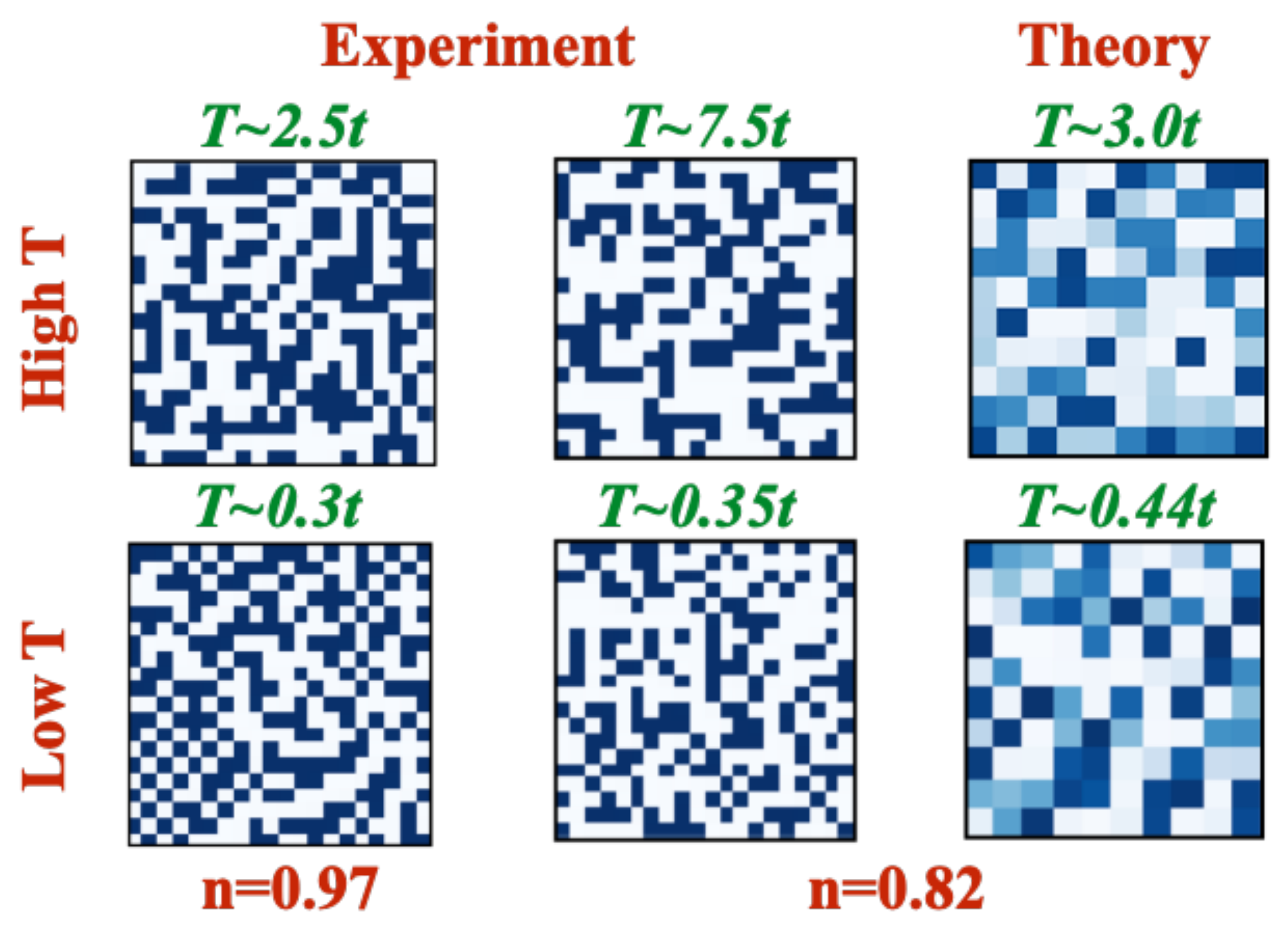}}
\caption{Sample snapshots used in the training of the CNN in Fig.~\ref{fig:CNN1}. Shown are four from the experiments for densities $n=0.97$ and $n=0.82$ at the two extreme temperatures available, and two from DQMC for $n=0.82$. Image partly reproduced from Ref.~\cite{e_khatami_20}. 
\label{fig:Snapshots}}
\end{figure}

Two types of experimental snapshots were available at the time, mainly around the strange metal phase of the Hubbard model, and were used in this study: (1) Those of a single species of fermions and (2) those of the two species together, minus the double occupations, which would show up as empty sites for technical reasons. (2) are often called ``singles'' snapshots and can also be thought of as snapshots of local moments. In addition, theory snapshots were generated by periodically pausing the DQMC simulations of the 2D Hubbard model and calculating average site occupations. The latter leads to non-binary snapshots (non-projective measurements), which nevertheless proved useful for theory comparisons. Fig.~\ref{fig:Snapshots} shows a sample of these snapshots at different temperatures. 

\begin{figure}[b]
\centerline {\includegraphics*[width=\columnwidth]{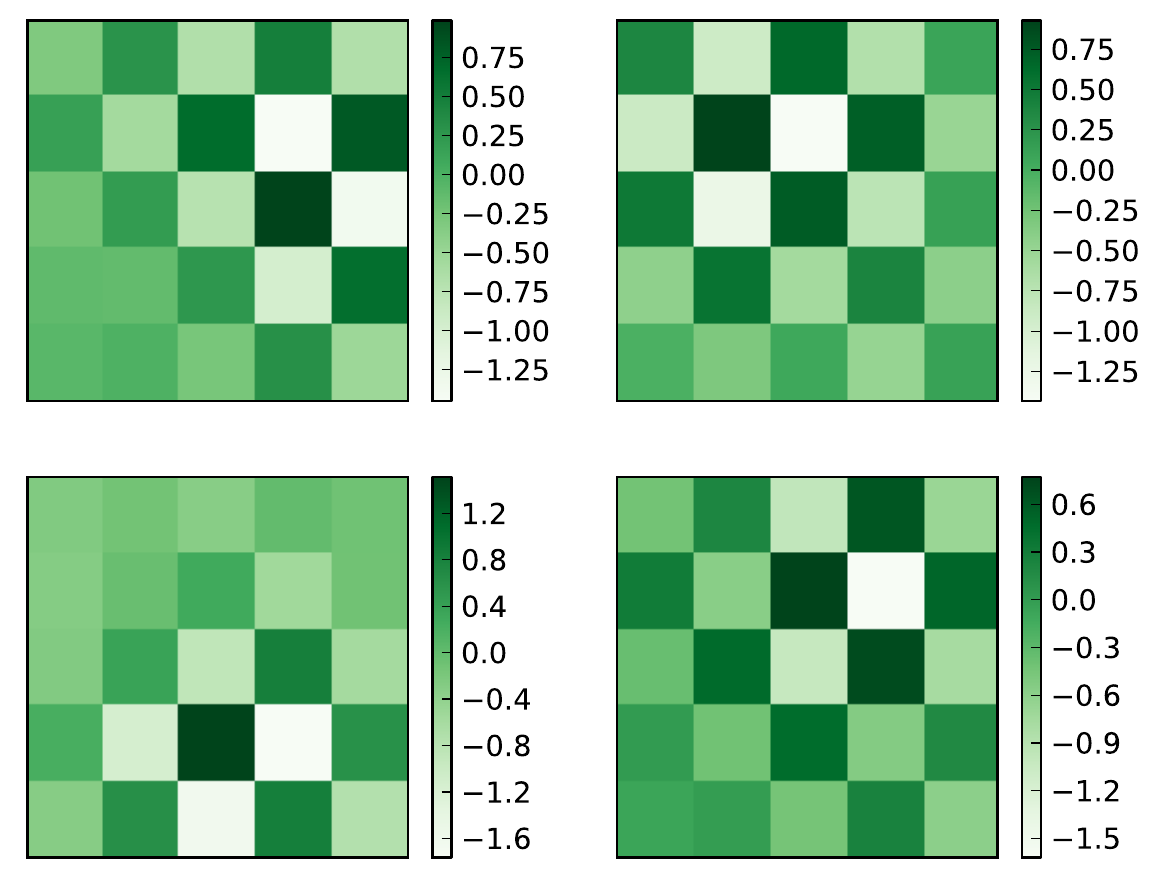}}
\caption{Sample filters after separate training using experimental single-species snapshots near half filling. The observed checkerboard patterns point to long-range antiferromagnetic correlations in the system at low temperatures.
\label{fig:AFM}}
\end{figure}

\begin{figure}[t]
\centerline {\includegraphics[width=\columnwidth]{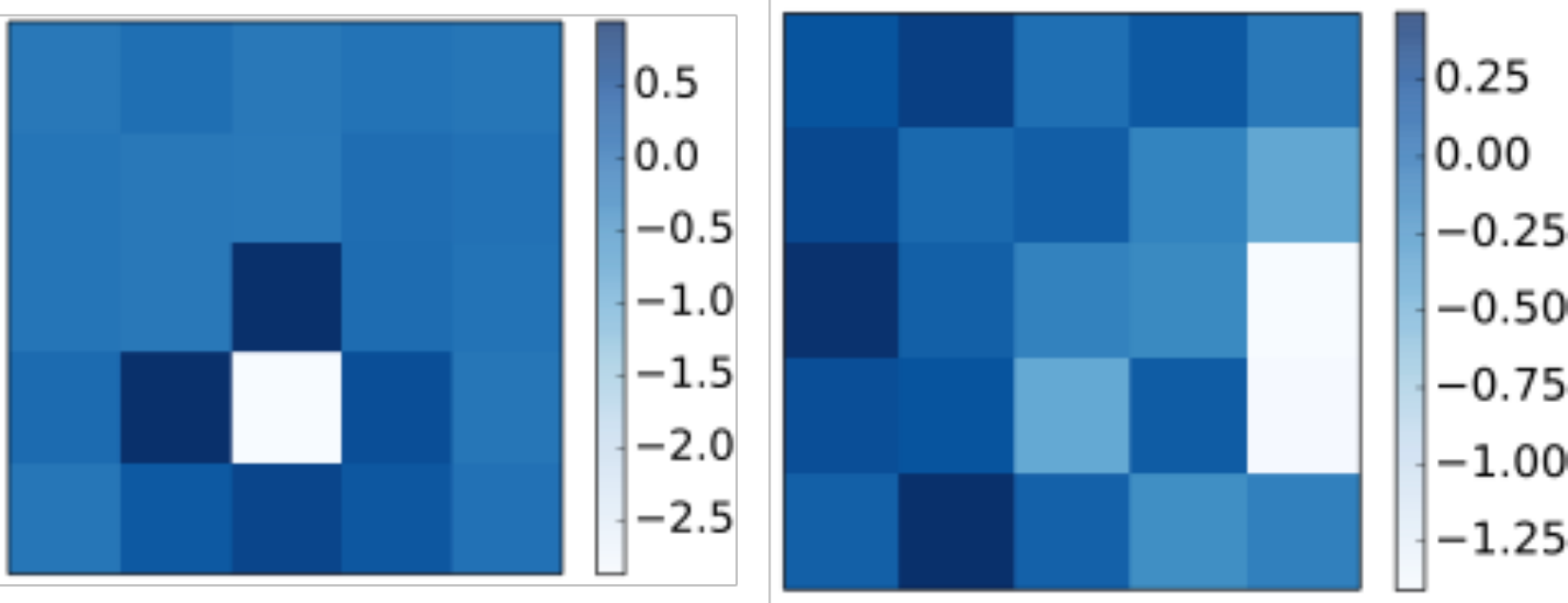}}
\caption{Example of two filters after training using experimental local moment snapshots at densities $n=0.58$ (left) and $0.97$ (right). Signatures of nearest-neighbor anti-correlation and doublon-hole fluctuations are seen in the corresponding filters, respectively. Image reproduced from Ref.~\cite{e_khatami_20}. 
\label{fig:LocMoment}}
\end{figure}

Figure~\ref{fig:AFM} shows a sample of $5$-pixel by $5$-pixel filters from four different CNNs that have been separately trained to distinguish single-species snapshots at half-filling at two different temperatures of $T = 0.35t$ and $2.5t$. The long-range checkerboard patterns in these filters hint at developing antiferromagnetic correlations at temperatures below $T = t$. Typical patterns drawn from other training sets using snapshots of local moments are shown in Fig.~\ref{fig:LocMoment} for densities $\langle n \rangle=0.58$ and $0.97$. They show the anti-correlation of nearest-neighbor fermions for a density close to quarter-filling as expected and a pattern that can be interpreted as significant nearest-neighbor doublon-hole fluctuations, which are known to be large near half-filling~\cite{l_cheuk_16}. In local moment snapshots, empty sites could represent holes or double occupancies. 

Similar results obtained at a density of $0.82$, where a strange metallic behavior was directly observed through the dynamical properties of the same system~\cite{p_brown_18}, point to short-range antiferromagnetic correlations when training with single-species snapshots, and more or less random patterns when training with snapshots of local moments. Training with theory snapshots leads to similar results. To shed some light on density correlations (or more accurately, local moment correlations) that might be specific to the strange metal phase, the authors designed a CNN with six larger $7$-pixel by $7$-pixel filters and were able to show that the CNN trained at the density of $0.82$ can as a whole act as an order parameter for the strange metal phase with a signal that decreases as one moves away from this density. They eliminated the density of particles as an obvious indicator by subtracting the network signal for the real snapshots of local moments at the other densities from that obtained for the same snapshots but with randomized pixels (fake snapshots). This work demonstrated that by studying the inner workings of neural networks, one could gain unbiased insight into the physics of the problem.

Until now, this perspective has focused squarely on solving low-energy effective Hamiltonians for strongly correlated materials. However, it is important to keep in mind that the right effective models are not known or are still being debated for many classes of models. In this context, an important task is to extract effective models from inelastic scattering data by solving the \emph{inverse} scattering problem, where the high-dimensional inelastic neutron or x-ray scattering data for a dynamical structure factor $S({\bf q},\omega)$ is used to determine the parameters of an effective model. \gls*{ML} methods can help with this task by formulating it as a supervised learning problem \cite{SamarakoonJPCM2021, ChenReview2021}. Essentially, one formulates a model with parameters $\boldsymbol{\theta} = [\theta_1,\theta_2,\dots,\theta_n]$ that can be used to generate a model spectra $S_\mathrm{model}({\bf q},\omega,\boldsymbol{\theta})$. One then adjusts the parameters to minimize 
\begin{equation}\label{eq:L2}
    \chi^2 = || S_\mathrm{model}({\bf q},\omega,\boldsymbol{\theta}) - S({\bf q},\omega)||^2,
\end{equation}
where $||\cdot||$ defines a norm~\cite{ChenReview2021}. Eq.~\eqref{eq:L2} is minimized using AI and ML-based optimization methods. In this context, autoencoders have been particularly useful for reducing the complexity of the optimization problem~\cite{SamarakoonJPCM2021, Samarakoon2020, ChenReview2021}. 
For this approach to be successful, however, one must rapidly generate dynamic structure factors for a given candidate model. Therefore, a crucial component here is the existence of fast solvers for the \emph{direct} scattering problem. In many cases, this component is the bottleneck; however, methods for treating a large class of quantum magnets have recently been developed that  generalize semiclassical Landau-Lifshitz dynamics to SU($N$) ~\cite{ZhangPRB2021, DahlbomPreprint, DahlbomPreprint2}. These kinds of solvers have enabled the successful extraction of model parameters for Dy$_2$Ti$_2$O$_7$ 
\cite{Samarakoon2020} and $\alpha$-RuCl$_3$~\cite{SamarakoonPRR2022} from inelastic neutron scattering data. Achieving a similar level of success for resonant inelastic x-ray scattering experiments, with its more complex and challenging cross-section, will require additional work. 

Another popular approach for obtaining model parameters relies on down-folding {\it ab initio} electronic structure calculations onto low-energy target subspaces. These calculations typically project the bands near the Fermi level onto a minimal set of maximally localized Wannier functions to obtain tight-binding parameters~\cite{MarzariRMP2012}. A set of constrained random phase approximation calculations can then be conducted to compute the corresponding interaction parameters~\cite{Aryasetiawan, EichstaedtPRB2019}. 
ML and AI methods can also play a role here~\cite{Kulik_2022}. For example, techniques from these fields have been used to construct functionals~\cite{Brockherde2017} and accelerate density functional theory simulations~\cite{Chandrasekaran2018}. They have also been applied towards extracting tight-binding~\cite{Wang2021} and interaction~\cite{Yu2020} parameters, as well as impurity and defect parameters \cite{Schattauer2022}. These methods thus provide an alternative route towards deriving the appropriate effective models for many-body simulations.

\section{Accelerated Monte Carlo simulations}\label{sec:SLMC}
The term ``self-learning'' Monte Carlo (SLMC) algorithms refers to a powerful class of \gls*{ML}--accelerated MCMC methods that have been developed and refined in recent years. These methods were first introduced by Liu \emph{et al.}~\cite{SLMC, LiuPRB2017} in the context of classical MC simulations of the Ising model. Since then, the method has been expanded to a much broader class of correlated electron models, different flavors of classical and quantum MC algorithms, and \gls*{ML} frameworks \cite{HuangPRE2017, NagaiPRB2017, XuPRB2017, ShenPRB2018, LiuPRB2018, ChenPRB2018, LiPRB2019, TanakaPreprint, a_bohrdt_18, Kohshiro2021, Monroe2022}. 

\subsection{Overview of self-learning Monte Carlo method}    
The objective of an SLMC algorithm is to learn an accurate proxy function for the transition probabilities between different MC configurations. In other words, the algorithm learns an effective energy $E_\mathrm{eff}$ such that  
\begin{equation*}
e^{-\beta \Delta E_\mathrm{eff}(\{X\},\{X^\prime\})} \approx  P(\{X\}\rightarrow \{X^\prime\}) = \frac{W(\{X^\prime\})}{W(\{X\})}, 
\end{equation*}
where $W(\{X\})$ is the true MC weight of the system for a given configuration $\{X\}$ (see Sec.~\ref{sec:MCMC}). 

If the proxy energy $E_\mathrm{eff}(\{X\})$ can be evaluated more efficiently than $W(\{X\})$, then it can be used to quickly evolve the Markov chain between largely \emph{uncorrelated} configurations. Most implementations to date perform this task through a series of local updates in the configuration space, following the standard Metropolis-Hastings scheme. Specifically, one sweeps through all sites in configuration space proposing local updates to the MC configuration that are accepted or rejected with a probability estimated by the learned effective model
\begin{equation*}
  p_l = \mathrm{min}\left[1, e^{-\beta \Delta E_\mathrm{eff}(\{X\},\{X^\prime\})} \right].
\end{equation*}
After many sweeps through configuration space, this procedure should produce a new MC configuration $\{X^\prime\}$ that is very far removed from the starting one. This procedure can thus be viewed as a means for constructing non-trivial \emph{global} updates of the MC configurations. However, to maintain detailed balance~\cite{SLMC}, a final (cumulative) acceptance step is required where the entire move $\{X\}\rightarrow \{X^\prime\}$ is accepted with probability
\begin{equation}\label{eq:cumulative}
    p_c = \mathrm{min}\left[1,\frac{W(\{X\})}{W(\{X^\prime\})}
   e^{-\beta \Delta E_\mathrm{eff}(\{X\},\{X^\prime\})} 
    \right].
\end{equation}
While this final step requires evaluating the proper MC weights, it occurs infrequently enough that a considerable algorithm speedup can often be obtained, assuming $p_c$ is not too small. A large $p_c$ will be achieved when 
$ e^{-\beta \Delta E_\mathrm{eff}(\{X\},\{X^\prime\})} $
is close to $W(\{X^\prime\})/W(\{X\})$.

To illustrate this procedure, consider an MC simulation of a spin-Fermion model, where itinerant fermions are coupled to a classical spin ${\bf S}_i$ \cite{LiuPRB2017}
\begin{equation}\label{eq:spinfermion}
    H = -\sum_{i,j,\sigma} t_{i,j}c^\dagger_{i,\sigma}c^{\phantom\dagger}_{j,\sigma} - J \sum_{i,\sigma,\sigma^\prime} {\bf S}_i \cdot c^\dagger_{i,\sigma} 
    \boldsymbol{\tau}_{\sigma,\sigma^\prime}^{\phantom{\dagger}}c^{\phantom\dagger}_{i,\sigma^\prime}, 
\end{equation}
where $\boldsymbol{\tau}$ is a vector of Pauli matrices. For a given configuration of the classical spins $\{{\bf S}_i\}$, Eq.~\eqref{eq:spinfermion} can be diagonalized exactly to obtain the energy $E(\{{\bf S}_i\})$, which is an order $O(N^3)$ operation.  One then samples the classical spin configurations by proposing updates of ${\bf S}_i \rightarrow {\bf S}_i^\prime$ at each site such that the total computational cost of a full Monte Carlo sweep is $O(N^4)$. 

In an SLMC implementation, one assumes that the itinerant electrons mediate an effective RKKY-like spin-spin interaction of the form 
\begin{equation}\label{eq:JHeff}
    H_\mathrm{eff} = J_0 + \sum_{\langle i,j\rangle_n} J_n{\bf S}_i\cdot{\bf S}_j + \dots,
\end{equation}
where $J_n$ is the effective coupling between pairs of $n$\textsuperscript{th}-neighbor spins. To determine the coefficients, one fits Eq.~\eqref{eq:JHeff} to a large set of training data obtained using standard MCMC simulations of the original Hamiltonian. Once trained, the proxy energy provided by Eq.~\eqref{eq:JHeff} can be evaluated in $O(N)$ operations as opposed to the original $O(N^3)$ operations that are necessary to evaluate 
the DQMC determinant ratios in ${W(\{X^\prime\})}/{W(\{X\})}$.
The effective Hamiltonian can also be easily extended to include more complicated three- and four-body interactions \cite{SLMC}. One can also enforce specific symmetries into the form of $H_\mathrm{eff}$ if these are known \cite{ChenPRB2018}.

The efficiency of the SLMC approach depends crucially on the accuracy of the underlying effective model. For example, if the effective model is not sufficiently flexible to capture the training data patterns, it can be challenging to train an accurate surrogate $H_\mathrm{eff}$. Moreover, a specified effective model's ability to accurately capture the MC weights can depend very strongly on the parameters of the full model or the simulation temperature. One might expect this limitation to some extent, as different low-energy effective models describe different ordered phases. When the effective model cannot describe the true model's physics, the SLMC algorithm will attempt to construct updates that the original MCMC algorithm would typically reject. Because of this, the final cumulative update given Eq.~\eqref{eq:cumulative} will begin to have a high rejection rate. 

In some cases, one can improve the quality of the surrogate model by including longer-range interactions, many-body interactions, or additional symmetries in the underlying model. However, no well-defined procedure exists for systematically deriving the correct effective models. To overcome this shortcoming, several research groups have attempted to implement deep learning frameworks where an artificial neural network tries to learn the form of the effective model \cite{ShenPRB2018, LiPRB2019}. 

\subsection{Self-learning Monte Carlo for the Holstein model}
To overcome the need to have a priori information about the underlying effective model in SLMC applications, some of the current authors implemented a set of artificial neural networks for performing SLMC simulations of lattice QMC simulations \cite{LiPRB2019}. As a proof-of-principle, we applied this approach to simulations of the \gls*{CDW} transition in the two-dimensional half-filled Holstein model. We will, therefore, first give a brief overview of DQMC simulations for this challenging model.

The integral over fields in Eq.~\eqref{Eq:ZHolstein} is accomplished using MCMC sampling; however, two different types of update procedures are required. The first is the standard \emph{local} updates, where the displacements are changed individually at each $(i,l)$ site. This update can be performed on a single location with a computational cost that scales like $O(N^2)$ using an efficient Sherman–Morrison updating scheme \cite{WhitePRB1989}. An entire sweep of single-site updates costs $O(N^3L)$ operations to perform. 
The second class of updates is so-called \emph{block} or \emph{global} updates, where the displacements at all time slices are shifted by the same amount in a single update \cite{JohnstonPRB2013}. These updates are nonlocal in imaginary time and require one to explicitly compute the weights in Eq.~\eqref{Eq:ZHolstein}. Since this task required evaluating a matrix determinant, the nominal cost of performing a block update at a single site is $O(N^3)$ while an entire sweep costs $O(N^4)$. Despite their higher computational costs, block updates are needed to achieve reasonably short autocorrelation times in simulations of the Holstein model. (The same is true for simulations of the Hubbard model whenever $U\beta$ is large~\cite{ScalettarPRB1991}). 

\begin{figure}[t]
    \centering
    \includegraphics[width=0.8\columnwidth]{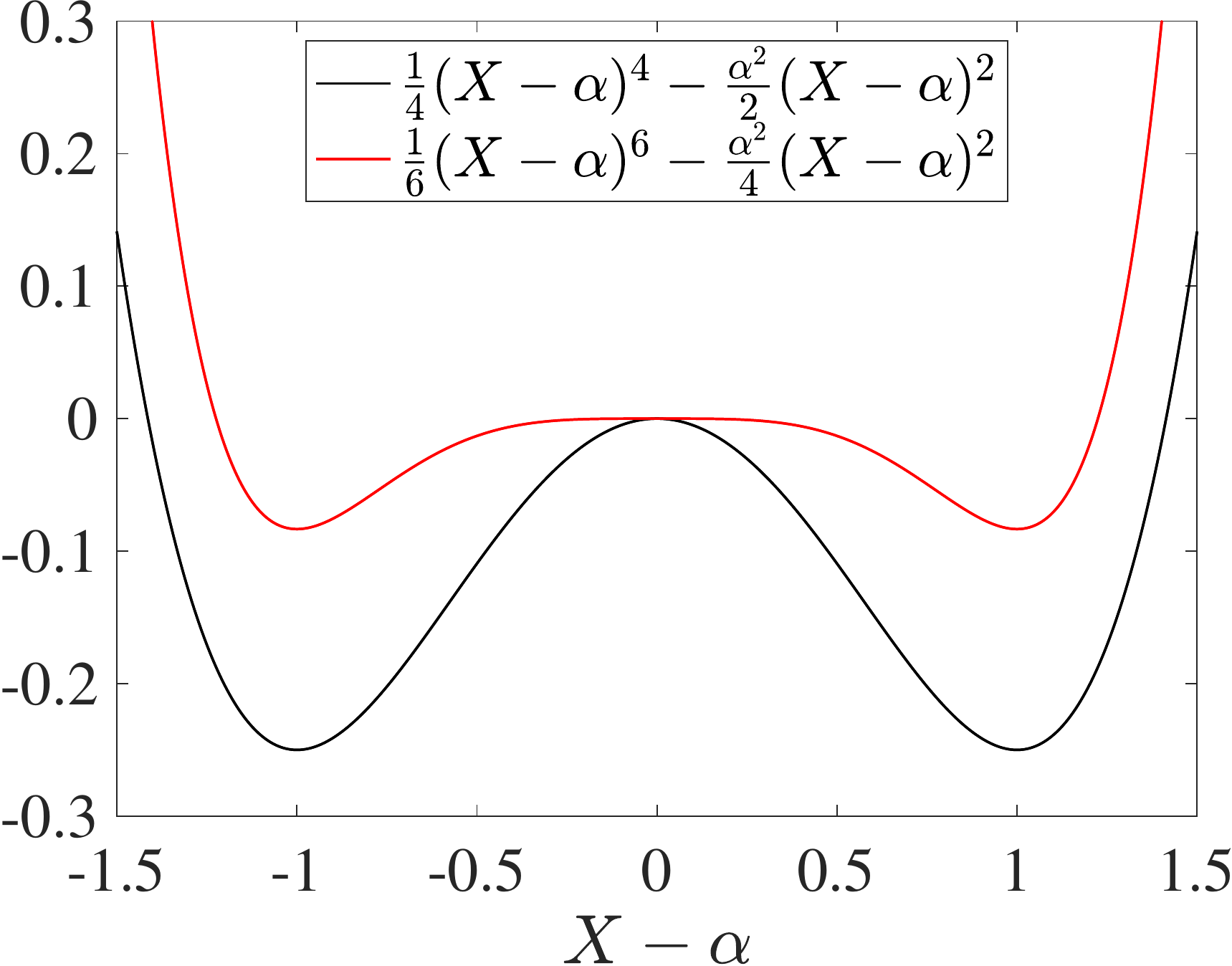}
    \caption{The symmetric functions used to enforce the double-well potential 
    present in the Holstein model near half-filling for the case $\alpha = -g/\omega_0^2 = 1$. Adapted from Ref.~\cite{ChenPRB2018}.}
    \label{fig:TwoWell}
\end{figure}

Reference~\cite{ChenPRB2018} was the first to attempt to conduct SLMC simulations of the Holstein model. In this approach, the authors defined an effective model in the spirit of Eq.~\eqref{eq:JHeff} but with terms designed to reflect the known $\mathbb{Z}_2$ ($X \rightarrow -X$) symmetry of the model. Specifically, they adopted
\begin{eqnarray}\label{eq:HolstEff}
    -\beta H_\mathrm{eff}&=&J_k\sum_{i,l} (X_{i,l+1}-X_{i,l})^2 \\\nonumber
    &+&J_p \sum_{i,l}\left(\frac{1}{4}(X-\alpha)^4-\frac{\alpha^2}{2}(X-\alpha)^2\right)\\\nonumber
    &+&J^\prime_p \sum_{i,l}\left(\frac{1}{6}(X-\alpha)^6-\frac{\alpha^2}{4}(X-\alpha)^4\right) \\\nonumber
    &+&J_{nn} \sum_{\langle i,j\rangle,l}\left(X_{i,l}-\alpha\right)\left(X_{j,l}-\alpha\right)\\\nonumber&+&J_{nn} \sum_{i,\langle l,l^\prime \rangle}\left(X_{i,l}-\alpha\right)\left(X_{i,l^\prime}-\alpha\right),
\end{eqnarray}
where $\alpha = -\frac{g}{\omega_0^2}$. The first term reflects the usual kinetic energy of the phonon fields, while the fourth and fifth terms reflect nearest-neighbor interactions between the lattice fields in both space and imaginary time. The second and third terms, plotted in Fig.~\ref{fig:TwoWell}, allow for the presence of a double-well potential in the Holstein model, which can become deep in the strong coupling limit. 

Chen \emph{et al}.~\cite{ChenPRB2018} used this effective model to study the CDW transition in the half-filled Holstein model for $\omega_0 = t/2$ on large $16\times 16$ lattices, large enough to perform reliable scaling analysis to estimate $T_\mathrm{CDW}$. The authors also reported a significant reduction in the autocorrelation time of the CDW structure factor using the SLMC method compared to the conventional DQMC algorithm. We note, however, that the authors defined a full MC sweep in the DQMC case as proposing single-site updates at all spacetime points together with four block updates at (presumably) randomly chosen sites. In the SLMC case, however, they defined an MC sweep as a sweep of local updates followed by a Wolff-cluster update. The sampling procedures differ, so it is unclear what the ultimate mechanisms of the autocorrelation reduction were.  

Li \emph{et al}. \cite{LiPRB2019} later employed artificial neural networks to perform SLMC simulations of the Holstein model, where the form of the effective Hamiltonian was learned directly from the training data. In this case, the authors trained two neural networks: a fully connected network to perform local single-site updates and a convolutional neural network to act as a proxy for global moves. Like the effective model of Chen \emph{et al}.~\cite{ChenPRB2018}, both networks used the proposed change in displacement $\Delta X_{i,l}$ (or $\Delta X$ for global updates) together with \emph{local} nearest neighbor information about the surrounding fields as input features. The networks could thus be trained using data generated using cheaper small clusters before being deployed on larger systems. For example, Li \emph{et al}. trained their networks on $6\times 6$ clusters before deploying them on system sizes up to $16\times 16$. 

\begin{figure}
    \centering
    \includegraphics[width=\columnwidth]{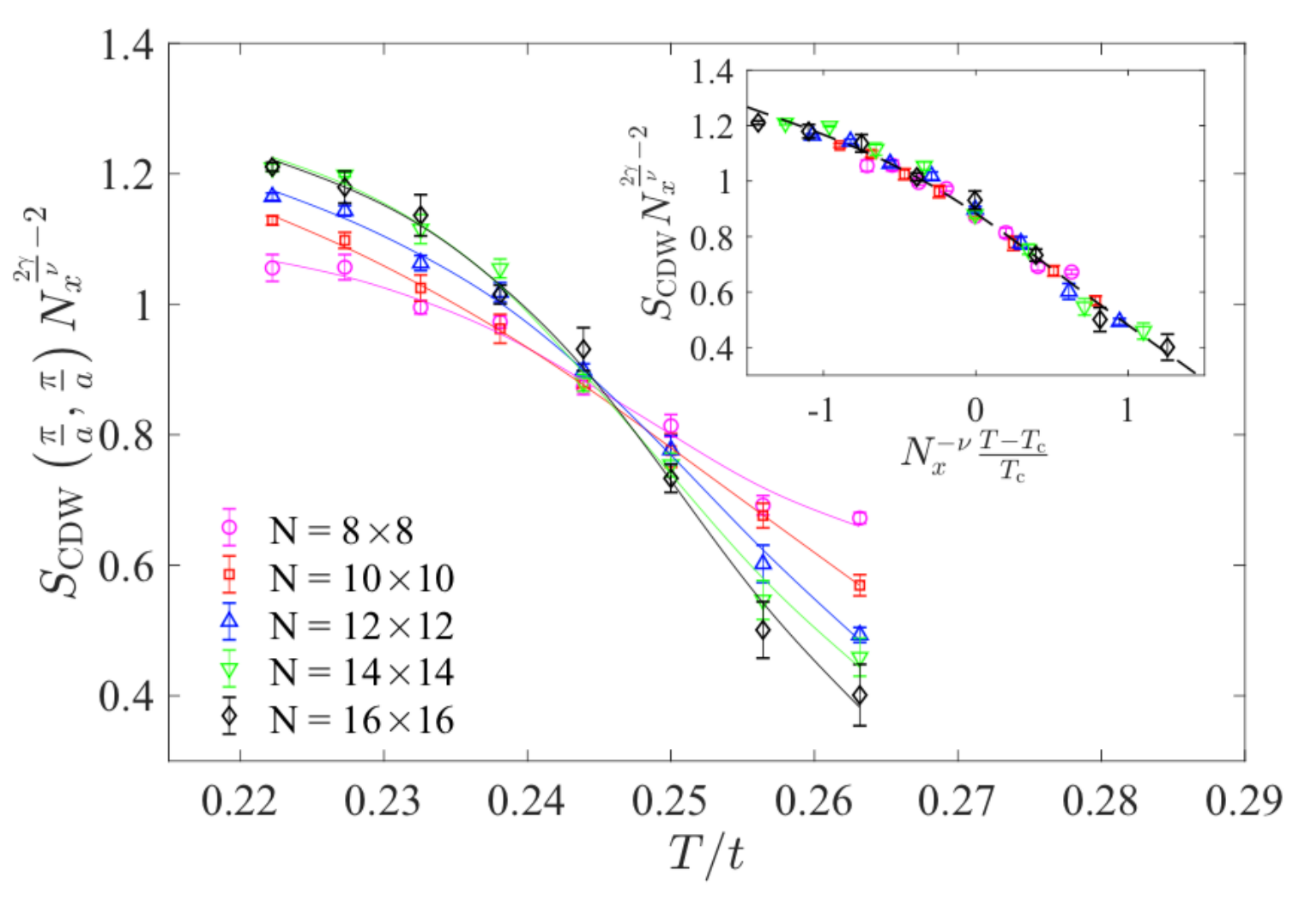}
    \caption{A finite-size scaling analysis of the ${\bf q}=(\pi,\pi)$ CDW 
    structure factor $S({\bf q})N_x^{2\gamma/\nu - 2}$ vs $T/t$. 
    The CDW transition is in the 2D Ising universality class with critical exponents $\gamma =\frac{1}{8}$ and $\nu = 1$. 
    The inset shows the collapse of the data for a critical temperature 
    $T_c = 0.244$. Reproduced from Ref.~\cite{LiPRB2019}}
    \label{fig:NNMC}
\end{figure}

Li \emph{et al}. \cite{LiPRB2019}, benchmarked their method for the challenging half-filled model with $\omega_0 = t/2$. They found that the neural networks were able to reproduce the results of the conventional DQMC algorithm accurately and obtained a thermodynamic value of $T_\mathrm{CDW}$ similar to Chen \emph{et al}.~\cite{ChenPRB2018} (see Fig.~\ref{fig:NNMC}). Li \emph{et al}. also examined the autocorrelation time for their approach. Because they were able to perform both global and local updates within the SLMC framework, the authors could make more meaningful comparisons with DQMC by defining an MC sweep in the same way for each case (a full sweep of single-site updates, followed by a block update performed at every site). In doing so, they found that the SLMC approach produced \emph{identical} autocorrelation times compared to the conventional algorithm. This result makes intuitive sense. Suppose the SLMC algorithm is accurately reproducing the accept/reject steps of the DQMC algorithm. In that case, it should produce the same autocorrelations in the Monte Carlo configurations, provided the SLMC and DQMC algorithms perform the same updates. 
This notion was reinforced by the observed behavior in the cumulative acceptance rates. For the neural networks, the cumulative acceptance rate remained near 100\%, independent of the temperature of the simulation. On the other hand, Chen \emph{et al}.~\cite{ChenPRB2018} obtained cumulative acceptance rates that fell off rapidly as the temperature was lowered. 

Despite these successes near half-filling, it is essential to note that the neural-network-based method has not yet been generalized to the metallic state that appears with doping. This failure occurs even though the networks often accurately reproduce the validation data sets. We speculate that this issue might be related to longer-range effective interactions in space or imaginary time that are not captured by the local input features used in the neural networks. For example, it is possible that no local models like 
Eq.~\eqref{eq:HolstEff} or Eq.~\eqref{eq:JHeff} exists for this problem. One possible solution would be to develop neural networks that use the entire set of auxiliary fields as input features. While costly, modern GPU resources would make this feasible. Additional research is needed to push this technology further.

\section{Outlook and conclusions}\label{sec:outlook}
From this short perspective, we hope it is clear that over the last several years, methods in artificial intelligence, machine learning, and data science have leaped from novel tools for studying {\it classical} phase transitions to providing powerful new means to solve {\it quantum} many-body problems. Moreover, these methods can provide a pathway forward to solving new and challenging forefront questions in strongly correlated materials. However, we also believe it is important to take a step back and remind ourselves that there remains work to be done in this area. 

The first issue concerns the `value added'. For example, it is clear that PCA can capture phase transitions of many interesting quantum models of interacting fermions. However, it is important to consider whether it (or other \gls*{ML} and \gls*{AI} methods) can do so {\it more effectively} than data analysis tools that have been refined over the last several decades centered on sophisticated finite-size scaling (FSS) of appropriate correlation functions. That is, in what ways do \gls*{ML} and \gls*{AI} `revolutionize' the field of computational quantum many-body physics? In the examples noted above, the characterizations of the critical points with \gls*{ML} and \gls*{AI} are {\it less precise}. Part of this is, of course, due to the time frames of the approaches; \gls*{ML} and \gls*{AI} have not yet matured in these investigations. There is roughly an order of magnitude difference in the time frames over which ``traditional'' and \gls*{ML} approaches have been under development. It seems likely that, given the opportunity, \gls*{ML} will approach the accuracy of FSS-based methods, but it is too soon to know if they will surpass them.

More promising, \gls*{ML} and \gls*{AI}-based methods can provide additional information, such as the ordering patterns in experimental images of fermions or identifying the presence of broken symmetries in MCMC simulations without having to know which observable to measure in advance. This problem is one area where these methods can provide new insights; similarly, one can envision hybrid approaches where a \gls*{ML} method quick surveys the model phase space to identify promising areas that are then expanded on using conventional methods. 

We believe that a particularly important set of opportunities involves deploying \gls*{ML} and \gls*{AI} to attack the bottlenecks of quantum simulations. These include 1) speeding up the generation of independent samples in regimes of long autocorrelation times, 2) identifying complex intertwined orders, and 3) (possibly) helping to mitigate the sign problem. \gls*{ML} applications to address the first are perhaps the most mature with the rapid development of the SLMC method. Nevertheless, SLMC methods have yet to be widely adopted by QMC practitioners. This is likely because the approach has only been validated on a small subset of problems and model parameters and often fails to generalize to other regimes. Additional research will be needed to develop robust design and deployment methods. For example, determining ways to quantify the level of trust in the model proxy's ability to sample the entire configuration space would be crucial. For example, it is entirely possible that the effective model used in an SLMC simulation can introduce more local minima in the MCMC configuration space that could introduce ergodicity issues. Another promising line of research involves the extension of matrix product states and tensor networks using CNNs or restricted Boltzmann machines~\cite{y_levine_19,o_sharir_22,PhysRevB.97.085104, PhysRevB.99.155129}.

Finally, using \gls*{ML} and \gls*{AI}-methods to solve the inverse scattering problem, where model parameters are extracted from inelastic neutron and x-ray scattering data, is a growing and promising avenue of research. This application, however, urgently needs fast methods for solving the direct scattering problem. Of course, such solvers can be \gls*{AI}-accelerated, but other avenues should also be explored ~\cite{DahlbomPreprint, DahlbomPreprint2, CohenSteadPRE2022}.  \\

\noindent{\bf Acknowledgements} --- We thank D. Agrawal, W. Bakr, K. Barros, K. Ch'ng, N. Costa, P. M. Dee, A. Del Maestro, Wenjian Hu, E. Kim, S. Li, Y. W. Li, J. Ostrowski, R. Singh, N. Vazquez, Bo Xiao, and Y. Zhang for their collaborations and discussions in this area over the years. This work was supported by the U.S. Department of Energy, Office of Science, Office of Basic Energy Sciences, under Award Number DE-SC0022311. 

\bibliography{references}

\begin{thebibliography}{218}%
\makeatletter
\providecommand \@ifxundefined [1]{%
 \@ifx{#1\undefined}
}%
\providecommand \@ifnum [1]{%
 \ifnum #1\expandafter \@firstoftwo
 \else \expandafter \@secondoftwo
 \fi
}%
\providecommand \@ifx [1]{%
 \ifx #1\expandafter \@firstoftwo
 \else \expandafter \@secondoftwo
 \fi
}%
\providecommand \natexlab [1]{#1}%
\providecommand \enquote  [1]{``#1''}%
\providecommand \bibnamefont  [1]{#1}%
\providecommand \bibfnamefont [1]{#1}%
\providecommand \citenamefont [1]{#1}%
\providecommand \href@noop [0]{\@secondoftwo}%
\providecommand \href [0]{\begingroup \@sanitize@url \@href}%
\providecommand \@href[1]{\@@startlink{#1}\@@href}%
\providecommand \@@href[1]{\endgroup#1\@@endlink}%
\providecommand \@sanitize@url [0]{\catcode `\\12\catcode `\$12\catcode
  `\&12\catcode `\#12\catcode `\^12\catcode `\_12\catcode `\%12\relax}%
\providecommand \@@startlink[1]{}%
\providecommand \@@endlink[0]{}%
\providecommand \url  [0]{\begingroup\@sanitize@url \@url }%
\providecommand \@url [1]{\endgroup\@href {#1}{\urlprefix }}%
\providecommand \urlprefix  [0]{URL }%
\providecommand \Eprint [0]{\href }%
\providecommand \doibase [0]{https://doi.org/}%
\providecommand \selectlanguage [0]{\@gobble}%
\providecommand \bibinfo  [0]{\@secondoftwo}%
\providecommand \bibfield  [0]{\@secondoftwo}%
\providecommand \translation [1]{[#1]}%
\providecommand \BibitemOpen [0]{}%
\providecommand \bibitemStop [0]{}%
\providecommand \bibitemNoStop [0]{.\EOS\space}%
\providecommand \EOS [0]{\spacefactor3000\relax}%
\providecommand \BibitemShut  [1]{\csname bibitem#1\endcsname}%
\let\auto@bib@innerbib\@empty
%</preamble>
\bibitem [{\citenamefont {White}\ \emph {et~al.}(1989)\citenamefont {White},
  \citenamefont {Scalapino}, \citenamefont {Sugar}, \citenamefont {Loh},
  \citenamefont {Gubernatis},\ and\ \citenamefont {Scalettar}}]{WhitePRB1989}%
  \BibitemOpen
  \bibfield  {author} {\bibinfo {author} {\bibfnamefont {S.~R.}\ \bibnamefont
  {White}}, \bibinfo {author} {\bibfnamefont {D.~J.}\ \bibnamefont
  {Scalapino}}, \bibinfo {author} {\bibfnamefont {R.~L.}\ \bibnamefont
  {Sugar}}, \bibinfo {author} {\bibfnamefont {E.~Y.}\ \bibnamefont {Loh}},
  \bibinfo {author} {\bibfnamefont {J.~E.}\ \bibnamefont {Gubernatis}},\ and\
  \bibinfo {author} {\bibfnamefont {R.~T.}\ \bibnamefont {Scalettar}},\
  }\bibfield  {title} {\bibinfo {title} {Numerical study of the two-dimensional
  {Hubbard} model},\ }\href {https://doi.org/10.1103/PhysRevB.40.506}
  {\bibfield  {journal} {\bibinfo  {journal} {Phys. Rev. B}\ }\textbf {\bibinfo
  {volume} {40}},\ \bibinfo {pages} {506} (\bibinfo {year} {1989})}\BibitemShut
  {NoStop}%
\bibitem [{\citenamefont {Park}\ \emph {et~al.}(2008)\citenamefont {Park},
  \citenamefont {Haule},\ and\ \citenamefont
  {Kotliar}}]{PhysRevLett.101.186403}%
  \BibitemOpen
  \bibfield  {author} {\bibinfo {author} {\bibfnamefont {H.}~\bibnamefont
  {Park}}, \bibinfo {author} {\bibfnamefont {K.}~\bibnamefont {Haule}},\ and\
  \bibinfo {author} {\bibfnamefont {G.}~\bibnamefont {Kotliar}},\ }\bibfield
  {title} {\bibinfo {title} {Cluster dynamical mean field theory of the {Mott}
  transition},\ }\href {https://doi.org/10.1103/PhysRevLett.101.186403}
  {\bibfield  {journal} {\bibinfo  {journal} {Phys. Rev. Lett.}\ }\textbf
  {\bibinfo {volume} {101}},\ \bibinfo {pages} {186403} (\bibinfo {year}
  {2008})}\BibitemShut {NoStop}%
\bibitem [{\citenamefont {Jarrell}\ \emph {et~al.}(2001)\citenamefont
  {Jarrell}, \citenamefont {Maier}, \citenamefont {Hettler},\ and\
  \citenamefont {Tahvildarzadeh}}]{Jarrell_2001}%
  \BibitemOpen
  \bibfield  {author} {\bibinfo {author} {\bibfnamefont {M.}~\bibnamefont
  {Jarrell}}, \bibinfo {author} {\bibfnamefont {T.}~\bibnamefont {Maier}},
  \bibinfo {author} {\bibfnamefont {M.~H.}\ \bibnamefont {Hettler}},\ and\
  \bibinfo {author} {\bibfnamefont {A.~N.}\ \bibnamefont {Tahvildarzadeh}},\
  }\bibfield  {title} {\bibinfo {title} {Phase diagram of the {Hubbard} model:
  Beyond the dynamical mean field},\ }\href
  {https://doi.org/10.1209/epl/i2001-00557-x} {\bibfield  {journal} {\bibinfo
  {journal} {Europhysics Letters ({EPL})}\ }\textbf {\bibinfo {volume} {56}},\
  \bibinfo {pages} {563} (\bibinfo {year} {2001})}\BibitemShut {NoStop}%
\bibitem [{\citenamefont {Ido}\ \emph {et~al.}(2018)\citenamefont {Ido},
  \citenamefont {Ohgoe},\ and\ \citenamefont {Imada}}]{IdoPRB2018}%
  \BibitemOpen
  \bibfield  {author} {\bibinfo {author} {\bibfnamefont {K.}~\bibnamefont
  {Ido}}, \bibinfo {author} {\bibfnamefont {T.}~\bibnamefont {Ohgoe}},\ and\
  \bibinfo {author} {\bibfnamefont {M.}~\bibnamefont {Imada}},\ }\bibfield
  {title} {\bibinfo {title} {Competition among various charge-inhomogeneous
  states and $d$-wave superconducting state in {Hubbard} models on square
  lattices},\ }\href {https://doi.org/10.1103/PhysRevB.97.045138} {\bibfield
  {journal} {\bibinfo  {journal} {Phys. Rev. B}\ }\textbf {\bibinfo {volume}
  {97}},\ \bibinfo {pages} {045138} (\bibinfo {year} {2018})}\BibitemShut
  {NoStop}%
\bibitem [{\citenamefont {White}\ and\ \citenamefont
  {Scalapino}(1998{\natexlab{a}})}]{WhitePRL1989a}%
  \BibitemOpen
  \bibfield  {author} {\bibinfo {author} {\bibfnamefont {S.~R.}\ \bibnamefont
  {White}}\ and\ \bibinfo {author} {\bibfnamefont {D.~J.}\ \bibnamefont
  {Scalapino}},\ }\bibfield  {title} {\bibinfo {title} {Density matrix
  renormalization group study of the striped phase in the {2D}
  $\mathit{t}\ensuremath{-}\mathit{J}$ model},\ }\href
  {https://doi.org/10.1103/PhysRevLett.80.1272} {\bibfield  {journal} {\bibinfo
   {journal} {Phys. Rev. Lett.}\ }\textbf {\bibinfo {volume} {80}},\ \bibinfo
  {pages} {1272} (\bibinfo {year} {1998}{\natexlab{a}})}\BibitemShut {NoStop}%
\bibitem [{\citenamefont {Macridin}\ and\ \citenamefont
  {Jarrell}(2008)}]{PhysRevB.78.241101}%
  \BibitemOpen
  \bibfield  {author} {\bibinfo {author} {\bibfnamefont {A.}~\bibnamefont
  {Macridin}}\ and\ \bibinfo {author} {\bibfnamefont {M.}~\bibnamefont
  {Jarrell}},\ }\bibfield  {title} {\bibinfo {title} {Bond excitations in the
  pseudogap phase of the {Hubbard} model},\ }\href
  {https://doi.org/10.1103/PhysRevB.78.241101} {\bibfield  {journal} {\bibinfo
  {journal} {Phys. Rev. B}\ }\textbf {\bibinfo {volume} {78}},\ \bibinfo
  {pages} {241101} (\bibinfo {year} {2008})}\BibitemShut {NoStop}%
\bibitem [{\citenamefont {White}\ and\ \citenamefont
  {Scalapino}(1998{\natexlab{b}})}]{WhitePRL1989b}%
  \BibitemOpen
  \bibfield  {author} {\bibinfo {author} {\bibfnamefont {S.~R.}\ \bibnamefont
  {White}}\ and\ \bibinfo {author} {\bibfnamefont {D.~J.}\ \bibnamefont
  {Scalapino}},\ }\bibfield  {title} {\bibinfo {title} {Energetics of domain
  walls in the {2D} $\mathit{t}\ensuremath{-}\mathit{J}$ model},\ }\href
  {https://doi.org/10.1103/PhysRevLett.81.3227} {\bibfield  {journal} {\bibinfo
   {journal} {Phys. Rev. Lett.}\ }\textbf {\bibinfo {volume} {81}},\ \bibinfo
  {pages} {3227} (\bibinfo {year} {1998}{\natexlab{b}})}\BibitemShut {NoStop}%
\bibitem [{\citenamefont {Huang}\ \emph {et~al.}(2018)\citenamefont {Huang},
  \citenamefont {Mendl}, \citenamefont {Jiang}, \citenamefont {Moritz},\ and\
  \citenamefont {Devereaux}}]{HuangQuantMat2018}%
  \BibitemOpen
  \bibfield  {author} {\bibinfo {author} {\bibfnamefont {E.~W.}\ \bibnamefont
  {Huang}}, \bibinfo {author} {\bibfnamefont {C.~B.}\ \bibnamefont {Mendl}},
  \bibinfo {author} {\bibfnamefont {H.-C.}\ \bibnamefont {Jiang}}, \bibinfo
  {author} {\bibfnamefont {B.}~\bibnamefont {Moritz}},\ and\ \bibinfo {author}
  {\bibfnamefont {T.~P.}\ \bibnamefont {Devereaux}},\ }\bibfield  {title}
  {\bibinfo {title} {Stripe order from the perspective of the {Hubbard}
  model},\ }\href {https://doi.org/10.1038/s41535-018-0097-0} {\bibfield
  {journal} {\bibinfo  {journal} {npj Quantum Mater.}\ }\textbf {\bibinfo
  {volume} {3}},\ \bibinfo {pages} {22} (\bibinfo {year} {2018})}\BibitemShut
  {NoStop}%
\bibitem [{\citenamefont {Maier}\ \emph
  {et~al.}(2005{\natexlab{a}})\citenamefont {Maier}, \citenamefont {Jarrell},
  \citenamefont {Schulthess}, \citenamefont {Kent},\ and\ \citenamefont
  {White}}]{MaierPRL2005}%
  \BibitemOpen
  \bibfield  {author} {\bibinfo {author} {\bibfnamefont {T.~A.}\ \bibnamefont
  {Maier}}, \bibinfo {author} {\bibfnamefont {M.}~\bibnamefont {Jarrell}},
  \bibinfo {author} {\bibfnamefont {T.~C.}\ \bibnamefont {Schulthess}},
  \bibinfo {author} {\bibfnamefont {P.~R.~C.}\ \bibnamefont {Kent}},\ and\
  \bibinfo {author} {\bibfnamefont {J.~B.}\ \bibnamefont {White}},\ }\bibfield
  {title} {\bibinfo {title} {Systematic study of $d$-wave superconductivity in
  the {2D} repulsive {Hubbard} model},\ }\href
  {https://link.aps.org/doi/10.1103/PhysRevLett.95.237001} {\bibfield
  {journal} {\bibinfo  {journal} {Phys. Rev. Lett.}\ }\textbf {\bibinfo
  {volume} {95}},\ \bibinfo {pages} {237001} (\bibinfo {year}
  {2005}{\natexlab{a}})}\BibitemShut {NoStop}%
\bibitem [{\citenamefont {Li}\ \emph {et~al.}(2021)\citenamefont {Li},
  \citenamefont {Nocera}, \citenamefont {Kumar},\ and\ \citenamefont
  {Johnston}}]{LiCommPhys2021}%
  \BibitemOpen
  \bibfield  {author} {\bibinfo {author} {\bibfnamefont {S.}~\bibnamefont
  {Li}}, \bibinfo {author} {\bibfnamefont {A.}~\bibnamefont {Nocera}}, \bibinfo
  {author} {\bibfnamefont {U.}~\bibnamefont {Kumar}},\ and\ \bibinfo {author}
  {\bibfnamefont {S.}~\bibnamefont {Johnston}},\ }\bibfield  {title} {\bibinfo
  {title} {Particle-hole asymmetry in the dynamical spin and charge responses
  of corner-shared {1D} cuprates},\ }\href
  {https://doi.org/10.1038/s42005-021-00718-w} {\bibfield  {journal} {\bibinfo
  {journal} {Communications Physics}\ }\textbf {\bibinfo {volume} {4}},\
  \bibinfo {pages} {217} (\bibinfo {year} {2021})}\BibitemShut {NoStop}%
\bibitem [{\citenamefont {Mai}\ \emph {et~al.}(2022)\citenamefont {Mai},
  \citenamefont {Karakuzu}, \citenamefont {Balduzzi}, \citenamefont
  {Johnston},\ and\ \citenamefont {Maier}}]{MaiPNAS2022}%
  \BibitemOpen
  \bibfield  {author} {\bibinfo {author} {\bibfnamefont {P.}~\bibnamefont
  {Mai}}, \bibinfo {author} {\bibfnamefont {S.}~\bibnamefont {Karakuzu}},
  \bibinfo {author} {\bibfnamefont {G.}~\bibnamefont {Balduzzi}}, \bibinfo
  {author} {\bibfnamefont {S.}~\bibnamefont {Johnston}},\ and\ \bibinfo
  {author} {\bibfnamefont {T.~A.}\ \bibnamefont {Maier}},\ }\bibfield  {title}
  {\bibinfo {title} {Intertwined spin, charge, and pair correlations in the
  two-dimensional {Hubbard} model in the thermodynamic limit},\ }\href
  {https://doi.org/10.1073/pnas.2112806119} {\bibfield  {journal} {\bibinfo
  {journal} {Proceedings of the National Academy of Sciences}\ }\textbf
  {\bibinfo {volume} {119}},\ \bibinfo {pages} {e2112806119} (\bibinfo {year}
  {2022})}\BibitemShut {NoStop}%
\bibitem [{\citenamefont {Huang}\ \emph {et~al.}(2022)\citenamefont {Huang},
  \citenamefont {Liu}, \citenamefont {Wang}, \citenamefont {Jiang},
  \citenamefont {Mai}, \citenamefont {Maier}, \citenamefont {Johnston},
  \citenamefont {Moritz},\ and\ \citenamefont {Devereaux}}]{HuangPreprint}%
  \BibitemOpen
  \bibfield  {author} {\bibinfo {author} {\bibfnamefont {E.~W.}\ \bibnamefont
  {Huang}}, \bibinfo {author} {\bibfnamefont {T.}~\bibnamefont {Liu}}, \bibinfo
  {author} {\bibfnamefont {W.~O.}\ \bibnamefont {Wang}}, \bibinfo {author}
  {\bibfnamefont {J.-C.}\ \bibnamefont {Jiang}}, \bibinfo {author}
  {\bibfnamefont {P.}~\bibnamefont {Mai}}, \bibinfo {author} {\bibfnamefont
  {T.~A.}\ \bibnamefont {Maier}}, \bibinfo {author} {\bibfnamefont
  {S.}~\bibnamefont {Johnston}}, \bibinfo {author} {\bibfnamefont
  {B.}~\bibnamefont {Moritz}},\ and\ \bibinfo {author} {\bibfnamefont {T.~P.}\
  \bibnamefont {Devereaux}},\ }\bibfield  {title} {\bibinfo {title}
  {Fluctuating intertwined stripes in the strange metal regime of the {Hubbard}
  model},\ }\href {https://arxiv.org/abs/2202.08845} {\bibfield  {journal}
  {\bibinfo  {journal} {arXiv:2202.08845}\ } (\bibinfo {year}
  {2022})}\BibitemShut {NoStop}%
\bibitem [{\citenamefont {Qin}\ \emph {et~al.}(2020)\citenamefont {Qin},
  \citenamefont {Chung}, \citenamefont {Shi}, \citenamefont {Vitali},
  \citenamefont {Hubig}, \citenamefont {Schollw\"ock}, \citenamefont {White},\
  and\ \citenamefont {Zhang}}]{Qin2020}%
  \BibitemOpen
  \bibfield  {author} {\bibinfo {author} {\bibfnamefont {M.}~\bibnamefont
  {Qin}}, \bibinfo {author} {\bibfnamefont {C.-M.}\ \bibnamefont {Chung}},
  \bibinfo {author} {\bibfnamefont {H.}~\bibnamefont {Shi}}, \bibinfo {author}
  {\bibfnamefont {E.}~\bibnamefont {Vitali}}, \bibinfo {author} {\bibfnamefont
  {C.}~\bibnamefont {Hubig}}, \bibinfo {author} {\bibfnamefont
  {U.}~\bibnamefont {Schollw\"ock}}, \bibinfo {author} {\bibfnamefont {S.~R.}\
  \bibnamefont {White}},\ and\ \bibinfo {author} {\bibfnamefont
  {S.}~\bibnamefont {Zhang}} (\bibinfo {collaboration} {Simons Collaboration on
  the Many-Electron Problem}),\ }\bibfield  {title} {\bibinfo {title} {Absence
  of superconductivity in the pure two-dimensional {Hubbard} model},\ }\href
  {https://doi.org/10.1103/PhysRevX.10.031016} {\bibfield  {journal} {\bibinfo
  {journal} {Phys. Rev. X}\ }\textbf {\bibinfo {volume} {10}},\ \bibinfo
  {pages} {031016} (\bibinfo {year} {2020})}\BibitemShut {NoStop}%
\bibitem [{\citenamefont {Jiang}\ \emph {et~al.}(2021)\citenamefont {Jiang},
  \citenamefont {Scalapino},\ and\ \citenamefont {White}}]{JiangPNAS2021}%
  \BibitemOpen
  \bibfield  {author} {\bibinfo {author} {\bibfnamefont {S.}~\bibnamefont
  {Jiang}}, \bibinfo {author} {\bibfnamefont {D.~J.}\ \bibnamefont
  {Scalapino}},\ and\ \bibinfo {author} {\bibfnamefont {S.~R.}\ \bibnamefont
  {White}},\ }\bibfield  {title} {\bibinfo {title} {Ground-state phase diagram
  of the $t-t^\prime-{J}$ model},\ }\href
  {https://doi.org/10.1073/pnas.2109978118} {\bibfield  {journal} {\bibinfo
  {journal} {Proceedings of the National Academy of Sciences}\ }\textbf
  {\bibinfo {volume} {118}},\ \bibinfo {pages} {e2109978118} (\bibinfo {year}
  {2021})}\BibitemShut {NoStop}%
\bibitem [{\citenamefont {Zheng}\ \emph
  {et~al.}(2017{\natexlab{a}})\citenamefont {Zheng}, \citenamefont {Chung},
  \citenamefont {Corboz}, \citenamefont {Ehlers}, \citenamefont {Qin},
  \citenamefont {Noack}, \citenamefont {Shi}, \citenamefont {White},
  \citenamefont {Zhang},\ and\ \citenamefont {Chan}}]{Zheng1155}%
  \BibitemOpen
  \bibfield  {author} {\bibinfo {author} {\bibfnamefont {B.-X.}\ \bibnamefont
  {Zheng}}, \bibinfo {author} {\bibfnamefont {C.-M.}\ \bibnamefont {Chung}},
  \bibinfo {author} {\bibfnamefont {P.}~\bibnamefont {Corboz}}, \bibinfo
  {author} {\bibfnamefont {G.}~\bibnamefont {Ehlers}}, \bibinfo {author}
  {\bibfnamefont {M.-P.}\ \bibnamefont {Qin}}, \bibinfo {author} {\bibfnamefont
  {R.~M.}\ \bibnamefont {Noack}}, \bibinfo {author} {\bibfnamefont
  {H.}~\bibnamefont {Shi}}, \bibinfo {author} {\bibfnamefont {S.~R.}\
  \bibnamefont {White}}, \bibinfo {author} {\bibfnamefont {S.}~\bibnamefont
  {Zhang}},\ and\ \bibinfo {author} {\bibfnamefont {G.~K.-L.}\ \bibnamefont
  {Chan}},\ }\bibfield  {title} {\bibinfo {title} {Stripe order in the
  underdoped region of the two-dimensional {Hubbard} model},\ }\href
  {https://doi.org/10.1126/science.aam7127} {\bibfield  {journal} {\bibinfo
  {journal} {Science}\ }\textbf {\bibinfo {volume} {358}},\ \bibinfo {pages}
  {1155} (\bibinfo {year} {2017}{\natexlab{a}})}\BibitemShut {NoStop}%
\bibitem [{\citenamefont {White}\ and\ \citenamefont
  {Scalapino}(2003)}]{WhitePRL2003}%
  \BibitemOpen
  \bibfield  {author} {\bibinfo {author} {\bibfnamefont {S.~R.}\ \bibnamefont
  {White}}\ and\ \bibinfo {author} {\bibfnamefont {D.~J.}\ \bibnamefont
  {Scalapino}},\ }\bibfield  {title} {\bibinfo {title} {Stripes on a 6-leg
  {{Hubbard}} ladder},\ }\href
  {https://link.aps.org/doi/10.1103/PhysRevLett.91.136403} {\bibfield
  {journal} {\bibinfo  {journal} {Phys. Rev. Lett.}\ }\textbf {\bibinfo
  {volume} {91}},\ \bibinfo {pages} {136403} (\bibinfo {year}
  {2003})}\BibitemShut {NoStop}%
\bibitem [{\citenamefont {Xiao}\ \emph {et~al.}(2022)\citenamefont {Xiao},
  \citenamefont {He}, \citenamefont {Georges},\ and\ \citenamefont
  {Zhang}}]{XiaoPreprint}%
  \BibitemOpen
  \bibfield  {author} {\bibinfo {author} {\bibfnamefont {B.}~\bibnamefont
  {Xiao}}, \bibinfo {author} {\bibfnamefont {Y.-Y.}\ \bibnamefont {He}},
  \bibinfo {author} {\bibfnamefont {A.}~\bibnamefont {Georges}},\ and\ \bibinfo
  {author} {\bibfnamefont {S.}~\bibnamefont {Zhang}},\ }\bibfield  {title}
  {\bibinfo {title} {Temperature dependence of spin and charge orders in the
  doped two-dimensional {Hubbard} model},\ }\href
  {https://arxiv.org/abs/2202.11741} {\bibfield  {journal} {\bibinfo  {journal}
  {arXiv:2202.11741}\ } (\bibinfo {year} {2022})}\BibitemShut {NoStop}%
\bibitem [{\citenamefont {Wietek}\ \emph {et~al.}(2021)\citenamefont {Wietek},
  \citenamefont {He}, \citenamefont {White}, \citenamefont {Georges},\ and\
  \citenamefont {Stoudenmire}}]{Wietek2021}%
  \BibitemOpen
  \bibfield  {author} {\bibinfo {author} {\bibfnamefont {A.}~\bibnamefont
  {Wietek}}, \bibinfo {author} {\bibfnamefont {Y.-Y.}\ \bibnamefont {He}},
  \bibinfo {author} {\bibfnamefont {S.~R.}\ \bibnamefont {White}}, \bibinfo
  {author} {\bibfnamefont {A.}~\bibnamefont {Georges}},\ and\ \bibinfo {author}
  {\bibfnamefont {E.~M.}\ \bibnamefont {Stoudenmire}},\ }\bibfield  {title}
  {\bibinfo {title} {Stripes, antiferromagnetism, and the pseudogap in the
  doped {Hubbard} model at finite temperature},\ }\href
  {https://doi.org/10.1103/PhysRevX.11.031007} {\bibfield  {journal} {\bibinfo
  {journal} {Phys. Rev. X}\ }\textbf {\bibinfo {volume} {11}},\ \bibinfo
  {pages} {031007} (\bibinfo {year} {2021})}\BibitemShut {NoStop}%
\bibitem [{\citenamefont {Scalettar}\ \emph {et~al.}(1989)\citenamefont
  {Scalettar}, \citenamefont {Bickers},\ and\ \citenamefont
  {Scalapino}}]{ScalettarPRB1989}%
  \BibitemOpen
  \bibfield  {author} {\bibinfo {author} {\bibfnamefont {R.~T.}\ \bibnamefont
  {Scalettar}}, \bibinfo {author} {\bibfnamefont {N.~E.}\ \bibnamefont
  {Bickers}},\ and\ \bibinfo {author} {\bibfnamefont {D.~J.}\ \bibnamefont
  {Scalapino}},\ }\bibfield  {title} {\bibinfo {title} {Competition of pairing
  and {Peierls}--charge-density-wave correlations in a two-dimensional
  electron-phonon model},\ }\href {https://doi.org/10.1103/PhysRevB.40.197}
  {\bibfield  {journal} {\bibinfo  {journal} {Phys. Rev. B}\ }\textbf {\bibinfo
  {volume} {40}},\ \bibinfo {pages} {197} (\bibinfo {year} {1989})}\BibitemShut
  {NoStop}%
\bibitem [{\citenamefont {Capone}\ \emph {et~al.}(2006)\citenamefont {Capone},
  \citenamefont {Carta},\ and\ \citenamefont {Ciuchi}}]{PhysRevB.74.045106}%
  \BibitemOpen
  \bibfield  {author} {\bibinfo {author} {\bibfnamefont {M.}~\bibnamefont
  {Capone}}, \bibinfo {author} {\bibfnamefont {P.}~\bibnamefont {Carta}},\ and\
  \bibinfo {author} {\bibfnamefont {S.}~\bibnamefont {Ciuchi}},\ }\bibfield
  {title} {\bibinfo {title} {Dynamical mean field theory of polarons and
  bipolarons in the half-filled {Holstein} model},\ }\href
  {https://doi.org/10.1103/PhysRevB.74.045106} {\bibfield  {journal} {\bibinfo
  {journal} {Phys. Rev. B}\ }\textbf {\bibinfo {volume} {74}},\ \bibinfo
  {pages} {045106} (\bibinfo {year} {2006})}\BibitemShut {NoStop}%
\bibitem [{\citenamefont {Hague}(2005)}]{Hague_2005}%
  \BibitemOpen
  \bibfield  {author} {\bibinfo {author} {\bibfnamefont {J.~P.}\ \bibnamefont
  {Hague}},\ }\bibfield  {title} {\bibinfo {title} {Superconducting states of
  the quasi-{2D} {H}olstein model: effects of vertex and non-local
  corrections},\ }\href {https://doi.org/10.1088/0953-8984/17/37/005}
  {\bibfield  {journal} {\bibinfo  {journal} {Journal of Physics: Condensed
  Matter}\ }\textbf {\bibinfo {volume} {17}},\ \bibinfo {pages} {5663}
  (\bibinfo {year} {2005})}\BibitemShut {NoStop}%
\bibitem [{\citenamefont {Dee}\ \emph {et~al.}(2020{\natexlab{a}})\citenamefont
  {Dee}, \citenamefont {Coulter}, \citenamefont {Kleiner},\ and\ \citenamefont
  {Johnston}}]{DeeCommPhys2021}%
  \BibitemOpen
  \bibfield  {author} {\bibinfo {author} {\bibfnamefont {P.~M.}\ \bibnamefont
  {Dee}}, \bibinfo {author} {\bibfnamefont {J.}~\bibnamefont {Coulter}},
  \bibinfo {author} {\bibfnamefont {K.~G.}\ \bibnamefont {Kleiner}},\ and\
  \bibinfo {author} {\bibfnamefont {S.}~\bibnamefont {Johnston}},\ }\bibfield
  {title} {\bibinfo {title} {Relative importance of nonlinear electron-phonon
  coupling and vertex corrections in the {Holstein} model},\ }\href
  {https://doi.org/10.1038/s42005-020-00413-2} {\bibfield  {journal} {\bibinfo
  {journal} {Communications Physics}\ }\textbf {\bibinfo {volume} {3}},\
  \bibinfo {pages} {145} (\bibinfo {year} {2020}{\natexlab{a}})}\BibitemShut
  {NoStop}%
\bibitem [{\citenamefont {Nosarzewski}\ \emph {et~al.}(2021)\citenamefont
  {Nosarzewski}, \citenamefont {Huang}, \citenamefont {Dee}, \citenamefont
  {Esterlis}, \citenamefont {Moritz}, \citenamefont {Kivelson}, \citenamefont
  {Johnston},\ and\ \citenamefont {Devereaux}}]{NosarzewskiPRB2021}%
  \BibitemOpen
  \bibfield  {author} {\bibinfo {author} {\bibfnamefont {B.}~\bibnamefont
  {Nosarzewski}}, \bibinfo {author} {\bibfnamefont {E.~W.}\ \bibnamefont
  {Huang}}, \bibinfo {author} {\bibfnamefont {P.~M.}\ \bibnamefont {Dee}},
  \bibinfo {author} {\bibfnamefont {I.}~\bibnamefont {Esterlis}}, \bibinfo
  {author} {\bibfnamefont {B.}~\bibnamefont {Moritz}}, \bibinfo {author}
  {\bibfnamefont {S.~A.}\ \bibnamefont {Kivelson}}, \bibinfo {author}
  {\bibfnamefont {S.}~\bibnamefont {Johnston}},\ and\ \bibinfo {author}
  {\bibfnamefont {T.~P.}\ \bibnamefont {Devereaux}},\ }\bibfield  {title}
  {\bibinfo {title} {Superconductivity, charge density waves, and bipolarons in
  the {Holstein} model},\ }\href {https://doi.org/10.1103/PhysRevB.103.235156}
  {\bibfield  {journal} {\bibinfo  {journal} {Phys. Rev. B}\ }\textbf {\bibinfo
  {volume} {103}},\ \bibinfo {pages} {235156} (\bibinfo {year}
  {2021})}\BibitemShut {NoStop}%
\bibitem [{\citenamefont {Bradley}\ \emph {et~al.}(2021)\citenamefont
  {Bradley}, \citenamefont {Batrouni},\ and\ \citenamefont
  {Scalettar}}]{BradleyPRB2021}%
  \BibitemOpen
  \bibfield  {author} {\bibinfo {author} {\bibfnamefont {O.}~\bibnamefont
  {Bradley}}, \bibinfo {author} {\bibfnamefont {G.~G.}\ \bibnamefont
  {Batrouni}},\ and\ \bibinfo {author} {\bibfnamefont {R.~T.}\ \bibnamefont
  {Scalettar}},\ }\bibfield  {title} {\bibinfo {title} {Superconductivity and
  charge density wave order in the two-dimensional {Holstein} model},\ }\href
  {https://doi.org/10.1103/PhysRevB.103.235104} {\bibfield  {journal} {\bibinfo
   {journal} {Phys. Rev. B}\ }\textbf {\bibinfo {volume} {103}},\ \bibinfo
  {pages} {235104} (\bibinfo {year} {2021})}\BibitemShut {NoStop}%
\bibitem [{\citenamefont {Cohen-Stead}\ \emph {et~al.}(2020)\citenamefont
  {Cohen-Stead}, \citenamefont {Barros}, \citenamefont {Meng}, \citenamefont
  {Chen}, \citenamefont {Scalettar},\ and\ \citenamefont
  {Batrouni}}]{CohenSteadPRB2021}%
  \BibitemOpen
  \bibfield  {author} {\bibinfo {author} {\bibfnamefont {B.}~\bibnamefont
  {Cohen-Stead}}, \bibinfo {author} {\bibfnamefont {K.}~\bibnamefont {Barros}},
  \bibinfo {author} {\bibfnamefont {Z.}~\bibnamefont {Meng}}, \bibinfo {author}
  {\bibfnamefont {C.}~\bibnamefont {Chen}}, \bibinfo {author} {\bibfnamefont
  {R.~T.}\ \bibnamefont {Scalettar}},\ and\ \bibinfo {author} {\bibfnamefont
  {G.~G.}\ \bibnamefont {Batrouni}},\ }\bibfield  {title} {\bibinfo {title}
  {Langevin simulations of the half-filled cubic {Holstein} model},\ }\href
  {https://doi.org/10.1103/PhysRevB.102.161108} {\bibfield  {journal} {\bibinfo
   {journal} {Phys. Rev. B}\ }\textbf {\bibinfo {volume} {102}},\ \bibinfo
  {pages} {161108} (\bibinfo {year} {2020})}\BibitemShut {NoStop}%
\bibitem [{\citenamefont {Marchand}\ \emph {et~al.}(2010)\citenamefont
  {Marchand}, \citenamefont {De~Filippis}, \citenamefont {Cataudella},
  \citenamefont {Berciu}, \citenamefont {Nagaosa}, \citenamefont {Prokof\'ev},
  \citenamefont {Mishchenko},\ and\ \citenamefont {Stamp}}]{marchand10}%
  \BibitemOpen
  \bibfield  {author} {\bibinfo {author} {\bibfnamefont {D.~J.~J.}\
  \bibnamefont {Marchand}}, \bibinfo {author} {\bibfnamefont {G.}~\bibnamefont
  {De~Filippis}}, \bibinfo {author} {\bibfnamefont {V.}~\bibnamefont
  {Cataudella}}, \bibinfo {author} {\bibfnamefont {M.}~\bibnamefont {Berciu}},
  \bibinfo {author} {\bibfnamefont {N.}~\bibnamefont {Nagaosa}}, \bibinfo
  {author} {\bibfnamefont {N.~V.}\ \bibnamefont {Prokof\'ev}}, \bibinfo
  {author} {\bibfnamefont {A.~S.}\ \bibnamefont {Mishchenko}},\ and\ \bibinfo
  {author} {\bibfnamefont {P.~C.~E.}\ \bibnamefont {Stamp}},\ }\bibfield
  {title} {\bibinfo {title} {Sharp transition for single polarons in the
  one-dimensional {Su-Schrieffer-Heeger} model},\ }\href
  {https://doi.org/10.1103/PhysRevLett.105.266605} {\bibfield  {journal}
  {\bibinfo  {journal} {Phys. Rev. Lett.}\ }\textbf {\bibinfo {volume} {105}},\
  \bibinfo {pages} {266605} (\bibinfo {year} {2010})}\BibitemShut {NoStop}%
\bibitem [{\citenamefont {Weber}\ \emph {et~al.}(2015)\citenamefont {Weber},
  \citenamefont {Assaad},\ and\ \citenamefont {Hohenadler}}]{weber15}%
  \BibitemOpen
  \bibfield  {author} {\bibinfo {author} {\bibfnamefont {M.}~\bibnamefont
  {Weber}}, \bibinfo {author} {\bibfnamefont {F.~F.}\ \bibnamefont {Assaad}},\
  and\ \bibinfo {author} {\bibfnamefont {M.}~\bibnamefont {Hohenadler}},\
  }\bibfield  {title} {\bibinfo {title} {Excitation spectra and correlation
  functions of quantum {Su-Schrieffer-Heeger} models},\ }\href
  {https://doi.org/10.1103/PhysRevB.91.245147} {\bibfield  {journal} {\bibinfo
  {journal} {Phys. Rev. B}\ }\textbf {\bibinfo {volume} {91}},\ \bibinfo
  {pages} {245147} (\bibinfo {year} {2015})}\BibitemShut {NoStop}%
\bibitem [{\citenamefont {Sous}\ \emph {et~al.}(2018)\citenamefont {Sous},
  \citenamefont {Chakraborty}, \citenamefont {Krems},\ and\ \citenamefont
  {Berciu}}]{johnsousmonaberciu2018}%
  \BibitemOpen
  \bibfield  {author} {\bibinfo {author} {\bibfnamefont {J.}~\bibnamefont
  {Sous}}, \bibinfo {author} {\bibfnamefont {M.}~\bibnamefont {Chakraborty}},
  \bibinfo {author} {\bibfnamefont {R.~V.}\ \bibnamefont {Krems}},\ and\
  \bibinfo {author} {\bibfnamefont {M.}~\bibnamefont {Berciu}},\ }\bibfield
  {title} {\bibinfo {title} {Light bipolarons stabilized by {Peierls}
  electron-phonon coupling},\ }\href
  {https://doi.org/10.1103/PhysRevLett.121.247001} {\bibfield  {journal}
  {\bibinfo  {journal} {Phys. Rev. Lett.}\ }\textbf {\bibinfo {volume} {121}},\
  \bibinfo {pages} {247001} (\bibinfo {year} {2018})}\BibitemShut {NoStop}%
\bibitem [{\citenamefont {Li}\ and\ \citenamefont {Johnston}(2020)}]{LiQM2020}%
  \BibitemOpen
  \bibfield  {author} {\bibinfo {author} {\bibfnamefont {S.}~\bibnamefont
  {Li}}\ and\ \bibinfo {author} {\bibfnamefont {S.}~\bibnamefont {Johnston}},\
  }\bibfield  {title} {\bibinfo {title} {Quantum {Monte Carlo} study of lattice
  polarons in the two-dimensional three-orbital {Su-Schrieffer-Heeger} model},\
  }\href {https://doi.org/10.1038/s41535-020-0243-3} {\bibfield  {journal}
  {\bibinfo  {journal} {npj Quantum Materials}\ }\textbf {\bibinfo {volume}
  {5}},\ \bibinfo {pages} {40} (\bibinfo {year} {2020})}\BibitemShut {NoStop}%
\bibitem [{\citenamefont {Cai}\ \emph {et~al.}(2021)\citenamefont {Cai},
  \citenamefont {Li},\ and\ \citenamefont {Yao}}]{cai_2021}%
  \BibitemOpen
  \bibfield  {author} {\bibinfo {author} {\bibfnamefont {X.}~\bibnamefont
  {Cai}}, \bibinfo {author} {\bibfnamefont {Z.-X.}\ \bibnamefont {Li}},\ and\
  \bibinfo {author} {\bibfnamefont {H.}~\bibnamefont {Yao}},\ }\bibfield
  {title} {\bibinfo {title} {Robustness of {Antiferromagnetism} in the
  {Su}-{Schrieffer}-{Heeger}-{{Hubbard}} model},\ }\href
  {http://arxiv.org/abs/2112.14744} {\bibfield  {journal} {\bibinfo  {journal}
  {arXiv:2112.14744}\ } (\bibinfo {year} {2021})}\BibitemShut {NoStop}%
\bibitem [{\citenamefont {G\"otz}\ \emph {et~al.}(2022)\citenamefont {G\"otz},
  \citenamefont {Beyl}, \citenamefont {Hohenadler},\ and\ \citenamefont
  {Assaad}}]{GoetzAssaad2021}%
  \BibitemOpen
  \bibfield  {author} {\bibinfo {author} {\bibfnamefont {A.}~\bibnamefont
  {G\"otz}}, \bibinfo {author} {\bibfnamefont {S.}~\bibnamefont {Beyl}},
  \bibinfo {author} {\bibfnamefont {M.}~\bibnamefont {Hohenadler}},\ and\
  \bibinfo {author} {\bibfnamefont {F.~F.}\ \bibnamefont {Assaad}},\ }\bibfield
   {title} {\bibinfo {title} {Valence-bond solid to antiferromagnet transition
  in the two-dimensional {Su-Schrieffer-Heeger} model by {Langevin} dynamics},\
  }\href {https://doi.org/10.1103/PhysRevB.105.085151} {\bibfield  {journal}
  {\bibinfo  {journal} {Phys. Rev. B}\ }\textbf {\bibinfo {volume} {105}},\
  \bibinfo {pages} {085151} (\bibinfo {year} {2022})}\BibitemShut {NoStop}%
\bibitem [{\citenamefont {Xing}\ \emph {et~al.}(2021)\citenamefont {Xing},
  \citenamefont {Chiu}, \citenamefont {Poletti}, \citenamefont {Scalettar},\
  and\ \citenamefont {Batrouni}}]{XingBatrouni2021}%
  \BibitemOpen
  \bibfield  {author} {\bibinfo {author} {\bibfnamefont {B.}~\bibnamefont
  {Xing}}, \bibinfo {author} {\bibfnamefont {W.-T.}\ \bibnamefont {Chiu}},
  \bibinfo {author} {\bibfnamefont {D.}~\bibnamefont {Poletti}}, \bibinfo
  {author} {\bibfnamefont {R.~T.}\ \bibnamefont {Scalettar}},\ and\ \bibinfo
  {author} {\bibfnamefont {G.}~\bibnamefont {Batrouni}},\ }\bibfield  {title}
  {\bibinfo {title} {{Quantum {Monte Carlo} Simulations of the {2D}
  {Su-Schrieffer-Heeger} Model}},\ }\href
  {https://doi.org/10.1103/PhysRevLett.126.017601} {\bibfield  {journal}
  {\bibinfo  {journal} {Phys. Rev. Lett.}\ }\textbf {\bibinfo {volume} {126}},\
  \bibinfo {pages} {017601} (\bibinfo {year} {2021})}\BibitemShut {NoStop}%
\bibitem [{\citenamefont {Zhang}\ \emph {et~al.}(2022)\citenamefont {Zhang},
  \citenamefont {Sous}, \citenamefont {Reichman}, \citenamefont {Berciu},
  \citenamefont {Millis}, \citenamefont {Prokof'ev},\ and\ \citenamefont
  {Svistunov}}]{Zhang_Sous2022}%
  \BibitemOpen
  \bibfield  {author} {\bibinfo {author} {\bibfnamefont {C.}~\bibnamefont
  {Zhang}}, \bibinfo {author} {\bibfnamefont {J.}~\bibnamefont {Sous}},
  \bibinfo {author} {\bibfnamefont {D.~R.}\ \bibnamefont {Reichman}}, \bibinfo
  {author} {\bibfnamefont {M.}~\bibnamefont {Berciu}}, \bibinfo {author}
  {\bibfnamefont {A.~J.}\ \bibnamefont {Millis}}, \bibinfo {author}
  {\bibfnamefont {N.~V.}\ \bibnamefont {Prokof'ev}},\ and\ \bibinfo {author}
  {\bibfnamefont {B.~V.}\ \bibnamefont {Svistunov}},\ }\bibfield  {title}
  {\bibinfo {title} {Bipolaronic high-temperature superconductivity},\ }\href
  {https://arxiv.org/abs/2203.07380} {\bibfield  {journal} {\bibinfo  {journal}
  {arXiv:2203.07380}\ } (\bibinfo {year} {2022})}\BibitemShut {NoStop}%
\bibitem [{\citenamefont {Jarrell}\ \emph {et~al.}(1993)\citenamefont
  {Jarrell}, \citenamefont {Akhlaghpour},\ and\ \citenamefont
  {Pruschke}}]{jarrell1993periodic}%
  \BibitemOpen
  \bibfield  {author} {\bibinfo {author} {\bibfnamefont {M.}~\bibnamefont
  {Jarrell}}, \bibinfo {author} {\bibfnamefont {H.}~\bibnamefont
  {Akhlaghpour}},\ and\ \bibinfo {author} {\bibfnamefont {T.}~\bibnamefont
  {Pruschke}},\ }\bibfield  {title} {\bibinfo {title} {Periodic {A}nderson
  model in infinite dimensions},\ }\href
  {https://doi.org/10.1103/PhysRevLett.70.1670} {\bibfield  {journal} {\bibinfo
   {journal} {Phys. Rev. Lett.}\ }\textbf {\bibinfo {volume} {70}},\ \bibinfo
  {pages} {1670} (\bibinfo {year} {1993})}\BibitemShut {NoStop}%
\bibitem [{\citenamefont {Veki{\'c}}\ \emph {et~al.}(1995)\citenamefont
  {Veki{\'c}}, \citenamefont {Cannon}, \citenamefont {Scalapino}, \citenamefont
  {Scalettar},\ and\ \citenamefont {Sugar}}]{vekic1995competition}%
  \BibitemOpen
  \bibfield  {author} {\bibinfo {author} {\bibfnamefont {M.}~\bibnamefont
  {Veki{\'c}}}, \bibinfo {author} {\bibfnamefont {J.}~\bibnamefont {Cannon}},
  \bibinfo {author} {\bibfnamefont {D.}~\bibnamefont {Scalapino}}, \bibinfo
  {author} {\bibfnamefont {R.}~\bibnamefont {Scalettar}},\ and\ \bibinfo
  {author} {\bibfnamefont {R.}~\bibnamefont {Sugar}},\ }\bibfield  {title}
  {\bibinfo {title} {Competition between antiferromagnetic order and
  spin-liquid behavior in the two-dimensional periodic {Anderson} model at half
  filling},\ }\href
  {https://journals.aps.org/prl/abstract/10.1103/PhysRevLett.74.2367}
  {\bibfield  {journal} {\bibinfo  {journal} {Phys. Rev. Lett.}\ }\textbf
  {\bibinfo {volume} {74}},\ \bibinfo {pages} {2367} (\bibinfo {year}
  {1995})}\BibitemShut {NoStop}%
\bibitem [{\citenamefont {Held}\ \emph {et~al.}(2000)\citenamefont {Held},
  \citenamefont {Huscroft}, \citenamefont {Scalettar},\ and\ \citenamefont
  {McMahan}}]{held2000similarities}%
  \BibitemOpen
  \bibfield  {author} {\bibinfo {author} {\bibfnamefont {K.}~\bibnamefont
  {Held}}, \bibinfo {author} {\bibfnamefont {C.}~\bibnamefont {Huscroft}},
  \bibinfo {author} {\bibfnamefont {R.}~\bibnamefont {Scalettar}},\ and\
  \bibinfo {author} {\bibfnamefont {A.}~\bibnamefont {McMahan}},\ }\bibfield
  {title} {\bibinfo {title} {Similarities between the {Hubbard} and periodic
  {Anderson} models at finite temperatures},\ }\href
  {https://journals.aps.org/prl/abstract/10.1103/PhysRevLett.85.373} {\bibfield
   {journal} {\bibinfo  {journal} {Phys. Rev. Lett.}\ }\textbf {\bibinfo
  {volume} {85}},\ \bibinfo {pages} {373} (\bibinfo {year} {2000})}\BibitemShut
  {NoStop}%
\bibitem [{\citenamefont {Sun}\ and\ \citenamefont
  {Kotliar}(2003)}]{sun2003extended}%
  \BibitemOpen
  \bibfield  {author} {\bibinfo {author} {\bibfnamefont {P.}~\bibnamefont
  {Sun}}\ and\ \bibinfo {author} {\bibfnamefont {G.}~\bibnamefont {Kotliar}},\
  }\bibfield  {title} {\bibinfo {title} {Extended dynamical mean field theory
  study of the periodic {A}nderson model},\ }\href
  {https://journals.aps.org/prl/pdf/10.1103/PhysRevLett.91.037209} {\bibfield
  {journal} {\bibinfo  {journal} {Phys. Rev. Lett.}\ }\textbf {\bibinfo
  {volume} {91}},\ \bibinfo {pages} {037209} (\bibinfo {year}
  {2003})}\BibitemShut {NoStop}%
\bibitem [{\citenamefont {Luitz}\ and\ \citenamefont
  {Assaad}(2010)}]{luitz2010weak}%
  \BibitemOpen
  \bibfield  {author} {\bibinfo {author} {\bibfnamefont {D.~J.}\ \bibnamefont
  {Luitz}}\ and\ \bibinfo {author} {\bibfnamefont {F.~F.}\ \bibnamefont
  {Assaad}},\ }\bibfield  {title} {\bibinfo {title} {Weak coupling continuous
  time quantum {M}onte {C}arlo study of the single impurity and periodic
  {A}nderson models with s-wave superconducting baths},\ }\href
  {https://journals.aps.org/prb/pdf/10.1103/PhysRevB.81.024509} {\bibfield
  {journal} {\bibinfo  {journal} {Phys. Rev. B}\ }\textbf {\bibinfo {volume}
  {81}},\ \bibinfo {pages} {024509} (\bibinfo {year} {2010})}\BibitemShut
  {NoStop}%
\bibitem [{\citenamefont {Wu}\ and\ \citenamefont {Tremblay}(2015)}]{wu2015d}%
  \BibitemOpen
  \bibfield  {author} {\bibinfo {author} {\bibfnamefont {W.}~\bibnamefont
  {Wu}}\ and\ \bibinfo {author} {\bibfnamefont {A.-M.-S.}\ \bibnamefont
  {Tremblay}},\ }\bibfield  {title} {\bibinfo {title} {d-wave superconductivity
  in the frustrated two-dimensional periodic {A}nderson model},\ }\href
  {https://journals.aps.org/prx/pdf/10.1103/PhysRevX.5.011019} {\bibfield
  {journal} {\bibinfo  {journal} {Phys. Rev. X}\ }\textbf {\bibinfo {volume}
  {5}},\ \bibinfo {pages} {011019} (\bibinfo {year} {2015})}\BibitemShut
  {NoStop}%
\bibitem [{\citenamefont {Werner}\ and\ \citenamefont
  {Millis}(2007)}]{PhysRevLett.99.146404}%
  \BibitemOpen
  \bibfield  {author} {\bibinfo {author} {\bibfnamefont {P.}~\bibnamefont
  {Werner}}\ and\ \bibinfo {author} {\bibfnamefont {A.~J.}\ \bibnamefont
  {Millis}},\ }\bibfield  {title} {\bibinfo {title} {Efficient dynamical mean
  field simulation of the {Holstein}-{Hubbard} model},\ }\href
  {https://doi.org/10.1103/PhysRevLett.99.146404} {\bibfield  {journal}
  {\bibinfo  {journal} {Phys. Rev. Lett.}\ }\textbf {\bibinfo {volume} {99}},\
  \bibinfo {pages} {146404} (\bibinfo {year} {2007})}\BibitemShut {NoStop}%
\bibitem [{\citenamefont {Fehske}\ \emph {et~al.}(2008)\citenamefont {Fehske},
  \citenamefont {Hager},\ and\ \citenamefont {Jeckelmann}}]{fehske08}%
  \BibitemOpen
  \bibfield  {author} {\bibinfo {author} {\bibfnamefont {H.}~\bibnamefont
  {Fehske}}, \bibinfo {author} {\bibfnamefont {G.}~\bibnamefont {Hager}},\ and\
  \bibinfo {author} {\bibfnamefont {E.}~\bibnamefont {Jeckelmann}},\ }\bibfield
   {title} {\bibinfo {title} {{Metallicity in the half-filled
  {Holstein}-{Hubbard} model}},\ }\href
  {https://doi.org/10.1209/0295-5075/84/57001} {\bibfield  {journal} {\bibinfo
  {journal} {Europhys. Lett.}\ }\textbf {\bibinfo {volume} {84}},\ \bibinfo
  {pages} {57001} (\bibinfo {year} {2008})}\BibitemShut {NoStop}%
\bibitem [{\citenamefont {Nocera}\ \emph {et~al.}(2014)\citenamefont {Nocera},
  \citenamefont {Soltanieh-ha}, \citenamefont {Perroni}, \citenamefont
  {Cataudella},\ and\ \citenamefont {Feiguin}}]{PhysRevB.90.195134}%
  \BibitemOpen
  \bibfield  {author} {\bibinfo {author} {\bibfnamefont {A.}~\bibnamefont
  {Nocera}}, \bibinfo {author} {\bibfnamefont {M.}~\bibnamefont
  {Soltanieh-ha}}, \bibinfo {author} {\bibfnamefont {C.~A.}\ \bibnamefont
  {Perroni}}, \bibinfo {author} {\bibfnamefont {V.}~\bibnamefont
  {Cataudella}},\ and\ \bibinfo {author} {\bibfnamefont {A.~E.}\ \bibnamefont
  {Feiguin}},\ }\bibfield  {title} {\bibinfo {title} {Interplay of charge,
  spin, and lattice degrees of freedom in the spectral properties of the
  one-dimensional {Hubbard}-{Holstein} model},\ }\href
  {https://doi.org/10.1103/PhysRevB.90.195134} {\bibfield  {journal} {\bibinfo
  {journal} {Phys. Rev. B}\ }\textbf {\bibinfo {volume} {90}},\ \bibinfo
  {pages} {195134} (\bibinfo {year} {2014})}\BibitemShut {NoStop}%
\bibitem [{\citenamefont {Greitemann}\ \emph {et~al.}(2015)\citenamefont
  {Greitemann}, \citenamefont {Hesselmann}, \citenamefont {Wessel},
  \citenamefont {Assaad},\ and\ \citenamefont {Hohenadler}}]{greitemann15}%
  \BibitemOpen
  \bibfield  {author} {\bibinfo {author} {\bibfnamefont {J.}~\bibnamefont
  {Greitemann}}, \bibinfo {author} {\bibfnamefont {S.}~\bibnamefont
  {Hesselmann}}, \bibinfo {author} {\bibfnamefont {S.}~\bibnamefont {Wessel}},
  \bibinfo {author} {\bibfnamefont {F.~F.}\ \bibnamefont {Assaad}},\ and\
  \bibinfo {author} {\bibfnamefont {M.}~\bibnamefont {Hohenadler}},\ }\bibfield
   {title} {\bibinfo {title} {Finite-size effects in {L}uther-{E}mery phases of
  {Holstein} and {Hubbard} models},\ }\href
  {https://doi.org/10.1103/PhysRevB.92.245132} {\bibfield  {journal} {\bibinfo
  {journal} {Phys. Rev. B}\ }\textbf {\bibinfo {volume} {92}},\ \bibinfo
  {pages} {245132} (\bibinfo {year} {2015})}\BibitemShut {NoStop}%
\bibitem [{\citenamefont {Hohenadler}(2016)}]{hohenadler16}%
  \BibitemOpen
  \bibfield  {author} {\bibinfo {author} {\bibfnamefont {M.}~\bibnamefont
  {Hohenadler}},\ }\bibfield  {title} {\bibinfo {title} {Interplay of site and
  bond electron-phonon coupling in one dimension},\ }\href
  {https://doi.org/10.1103/PhysRevLett.117.206404} {\bibfield  {journal}
  {\bibinfo  {journal} {Phys. Rev. Lett.}\ }\textbf {\bibinfo {volume} {117}},\
  \bibinfo {pages} {206404} (\bibinfo {year} {2016})}\BibitemShut {NoStop}%
\bibitem [{\citenamefont {Wang}\ \emph {et~al.}(2020)\citenamefont {Wang},
  \citenamefont {Esterlis}, \citenamefont {Shi}, \citenamefont {Cirac},\ and\
  \citenamefont {Demler}}]{WangDemler2020}%
  \BibitemOpen
  \bibfield  {author} {\bibinfo {author} {\bibfnamefont {Y.}~\bibnamefont
  {Wang}}, \bibinfo {author} {\bibfnamefont {I.}~\bibnamefont {Esterlis}},
  \bibinfo {author} {\bibfnamefont {T.}~\bibnamefont {Shi}}, \bibinfo {author}
  {\bibfnamefont {J.~I.}\ \bibnamefont {Cirac}},\ and\ \bibinfo {author}
  {\bibfnamefont {E.}~\bibnamefont {Demler}},\ }\bibfield  {title} {\bibinfo
  {title} {{Zero-temperature phases of the two-dimensional {Hubbard}-{Holstein}
  model: A non-Gaussian exact diagonalization study}},\ }\href
  {https://doi.org/10.1103/PhysRevResearch.2.043258} {\bibfield  {journal}
  {\bibinfo  {journal} {Phys. Rev. Research}\ }\textbf {\bibinfo {volume}
  {2}},\ \bibinfo {pages} {043258} (\bibinfo {year} {2020})}\BibitemShut
  {NoStop}%
\bibitem [{\citenamefont {Costa}\ \emph
  {et~al.}(2021{\natexlab{a}})\citenamefont {Costa}, \citenamefont {Seki},\
  and\ \citenamefont {Sorella}}]{costa2020}%
  \BibitemOpen
  \bibfield  {author} {\bibinfo {author} {\bibfnamefont {N.~C.}\ \bibnamefont
  {Costa}}, \bibinfo {author} {\bibfnamefont {K.}~\bibnamefont {Seki}},\ and\
  \bibinfo {author} {\bibfnamefont {S.}~\bibnamefont {Sorella}},\ }\bibfield
  {title} {\bibinfo {title} {Magnetism and charge order in the honeycomb
  lattice},\ }\href {https://doi.org/10.1103/PhysRevLett.126.107205} {\bibfield
   {journal} {\bibinfo  {journal} {Phys. Rev. Lett.}\ }\textbf {\bibinfo
  {volume} {126}},\ \bibinfo {pages} {107205} (\bibinfo {year}
  {2021}{\natexlab{a}})}\BibitemShut {NoStop}%
\bibitem [{\citenamefont {Johnston}\ \emph {et~al.}(2013)\citenamefont
  {Johnston}, \citenamefont {Nowadnick}, \citenamefont {Kung}, \citenamefont
  {Moritz}, \citenamefont {Scalettar},\ and\ \citenamefont
  {Devereaux}}]{JohnstonPRB2013}%
  \BibitemOpen
  \bibfield  {author} {\bibinfo {author} {\bibfnamefont {S.}~\bibnamefont
  {Johnston}}, \bibinfo {author} {\bibfnamefont {E.~A.}\ \bibnamefont
  {Nowadnick}}, \bibinfo {author} {\bibfnamefont {Y.~F.}\ \bibnamefont {Kung}},
  \bibinfo {author} {\bibfnamefont {B.}~\bibnamefont {Moritz}}, \bibinfo
  {author} {\bibfnamefont {R.~T.}\ \bibnamefont {Scalettar}},\ and\ \bibinfo
  {author} {\bibfnamefont {T.~P.}\ \bibnamefont {Devereaux}},\ }\bibfield
  {title} {\bibinfo {title} {Determinant quantum {Monte Carlo} study of the
  two-dimensional single-band {Hubbard}-{Holstein} model},\ }\href
  {https://doi.org/10.1103/PhysRevB.87.235133} {\bibfield  {journal} {\bibinfo
  {journal} {Phys. Rev. B}\ }\textbf {\bibinfo {volume} {87}},\ \bibinfo
  {pages} {235133} (\bibinfo {year} {2013})}\BibitemShut {NoStop}%
\bibitem [{\citenamefont {Karakuzu}\ \emph {et~al.}(2022)\citenamefont
  {Karakuzu}, \citenamefont {Ly}, \citenamefont {Mai}, \citenamefont {Neuhaus},
  \citenamefont {Maier},\ and\ \citenamefont
  {Johnston}}]{KarakuzuPreprint2022}%
  \BibitemOpen
  \bibfield  {author} {\bibinfo {author} {\bibfnamefont {S.}~\bibnamefont
  {Karakuzu}}, \bibinfo {author} {\bibfnamefont {A.~T.}\ \bibnamefont {Ly}},
  \bibinfo {author} {\bibfnamefont {P.}~\bibnamefont {Mai}}, \bibinfo {author}
  {\bibfnamefont {J.}~\bibnamefont {Neuhaus}}, \bibinfo {author} {\bibfnamefont
  {T.~A.}\ \bibnamefont {Maier}},\ and\ \bibinfo {author} {\bibfnamefont
  {S.}~\bibnamefont {Johnston}},\ }\bibfield  {title} {\bibinfo {title} {Stripe
  correlations in the two-dimensional {Hubbard}-{Holstein} model},\ }\href
  {https://arxiv.org/abs/2205.15464} {\bibfield  {journal} {\bibinfo  {journal}
  {arXiv:2205.15464}\ } (\bibinfo {year} {2022})}\BibitemShut {NoStop}%
\bibitem [{\citenamefont {Li}\ and\ \citenamefont
  {Johnston}(2022)}]{LiPreprint2022}%
  \BibitemOpen
  \bibfield  {author} {\bibinfo {author} {\bibfnamefont {S.}~\bibnamefont
  {Li}}\ and\ \bibinfo {author} {\bibfnamefont {S.}~\bibnamefont {Johnston}},\
  }\bibfield  {title} {\bibinfo {title} {Suppressed superexchange interactions
  in the cuprates by bond-stretching oxygen phonons},\ }\href
  {https://arxiv.org/abs/2205.12678} {\bibfield  {journal} {\bibinfo  {journal}
  {arXiv:2205.12678}\ } (\bibinfo {year} {2022})}\BibitemShut {NoStop}%
\bibitem [{\citenamefont {Hirsch}(2002)}]{PhysRevB.65.214510}%
  \BibitemOpen
  \bibfield  {author} {\bibinfo {author} {\bibfnamefont {J.~E.}\ \bibnamefont
  {Hirsch}},\ }\bibfield  {title} {\bibinfo {title} {Quantum {Monte Carlo} and
  exact diagonalization study of a dynamic {Hubbard} model},\ }\href
  {https://doi.org/10.1103/PhysRevB.65.214510} {\bibfield  {journal} {\bibinfo
  {journal} {Phys. Rev. B}\ }\textbf {\bibinfo {volume} {65}},\ \bibinfo
  {pages} {214510} (\bibinfo {year} {2002})}\BibitemShut {NoStop}%
\bibitem [{\citenamefont {Li}\ \emph {et~al.}(2015{\natexlab{a}})\citenamefont
  {Li}, \citenamefont {Nowadnick},\ and\ \citenamefont {Johnston}}]{LiPRB2015}%
  \BibitemOpen
  \bibfield  {author} {\bibinfo {author} {\bibfnamefont {S.}~\bibnamefont
  {Li}}, \bibinfo {author} {\bibfnamefont {E.~A.}\ \bibnamefont {Nowadnick}},\
  and\ \bibinfo {author} {\bibfnamefont {S.}~\bibnamefont {Johnston}},\
  }\bibfield  {title} {\bibinfo {title} {Quasiparticle properties of the
  nonlinear {Holstein} model at finite doping and temperature},\ }\href
  {https://doi.org/10.1103/PhysRevB.92.064301} {\bibfield  {journal} {\bibinfo
  {journal} {Phys. Rev. B}\ }\textbf {\bibinfo {volume} {92}},\ \bibinfo
  {pages} {064301} (\bibinfo {year} {2015}{\natexlab{a}})}\BibitemShut
  {NoStop}%
\bibitem [{\citenamefont {Ayral}\ \emph {et~al.}(2013)\citenamefont {Ayral},
  \citenamefont {Biermann},\ and\ \citenamefont {Werner}}]{PhysRevB.87.125149}%
  \BibitemOpen
  \bibfield  {author} {\bibinfo {author} {\bibfnamefont {T.}~\bibnamefont
  {Ayral}}, \bibinfo {author} {\bibfnamefont {S.}~\bibnamefont {Biermann}},\
  and\ \bibinfo {author} {\bibfnamefont {P.}~\bibnamefont {Werner}},\
  }\bibfield  {title} {\bibinfo {title} {Screening and nonlocal correlations in
  the extended {H}ubbard model from self-consistent combined {$GW$} and
  dynamical mean field theory},\ }\href
  {https://doi.org/10.1103/PhysRevB.87.125149} {\bibfield  {journal} {\bibinfo
  {journal} {Phys. Rev. B}\ }\textbf {\bibinfo {volume} {87}},\ \bibinfo
  {pages} {125149} (\bibinfo {year} {2013})}\BibitemShut {NoStop}%
\bibitem [{\citenamefont {Paki}\ \emph {et~al.}(2019)\citenamefont {Paki},
  \citenamefont {Terletska}, \citenamefont {Iskakov},\ and\ \citenamefont
  {Gull}}]{PakiPRB2019}%
  \BibitemOpen
  \bibfield  {author} {\bibinfo {author} {\bibfnamefont {J.}~\bibnamefont
  {Paki}}, \bibinfo {author} {\bibfnamefont {H.}~\bibnamefont {Terletska}},
  \bibinfo {author} {\bibfnamefont {S.}~\bibnamefont {Iskakov}},\ and\ \bibinfo
  {author} {\bibfnamefont {E.}~\bibnamefont {Gull}},\ }\bibfield  {title}
  {\bibinfo {title} {Charge order and antiferromagnetism in the extended
  {Hubbard} model},\ }\href {https://doi.org/10.1103/PhysRevB.99.245146}
  {\bibfield  {journal} {\bibinfo  {journal} {Phys. Rev. B}\ }\textbf {\bibinfo
  {volume} {99}},\ \bibinfo {pages} {245146} (\bibinfo {year}
  {2019})}\BibitemShut {NoStop}%
\bibitem [{\citenamefont {Dee}\ \emph {et~al.}(2020{\natexlab{b}})\citenamefont
  {Dee}, \citenamefont {Coulter}, \citenamefont {Kleiner},\ and\ \citenamefont
  {Johnston}}]{dee2020relative}%
  \BibitemOpen
  \bibfield  {author} {\bibinfo {author} {\bibfnamefont {P.~M.}\ \bibnamefont
  {Dee}}, \bibinfo {author} {\bibfnamefont {J.}~\bibnamefont {Coulter}},
  \bibinfo {author} {\bibfnamefont {K.~G.}\ \bibnamefont {Kleiner}},\ and\
  \bibinfo {author} {\bibfnamefont {S.}~\bibnamefont {Johnston}},\ }\bibfield
  {title} {\bibinfo {title} {{Relative importance of nonlinear electron-phonon
  coupling and vertex corrections in the {Holstein} model}},\ }\href
  {https://doi.org/10.1038/s42005-020-00413-2} {\bibfield  {journal} {\bibinfo
  {journal} {Communications Physics}\ }\textbf {\bibinfo {volume} {3}},\
  \bibinfo {pages} {1} (\bibinfo {year} {2020}{\natexlab{b}})}\BibitemShut
  {NoStop}%
\bibitem [{\citenamefont {Chen}\ \emph
  {et~al.}(2021{\natexlab{a}})\citenamefont {Chen}, \citenamefont {Wang},
  \citenamefont {Rebec}, \citenamefont {Jia}, \citenamefont {Hashimoto},
  \citenamefont {Lu}, \citenamefont {Moritz}, \citenamefont {Moore},
  \citenamefont {Devereaux},\ and\ \citenamefont {Shen}}]{ChenScience2021}%
  \BibitemOpen
  \bibfield  {author} {\bibinfo {author} {\bibfnamefont {Z.}~\bibnamefont
  {Chen}}, \bibinfo {author} {\bibfnamefont {Y.}~\bibnamefont {Wang}}, \bibinfo
  {author} {\bibfnamefont {S.~N.}\ \bibnamefont {Rebec}}, \bibinfo {author}
  {\bibfnamefont {T.}~\bibnamefont {Jia}}, \bibinfo {author} {\bibfnamefont
  {M.}~\bibnamefont {Hashimoto}}, \bibinfo {author} {\bibfnamefont
  {D.}~\bibnamefont {Lu}}, \bibinfo {author} {\bibfnamefont {B.}~\bibnamefont
  {Moritz}}, \bibinfo {author} {\bibfnamefont {R.~G.}\ \bibnamefont {Moore}},
  \bibinfo {author} {\bibfnamefont {T.~P.}\ \bibnamefont {Devereaux}},\ and\
  \bibinfo {author} {\bibfnamefont {Z.-X.}\ \bibnamefont {Shen}},\ }\bibfield
  {title} {\bibinfo {title} {Anomalously strong near-neighbor attraction in
  doped {1D} cuprate chains},\ }\href {https://doi.org/10.1126/science.abf5174}
  {\bibfield  {journal} {\bibinfo  {journal} {Science}\ }\textbf {\bibinfo
  {volume} {373}},\ \bibinfo {pages} {1235} (\bibinfo {year}
  {2021}{\natexlab{a}})}\BibitemShut {NoStop}%
\bibitem [{\citenamefont {Huang}\ \emph
  {et~al.}(2017{\natexlab{a}})\citenamefont {Huang}, \citenamefont {Mendl},
  \citenamefont {Liu}, \citenamefont {Johnston}, \citenamefont {Jiang},
  \citenamefont {Moritz},\ and\ \citenamefont {Devereaux}}]{HuangScience2017}%
  \BibitemOpen
  \bibfield  {author} {\bibinfo {author} {\bibfnamefont {E.~W.}\ \bibnamefont
  {Huang}}, \bibinfo {author} {\bibfnamefont {C.~B.}\ \bibnamefont {Mendl}},
  \bibinfo {author} {\bibfnamefont {S.}~\bibnamefont {Liu}}, \bibinfo {author}
  {\bibfnamefont {S.}~\bibnamefont {Johnston}}, \bibinfo {author}
  {\bibfnamefont {H.-C.}\ \bibnamefont {Jiang}}, \bibinfo {author}
  {\bibfnamefont {B.}~\bibnamefont {Moritz}},\ and\ \bibinfo {author}
  {\bibfnamefont {T.~P.}\ \bibnamefont {Devereaux}},\ }\bibfield  {title}
  {\bibinfo {title} {Numerical evidence of fluctuating stripes in the normal
  state of high-{T$_c$} cuprate superconductors},\ }\href
  {https://doi.org/10.1126/science.aak9546} {\bibfield  {journal} {\bibinfo
  {journal} {Science}\ }\textbf {\bibinfo {volume} {358}},\ \bibinfo {pages}
  {1161} (\bibinfo {year} {2017}{\natexlab{a}})}\BibitemShut {NoStop}%
\bibitem [{\citenamefont {Loh}\ \emph {et~al.}(1990)\citenamefont {Loh},
  \citenamefont {Gubernatis}, \citenamefont {Scalettar}, \citenamefont {White},
  \citenamefont {Scalapino},\ and\ \citenamefont {Sugar}}]{loh90}%
  \BibitemOpen
  \bibfield  {author} {\bibinfo {author} {\bibfnamefont {E.}~\bibnamefont
  {Loh}}, \bibinfo {author} {\bibfnamefont {J.}~\bibnamefont {Gubernatis}},
  \bibinfo {author} {\bibfnamefont {R.}~\bibnamefont {Scalettar}}, \bibinfo
  {author} {\bibfnamefont {S.}~\bibnamefont {White}}, \bibinfo {author}
  {\bibfnamefont {D.}~\bibnamefont {Scalapino}},\ and\ \bibinfo {author}
  {\bibfnamefont {R.}~\bibnamefont {Sugar}},\ }\bibfield  {title} {\bibinfo
  {title} {Sign problem in the numerical simulation of many-electron systems},\
  }\href {https://doi.org/10.1103/PhysRevB.41.9301} {\bibfield  {journal}
  {\bibinfo  {journal} {Phys. Rev. B}\ }\textbf {\bibinfo {volume} {41}},\
  \bibinfo {pages} {9301} (\bibinfo {year} {1990})}\BibitemShut {NoStop}%
\bibitem [{\citenamefont {Wu}\ and\ \citenamefont {Zhang}(2005)}]{wu05}%
  \BibitemOpen
  \bibfield  {author} {\bibinfo {author} {\bibfnamefont {C.}~\bibnamefont
  {Wu}}\ and\ \bibinfo {author} {\bibfnamefont {S.-C.}\ \bibnamefont {Zhang}},\
  }\bibfield  {title} {\bibinfo {title} {{Sufficient condition for absence of
  the sign problem in the fermionic quantum {Monte Carlo} algorithm}},\ }\href
  {https://doi.org/10.1103/PhysRevB.71.155115} {\bibfield  {journal} {\bibinfo
  {journal} {Phys. Rev. B}\ }\textbf {\bibinfo {volume} {71}},\ \bibinfo
  {pages} {155115} (\bibinfo {year} {2005})}\BibitemShut {NoStop}%
\bibitem [{\citenamefont {Troyer}\ and\ \citenamefont
  {Wiese}(2005)}]{troyer05}%
  \BibitemOpen
  \bibfield  {author} {\bibinfo {author} {\bibfnamefont {M.}~\bibnamefont
  {Troyer}}\ and\ \bibinfo {author} {\bibfnamefont {U.-J.}\ \bibnamefont
  {Wiese}},\ }\bibfield  {title} {\bibinfo {title} {Computational complexity
  and fundamental limitations to fermionic quantum {M}onte {C}arlo
  simulations},\ }\href {https://doi.org/10.1103/PhysRevLett.94.170201}
  {\bibfield  {journal} {\bibinfo  {journal} {Phys. Rev. Lett.}\ }\textbf
  {\bibinfo {volume} {94}},\ \bibinfo {pages} {170201} (\bibinfo {year}
  {2005})}\BibitemShut {NoStop}%
\bibitem [{\citenamefont {Chandrasekharan}(2010)}]{chandrasekharan10}%
  \BibitemOpen
  \bibfield  {author} {\bibinfo {author} {\bibfnamefont {S.}~\bibnamefont
  {Chandrasekharan}},\ }\bibfield  {title} {\bibinfo {title} {Fermion bag
  approach to lattice field theories},\ }\href
  {https://doi.org/10.1103/PhysRevD.82.025007} {\bibfield  {journal} {\bibinfo
  {journal} {Phys. Rev. D}\ }\textbf {\bibinfo {volume} {82}},\ \bibinfo
  {pages} {025007} (\bibinfo {year} {2010})}\BibitemShut {NoStop}%
\bibitem [{\citenamefont {Li}\ \emph {et~al.}(2016)\citenamefont {Li},
  \citenamefont {Jiang},\ and\ \citenamefont {Yao}}]{li16}%
  \BibitemOpen
  \bibfield  {author} {\bibinfo {author} {\bibfnamefont {Z.-X.}\ \bibnamefont
  {Li}}, \bibinfo {author} {\bibfnamefont {Y.-F.}\ \bibnamefont {Jiang}},\ and\
  \bibinfo {author} {\bibfnamefont {H.}~\bibnamefont {Yao}},\ }\bibfield
  {title} {\bibinfo {title} {Majorana-time-reversal symmetries: A fundamental
  principle for sign-problem-free quantum {M}onte {C}arlo simulations},\ }\href
  {https://doi.org/10.1103/PhysRevLett.117.267002} {\bibfield  {journal}
  {\bibinfo  {journal} {Phys. Rev. Lett.}\ }\textbf {\bibinfo {volume} {117}},\
  \bibinfo {pages} {267002} (\bibinfo {year} {2016})}\BibitemShut {NoStop}%
\bibitem [{\citenamefont {Iazzi}\ \emph {et~al.}(2016)\citenamefont {Iazzi},
  \citenamefont {Soluyanov},\ and\ \citenamefont {Troyer}}]{Iazzi2016}%
  \BibitemOpen
  \bibfield  {author} {\bibinfo {author} {\bibfnamefont {M.}~\bibnamefont
  {Iazzi}}, \bibinfo {author} {\bibfnamefont {A.~A.}\ \bibnamefont
  {Soluyanov}},\ and\ \bibinfo {author} {\bibfnamefont {M.}~\bibnamefont
  {Troyer}},\ }\bibfield  {title} {\bibinfo {title} {Topological origin of the
  fermion sign problem},\ }\href {https://doi.org/10.1103/PhysRevB.93.115102}
  {\bibfield  {journal} {\bibinfo  {journal} {Phys. Rev. B}\ }\textbf {\bibinfo
  {volume} {93}},\ \bibinfo {pages} {115102} (\bibinfo {year}
  {2016})}\BibitemShut {NoStop}%
\bibitem [{\citenamefont {Hangleiter}\ \emph {et~al.}(2020)\citenamefont
  {Hangleiter}, \citenamefont {Roth}, \citenamefont {Nagaj},\ and\
  \citenamefont {Eisert}}]{Hangleiter2020}%
  \BibitemOpen
  \bibfield  {author} {\bibinfo {author} {\bibfnamefont {D.}~\bibnamefont
  {Hangleiter}}, \bibinfo {author} {\bibfnamefont {I.}~\bibnamefont {Roth}},
  \bibinfo {author} {\bibfnamefont {D.}~\bibnamefont {Nagaj}},\ and\ \bibinfo
  {author} {\bibfnamefont {J.}~\bibnamefont {Eisert}},\ }\bibfield  {title}
  {\bibinfo {title} {Easing the {M}onte {C}arlo sign problem},\ }\bibfield
  {journal} {\bibinfo  {journal} {Science Advances}\ }\textbf {\bibinfo
  {volume} {6}},\ \href {https://doi.org/10.1126/sciadv.abb8341}
  {10.1126/sciadv.abb8341} (\bibinfo {year} {2020})\BibitemShut {NoStop}%
\bibitem [{\citenamefont {Mondaini}\ \emph {et~al.}(2022)\citenamefont
  {Mondaini}, \citenamefont {Tarat},\ and\ \citenamefont
  {Scalettar}}]{MondainiScience2022}%
  \BibitemOpen
  \bibfield  {author} {\bibinfo {author} {\bibfnamefont {R.}~\bibnamefont
  {Mondaini}}, \bibinfo {author} {\bibfnamefont {S.}~\bibnamefont {Tarat}},\
  and\ \bibinfo {author} {\bibfnamefont {R.~T.}\ \bibnamefont {Scalettar}},\
  }\bibfield  {title} {\bibinfo {title} {Quantum critical points and the sign
  problem},\ }\href {https://doi.org/10.1126/science.abg9299} {\bibfield
  {journal} {\bibinfo  {journal} {Science}\ }\textbf {\bibinfo {volume}
  {375}},\ \bibinfo {pages} {418} (\bibinfo {year} {2022})}\BibitemShut
  {NoStop}%
\bibitem [{\citenamefont {Swendsen}\ and\ \citenamefont
  {Wang}(1987)}]{swendsen87}%
  \BibitemOpen
  \bibfield  {author} {\bibinfo {author} {\bibfnamefont {R.~H.}\ \bibnamefont
  {Swendsen}}\ and\ \bibinfo {author} {\bibfnamefont {J.-S.}\ \bibnamefont
  {Wang}},\ }\bibfield  {title} {\bibinfo {title} {Nonuniversal critical
  dynamics in {Monte Carlo} simulations},\ }\href
  {https://doi.org/10.1103/PhysRevLett.58.86} {\bibfield  {journal} {\bibinfo
  {journal} {Phys. Rev. Lett.}\ }\textbf {\bibinfo {volume} {58}},\ \bibinfo
  {pages} {86} (\bibinfo {year} {1987})}\BibitemShut {NoStop}%
\bibitem [{\citenamefont {Wolff}(1988)}]{wolff88}%
  \BibitemOpen
  \bibfield  {author} {\bibinfo {author} {\bibfnamefont {U.}~\bibnamefont
  {Wolff}},\ }\bibfield  {title} {\bibinfo {title} {Lattice field theory as a
  percolation process},\ }\href {https://doi.org/10.1103/PhysRevLett.60.1461}
  {\bibfield  {journal} {\bibinfo  {journal} {Phys. Rev. Lett.}\ }\textbf
  {\bibinfo {volume} {60}},\ \bibinfo {pages} {1461} (\bibinfo {year}
  {1988})}\BibitemShut {NoStop}%
\bibitem [{\citenamefont {Edwards}\ and\ \citenamefont
  {Sokal}(1988)}]{edwards88}%
  \BibitemOpen
  \bibfield  {author} {\bibinfo {author} {\bibfnamefont {R.~G.}\ \bibnamefont
  {Edwards}}\ and\ \bibinfo {author} {\bibfnamefont {A.~D.}\ \bibnamefont
  {Sokal}},\ }\bibfield  {title} {\bibinfo {title} {Generalization of the
  {Fortuin-Kasteleyn-Swendsen-Wang} representation and {Monte Carlo}
  algorithm},\ }\href {https://doi.org/10.1103/PhysRevD.38.2009} {\bibfield
  {journal} {\bibinfo  {journal} {Phys. Rev. D}\ }\textbf {\bibinfo {volume}
  {38}},\ \bibinfo {pages} {2009} (\bibinfo {year} {1988})}\BibitemShut
  {NoStop}%
\bibitem [{\citenamefont {Carleo}\ \emph {et~al.}(2019)\citenamefont {Carleo},
  \citenamefont {Cirac}, \citenamefont {Cranmer}, \citenamefont {Daudet},
  \citenamefont {Schuld}, \citenamefont {Tishby}, \citenamefont
  {Vogt-Maranto},\ and\ \citenamefont {Zdeborov\'a}}]{CarleoRMP2019}%
  \BibitemOpen
  \bibfield  {author} {\bibinfo {author} {\bibfnamefont {G.}~\bibnamefont
  {Carleo}}, \bibinfo {author} {\bibfnamefont {I.}~\bibnamefont {Cirac}},
  \bibinfo {author} {\bibfnamefont {K.}~\bibnamefont {Cranmer}}, \bibinfo
  {author} {\bibfnamefont {L.}~\bibnamefont {Daudet}}, \bibinfo {author}
  {\bibfnamefont {M.}~\bibnamefont {Schuld}}, \bibinfo {author} {\bibfnamefont
  {N.}~\bibnamefont {Tishby}}, \bibinfo {author} {\bibfnamefont
  {L.}~\bibnamefont {Vogt-Maranto}},\ and\ \bibinfo {author} {\bibfnamefont
  {L.}~\bibnamefont {Zdeborov\'a}},\ }\bibfield  {title} {\bibinfo {title}
  {Machine learning and the physical sciences},\ }\href
  {https://doi.org/10.1103/RevModPhys.91.045002} {\bibfield  {journal}
  {\bibinfo  {journal} {Rev. Mod. Phys.}\ }\textbf {\bibinfo {volume} {91}},\
  \bibinfo {pages} {045002} (\bibinfo {year} {2019})}\BibitemShut {NoStop}%
\bibitem [{\citenamefont {Carrasquilla}(2020)}]{Carrasquilla2020}%
  \BibitemOpen
  \bibfield  {author} {\bibinfo {author} {\bibfnamefont {J.}~\bibnamefont
  {Carrasquilla}},\ }\bibfield  {title} {\bibinfo {title} {Machine learning for
  quantum matter},\ }\href {https://doi.org/10.1080/23746149.2020.1797528}
  {\bibfield  {journal} {\bibinfo  {journal} {Advances in Physics: X}\ }\textbf
  {\bibinfo {volume} {5}},\ \bibinfo {pages} {1797528} (\bibinfo {year}
  {2020})}\BibitemShut {NoStop}%
\bibitem [{\citenamefont {Feickert}\ and\ \citenamefont
  {Nachman}(2021)}]{FeickertPreprint}%
  \BibitemOpen
  \bibfield  {author} {\bibinfo {author} {\bibfnamefont {M.}~\bibnamefont
  {Feickert}}\ and\ \bibinfo {author} {\bibfnamefont {B.}~\bibnamefont
  {Nachman}},\ }\bibfield  {title} {\bibinfo {title} {A living review of
  machine learning for particle physics},\ }\href
  {https://arxiv.org/abs/2102.02770} {\bibfield  {journal} {\bibinfo  {journal}
  {arXiv:2102.02770}\ } (\bibinfo {year} {2021})}\BibitemShut {NoStop}%
\bibitem [{\citenamefont {Karniadakis}\ \emph {et~al.}(2021)\citenamefont
  {Karniadakis}, \citenamefont {Kevrekidis}, \citenamefont {Lu}, \citenamefont
  {Perdikaris}, \citenamefont {Wang},\ and\ \citenamefont
  {Yang}}]{Karniadakis2021}%
  \BibitemOpen
  \bibfield  {author} {\bibinfo {author} {\bibfnamefont {G.~E.}\ \bibnamefont
  {Karniadakis}}, \bibinfo {author} {\bibfnamefont {I.~G.}\ \bibnamefont
  {Kevrekidis}}, \bibinfo {author} {\bibfnamefont {L.}~\bibnamefont {Lu}},
  \bibinfo {author} {\bibfnamefont {P.}~\bibnamefont {Perdikaris}}, \bibinfo
  {author} {\bibfnamefont {S.}~\bibnamefont {Wang}},\ and\ \bibinfo {author}
  {\bibfnamefont {L.}~\bibnamefont {Yang}},\ }\bibfield  {title} {\bibinfo
  {title} {Physics-informed machine learning},\ }\href
  {https://doi.org/10.1038/s42254-021-00314-5} {\bibfield  {journal} {\bibinfo
  {journal} {Nature Reviews Physics}\ }\textbf {\bibinfo {volume} {3}},\
  \bibinfo {pages} {422} (\bibinfo {year} {2021})}\BibitemShut {NoStop}%
\bibitem [{\citenamefont {Chen}\ \emph
  {et~al.}(2021{\natexlab{b}})\citenamefont {Chen}, \citenamefont {Andrejevic},
  \citenamefont {Drucker}, \citenamefont {Nguyen}, \citenamefont {Xian},
  \citenamefont {Smidt}, \citenamefont {Wang}, \citenamefont {Ernstorfer},
  \citenamefont {Tennant}, \citenamefont {Chan},\ and\ \citenamefont
  {Li}}]{ChenReview2021}%
  \BibitemOpen
  \bibfield  {author} {\bibinfo {author} {\bibfnamefont {Z.}~\bibnamefont
  {Chen}}, \bibinfo {author} {\bibfnamefont {N.}~\bibnamefont {Andrejevic}},
  \bibinfo {author} {\bibfnamefont {N.~C.}\ \bibnamefont {Drucker}}, \bibinfo
  {author} {\bibfnamefont {T.}~\bibnamefont {Nguyen}}, \bibinfo {author}
  {\bibfnamefont {R.~P.}\ \bibnamefont {Xian}}, \bibinfo {author}
  {\bibfnamefont {T.}~\bibnamefont {Smidt}}, \bibinfo {author} {\bibfnamefont
  {Y.}~\bibnamefont {Wang}}, \bibinfo {author} {\bibfnamefont {R.}~\bibnamefont
  {Ernstorfer}}, \bibinfo {author} {\bibfnamefont {D.~A.}\ \bibnamefont
  {Tennant}}, \bibinfo {author} {\bibfnamefont {M.}~\bibnamefont {Chan}},\ and\
  \bibinfo {author} {\bibfnamefont {M.}~\bibnamefont {Li}},\ }\bibfield
  {title} {\bibinfo {title} {Machine learning on neutron and x-ray scattering
  and spectroscopies},\ }\href {https://doi.org/10.1063/5.0049111} {\bibfield
  {journal} {\bibinfo  {journal} {Chemical Physics Reviews}\ }\textbf {\bibinfo
  {volume} {2}},\ \bibinfo {pages} {031301} (\bibinfo {year}
  {2021}{\natexlab{b}})}\BibitemShut {NoStop}%
\bibitem [{\citenamefont {Roberts}\ \emph {et~al.}(2022)\citenamefont
  {Roberts}, \citenamefont {Yaida},\ and\ \citenamefont {Hanin}}]{PDLT-2022}%
  \BibitemOpen
  \bibfield  {author} {\bibinfo {author} {\bibfnamefont {D.~A.}\ \bibnamefont
  {Roberts}}, \bibinfo {author} {\bibfnamefont {S.}~\bibnamefont {Yaida}},\
  and\ \bibinfo {author} {\bibfnamefont {B.}~\bibnamefont {Hanin}},\
  }\href@noop {} {\emph {\bibinfo {title} {The Principles of Deep Learning
  Theory}}}\ (\bibinfo  {publisher} {Cambridge University Press},\ \bibinfo
  {year} {2022})\ \bibinfo {note} {\url{https://deeplearningtheory.com}},\
  \Eprint {https://arxiv.org/abs/2106.10165} {arXiv:2106.10165 [cs.LG]}
  \BibitemShut {NoStop}%
\bibitem [{\citenamefont {Dawid}\ \emph {et~al.}(2022)\citenamefont {Dawid},
  \citenamefont {Arnold}, \citenamefont {Requena}, \citenamefont {Gresch},
  \citenamefont {Płodzień}, \citenamefont {Donatella}, \citenamefont
  {Nicoli}, \citenamefont {Stornati}, \citenamefont {Koch}, \citenamefont
  {Büttner}, \citenamefont {Okuła}, \citenamefont {Muñoz-Gil}, \citenamefont
  {Vargas-Hernández}, \citenamefont {Cervera-Lierta}, \citenamefont
  {Carrasquilla}, \citenamefont {Dunjko}, \citenamefont {Gabrié},
  \citenamefont {Huembeli}, \citenamefont {van Nieuwenburg}, \citenamefont
  {Vicentini}, \citenamefont {Wang}, \citenamefont {Wetzel}, \citenamefont
  {Carleo}, \citenamefont {Greplová}, \citenamefont {Krems}, \citenamefont
  {Marquardt}, \citenamefont {Tomza}, \citenamefont {Lewenstein},\ and\
  \citenamefont {Dauphin}}]{DawidPreprint}%
  \BibitemOpen
  \bibfield  {author} {\bibinfo {author} {\bibfnamefont {A.}~\bibnamefont
  {Dawid}}, \bibinfo {author} {\bibfnamefont {J.}~\bibnamefont {Arnold}},
  \bibinfo {author} {\bibfnamefont {B.}~\bibnamefont {Requena}}, \bibinfo
  {author} {\bibfnamefont {A.}~\bibnamefont {Gresch}}, \bibinfo {author}
  {\bibfnamefont {M.}~\bibnamefont {Płodzień}}, \bibinfo {author}
  {\bibfnamefont {K.}~\bibnamefont {Donatella}}, \bibinfo {author}
  {\bibfnamefont {K.~A.}\ \bibnamefont {Nicoli}}, \bibinfo {author}
  {\bibfnamefont {P.}~\bibnamefont {Stornati}}, \bibinfo {author}
  {\bibfnamefont {R.}~\bibnamefont {Koch}}, \bibinfo {author} {\bibfnamefont
  {M.}~\bibnamefont {Büttner}}, \bibinfo {author} {\bibfnamefont
  {R.}~\bibnamefont {Okuła}}, \bibinfo {author} {\bibfnamefont
  {G.}~\bibnamefont {Muñoz-Gil}}, \bibinfo {author} {\bibfnamefont {R.~A.}\
  \bibnamefont {Vargas-Hernández}}, \bibinfo {author} {\bibfnamefont
  {A.}~\bibnamefont {Cervera-Lierta}}, \bibinfo {author} {\bibfnamefont
  {J.}~\bibnamefont {Carrasquilla}}, \bibinfo {author} {\bibfnamefont
  {V.}~\bibnamefont {Dunjko}}, \bibinfo {author} {\bibfnamefont
  {M.}~\bibnamefont {Gabrié}}, \bibinfo {author} {\bibfnamefont
  {P.}~\bibnamefont {Huembeli}}, \bibinfo {author} {\bibfnamefont
  {E.}~\bibnamefont {van Nieuwenburg}}, \bibinfo {author} {\bibfnamefont
  {F.}~\bibnamefont {Vicentini}}, \bibinfo {author} {\bibfnamefont
  {L.}~\bibnamefont {Wang}}, \bibinfo {author} {\bibfnamefont {S.~J.}\
  \bibnamefont {Wetzel}}, \bibinfo {author} {\bibfnamefont {G.}~\bibnamefont
  {Carleo}}, \bibinfo {author} {\bibfnamefont {E.}~\bibnamefont {Greplová}},
  \bibinfo {author} {\bibfnamefont {R.}~\bibnamefont {Krems}}, \bibinfo
  {author} {\bibfnamefont {F.}~\bibnamefont {Marquardt}}, \bibinfo {author}
  {\bibfnamefont {M.}~\bibnamefont {Tomza}}, \bibinfo {author} {\bibfnamefont
  {M.}~\bibnamefont {Lewenstein}},\ and\ \bibinfo {author} {\bibfnamefont
  {A.}~\bibnamefont {Dauphin}},\ }\bibfield  {title} {\bibinfo {title} {Modern
  applications of machine learning in quantum sciences},\ }\href
  {https://arxiv.org/abs/2204.04198} {\bibfield  {journal} {\bibinfo  {journal}
  {arXiv:2204.04198}\ } (\bibinfo {year} {2022})}\BibitemShut {NoStop}%
\bibitem [{\citenamefont {Onsager}(1944)}]{Onsager}%
  \BibitemOpen
  \bibfield  {author} {\bibinfo {author} {\bibfnamefont {L.}~\bibnamefont
  {Onsager}},\ }\bibfield  {title} {\bibinfo {title} {Crystal statistics.
  {{I}}. {{A}} two-dimensional model with an order-disorder transition},\
  }\href {https://doi.org/10.1103/PhysRev.65.117} {\bibfield  {journal}
  {\bibinfo  {journal} {Phys. Rev.}\ }\textbf {\bibinfo {volume} {65}},\
  \bibinfo {pages} {117} (\bibinfo {year} {1944})}\BibitemShut {NoStop}%
\bibitem [{\citenamefont {Liu}\ \emph {et~al.}(2017{\natexlab{a}})\citenamefont
  {Liu}, \citenamefont {Qi}, \citenamefont {Meng},\ and\ \citenamefont
  {Fu}}]{SLMC}%
  \BibitemOpen
  \bibfield  {author} {\bibinfo {author} {\bibfnamefont {J.}~\bibnamefont
  {Liu}}, \bibinfo {author} {\bibfnamefont {Y.}~\bibnamefont {Qi}}, \bibinfo
  {author} {\bibfnamefont {Z.~Y.}\ \bibnamefont {Meng}},\ and\ \bibinfo
  {author} {\bibfnamefont {L.}~\bibnamefont {Fu}},\ }\bibfield  {title}
  {\bibinfo {title} {Self-learning {Monte Carlo} method},\ }\href
  {https://doi.org/10.1103/PhysRevB.95.041101} {\bibfield  {journal} {\bibinfo
  {journal} {Phys. Rev. B}\ }\textbf {\bibinfo {volume} {95}},\ \bibinfo
  {pages} {041101} (\bibinfo {year} {2017}{\natexlab{a}})}\BibitemShut
  {NoStop}%
\bibitem [{\citenamefont {Wetzel}\ and\ \citenamefont
  {Scherzer}(2017)}]{wetzel2017machine}%
  \BibitemOpen
  \bibfield  {author} {\bibinfo {author} {\bibfnamefont {S.~J.}\ \bibnamefont
  {Wetzel}}\ and\ \bibinfo {author} {\bibfnamefont {M.}~\bibnamefont
  {Scherzer}},\ }\bibfield  {title} {\bibinfo {title} {Machine learning of
  explicit order parameters: From the {Ising} model to {SU(2)} lattice gauge
  theory},\ }\href {https://doi.org/10.1103/PhysRevB.96.184410} {\bibfield
  {journal} {\bibinfo  {journal} {Phys. Rev. B}\ }\textbf {\bibinfo {volume}
  {96}},\ \bibinfo {pages} {184410} (\bibinfo {year} {2017})}\BibitemShut
  {NoStop}%
\bibitem [{\citenamefont {Kim}\ and\ \citenamefont
  {Kim}(2018)}]{kim2018smallest}%
  \BibitemOpen
  \bibfield  {author} {\bibinfo {author} {\bibfnamefont {D.}~\bibnamefont
  {Kim}}\ and\ \bibinfo {author} {\bibfnamefont {D.-H.}\ \bibnamefont {Kim}},\
  }\bibfield  {title} {\bibinfo {title} {Smallest neural network to learn the
  {Ising} criticality},\ }\href {https://doi.org/10.1103/PhysRevE.98.022138}
  {\bibfield  {journal} {\bibinfo  {journal} {Phys. Rev. E}\ }\textbf {\bibinfo
  {volume} {98}},\ \bibinfo {pages} {022138} (\bibinfo {year}
  {2018})}\BibitemShut {NoStop}%
\bibitem [{\citenamefont {Morningstar}\ and\ \citenamefont
  {Melko}(2018)}]{Morningstar2018}%
  \BibitemOpen
  \bibfield  {author} {\bibinfo {author} {\bibfnamefont {A.}~\bibnamefont
  {Morningstar}}\ and\ \bibinfo {author} {\bibfnamefont {R.~G.}\ \bibnamefont
  {Melko}},\ }\bibfield  {title} {\bibinfo {title} {Deep learning the {Ising}
  model near criticality},\ }\href {http://jmlr.org/papers/v18/17-527.html}
  {\bibfield  {journal} {\bibinfo  {journal} {J. Mach. Learn. Res.}\ }\textbf
  {\bibinfo {volume} {18}},\ \bibinfo {pages} {1} (\bibinfo {year}
  {2018})}\BibitemShut {NoStop}%
\bibitem [{\citenamefont {Alexandrou}\ \emph {et~al.}(2020)\citenamefont
  {Alexandrou}, \citenamefont {Athenodorou}, \citenamefont {Chrysostomou},\
  and\ \citenamefont {Paul}}]{alexandrou2020critical}%
  \BibitemOpen
  \bibfield  {author} {\bibinfo {author} {\bibfnamefont {C.}~\bibnamefont
  {Alexandrou}}, \bibinfo {author} {\bibfnamefont {A.}~\bibnamefont
  {Athenodorou}}, \bibinfo {author} {\bibfnamefont {C.}~\bibnamefont
  {Chrysostomou}},\ and\ \bibinfo {author} {\bibfnamefont {S.}~\bibnamefont
  {Paul}},\ }\bibfield  {title} {\bibinfo {title} {{T}he critical temperature
  of the 2{D}-{Ising} model through deep learning autoencoders},\ }\href
  {https://doi.org/10.1140/epjb/e2020-100506-5} {\bibfield  {journal} {\bibinfo
   {journal} {The European Physical Journal B}\ }\textbf {\bibinfo {volume}
  {93}},\ \bibinfo {pages} {1140} (\bibinfo {year} {2020})}\BibitemShut
  {NoStop}%
\bibitem [{\citenamefont {Yevick}(2021)}]{yevick2021variational}%
  \BibitemOpen
  \bibfield  {author} {\bibinfo {author} {\bibfnamefont {D.}~\bibnamefont
  {Yevick}},\ }\bibfield  {title} {\bibinfo {title} {Variational autoencoder
  analysis of {Ising} model statistical distributions and phase transitions},\
  }\href {https://arxiv.org/abs/2104.06368} {\bibfield  {journal} {\bibinfo
  {journal} {arXiv:2104.06368}\ } (\bibinfo {year} {2021})}\BibitemShut
  {NoStop}%
\bibitem [{\citenamefont {Agrawal}\ \emph {et~al.}(2022)\citenamefont
  {Agrawal}, \citenamefont {Del~Maestro}, \citenamefont {Johnston},\ and\
  \citenamefont {Ostrowski}}]{Agrawal2022}%
  \BibitemOpen
  \bibfield  {author} {\bibinfo {author} {\bibfnamefont {D.}~\bibnamefont
  {Agrawal}}, \bibinfo {author} {\bibfnamefont {A.}~\bibnamefont
  {Del~Maestro}}, \bibinfo {author} {\bibfnamefont {S.}~\bibnamefont
  {Johnston}},\ and\ \bibinfo {author} {\bibfnamefont {J.}~\bibnamefont
  {Ostrowski}},\ }\bibfield  {title} {\bibinfo {title} {A group-equivariant
  autoencoder for identifying spontaneously broken symmetries in the {Ising}
  model},\ }\href {https://doi.org/10.48550/arXiv.2202.06319} {\bibfield
  {journal} {\bibinfo  {journal} {arXiv:2202.06319}\ } (\bibinfo {year}
  {2022})}\BibitemShut {NoStop}%
\bibitem [{\citenamefont {Stephenson}(1964)}]{stephenson1964ising}%
  \BibitemOpen
  \bibfield  {author} {\bibinfo {author} {\bibfnamefont {J.}~\bibnamefont
  {Stephenson}},\ }\bibfield  {title} {\bibinfo {title} {Ising-model spin
  correlations on the triangular lattice},\ }\href
  {https://aip.scitation.org/doi/abs/10.1063/1.1704202} {\bibfield  {journal}
  {\bibinfo  {journal} {Journal of Mathematical Physics}\ }\textbf {\bibinfo
  {volume} {5}},\ \bibinfo {pages} {1009} (\bibinfo {year} {1964})}\BibitemShut
  {NoStop}%
\bibitem [{\citenamefont {Landau}(1983)}]{landau1983critical}%
  \BibitemOpen
  \bibfield  {author} {\bibinfo {author} {\bibfnamefont {D.}~\bibnamefont
  {Landau}},\ }\bibfield  {title} {\bibinfo {title} {Critical and multicritical
  behavior in a triangular-lattice-gas {I}sing model: Repulsive
  nearest-neighbor and attractive next-nearest-neighbor coupling},\ }\href
  {https://journals.aps.org/prb/pdf/10.1103/PhysRevB.27.5604} {\bibfield
  {journal} {\bibinfo  {journal} {Phys. Rev. B}\ }\textbf {\bibinfo {volume}
  {27}},\ \bibinfo {pages} {5604} (\bibinfo {year} {1983})}\BibitemShut
  {NoStop}%
\bibitem [{\citenamefont {Moessner}\ and\ \citenamefont
  {Sondhi}(2001)}]{moessner2001ising}%
  \BibitemOpen
  \bibfield  {author} {\bibinfo {author} {\bibfnamefont {R.}~\bibnamefont
  {Moessner}}\ and\ \bibinfo {author} {\bibfnamefont {S.~L.}\ \bibnamefont
  {Sondhi}},\ }\bibfield  {title} {\bibinfo {title} {Ising models of quantum
  frustration},\ }\href
  {https://journals.aps.org/prb/pdf/10.1103/PhysRevB.63.22440} {\bibfield
  {journal} {\bibinfo  {journal} {Phys. Rev. B}\ }\textbf {\bibinfo {volume}
  {63}},\ \bibinfo {pages} {224401} (\bibinfo {year} {2001})}\BibitemShut
  {NoStop}%
\bibitem [{\citenamefont {Sherrington}\ and\ \citenamefont
  {Kirkpatrick}(1975)}]{sherrington1975solvable}%
  \BibitemOpen
  \bibfield  {author} {\bibinfo {author} {\bibfnamefont {D.}~\bibnamefont
  {Sherrington}}\ and\ \bibinfo {author} {\bibfnamefont {S.}~\bibnamefont
  {Kirkpatrick}},\ }\bibfield  {title} {\bibinfo {title} {Solvable model of a
  spin-glass},\ }\href
  {https://journals.aps.org/prl/pdf/10.1103/PhysRevLett.35.1792} {\bibfield
  {journal} {\bibinfo  {journal} {Phys. Rev. Letters}\ }\textbf {\bibinfo
  {volume} {35}},\ \bibinfo {pages} {1792} (\bibinfo {year}
  {1975})}\BibitemShut {NoStop}%
\bibitem [{\citenamefont {Singh}\ and\ \citenamefont
  {Chakravarty}(1986)}]{singh1986critical}%
  \BibitemOpen
  \bibfield  {author} {\bibinfo {author} {\bibfnamefont {R.~R.}\ \bibnamefont
  {Singh}}\ and\ \bibinfo {author} {\bibfnamefont {S.}~\bibnamefont
  {Chakravarty}},\ }\bibfield  {title} {\bibinfo {title} {Critical behavior of
  an {I}sing spin-glass},\ }\href
  {https://journals.aps.org/prl/pdf/10.1103/PhysRevLett.57.245} {\bibfield
  {journal} {\bibinfo  {journal} {Phys. Rev. Lett.}\ }\textbf {\bibinfo
  {volume} {57}},\ \bibinfo {pages} {245} (\bibinfo {year} {1986})}\BibitemShut
  {NoStop}%
\bibitem [{\citenamefont {McMillan}(1984)}]{mcmillan1984scaling}%
  \BibitemOpen
  \bibfield  {author} {\bibinfo {author} {\bibfnamefont {W.}~\bibnamefont
  {McMillan}},\ }\bibfield  {title} {\bibinfo {title} {Scaling theory of
  {I}sing spin glasses},\ }\href
  {https://iopscience.iop.org/article/10.1088/0022-3719/17/18/010/pdf}
  {\bibfield  {journal} {\bibinfo  {journal} {Journal of Physics C: Solid State
  Physics}\ }\textbf {\bibinfo {volume} {17}},\ \bibinfo {pages} {3179}
  (\bibinfo {year} {1984})}\BibitemShut {NoStop}%
\bibitem [{\citenamefont {Blume}(1966)}]{Blume}%
  \BibitemOpen
  \bibfield  {author} {\bibinfo {author} {\bibfnamefont {M.}~\bibnamefont
  {Blume}},\ }\bibfield  {title} {\bibinfo {title} {Theory of the first-order
  magnetic phase change in {U${\mathrm{O}}_{2}$}},\ }\href
  {https://doi.org/10.1103/PhysRev.141.517} {\bibfield  {journal} {\bibinfo
  {journal} {Phys. Rev.}\ }\textbf {\bibinfo {volume} {141}},\ \bibinfo {pages}
  {517} (\bibinfo {year} {1966})}\BibitemShut {NoStop}%
\bibitem [{\citenamefont {Capel}(1966)}]{Capel}%
  \BibitemOpen
  \bibfield  {author} {\bibinfo {author} {\bibfnamefont {H.}~\bibnamefont
  {Capel}},\ }\bibfield  {title} {\bibinfo {title} {On the possibility of
  first-order phase transitions in {Ising} systems of triplet ions with
  zero-field splitting},\ }\href
  {https://doi.org/https://doi.org/10.1016/0031-8914(66)90027-9} {\bibfield
  {journal} {\bibinfo  {journal} {Physica}\ }\textbf {\bibinfo {volume} {32}},\
  \bibinfo {pages} {966} (\bibinfo {year} {1966})}\BibitemShut {NoStop}%
\bibitem [{\citenamefont {Hubbard}\ and\ \citenamefont
  {Flowers}(1963)}]{Hubbard}%
  \BibitemOpen
  \bibfield  {author} {\bibinfo {author} {\bibfnamefont {J.}~\bibnamefont
  {Hubbard}}\ and\ \bibinfo {author} {\bibfnamefont {B.~H.}\ \bibnamefont
  {Flowers}},\ }\bibfield  {title} {\bibinfo {title} {Electron correlations in
  narrow energy bands},\ }\href {https://doi.org/10.1098/rspa.1963.0204}
  {\bibfield  {journal} {\bibinfo  {journal} {Proceedings of the Royal Society
  of London. Series A. Mathematical and Physical Sciences}\ }\textbf {\bibinfo
  {volume} {276}},\ \bibinfo {pages} {238} (\bibinfo {year}
  {1963})}\BibitemShut {NoStop}%
\bibitem [{\citenamefont {Lieb}\ and\ \citenamefont {Wu}(1968)}]{LiebPRL1968}%
  \BibitemOpen
  \bibfield  {author} {\bibinfo {author} {\bibfnamefont {E.~H.}\ \bibnamefont
  {Lieb}}\ and\ \bibinfo {author} {\bibfnamefont {F.~Y.}\ \bibnamefont {Wu}},\
  }\bibfield  {title} {\bibinfo {title} {Absence of {Mott} transition in an
  exact solution of the short-range, one-band model in one dimension},\ }\href
  {https://doi.org/10.1103/PhysRevLett.21.192.2} {\bibfield  {journal}
  {\bibinfo  {journal} {Phys. Rev. Lett.}\ }\textbf {\bibinfo {volume} {21}},\
  \bibinfo {pages} {192} (\bibinfo {year} {1968})}\BibitemShut {NoStop}%
\bibitem [{\citenamefont {Sordi}\ \emph {et~al.}(2012)\citenamefont {Sordi},
  \citenamefont {S\'emon}, \citenamefont {Haule},\ and\ \citenamefont
  {Tremblay}}]{SordiPRL2012}%
  \BibitemOpen
  \bibfield  {author} {\bibinfo {author} {\bibfnamefont {G.}~\bibnamefont
  {Sordi}}, \bibinfo {author} {\bibfnamefont {P.}~\bibnamefont {S\'emon}},
  \bibinfo {author} {\bibfnamefont {K.}~\bibnamefont {Haule}},\ and\ \bibinfo
  {author} {\bibfnamefont {A.-M.~S.}\ \bibnamefont {Tremblay}},\ }\bibfield
  {title} {\bibinfo {title} {Strong coupling superconductivity, pseudogap, and
  {Mott} transition},\ }\href
  {https://link.aps.org/doi/10.1103/PhysRevLett.108.216401} {\bibfield
  {journal} {\bibinfo  {journal} {Phys. Rev. Lett.}\ }\textbf {\bibinfo
  {volume} {108}},\ \bibinfo {pages} {216401} (\bibinfo {year}
  {2012})}\BibitemShut {NoStop}%
\bibitem [{\citenamefont {Peters}\ and\ \citenamefont
  {Kawakami}(2014)}]{Robert2014}%
  \BibitemOpen
  \bibfield  {author} {\bibinfo {author} {\bibfnamefont {R.}~\bibnamefont
  {Peters}}\ and\ \bibinfo {author} {\bibfnamefont {N.}~\bibnamefont
  {Kawakami}},\ }\bibfield  {title} {\bibinfo {title} {Spin density waves in
  the {{Hubbard}} model: A {DMFT} approach},\ }\href
  {https://doi.org/10.1103/PhysRevB.89.155134} {\bibfield  {journal} {\bibinfo
  {journal} {Phys. Rev. B}\ }\textbf {\bibinfo {volume} {89}},\ \bibinfo
  {pages} {155134} (\bibinfo {year} {2014})}\BibitemShut {NoStop}%
\bibitem [{\citenamefont {Zheng}\ \emph
  {et~al.}(2017{\natexlab{b}})\citenamefont {Zheng}, \citenamefont {Chung},
  \citenamefont {Corboz}, \citenamefont {Ehlers}, \citenamefont {Qin},
  \citenamefont {Noack}, \citenamefont {Shi}, \citenamefont {White},
  \citenamefont {Zhang},\ and\ \citenamefont {Chan}}]{ZhengScience2017}%
  \BibitemOpen
  \bibfield  {author} {\bibinfo {author} {\bibfnamefont {B.-X.}\ \bibnamefont
  {Zheng}}, \bibinfo {author} {\bibfnamefont {C.-M.}\ \bibnamefont {Chung}},
  \bibinfo {author} {\bibfnamefont {P.}~\bibnamefont {Corboz}}, \bibinfo
  {author} {\bibfnamefont {G.}~\bibnamefont {Ehlers}}, \bibinfo {author}
  {\bibfnamefont {M.-P.}\ \bibnamefont {Qin}}, \bibinfo {author} {\bibfnamefont
  {R.~M.}\ \bibnamefont {Noack}}, \bibinfo {author} {\bibfnamefont
  {H.}~\bibnamefont {Shi}}, \bibinfo {author} {\bibfnamefont {S.~R.}\
  \bibnamefont {White}}, \bibinfo {author} {\bibfnamefont {S.}~\bibnamefont
  {Zhang}},\ and\ \bibinfo {author} {\bibfnamefont {G.~K.-L.}\ \bibnamefont
  {Chan}},\ }\bibfield  {title} {\bibinfo {title} {Stripe order in the
  underdoped region of the two-dimensional {Hubbard} model},\ }\href
  {https://science.sciencemag.org/content/358/6367/1155} {\bibfield  {journal}
  {\bibinfo  {journal} {Science}\ }\textbf {\bibinfo {volume} {358}},\ \bibinfo
  {pages} {1155} (\bibinfo {year} {2017}{\natexlab{b}})}\BibitemShut {NoStop}%
\bibitem [{\citenamefont {Huang}\ \emph {et~al.}(2019)\citenamefont {Huang},
  \citenamefont {Sheppard}, \citenamefont {Moritz},\ and\ \citenamefont
  {Devereaux}}]{HuangScience2019}%
  \BibitemOpen
  \bibfield  {author} {\bibinfo {author} {\bibfnamefont {E.~W.}\ \bibnamefont
  {Huang}}, \bibinfo {author} {\bibfnamefont {R.}~\bibnamefont {Sheppard}},
  \bibinfo {author} {\bibfnamefont {B.}~\bibnamefont {Moritz}},\ and\ \bibinfo
  {author} {\bibfnamefont {T.~P.}\ \bibnamefont {Devereaux}},\ }\bibfield
  {title} {\bibinfo {title} {Strange metallicity in the doped {Hubbard}
  model},\ }\href {https://doi.org/10.1126/science.aau7063} {\bibfield
  {journal} {\bibinfo  {journal} {Science}\ }\textbf {\bibinfo {volume}
  {366}},\ \bibinfo {pages} {987} (\bibinfo {year} {2019})}\BibitemShut
  {NoStop}%
\bibitem [{\citenamefont {Gull}\ \emph {et~al.}(2013)\citenamefont {Gull},
  \citenamefont {Parcollet},\ and\ \citenamefont {Millis}}]{GullPRL2013}%
  \BibitemOpen
  \bibfield  {author} {\bibinfo {author} {\bibfnamefont {E.}~\bibnamefont
  {Gull}}, \bibinfo {author} {\bibfnamefont {O.}~\bibnamefont {Parcollet}},\
  and\ \bibinfo {author} {\bibfnamefont {A.~J.}\ \bibnamefont {Millis}},\
  }\bibfield  {title} {\bibinfo {title} {Superconductivity and the pseudogap in
  the two-dimensional {Hubbard} model},\ }\href
  {https://doi.org/10.1103/PhysRevLett.110.216405} {\bibfield  {journal}
  {\bibinfo  {journal} {Phys. Rev. Lett.}\ }\textbf {\bibinfo {volume} {110}},\
  \bibinfo {pages} {216405} (\bibinfo {year} {2013})}\BibitemShut {NoStop}%
\bibitem [{\citenamefont {Chung}\ \emph {et~al.}(2020)\citenamefont {Chung},
  \citenamefont {Qin}, \citenamefont {Zhang}, \citenamefont {Schollw\"ock},\
  and\ \citenamefont {White}}]{ChungPRB2020}%
  \BibitemOpen
  \bibfield  {author} {\bibinfo {author} {\bibfnamefont {C.-M.}\ \bibnamefont
  {Chung}}, \bibinfo {author} {\bibfnamefont {M.}~\bibnamefont {Qin}}, \bibinfo
  {author} {\bibfnamefont {S.}~\bibnamefont {Zhang}}, \bibinfo {author}
  {\bibfnamefont {U.}~\bibnamefont {Schollw\"ock}},\ and\ \bibinfo {author}
  {\bibfnamefont {S.~R.}\ \bibnamefont {White}} (\bibinfo {collaboration} {The
  Simons Collaboration on the Many-Electron Problem}),\ }\bibfield  {title}
  {\bibinfo {title} {Plaquette versus ordinary $d$-wave pairing in the
  ${t}^\prime$-{{Hubbard}} model on a width-4 cylinder},\ }\href
  {https://link.aps.org/doi/10.1103/PhysRevB.102.041106} {\bibfield  {journal}
  {\bibinfo  {journal} {Phys. Rev. B}\ }\textbf {\bibinfo {volume} {102}},\
  \bibinfo {pages} {041106} (\bibinfo {year} {2020})}\BibitemShut {NoStop}%
\bibitem [{\citenamefont {Mai}\ \emph {et~al.}(2021)\citenamefont {Mai},
  \citenamefont {Balduzzi}, \citenamefont {Johnston},\ and\ \citenamefont
  {Maier}}]{Mai2021}%
  \BibitemOpen
  \bibfield  {author} {\bibinfo {author} {\bibfnamefont {P.}~\bibnamefont
  {Mai}}, \bibinfo {author} {\bibfnamefont {G.}~\bibnamefont {Balduzzi}},
  \bibinfo {author} {\bibfnamefont {S.}~\bibnamefont {Johnston}},\ and\
  \bibinfo {author} {\bibfnamefont {T.}~\bibnamefont {Maier}},\ }\bibfield
  {title} {\bibinfo {title} {Orbital structure of the effective pairing
  interaction in the high-temperature superconducting cuprates},\ }\href
  {https://doi.org/10.1038/s41535-021-00326-5} {\bibfield  {journal} {\bibinfo
  {journal} {npj Quantum Mater.}\ }\textbf {\bibinfo {volume} {6}},\ \bibinfo
  {pages} {26} (\bibinfo {year} {2021})}\BibitemShut {NoStop}%
\bibitem [{\citenamefont {Holstein}(1959)}]{Holstein}%
  \BibitemOpen
  \bibfield  {author} {\bibinfo {author} {\bibfnamefont {T.}~\bibnamefont
  {Holstein}},\ }\bibfield  {title} {\bibinfo {title} {Studies of polaron
  motion: Part {I}. the molecular-crystal model},\ }\href
  {https://doi.org/10.1016/0003-4916(59)90002-8} {\bibfield  {journal}
  {\bibinfo  {journal} {Annals of physics}\ }\textbf {\bibinfo {volume} {8}},\
  \bibinfo {pages} {325} (\bibinfo {year} {1959})}\BibitemShut {NoStop}%
\bibitem [{\citenamefont {Scalettar}\ \emph {et~al.}(1991)\citenamefont
  {Scalettar}, \citenamefont {Noack},\ and\ \citenamefont
  {Singh}}]{ScalettarPRB1991}%
  \BibitemOpen
  \bibfield  {author} {\bibinfo {author} {\bibfnamefont {R.~T.}\ \bibnamefont
  {Scalettar}}, \bibinfo {author} {\bibfnamefont {R.~M.}\ \bibnamefont
  {Noack}},\ and\ \bibinfo {author} {\bibfnamefont {R.~R.~P.}\ \bibnamefont
  {Singh}},\ }\bibfield  {title} {\bibinfo {title} {Ergodicity at large
  couplings with the determinant {Monte Carlo} algorithm},\ }\href
  {https://doi.org/10.1103/PhysRevB.44.10502} {\bibfield  {journal} {\bibinfo
  {journal} {Phys. Rev. B}\ }\textbf {\bibinfo {volume} {44}},\ \bibinfo
  {pages} {10502} (\bibinfo {year} {1991})}\BibitemShut {NoStop}%
\bibitem [{\citenamefont {Dee}\ \emph {et~al.}(2019)\citenamefont {Dee},
  \citenamefont {Nakatsukasa}, \citenamefont {Wang},\ and\ \citenamefont
  {Johnston}}]{DeePRB2019}%
  \BibitemOpen
  \bibfield  {author} {\bibinfo {author} {\bibfnamefont {P.~M.}\ \bibnamefont
  {Dee}}, \bibinfo {author} {\bibfnamefont {K.}~\bibnamefont {Nakatsukasa}},
  \bibinfo {author} {\bibfnamefont {Y.}~\bibnamefont {Wang}},\ and\ \bibinfo
  {author} {\bibfnamefont {S.}~\bibnamefont {Johnston}},\ }\bibfield  {title}
  {\bibinfo {title} {Temperature-filling phase diagram of the two-dimensional
  {Holstein} model in the thermodynamic limit by self-consistent {Migdal}
  approximation},\ }\href {https://doi.org/10.1103/PhysRevB.99.024514}
  {\bibfield  {journal} {\bibinfo  {journal} {Phys. Rev. B}\ }\textbf {\bibinfo
  {volume} {99}},\ \bibinfo {pages} {024514} (\bibinfo {year}
  {2019})}\BibitemShut {NoStop}%
\bibitem [{\citenamefont {Esterlis}\ \emph {et~al.}(2018)\citenamefont
  {Esterlis}, \citenamefont {Nosarzewski}, \citenamefont {Huang}, \citenamefont
  {Moritz}, \citenamefont {Devereaux}, \citenamefont {Scalapino},\ and\
  \citenamefont {Kivelson}}]{Esterlis2018}%
  \BibitemOpen
  \bibfield  {author} {\bibinfo {author} {\bibfnamefont {I.}~\bibnamefont
  {Esterlis}}, \bibinfo {author} {\bibfnamefont {B.}~\bibnamefont
  {Nosarzewski}}, \bibinfo {author} {\bibfnamefont {E.~W.}\ \bibnamefont
  {Huang}}, \bibinfo {author} {\bibfnamefont {B.}~\bibnamefont {Moritz}},
  \bibinfo {author} {\bibfnamefont {T.~P.}\ \bibnamefont {Devereaux}}, \bibinfo
  {author} {\bibfnamefont {D.~J.}\ \bibnamefont {Scalapino}},\ and\ \bibinfo
  {author} {\bibfnamefont {S.~A.}\ \bibnamefont {Kivelson}},\ }\bibfield
  {title} {\bibinfo {title} {Breakdown of the {Migdal}-{Eliashberg} theory: A
  determinant quantum {Monte Carlo} study},\ }\href
  {https://doi.org/10.1103/PhysRevB.97.140501} {\bibfield  {journal} {\bibinfo
  {journal} {Phys. Rev. B}\ }\textbf {\bibinfo {volume} {97}},\ \bibinfo
  {pages} {140501} (\bibinfo {year} {2018})}\BibitemShut {NoStop}%
\bibitem [{\citenamefont {Cohen-Stead}\ \emph {et~al.}(2022)\citenamefont
  {Cohen-Stead}, \citenamefont {Bradley}, \citenamefont {Miles}, \citenamefont
  {Batrouni}, \citenamefont {Scalettar},\ and\ \citenamefont
  {Barros}}]{CohenSteadPRE2022}%
  \BibitemOpen
  \bibfield  {author} {\bibinfo {author} {\bibfnamefont {B.}~\bibnamefont
  {Cohen-Stead}}, \bibinfo {author} {\bibfnamefont {O.}~\bibnamefont
  {Bradley}}, \bibinfo {author} {\bibfnamefont {C.}~\bibnamefont {Miles}},
  \bibinfo {author} {\bibfnamefont {G.}~\bibnamefont {Batrouni}}, \bibinfo
  {author} {\bibfnamefont {R.}~\bibnamefont {Scalettar}},\ and\ \bibinfo
  {author} {\bibfnamefont {K.}~\bibnamefont {Barros}},\ }\bibfield  {title}
  {\bibinfo {title} {Fast and scalable quantum monte carlo simulations of
  electron-phonon models},\ }\href
  {https://doi.org/10.1103/PhysRevE.105.065302} {\bibfield  {journal} {\bibinfo
   {journal} {Phys. Rev. E}\ }\textbf {\bibinfo {volume} {105}},\ \bibinfo
  {pages} {065302} (\bibinfo {year} {2022})}\BibitemShut {NoStop}%
\bibitem [{\citenamefont {Gubernatis}\ \emph {et~al.}(2016)\citenamefont
  {Gubernatis}, \citenamefont {Kawashima},\ and\ \citenamefont
  {Werner}}]{gubernatis_kawashima_werner_2016}%
  \BibitemOpen
  \bibfield  {author} {\bibinfo {author} {\bibfnamefont {J.}~\bibnamefont
  {Gubernatis}}, \bibinfo {author} {\bibfnamefont {N.}~\bibnamefont
  {Kawashima}},\ and\ \bibinfo {author} {\bibfnamefont {P.}~\bibnamefont
  {Werner}},\ }\href {https://doi.org/10.1017/CBO9780511902581} {\emph
  {\bibinfo {title} {Quantum {Monte Carlo} Methods: Algorithms for Lattice
  Models}}}\ (\bibinfo  {publisher} {Cambridge University Press},\ \bibinfo
  {year} {2016})\BibitemShut {NoStop}%
\bibitem [{\citenamefont {Becca}\ and\ \citenamefont
  {Sorella}(2017)}]{becca_sorella_2017}%
  \BibitemOpen
  \bibfield  {author} {\bibinfo {author} {\bibfnamefont {F.}~\bibnamefont
  {Becca}}\ and\ \bibinfo {author} {\bibfnamefont {S.}~\bibnamefont
  {Sorella}},\ }\href {https://doi.org/10.1017/9781316417041} {\emph {\bibinfo
  {title} {Quantum {Monte Carlo} Approaches for Correlated Systems}}}\
  (\bibinfo  {publisher} {Cambridge University Press},\ \bibinfo {year}
  {2017})\BibitemShut {NoStop}%
\bibitem [{\citenamefont {Hastings}(1970)}]{Hastings}%
  \BibitemOpen
  \bibfield  {author} {\bibinfo {author} {\bibfnamefont {W.~K.}\ \bibnamefont
  {Hastings}},\ }\bibfield  {title} {\bibinfo {title} {{{Monte Carlo} sampling
  methods using Markov chains and their applications}},\ }\href
  {https://doi.org/10.1093/biomet/57.1.97} {\bibfield  {journal} {\bibinfo
  {journal} {Biometrika}\ }\textbf {\bibinfo {volume} {57}},\ \bibinfo {pages}
  {97} (\bibinfo {year} {1970})}\BibitemShut {NoStop}%
\bibitem [{\citenamefont {Binder}\ \emph {et~al.}(1993)\citenamefont {Binder},
  \citenamefont {Heermann}, \citenamefont {Roelofs}, \citenamefont
  {Mallinckrodt},\ and\ \citenamefont {McKay}}]{binder1993monte}%
  \BibitemOpen
  \bibfield  {author} {\bibinfo {author} {\bibfnamefont {K.}~\bibnamefont
  {Binder}}, \bibinfo {author} {\bibfnamefont {D.}~\bibnamefont {Heermann}},
  \bibinfo {author} {\bibfnamefont {L.}~\bibnamefont {Roelofs}}, \bibinfo
  {author} {\bibfnamefont {A.~J.}\ \bibnamefont {Mallinckrodt}},\ and\ \bibinfo
  {author} {\bibfnamefont {S.}~\bibnamefont {McKay}},\ }\bibfield  {title}
  {\bibinfo {title} {{Monte Carlo} simulation in statistical physics},\ }\href
  {https://aip.scitation.org/doi/10.1063/1.4823159} {\bibfield  {journal}
  {\bibinfo  {journal} {Computers in Physics}\ }\textbf {\bibinfo {volume}
  {7}},\ \bibinfo {pages} {156} (\bibinfo {year} {1993})}\BibitemShut {NoStop}%
\bibitem [{\citenamefont {Assaad}(2002)}]{assaad2002quantum}%
  \BibitemOpen
  \bibfield  {author} {\bibinfo {author} {\bibfnamefont {F.~F.}\ \bibnamefont
  {Assaad}},\ }\bibfield  {title} {\bibinfo {title} {Quantum {Monte Carlo}
  methods on lattices: The determinantal approach},\ }\href
  {https://www.researchgate.net/profile/Fakher-Assaad/publication/242448168_Quantum_Monte_Carlo_Methods_on_Lattices_The_Determinantal_Approach/links/54748a190cf2778985abe630/Quantum-Monte-Carlo-Methods-on-Lattices-The-Determinantal-Approach.pdf}
  {\bibfield  {journal} {\bibinfo  {journal} {Quantum Simulations of Complex
  Many-Body Systems: From Theory to Algorithms}\ }\textbf {\bibinfo {volume}
  {10}},\ \bibinfo {pages} {99} (\bibinfo {year} {2002})}\BibitemShut {NoStop}%
\bibitem [{\citenamefont {{Santos, Raimundo R. dos}}(2003)}]{dossantos03}%
  \BibitemOpen
  \bibfield  {author} {\bibinfo {author} {\bibnamefont {{Santos, Raimundo R.
  dos}}},\ }\bibfield  {title} {\bibinfo {title} {{Introduction to quantum
  Monte Carlo simulations for fermionic systems}},\ }\href
  {http://www.scielo.br/scielo.php?script=sci_arttext&pid=S0103-97332003000100003&nrm=iso}
  {\bibfield  {journal} {\bibinfo  {journal} {{Brazilian Journal of Physics}}\
  }\textbf {\bibinfo {volume} {33}},\ \bibinfo {pages} {36} (\bibinfo {year}
  {2003})}\BibitemShut {NoStop}%
\bibitem [{\citenamefont {Assaad}\ and\ \citenamefont
  {Evertz}(2008)}]{assaad2008world}%
  \BibitemOpen
  \bibfield  {author} {\bibinfo {author} {\bibfnamefont {F.}~\bibnamefont
  {Assaad}}\ and\ \bibinfo {author} {\bibfnamefont {H.}~\bibnamefont
  {Evertz}},\ }\bibfield  {title} {\bibinfo {title} {World-line and
  determinantal quantum {M}onte {C}arlo methods for spins, phonons and
  electrons},\ }in\ \href
  {https://itp.tugraz.at/~evertz/Pubs/2007_Greifswald_QMC_Methods_chap10.pdf}
  {\emph {\bibinfo {booktitle} {Computational Many-Particle Physics}}}\
  (\bibinfo  {publisher} {Springer},\ \bibinfo {year} {2008})\ pp.\ \bibinfo
  {pages} {277--356}\BibitemShut {NoStop}%
\bibitem [{\citenamefont {Trotter}(1959)}]{trotter1959product}%
  \BibitemOpen
  \bibfield  {author} {\bibinfo {author} {\bibfnamefont {H.~F.}\ \bibnamefont
  {Trotter}},\ }\bibfield  {title} {\bibinfo {title} {On the product of
  semi-groups of operators},\ }\href
  {https://www.jstor.org/stable/pdf/2033649.pdf} {\bibfield  {journal}
  {\bibinfo  {journal} {Proceedings of the American Mathematical Society}\
  }\textbf {\bibinfo {volume} {10}},\ \bibinfo {pages} {545} (\bibinfo {year}
  {1959})}\BibitemShut {NoStop}%
\bibitem [{\citenamefont {Suzuki}(1976)}]{suzuki1976relationship}%
  \BibitemOpen
  \bibfield  {author} {\bibinfo {author} {\bibfnamefont {M.}~\bibnamefont
  {Suzuki}},\ }\bibfield  {title} {\bibinfo {title} {Relationship between
  d-dimensional quantal spin systems and (d+1)-dimensional {I}sing systems:
  Equivalence, critical exponents and systematic approximants of the partition
  function and spin correlations},\ }\href
  {https://academic.oup.com/ptp/article/56/5/1454/1860476?login=true}
  {\bibfield  {journal} {\bibinfo  {journal} {Progress of theoretical physics}\
  }\textbf {\bibinfo {volume} {56}},\ \bibinfo {pages} {1454} (\bibinfo {year}
  {1976})}\BibitemShut {NoStop}%
\bibitem [{\citenamefont {Fye}(1986)}]{fye1986new}%
  \BibitemOpen
  \bibfield  {author} {\bibinfo {author} {\bibfnamefont {R.}~\bibnamefont
  {Fye}},\ }\bibfield  {title} {\bibinfo {title} {New results on {T}rotter-like
  approximations},\ }\href
  {https://journals.aps.org/prb/pdf/10.1103/PhysRevB.33.6271} {\bibfield
  {journal} {\bibinfo  {journal} {Phys. Rev. B}\ }\textbf {\bibinfo {volume}
  {33}},\ \bibinfo {pages} {6271} (\bibinfo {year} {1986})}\BibitemShut
  {NoStop}%
\bibitem [{\citenamefont {Hirsch}(1983)}]{HirschPRB1983}%
  \BibitemOpen
  \bibfield  {author} {\bibinfo {author} {\bibfnamefont {J.~E.}\ \bibnamefont
  {Hirsch}},\ }\bibfield  {title} {\bibinfo {title} {Discrete
  {Hubbard-Stratonovich} transformation for fermion lattice models},\ }\href
  {https://doi.org/10.1103/PhysRevB.28.4059} {\bibfield  {journal} {\bibinfo
  {journal} {Phys. Rev. B}\ }\textbf {\bibinfo {volume} {28}},\ \bibinfo
  {pages} {4059} (\bibinfo {year} {1983})}\BibitemShut {NoStop}%
\bibitem [{\citenamefont {Blankenbecler}\ \emph {et~al.}(1981)\citenamefont
  {Blankenbecler}, \citenamefont {Scalapino},\ and\ \citenamefont
  {Sugar}}]{BlankenbeclerPRD1981}%
  \BibitemOpen
  \bibfield  {author} {\bibinfo {author} {\bibfnamefont {R.}~\bibnamefont
  {Blankenbecler}}, \bibinfo {author} {\bibfnamefont {D.~J.}\ \bibnamefont
  {Scalapino}},\ and\ \bibinfo {author} {\bibfnamefont {R.~L.}\ \bibnamefont
  {Sugar}},\ }\bibfield  {title} {\bibinfo {title} {Monte {C}arlo calculations
  of coupled boson-fermion systems. {I}},\ }\href
  {https://doi.org/10.1103/PhysRevD.24.2278} {\bibfield  {journal} {\bibinfo
  {journal} {Phys. Rev. D}\ }\textbf {\bibinfo {volume} {24}},\ \bibinfo
  {pages} {2278} (\bibinfo {year} {1981})}\BibitemShut {NoStop}%
\bibitem [{\citenamefont {Tomas}\ \emph {et~al.}(2012)\citenamefont {Tomas},
  \citenamefont {Chang}, \citenamefont {Scalettar},\ and\ \citenamefont
  {Bai}}]{Tomas}%
  \BibitemOpen
  \bibfield  {author} {\bibinfo {author} {\bibfnamefont {A.}~\bibnamefont
  {Tomas}}, \bibinfo {author} {\bibfnamefont {C.-C.}\ \bibnamefont {Chang}},
  \bibinfo {author} {\bibfnamefont {R.}~\bibnamefont {Scalettar}},\ and\
  \bibinfo {author} {\bibfnamefont {Z.}~\bibnamefont {Bai}},\ }\bibfield
  {title} {\bibinfo {title} {Advancing large scale many-body {QMC} simulations
  on {GPU} accelerated multicore systems},\ }in\ \href
  {https://doi.org/10.1109/IPDPS.2012.37} {\emph {\bibinfo {booktitle} {2012
  IEEE 26th International Parallel and Distributed Processing Symposium}}}\
  (\bibinfo {year} {2012})\ pp.\ \bibinfo {pages} {308--319}\BibitemShut
  {NoStop}%
\bibitem [{\citenamefont {Henelius}\ and\ \citenamefont
  {Sandvik}(2000)}]{henelius00}%
  \BibitemOpen
  \bibfield  {author} {\bibinfo {author} {\bibfnamefont {P.}~\bibnamefont
  {Henelius}}\ and\ \bibinfo {author} {\bibfnamefont {A.~W.}\ \bibnamefont
  {Sandvik}},\ }\bibfield  {title} {\bibinfo {title} {Sign problem in {M}onte
  {C}arlo simulations of frustrated quantum spin systems},\ }\href
  {https://doi.org/10.1103/PhysRevB.62.1102} {\bibfield  {journal} {\bibinfo
  {journal} {Phys. Rev. B}\ }\textbf {\bibinfo {volume} {62}},\ \bibinfo
  {pages} {1102} (\bibinfo {year} {2000})}\BibitemShut {NoStop}%
\bibitem [{\citenamefont {Chandrasekharan}\ and\ \citenamefont
  {Wiese}(1999)}]{chandrasekharan99}%
  \BibitemOpen
  \bibfield  {author} {\bibinfo {author} {\bibfnamefont {S.}~\bibnamefont
  {Chandrasekharan}}\ and\ \bibinfo {author} {\bibfnamefont {U.-J.}\
  \bibnamefont {Wiese}},\ }\bibfield  {title} {\bibinfo {title} {Meron-cluster
  solution of fermion sign problems},\ }\href
  {https://doi.org/10.1103/PhysRevLett.83.3116} {\bibfield  {journal} {\bibinfo
   {journal} {Phys. Rev. Lett.}\ }\textbf {\bibinfo {volume} {83}},\ \bibinfo
  {pages} {3116} (\bibinfo {year} {1999})}\BibitemShut {NoStop}%
\bibitem [{\citenamefont {{Wu, Congjun and Hu, Jiang-ping and Zhang,
  Shou-cheng}}(2003)}]{wu03}%
  \BibitemOpen
  \bibfield  {author} {\bibinfo {author} {\bibnamefont {{Wu, Congjun and Hu,
  Jiang-ping and Zhang, Shou-cheng}}},\ }\bibfield  {title} {\bibinfo {title}
  {Exact {SO(5)} symmetry in the spin-$3/2$ fermionic system},\ }\href
  {https://doi.org/10.1103/PhysRevLett.91.186402} {\bibfield  {journal}
  {\bibinfo  {journal} {Phys. Rev. Lett.}\ }\textbf {\bibinfo {volume} {91}},\
  \bibinfo {pages} {186402} (\bibinfo {year} {2003})}\BibitemShut {NoStop}%
\bibitem [{\citenamefont {Berg}\ \emph {et~al.}(2012)\citenamefont {Berg},
  \citenamefont {Metlitski},\ and\ \citenamefont {Sachdev}}]{berg12}%
  \BibitemOpen
  \bibfield  {author} {\bibinfo {author} {\bibfnamefont {E.}~\bibnamefont
  {Berg}}, \bibinfo {author} {\bibfnamefont {M.~A.}\ \bibnamefont
  {Metlitski}},\ and\ \bibinfo {author} {\bibfnamefont {S.}~\bibnamefont
  {Sachdev}},\ }\bibfield  {title} {\bibinfo {title}
  {Sign-problem{\textendash}free quantum {M}onte {C}arlo of the onset of
  antiferromagnetism in metals},\ }\href
  {https://doi.org/10.1126/science.1227769} {\bibfield  {journal} {\bibinfo
  {journal} {Science}\ }\textbf {\bibinfo {volume} {338}},\ \bibinfo {pages}
  {1606} (\bibinfo {year} {2012})}\BibitemShut {NoStop}%
\bibitem [{\citenamefont {Chandrasekharan}(2012)}]{chandrasekharan12}%
  \BibitemOpen
  \bibfield  {author} {\bibinfo {author} {\bibfnamefont {S.}~\bibnamefont
  {Chandrasekharan}},\ }\bibfield  {title} {\bibinfo {title} {{Solutions to
  sign problems in lattice Yukawa models}},\ }\href
  {https://doi.org/10.1103/PhysRevD.86.021701} {\bibfield  {journal} {\bibinfo
  {journal} {Phys. Rev. D}\ }\textbf {\bibinfo {volume} {86}},\ \bibinfo
  {pages} {021701} (\bibinfo {year} {2012})}\BibitemShut {NoStop}%
\bibitem [{\citenamefont {Cai}\ \emph {et~al.}(2013)\citenamefont {Cai},
  \citenamefont {Hung}, \citenamefont {Wang},\ and\ \citenamefont
  {Wu}}]{cai13}%
  \BibitemOpen
  \bibfield  {author} {\bibinfo {author} {\bibfnamefont {Z.}~\bibnamefont
  {Cai}}, \bibinfo {author} {\bibfnamefont {H.-H.}\ \bibnamefont {Hung}},
  \bibinfo {author} {\bibfnamefont {L.}~\bibnamefont {Wang}},\ and\ \bibinfo
  {author} {\bibfnamefont {C.}~\bibnamefont {Wu}},\ }\bibfield  {title}
  {\bibinfo {title} {{Quantum magnetic properties of the $SU(2N)$ Hubbard model
  in the square lattice: A quantum Monte Carlo study}},\ }\href
  {https://doi.org/10.1103/PhysRevB.88.125108} {\bibfield  {journal} {\bibinfo
  {journal} {Phys. Rev. B}\ }\textbf {\bibinfo {volume} {88}},\ \bibinfo
  {pages} {125108} (\bibinfo {year} {2013})}\BibitemShut {NoStop}%
\bibitem [{\citenamefont {Huffman}\ and\ \citenamefont
  {Chandrasekharan}(2014)}]{huffman14}%
  \BibitemOpen
  \bibfield  {author} {\bibinfo {author} {\bibfnamefont {E.~F.}\ \bibnamefont
  {Huffman}}\ and\ \bibinfo {author} {\bibfnamefont {S.}~\bibnamefont
  {Chandrasekharan}},\ }\bibfield  {title} {\bibinfo {title} {Solution to sign
  problems in half-filled spin-polarized electronic systems},\ }\href
  {https://doi.org/10.1103/PhysRevB.89.111101} {\bibfield  {journal} {\bibinfo
  {journal} {Phys. Rev. B}\ }\textbf {\bibinfo {volume} {89}},\ \bibinfo
  {pages} {111101} (\bibinfo {year} {2014})}\BibitemShut {NoStop}%
\bibitem [{\citenamefont {Wang}\ \emph {et~al.}(2015)\citenamefont {Wang},
  \citenamefont {Liu}, \citenamefont {Iazzi}, \citenamefont {Troyer},\ and\
  \citenamefont {Harcos}}]{wang15}%
  \BibitemOpen
  \bibfield  {author} {\bibinfo {author} {\bibfnamefont {L.}~\bibnamefont
  {Wang}}, \bibinfo {author} {\bibfnamefont {Y.-H.}\ \bibnamefont {Liu}},
  \bibinfo {author} {\bibfnamefont {M.}~\bibnamefont {Iazzi}}, \bibinfo
  {author} {\bibfnamefont {M.}~\bibnamefont {Troyer}},\ and\ \bibinfo {author}
  {\bibfnamefont {G.}~\bibnamefont {Harcos}},\ }\bibfield  {title} {\bibinfo
  {title} {Split orthogonal group: A guiding principle for sign-problem-free
  fermionic simulations},\ }\href
  {https://doi.org/10.1103/PhysRevLett.115.250601} {\bibfield  {journal}
  {\bibinfo  {journal} {Phys. Rev. Lett.}\ }\textbf {\bibinfo {volume} {115}},\
  \bibinfo {pages} {250601} (\bibinfo {year} {2015})}\BibitemShut {NoStop}%
\bibitem [{\citenamefont {Kaul}(2015)}]{kaul15}%
  \BibitemOpen
  \bibfield  {author} {\bibinfo {author} {\bibfnamefont {R.~K.}\ \bibnamefont
  {Kaul}},\ }\bibfield  {title} {\bibinfo {title} {Marshall-positive
  {$\mathrm{SU}(N)$} quantum spin systems and classical loop models: A
  practical strategy to design sign-problem-free spin {H}amiltonians},\ }\href
  {https://doi.org/10.1103/PhysRevB.91.054413} {\bibfield  {journal} {\bibinfo
  {journal} {Phys. Rev. B}\ }\textbf {\bibinfo {volume} {91}},\ \bibinfo
  {pages} {054413} (\bibinfo {year} {2015})}\BibitemShut {NoStop}%
\bibitem [{\citenamefont {Li}\ \emph {et~al.}(2015{\natexlab{b}})\citenamefont
  {Li}, \citenamefont {Jiang},\ and\ \citenamefont {Yao}}]{ZiXiang2015}%
  \BibitemOpen
  \bibfield  {author} {\bibinfo {author} {\bibfnamefont {Z.-X.}\ \bibnamefont
  {Li}}, \bibinfo {author} {\bibfnamefont {Y.-F.}\ \bibnamefont {Jiang}},\ and\
  \bibinfo {author} {\bibfnamefont {H.}~\bibnamefont {Yao}},\ }\bibfield
  {title} {\bibinfo {title} {{Solving the fermion sign problem in quantum Monte
  Carlo simulations by Majorana representation}},\ }\href
  {https://doi.org/10.1103/PhysRevB.91.241117} {\bibfield  {journal} {\bibinfo
  {journal} {Phys. Rev. B}\ }\textbf {\bibinfo {volume} {91}},\ \bibinfo
  {pages} {241117} (\bibinfo {year} {2015}{\natexlab{b}})}\BibitemShut
  {NoStop}%
\bibitem [{\citenamefont {Wei}\ \emph {et~al.}(2016)\citenamefont {Wei},
  \citenamefont {Wu}, \citenamefont {Li}, \citenamefont {Zhang},\ and\
  \citenamefont {Xiang}}]{Wei2016}%
  \BibitemOpen
  \bibfield  {author} {\bibinfo {author} {\bibfnamefont {Z.~C.}\ \bibnamefont
  {Wei}}, \bibinfo {author} {\bibfnamefont {C.}~\bibnamefont {Wu}}, \bibinfo
  {author} {\bibfnamefont {Y.}~\bibnamefont {Li}}, \bibinfo {author}
  {\bibfnamefont {S.}~\bibnamefont {Zhang}},\ and\ \bibinfo {author}
  {\bibfnamefont {T.}~\bibnamefont {Xiang}},\ }\bibfield  {title} {\bibinfo
  {title} {Majorana positivity and the fermion sign problem of quantum {M}onte
  {C}arlo simulations},\ }\href
  {https://doi.org/10.1103/PhysRevLett.116.250601} {\bibfield  {journal}
  {\bibinfo  {journal} {Phys. Rev. Lett.}\ }\textbf {\bibinfo {volume} {116}},\
  \bibinfo {pages} {250601} (\bibinfo {year} {2016})}\BibitemShut {NoStop}%
\bibitem [{\citenamefont {Li}\ and\ \citenamefont {Yao}(2019)}]{li19}%
  \BibitemOpen
  \bibfield  {author} {\bibinfo {author} {\bibfnamefont {Z.-X.}\ \bibnamefont
  {Li}}\ and\ \bibinfo {author} {\bibfnamefont {H.}~\bibnamefont {Yao}},\
  }\bibfield  {title} {\bibinfo {title} {Sign-problem-free fermionic quantum
  {M}onte {C}arlo: Developments and applications},\ }\href
  {https://doi.org/10.1146/annurev-conmatphys-033117-054307} {\bibfield
  {journal} {\bibinfo  {journal} {Annu. Rev. Condens. Matter Phys.}\ }\textbf
  {\bibinfo {volume} {10}},\ \bibinfo {pages} {337} (\bibinfo {year}
  {2019})}\BibitemShut {NoStop}%
\bibitem [{\citenamefont {Kim}\ \emph {et~al.}(2020)\citenamefont {Kim},
  \citenamefont {Werner},\ and\ \citenamefont {Valent\'{\i}}}]{kim20}%
  \BibitemOpen
  \bibfield  {author} {\bibinfo {author} {\bibfnamefont {A.~J.}\ \bibnamefont
  {Kim}}, \bibinfo {author} {\bibfnamefont {P.}~\bibnamefont {Werner}},\ and\
  \bibinfo {author} {\bibfnamefont {R.}~\bibnamefont {Valent\'{\i}}},\
  }\bibfield  {title} {\bibinfo {title} {{Alleviating the sign problem in
  quantum Monte Carlo simulations of spin-orbit-coupled multiorbital Hubbard
  models}},\ }\href {https://doi.org/10.1103/PhysRevB.101.045108} {\bibfield
  {journal} {\bibinfo  {journal} {Phys. Rev. B}\ }\textbf {\bibinfo {volume}
  {101}},\ \bibinfo {pages} {045108} (\bibinfo {year} {2020})}\BibitemShut
  {NoStop}%
\bibitem [{\citenamefont {Levy}\ and\ \citenamefont {Clark}(2021)}]{Levy2021}%
  \BibitemOpen
  \bibfield  {author} {\bibinfo {author} {\bibfnamefont {R.}~\bibnamefont
  {Levy}}\ and\ \bibinfo {author} {\bibfnamefont {B.~K.}\ \bibnamefont
  {Clark}},\ }\bibfield  {title} {\bibinfo {title} {Mitigating the sign problem
  through basis rotations},\ }\href
  {https://doi.org/10.1103/PhysRevLett.126.216401} {\bibfield  {journal}
  {\bibinfo  {journal} {Phys. Rev. Lett.}\ }\textbf {\bibinfo {volume} {126}},\
  \bibinfo {pages} {216401} (\bibinfo {year} {2021})}\BibitemShut {NoStop}%
\bibitem [{\citenamefont {Zhang}\ \emph {et~al.}(2021)\citenamefont {Zhang},
  \citenamefont {Pan}, \citenamefont {Xu},\ and\ \citenamefont
  {Meng}}]{zhang2021sign}%
  \BibitemOpen
  \bibfield  {author} {\bibinfo {author} {\bibfnamefont {X.}~\bibnamefont
  {Zhang}}, \bibinfo {author} {\bibfnamefont {G.}~\bibnamefont {Pan}}, \bibinfo
  {author} {\bibfnamefont {X.~Y.}\ \bibnamefont {Xu}},\ and\ \bibinfo {author}
  {\bibfnamefont {Z.~Y.}\ \bibnamefont {Meng}},\ }\bibfield  {title} {\bibinfo
  {title} {Sign problem finds its bounds},\ }\href
  {https://arxiv.org/pdf/2112.06139.pdf} {\bibfield  {journal} {\bibinfo
  {journal} {arXiv preprint arXiv:2112.06139}\ } (\bibinfo {year}
  {2021})}\BibitemShut {NoStop}%
\bibitem [{\citenamefont {Maier}\ \emph
  {et~al.}(2005{\natexlab{b}})\citenamefont {Maier}, \citenamefont {Jarrell},
  \citenamefont {Pruschke},\ and\ \citenamefont {Hettler}}]{MaierRMP2005}%
  \BibitemOpen
  \bibfield  {author} {\bibinfo {author} {\bibfnamefont {T.}~\bibnamefont
  {Maier}}, \bibinfo {author} {\bibfnamefont {M.}~\bibnamefont {Jarrell}},
  \bibinfo {author} {\bibfnamefont {T.}~\bibnamefont {Pruschke}},\ and\
  \bibinfo {author} {\bibfnamefont {M.~H.}\ \bibnamefont {Hettler}},\
  }\bibfield  {title} {\bibinfo {title} {Quantum cluster theories},\ }\href
  {https://doi.org/10.1103/RevModPhys.77.1027} {\bibfield  {journal} {\bibinfo
  {journal} {Rev. Mod. Phys.}\ }\textbf {\bibinfo {volume} {77}},\ \bibinfo
  {pages} {1027} (\bibinfo {year} {2005}{\natexlab{b}})}\BibitemShut {NoStop}%
\bibitem [{\citenamefont {Georges}\ \emph {et~al.}(1996)\citenamefont
  {Georges}, \citenamefont {Kotliar}, \citenamefont {Krauth},\ and\
  \citenamefont {Rozenberg}}]{GeorgesRMP1996}%
  \BibitemOpen
  \bibfield  {author} {\bibinfo {author} {\bibfnamefont {A.}~\bibnamefont
  {Georges}}, \bibinfo {author} {\bibfnamefont {G.}~\bibnamefont {Kotliar}},
  \bibinfo {author} {\bibfnamefont {W.}~\bibnamefont {Krauth}},\ and\ \bibinfo
  {author} {\bibfnamefont {M.~J.}\ \bibnamefont {Rozenberg}},\ }\bibfield
  {title} {\bibinfo {title} {Dynamical mean-field theory of strongly correlated
  fermion systems and the limit of infinite dimensions},\ }\href
  {https://doi.org/10.1103/RevModPhys.68.13} {\bibfield  {journal} {\bibinfo
  {journal} {Rev. Mod. Phys.}\ }\textbf {\bibinfo {volume} {68}},\ \bibinfo
  {pages} {13} (\bibinfo {year} {1996})}\BibitemShut {NoStop}%
\bibitem [{\citenamefont {Wolff}(1989)}]{Wolff}%
  \BibitemOpen
  \bibfield  {author} {\bibinfo {author} {\bibfnamefont {U.}~\bibnamefont
  {Wolff}},\ }\bibfield  {title} {\bibinfo {title} {Collective {Monte Carlo}
  updating for spin systems},\ }\href
  {https://doi.org/10.1103/PhysRevLett.62.361} {\bibfield  {journal} {\bibinfo
  {journal} {Phys. Rev. Lett.}\ }\textbf {\bibinfo {volume} {62}},\ \bibinfo
  {pages} {361} (\bibinfo {year} {1989})}\BibitemShut {NoStop}%
\bibitem [{ML()}]{ML}%
  \BibitemOpen
  \href@noop {} {}\bibinfo {note} {Michael Neilsen, Neural Networks and Deep
  Learning, \url{http://neuralnetworksanddeeplearning.com/}}\BibitemShut
  {NoStop}%
\bibitem [{You()}]{Youtube}%
  \BibitemOpen
  \href@noop {} {}\bibinfo {note} {3Blue1Brown Youtube Channel, But what is a
  neural network? | Chapter 1, Deep learning,
  \url{https://www.youtube.com/watch?v=aircAruvnKk}}\BibitemShut {NoStop}%
\bibitem [{\citenamefont {van~der Maaten}\ and\ \citenamefont
  {Hinton}(2008)}]{tSNE}%
  \BibitemOpen
  \bibfield  {author} {\bibinfo {author} {\bibfnamefont {L.}~\bibnamefont
  {van~der Maaten}}\ and\ \bibinfo {author} {\bibfnamefont {G.}~\bibnamefont
  {Hinton}},\ }\bibfield  {title} {\bibinfo {title} {Visualizing data using
  t-{SNE}},\ }\href {http://jmlr.org/papers/v9/vandermaaten08a.html} {\bibfield
   {journal} {\bibinfo  {journal} {Journal of Machine Learning Research}\
  }\textbf {\bibinfo {volume} {9}},\ \bibinfo {pages} {2579} (\bibinfo {year}
  {2008})}\BibitemShut {NoStop}%
\bibitem [{tSN()}]{tSNEHowTo}%
  \BibitemOpen
  \href@noop {} {}\bibinfo {note} {A tutorial on how to use t-SNE effectively
  can be found at \url{https://distill.pub/2016/misread-tsne}.}\BibitemShut
  {Stop}%
\bibitem [{\citenamefont {Moosmann}\ \emph {et~al.}(2006)\citenamefont
  {Moosmann}, \citenamefont {Triggs},\ and\ \citenamefont
  {Jurie}}]{f_moosmann_06}%
  \BibitemOpen
  \bibfield  {author} {\bibinfo {author} {\bibfnamefont {F.}~\bibnamefont
  {Moosmann}}, \bibinfo {author} {\bibfnamefont {B.}~\bibnamefont {Triggs}},\
  and\ \bibinfo {author} {\bibfnamefont {F.}~\bibnamefont {Jurie}},\ }\bibfield
   {title} {\bibinfo {title} {Fast discriminative visual codebooks using
  randomized clustering forests},\ }in\ \href
  {https://proceedings.neurips.cc/paper/2006/file/d3157f2f0212a80a5d042c127522a2d5-Paper.pdf}
  {\emph {\bibinfo {booktitle} {Advances in Neural Information Processing
  Systems}}},\ Vol.~\bibinfo {volume} {19},\ \bibinfo {editor} {edited by\
  \bibinfo {editor} {\bibfnamefont {B.}~\bibnamefont {Sch\"{o}lkopf}}, \bibinfo
  {editor} {\bibfnamefont {J.}~\bibnamefont {Platt}},\ and\ \bibinfo {editor}
  {\bibfnamefont {T.}~\bibnamefont {Hoffman}}}\ (\bibinfo  {publisher} {MIT
  Press},\ \bibinfo {year} {2006})\BibitemShut {NoStop}%
\bibitem [{Ran()}]{RandomTrees}%
  \BibitemOpen
  \href@noop {} {}\bibinfo {note} {For a code, examples, and more references
  see
  \url{https://scikit-learn.org/stable/modules/ensemble.html\#random-trees-embedding}.}\BibitemShut
  {Stop}%
\bibitem [{\citenamefont {Carrasquilla}\ and\ \citenamefont
  {Melko}(2017)}]{j_carrasquilla_16}%
  \BibitemOpen
  \bibfield  {author} {\bibinfo {author} {\bibfnamefont {J.}~\bibnamefont
  {Carrasquilla}}\ and\ \bibinfo {author} {\bibfnamefont {R.~G.}\ \bibnamefont
  {Melko}},\ }\bibfield  {title} {\bibinfo {title} {Machine learning phases of
  matter},\ }\href {http://dx.doi.org/10.1038/nphys4035} {\bibfield  {journal}
  {\bibinfo  {journal} {Nat Phys}\ }\textbf {\bibinfo {volume} {13}},\ \bibinfo
  {pages} {431} (\bibinfo {year} {2017})}\BibitemShut {NoStop}%
\bibitem [{\citenamefont {Wang}(2016)}]{Wang2016}%
  \BibitemOpen
  \bibfield  {author} {\bibinfo {author} {\bibfnamefont {L.}~\bibnamefont
  {Wang}},\ }\bibfield  {title} {\bibinfo {title} {Discovering phase
  transitions with unsupervised learning},\ }\href
  {https://doi.org/10.1103/PhysRevB.94.195105} {\bibfield  {journal} {\bibinfo
  {journal} {Phys. Rev. B}\ }\textbf {\bibinfo {volume} {94}},\ \bibinfo
  {pages} {195105} (\bibinfo {year} {2016})}\BibitemShut {NoStop}%
\bibitem [{\citenamefont {Carleo}\ and\ \citenamefont
  {Troyer}(2017)}]{Carleo2016}%
  \BibitemOpen
  \bibfield  {author} {\bibinfo {author} {\bibfnamefont {G.}~\bibnamefont
  {Carleo}}\ and\ \bibinfo {author} {\bibfnamefont {M.}~\bibnamefont
  {Troyer}},\ }\bibfield  {title} {\bibinfo {title} {Solving the quantum
  many-body problem with artificial neural networks},\ }\href
  {https://doi.org/10.1126/science.aag2302} {\bibfield  {journal} {\bibinfo
  {journal} {Science}\ }\textbf {\bibinfo {volume} {355}},\ \bibinfo {pages}
  {602} (\bibinfo {year} {2017})}\BibitemShut {NoStop}%
\bibitem [{\citenamefont {Torlai}\ and\ \citenamefont
  {Melko}(2016)}]{g_torlai_16}%
  \BibitemOpen
  \bibfield  {author} {\bibinfo {author} {\bibfnamefont {G.}~\bibnamefont
  {Torlai}}\ and\ \bibinfo {author} {\bibfnamefont {R.~G.}\ \bibnamefont
  {Melko}},\ }\bibfield  {title} {\bibinfo {title} {Learning thermodynamics
  with {Boltzmann} machines},\ }\href
  {https://doi.org/10.1103/PhysRevB.94.165134} {\bibfield  {journal} {\bibinfo
  {journal} {Phys. Rev. B}\ }\textbf {\bibinfo {volume} {94}},\ \bibinfo
  {pages} {165134} (\bibinfo {year} {2016})}\BibitemShut {NoStop}%
\bibitem [{\citenamefont {Hu}\ \emph {et~al.}(2017)\citenamefont {Hu},
  \citenamefont {Singh},\ and\ \citenamefont {Scalettar}}]{HuPRE2017}%
  \BibitemOpen
  \bibfield  {author} {\bibinfo {author} {\bibfnamefont {W.}~\bibnamefont
  {Hu}}, \bibinfo {author} {\bibfnamefont {R.~R.~P.}\ \bibnamefont {Singh}},\
  and\ \bibinfo {author} {\bibfnamefont {R.~T.}\ \bibnamefont {Scalettar}},\
  }\bibfield  {title} {\bibinfo {title} {Discovering phases, phase transitions,
  and crossovers through unsupervised machine learning: A critical
  examination},\ }\href {https://doi.org/10.1103/PhysRevE.95.062122} {\bibfield
   {journal} {\bibinfo  {journal} {Phys. Rev. E}\ }\textbf {\bibinfo {volume}
  {95}},\ \bibinfo {pages} {062122} (\bibinfo {year} {2017})}\BibitemShut
  {NoStop}%
\bibitem [{\citenamefont {Wang}\ and\ \citenamefont {Zhai}(2017)}]{c_wang_17}%
  \BibitemOpen
  \bibfield  {author} {\bibinfo {author} {\bibfnamefont {C.}~\bibnamefont
  {Wang}}\ and\ \bibinfo {author} {\bibfnamefont {H.}~\bibnamefont {Zhai}},\
  }\bibfield  {title} {\bibinfo {title} {Machine learning of frustrated
  classical spin models. {I}. principal component analysis},\ }\href
  {https://doi.org/10.1103/PhysRevB.96.144432} {\bibfield  {journal} {\bibinfo
  {journal} {Phys. Rev. B}\ }\textbf {\bibinfo {volume} {96}},\ \bibinfo
  {pages} {144432} (\bibinfo {year} {2017})}\BibitemShut {NoStop}%
\bibitem [{\citenamefont {Wetzel}(2017)}]{wetzel2017unsupervised}%
  \BibitemOpen
  \bibfield  {author} {\bibinfo {author} {\bibfnamefont {S.~J.}\ \bibnamefont
  {Wetzel}},\ }\bibfield  {title} {\bibinfo {title} {Unsupervised learning of
  phase transitions: From principal component analysis to variational
  autoencoders},\ }\href
  {https://journals.aps.org/pre/pdf/10.1103/PhysRevE.96.022140} {\bibfield
  {journal} {\bibinfo  {journal} {Phys. Rev. E}\ }\textbf {\bibinfo {volume}
  {96}},\ \bibinfo {pages} {022140} (\bibinfo {year} {2017})}\BibitemShut
  {NoStop}%
\bibitem [{\citenamefont {Tanaka}\ and\ \citenamefont
  {Tomiya}(2017{\natexlab{a}})}]{tanaka2017detection}%
  \BibitemOpen
  \bibfield  {author} {\bibinfo {author} {\bibfnamefont {A.}~\bibnamefont
  {Tanaka}}\ and\ \bibinfo {author} {\bibfnamefont {A.}~\bibnamefont
  {Tomiya}},\ }\bibfield  {title} {\bibinfo {title} {Detection of phase
  transition via convolutional neural networks},\ }\href
  {https://journals.jps.jp/doi/pdf/10.7566/JPSJ.86.063001} {\bibfield
  {journal} {\bibinfo  {journal} {Journal of the Physical Society of Japan}\
  }\textbf {\bibinfo {volume} {86}},\ \bibinfo {pages} {063001} (\bibinfo
  {year} {2017}{\natexlab{a}})}\BibitemShut {NoStop}%
\bibitem [{\citenamefont {Ch'ng}\ \emph {et~al.}(2018)\citenamefont {Ch'ng},
  \citenamefont {Vazquez},\ and\ \citenamefont {Khatami}}]{k_chng_18}%
  \BibitemOpen
  \bibfield  {author} {\bibinfo {author} {\bibfnamefont {K.}~\bibnamefont
  {Ch'ng}}, \bibinfo {author} {\bibfnamefont {N.}~\bibnamefont {Vazquez}},\
  and\ \bibinfo {author} {\bibfnamefont {E.}~\bibnamefont {Khatami}},\
  }\bibfield  {title} {\bibinfo {title} {Unsupervised machine learning account
  of magnetic transitions in the {Hubbard} model},\ }\href
  {https://doi.org/10.1103/PhysRevE.97.013306} {\bibfield  {journal} {\bibinfo
  {journal} {Phys. Rev. E}\ }\textbf {\bibinfo {volume} {97}},\ \bibinfo
  {pages} {013306} (\bibinfo {year} {2018})}\BibitemShut {NoStop}%
\bibitem [{\citenamefont {Binder}\ and\ \citenamefont
  {Luijten}(2001)}]{binder2001monte}%
  \BibitemOpen
  \bibfield  {author} {\bibinfo {author} {\bibfnamefont {K.}~\bibnamefont
  {Binder}}\ and\ \bibinfo {author} {\bibfnamefont {E.}~\bibnamefont
  {Luijten}},\ }\bibfield  {title} {\bibinfo {title} {Monte carlo tests of
  renormalization-group predictions for critical phenomena in {I}sing models},\
  }\href {https://www.sciencedirect.com/science/article/pii/S0370157300001277}
  {\bibfield  {journal} {\bibinfo  {journal} {Physics Reports}\ }\textbf
  {\bibinfo {volume} {344}},\ \bibinfo {pages} {179} (\bibinfo {year}
  {2001})}\BibitemShut {NoStop}%
\bibitem [{\citenamefont {Blume}\ \emph {et~al.}(1971)\citenamefont {Blume},
  \citenamefont {Emery},\ and\ \citenamefont {Griffiths}}]{Blume71}%
  \BibitemOpen
  \bibfield  {author} {\bibinfo {author} {\bibfnamefont {M.}~\bibnamefont
  {Blume}}, \bibinfo {author} {\bibfnamefont {V.~J.}\ \bibnamefont {Emery}},\
  and\ \bibinfo {author} {\bibfnamefont {R.~B.}\ \bibnamefont {Griffiths}},\
  }\bibfield  {title} {\bibinfo {title} {{Ising} model for the
  $\ensuremath{\lambda}$ transition and phase separation in
  {${\mathrm{He}}^{3}$-${\mathrm{He}}^{4}$} mixtures},\ }\href
  {https://doi.org/10.1103/PhysRevA.4.1071} {\bibfield  {journal} {\bibinfo
  {journal} {Phys. Rev. A}\ }\textbf {\bibinfo {volume} {4}},\ \bibinfo {pages}
  {1071} (\bibinfo {year} {1971})}\BibitemShut {NoStop}%
\bibitem [{\citenamefont {Wang}\ and\ \citenamefont
  {Zhai}(2018)}]{wang2018machine}%
  \BibitemOpen
  \bibfield  {author} {\bibinfo {author} {\bibfnamefont {C.}~\bibnamefont
  {Wang}}\ and\ \bibinfo {author} {\bibfnamefont {H.}~\bibnamefont {Zhai}},\
  }\bibfield  {title} {\bibinfo {title} {Machine learning of frustrated
  classical spin models ({II}): Kernel principal component analysis},\ }\href
  {https://doi.org/10.1007/s11467-018-0798-7} {\bibfield  {journal} {\bibinfo
  {journal} {Frontiers of Physics}\ }\textbf {\bibinfo {volume} {13}},\
  \bibinfo {pages} {1} (\bibinfo {year} {2018})}\BibitemShut {NoStop}%
\bibitem [{\citenamefont {Beach}\ \emph {et~al.}(2018)\citenamefont {Beach},
  \citenamefont {Golubeva},\ and\ \citenamefont {Melko}}]{beach2018machine}%
  \BibitemOpen
  \bibfield  {author} {\bibinfo {author} {\bibfnamefont {M.~J.~S.}\
  \bibnamefont {Beach}}, \bibinfo {author} {\bibfnamefont {A.}~\bibnamefont
  {Golubeva}},\ and\ \bibinfo {author} {\bibfnamefont {R.~G.}\ \bibnamefont
  {Melko}},\ }\bibfield  {title} {\bibinfo {title} {Machine learning vortices
  at the {K}osterlitz-{T}houless transition},\ }\href
  {https://doi.org/10.1103/PhysRevB.97.045207} {\bibfield  {journal} {\bibinfo
  {journal} {Phys. Rev. B}\ }\textbf {\bibinfo {volume} {97}},\ \bibinfo
  {pages} {045207} (\bibinfo {year} {2018})}\BibitemShut {NoStop}%
\bibitem [{\citenamefont {Rodriguez-Nieva}\ and\ \citenamefont
  {Scheurer}(2019)}]{rodriguez2019identifying}%
  \BibitemOpen
  \bibfield  {author} {\bibinfo {author} {\bibfnamefont {J.~F.}\ \bibnamefont
  {Rodriguez-Nieva}}\ and\ \bibinfo {author} {\bibfnamefont {M.~S.}\
  \bibnamefont {Scheurer}},\ }\bibfield  {title} {\bibinfo {title} {Identifying
  topological order through unsupervised machine learning},\ }\href
  {https://www.nature.com/articles/s41567-019-0512-x} {\bibfield  {journal}
  {\bibinfo  {journal} {Nature Physics}\ }\textbf {\bibinfo {volume} {15}},\
  \bibinfo {pages} {790} (\bibinfo {year} {2019})}\BibitemShut {NoStop}%
\bibitem [{Ann()}]{AnnasList}%
  \BibitemOpen
  \href@noop {} {}\bibinfo {note} {For an incomplete early list, see ``Applying
  Machine Learning to Physics'' in Dr. Anna Golubeva's website at
  \url{https://github.com/AnnaGolubeva/physicsml.github.io/blob/master/develop/content/pages/papers.md}}\BibitemShut
  {NoStop}%
\bibitem [{\citenamefont {Broecker}\ \emph
  {et~al.}(2017{\natexlab{a}})\citenamefont {Broecker}, \citenamefont
  {Carrasquilla}, \citenamefont {Melko},\ and\ \citenamefont
  {Trebst}}]{p_broecker_16}%
  \BibitemOpen
  \bibfield  {author} {\bibinfo {author} {\bibfnamefont {P.}~\bibnamefont
  {Broecker}}, \bibinfo {author} {\bibfnamefont {J.}~\bibnamefont
  {Carrasquilla}}, \bibinfo {author} {\bibfnamefont {R.~G.}\ \bibnamefont
  {Melko}},\ and\ \bibinfo {author} {\bibfnamefont {S.}~\bibnamefont
  {Trebst}},\ }\bibfield  {title} {\bibinfo {title} {Machine learning quantum
  phases of matter beyond the fermion sign problem},\ }\href
  {https://doi.org/10.1038/s41598-017-09098-0} {\bibfield  {journal} {\bibinfo
  {journal} {Scientific Reports}\ }\textbf {\bibinfo {volume} {7}},\ \bibinfo
  {pages} {8823} (\bibinfo {year} {2017}{\natexlab{a}})}\BibitemShut {NoStop}%
\bibitem [{\citenamefont {Ch'ng}\ \emph {et~al.}(2017)\citenamefont {Ch'ng},
  \citenamefont {Carrasquilla}, \citenamefont {Melko},\ and\ \citenamefont
  {Khatami}}]{k_chng_17}%
  \BibitemOpen
  \bibfield  {author} {\bibinfo {author} {\bibfnamefont {K.}~\bibnamefont
  {Ch'ng}}, \bibinfo {author} {\bibfnamefont {J.}~\bibnamefont {Carrasquilla}},
  \bibinfo {author} {\bibfnamefont {R.~G.}\ \bibnamefont {Melko}},\ and\
  \bibinfo {author} {\bibfnamefont {E.}~\bibnamefont {Khatami}},\ }\bibfield
  {title} {\bibinfo {title} {Machine learning phases of strongly correlated
  fermions},\ }\href {https://doi.org/10.1103/PhysRevX.7.031038} {\bibfield
  {journal} {\bibinfo  {journal} {Phys. Rev. X}\ }\textbf {\bibinfo {volume}
  {7}},\ \bibinfo {pages} {031038} (\bibinfo {year} {2017})}\BibitemShut
  {NoStop}%
\bibitem [{\citenamefont {Deng}\ \emph {et~al.}(2017)\citenamefont {Deng},
  \citenamefont {Li},\ and\ \citenamefont {Das~Sarma}}]{d_deng_17}%
  \BibitemOpen
  \bibfield  {author} {\bibinfo {author} {\bibfnamefont {D.-L.}\ \bibnamefont
  {Deng}}, \bibinfo {author} {\bibfnamefont {X.}~\bibnamefont {Li}},\ and\
  \bibinfo {author} {\bibfnamefont {S.}~\bibnamefont {Das~Sarma}},\ }\bibfield
  {title} {\bibinfo {title} {Machine learning topological states},\ }\href
  {https://doi.org/10.1103/PhysRevB.96.195145} {\bibfield  {journal} {\bibinfo
  {journal} {Phys. Rev. B}\ }\textbf {\bibinfo {volume} {96}},\ \bibinfo
  {pages} {195145} (\bibinfo {year} {2017})}\BibitemShut {NoStop}%
\bibitem [{\citenamefont {Torlai}\ and\ \citenamefont
  {Melko}(2017)}]{g_torlai_17}%
  \BibitemOpen
  \bibfield  {author} {\bibinfo {author} {\bibfnamefont {G.}~\bibnamefont
  {Torlai}}\ and\ \bibinfo {author} {\bibfnamefont {R.~G.}\ \bibnamefont
  {Melko}},\ }\bibfield  {title} {\bibinfo {title} {Neural decoder for
  topological codes},\ }\href {https://doi.org/10.1103/PhysRevLett.119.030501}
  {\bibfield  {journal} {\bibinfo  {journal} {Phys. Rev. Lett.}\ }\textbf
  {\bibinfo {volume} {119}},\ \bibinfo {pages} {030501} (\bibinfo {year}
  {2017})}\BibitemShut {NoStop}%
\bibitem [{\citenamefont {Zhang}\ and\ \citenamefont {Kim}(2017)}]{y_zhang_17}%
  \BibitemOpen
  \bibfield  {author} {\bibinfo {author} {\bibfnamefont {Y.}~\bibnamefont
  {Zhang}}\ and\ \bibinfo {author} {\bibfnamefont {E.-A.}\ \bibnamefont
  {Kim}},\ }\bibfield  {title} {\bibinfo {title} {Quantum loop topography for
  machine learning},\ }\href {https://doi.org/10.1103/PhysRevLett.118.216401}
  {\bibfield  {journal} {\bibinfo  {journal} {Phys. Rev. Lett.}\ }\textbf
  {\bibinfo {volume} {118}},\ \bibinfo {pages} {216401} (\bibinfo {year}
  {2017})}\BibitemShut {NoStop}%
\bibitem [{\citenamefont {van Nieuwenburg}\ \emph {et~al.}(2017)\citenamefont
  {van Nieuwenburg}, \citenamefont {Liu},\ and\ \citenamefont
  {Huber}}]{e_vanNieuwenburg_17}%
  \BibitemOpen
  \bibfield  {author} {\bibinfo {author} {\bibfnamefont {E.~P.~L.}\
  \bibnamefont {van Nieuwenburg}}, \bibinfo {author} {\bibfnamefont {Y.-H.}\
  \bibnamefont {Liu}},\ and\ \bibinfo {author} {\bibfnamefont {S.~D.}\
  \bibnamefont {Huber}},\ }\bibfield  {title} {\bibinfo {title} {Learning phase
  transitions by confusion},\ }\href {https://doi.org/10.1038/nphys4037}
  {\bibfield  {journal} {\bibinfo  {journal} {Nature Physics}\ }\textbf
  {\bibinfo {volume} {13}},\ \bibinfo {pages} {435} (\bibinfo {year}
  {2017})}\BibitemShut {NoStop}%
\bibitem [{\citenamefont {Broecker}\ \emph
  {et~al.}(2017{\natexlab{b}})\citenamefont {Broecker}, \citenamefont
  {Assaad},\ and\ \citenamefont {Trebst}}]{p_broecker_17}%
  \BibitemOpen
  \bibfield  {author} {\bibinfo {author} {\bibfnamefont {P.}~\bibnamefont
  {Broecker}}, \bibinfo {author} {\bibfnamefont {F.~F.}\ \bibnamefont
  {Assaad}},\ and\ \bibinfo {author} {\bibfnamefont {S.}~\bibnamefont
  {Trebst}},\ }\bibfield  {title} {\bibinfo {title} {Quantum phase recognition
  via unsupervised machine learning},\ }\href
  {https://arxiv.org/abs/1707.00663} {\bibfield  {journal} {\bibinfo  {journal}
  {arXiv:1707.00663}\ } (\bibinfo {year} {2017}{\natexlab{b}})}\BibitemShut
  {NoStop}%
\bibitem [{\citenamefont {Tirelli}\ and\ \citenamefont
  {Costa}(2021)}]{tirelli2021learning}%
  \BibitemOpen
  \bibfield  {author} {\bibinfo {author} {\bibfnamefont {A.}~\bibnamefont
  {Tirelli}}\ and\ \bibinfo {author} {\bibfnamefont {N.~C.}\ \bibnamefont
  {Costa}},\ }\bibfield  {title} {\bibinfo {title} {Learning quantum phase
  transitions through topological data analysis},\ }\href
  {https://doi.org/10.1103/PhysRevB.104.235146} {\bibfield  {journal} {\bibinfo
   {journal} {Phys. Rev. B}\ }\textbf {\bibinfo {volume} {104}},\ \bibinfo
  {pages} {235146} (\bibinfo {year} {2021})}\BibitemShut {NoStop}%
\bibitem [{\citenamefont {Fontenele}\ \emph {et~al.}(2022)\citenamefont
  {Fontenele}, \citenamefont {Costa}, \citenamefont {dos Santos},\ and\
  \citenamefont {Paiva}}]{Fontenele2022}%
  \BibitemOpen
  \bibfield  {author} {\bibinfo {author} {\bibfnamefont {R.~A.}\ \bibnamefont
  {Fontenele}}, \bibinfo {author} {\bibfnamefont {N.~C.}\ \bibnamefont
  {Costa}}, \bibinfo {author} {\bibfnamefont {R.~R.}\ \bibnamefont {dos
  Santos}},\ and\ \bibinfo {author} {\bibfnamefont {T.}~\bibnamefont {Paiva}},\
  }\bibfield  {title} {\bibinfo {title} {Two-dimensional attractive {Hubbard}
  model and the {BCS-BEC} crossover},\ }\href
  {https://doi.org/10.1103/PhysRevB.105.184502} {\bibfield  {journal} {\bibinfo
   {journal} {Phys. Rev. B}\ }\textbf {\bibinfo {volume} {105}},\ \bibinfo
  {pages} {184502} (\bibinfo {year} {2022})}\BibitemShut {NoStop}%
\bibitem [{\citenamefont {Costa}\ \emph {et~al.}(2017)\citenamefont {Costa},
  \citenamefont {Hu}, \citenamefont {Bai}, \citenamefont {Scalettar},\ and\
  \citenamefont {Singh}}]{costa17}%
  \BibitemOpen
  \bibfield  {author} {\bibinfo {author} {\bibfnamefont {N.~C.}\ \bibnamefont
  {Costa}}, \bibinfo {author} {\bibfnamefont {W.}~\bibnamefont {Hu}}, \bibinfo
  {author} {\bibfnamefont {Z.~J.}\ \bibnamefont {Bai}}, \bibinfo {author}
  {\bibfnamefont {R.~T.}\ \bibnamefont {Scalettar}},\ and\ \bibinfo {author}
  {\bibfnamefont {R.~R.~P.}\ \bibnamefont {Singh}},\ }\bibfield  {title}
  {\bibinfo {title} {Principal component analysis for fermionic critical
  points},\ }\href {https://doi.org/10.1103/PhysRevB.96.195138} {\bibfield
  {journal} {\bibinfo  {journal} {Phys. Rev. B}\ }\textbf {\bibinfo {volume}
  {96}},\ \bibinfo {pages} {195138} (\bibinfo {year} {2017})}\BibitemShut
  {NoStop}%
\bibitem [{\citenamefont {Hirsch}(1985)}]{hirsch1985two}%
  \BibitemOpen
  \bibfield  {author} {\bibinfo {author} {\bibfnamefont {J.~E.}\ \bibnamefont
  {Hirsch}},\ }\bibfield  {title} {\bibinfo {title} {Two-dimensional {H}ubbard
  model: Numerical simulation study},\ }\href
  {https://doi.org/10.1103/PhysRevB.31.4403} {\bibfield  {journal} {\bibinfo
  {journal} {Phys. Rev. B}\ }\textbf {\bibinfo {volume} {31}},\ \bibinfo
  {pages} {4403} (\bibinfo {year} {1985})}\BibitemShut {NoStop}%
\bibitem [{\citenamefont {Sorella}\ and\ \citenamefont
  {Tosatti}(1992)}]{Sorella_1992}%
  \BibitemOpen
  \bibfield  {author} {\bibinfo {author} {\bibfnamefont {S.}~\bibnamefont
  {Sorella}}\ and\ \bibinfo {author} {\bibfnamefont {E.}~\bibnamefont
  {Tosatti}},\ }\bibfield  {title} {\bibinfo {title} {Semi-metal-insulator
  transition of the {Hubbard} model in the honeycomb lattice},\ }\href
  {https://doi.org/10.1209/0295-5075/19/8/007} {\bibfield  {journal} {\bibinfo
  {journal} {Europhysics Letters ({EPL})}\ }\textbf {\bibinfo {volume} {19}},\
  \bibinfo {pages} {699} (\bibinfo {year} {1992})}\BibitemShut {NoStop}%
\bibitem [{\citenamefont {Paiva}\ \emph {et~al.}(2005)\citenamefont {Paiva},
  \citenamefont {Scalettar}, \citenamefont {Zheng}, \citenamefont {Singh},\
  and\ \citenamefont {Oitmaa}}]{paiva05}%
  \BibitemOpen
  \bibfield  {author} {\bibinfo {author} {\bibfnamefont {T.}~\bibnamefont
  {Paiva}}, \bibinfo {author} {\bibfnamefont {R.}~\bibnamefont {Scalettar}},
  \bibinfo {author} {\bibfnamefont {W.}~\bibnamefont {Zheng}}, \bibinfo
  {author} {\bibfnamefont {R.}~\bibnamefont {Singh}},\ and\ \bibinfo {author}
  {\bibfnamefont {J.}~\bibnamefont {Oitmaa}},\ }\bibfield  {title} {\bibinfo
  {title} {Ground-state and finite-temperature signatures of quantum phase
  transitions in the half-filled {H}ubbard model on a honeycomb lattice},\
  }\href {https://doi.org/10.1103/PhysRevB.72.085123} {\bibfield  {journal}
  {\bibinfo  {journal} {Phys. Rev. B}\ }\textbf {\bibinfo {volume} {72}},\
  \bibinfo {pages} {085123} (\bibinfo {year} {2005})}\BibitemShut {NoStop}%
\bibitem [{\citenamefont {Otsuka}\ \emph {et~al.}(2016)\citenamefont {Otsuka},
  \citenamefont {Yunoki},\ and\ \citenamefont {Sorella}}]{Otsuka2016}%
  \BibitemOpen
  \bibfield  {author} {\bibinfo {author} {\bibfnamefont {Y.}~\bibnamefont
  {Otsuka}}, \bibinfo {author} {\bibfnamefont {S.}~\bibnamefont {Yunoki}},\
  and\ \bibinfo {author} {\bibfnamefont {S.}~\bibnamefont {Sorella}},\
  }\bibfield  {title} {\bibinfo {title} {Universal quantum criticality in the
  metal-insulator transition of two-dimensional interacting {D}irac
  electrons},\ }\href {https://doi.org/10.1103/PhysRevX.6.011029} {\bibfield
  {journal} {\bibinfo  {journal} {Phys. Rev. X}\ }\textbf {\bibinfo {volume}
  {6}},\ \bibinfo {pages} {011029} (\bibinfo {year} {2016})}\BibitemShut
  {NoStop}%
\bibitem [{\citenamefont {Raczkowski}\ \emph {et~al.}(2020)\citenamefont
  {Raczkowski}, \citenamefont {Peters}, \citenamefont {Ph{\`u}ng},
  \citenamefont {Takemori}, \citenamefont {Assaad}, \citenamefont {Honecker},\
  and\ \citenamefont {Vahedi}}]{raczkowski2020hubbard}%
  \BibitemOpen
  \bibfield  {author} {\bibinfo {author} {\bibfnamefont {M.}~\bibnamefont
  {Raczkowski}}, \bibinfo {author} {\bibfnamefont {R.}~\bibnamefont {Peters}},
  \bibinfo {author} {\bibfnamefont {T.~T.}\ \bibnamefont {Ph{\`u}ng}}, \bibinfo
  {author} {\bibfnamefont {N.}~\bibnamefont {Takemori}}, \bibinfo {author}
  {\bibfnamefont {F.~F.}\ \bibnamefont {Assaad}}, \bibinfo {author}
  {\bibfnamefont {A.}~\bibnamefont {Honecker}},\ and\ \bibinfo {author}
  {\bibfnamefont {J.}~\bibnamefont {Vahedi}},\ }\bibfield  {title} {\bibinfo
  {title} {{H}ubbard model on the honeycomb lattice: From static and dynamical
  mean-field theories to lattice quantum {Monte Carlo} simulations},\ }\href
  {https://journals.aps.org/prb/abstract/10.1103/PhysRevB.101.125103}
  {\bibfield  {journal} {\bibinfo  {journal} {Phys. Rev. B}\ }\textbf {\bibinfo
  {volume} {101}},\ \bibinfo {pages} {125103} (\bibinfo {year}
  {2020})}\BibitemShut {NoStop}%
\bibitem [{\citenamefont {Costa}\ \emph
  {et~al.}(2021{\natexlab{b}})\citenamefont {Costa}, \citenamefont {Seki},\
  and\ \citenamefont {Sorella}}]{costa2021magnetism}%
  \BibitemOpen
  \bibfield  {author} {\bibinfo {author} {\bibfnamefont {N.~C.}\ \bibnamefont
  {Costa}}, \bibinfo {author} {\bibfnamefont {K.}~\bibnamefont {Seki}},\ and\
  \bibinfo {author} {\bibfnamefont {S.}~\bibnamefont {Sorella}},\ }\bibfield
  {title} {\bibinfo {title} {Magnetism and charge order in the honeycomb
  lattice},\ }\href {https://doi.org/10.1103/PhysRevLett.126.107205} {\bibfield
   {journal} {\bibinfo  {journal} {Phys. Rev. Lett.}\ }\textbf {\bibinfo
  {volume} {126}},\ \bibinfo {pages} {107205} (\bibinfo {year}
  {2021}{\natexlab{b}})}\BibitemShut {NoStop}%
\bibitem [{\citenamefont {Meng}\ \emph {et~al.}(2010)\citenamefont {Meng},
  \citenamefont {Wessel}, \citenamefont {Muramatsu}, \citenamefont {Lang},\
  and\ \citenamefont {Assaad}}]{meng10}%
  \BibitemOpen
  \bibfield  {author} {\bibinfo {author} {\bibfnamefont {Z.}~\bibnamefont
  {Meng}}, \bibinfo {author} {\bibfnamefont {S.}~\bibnamefont {Wessel}},
  \bibinfo {author} {\bibfnamefont {A.}~\bibnamefont {Muramatsu}}, \bibinfo
  {author} {\bibfnamefont {T.}~\bibnamefont {Lang}},\ and\ \bibinfo {author}
  {\bibfnamefont {F.}~\bibnamefont {Assaad}},\ }\bibfield  {title} {\bibinfo
  {title} {{Quantum spin liquid emerging in two-dimensional correlated Dirac
  fermions}},\ }\href {https://doi.org/https://doi.org/10.1038/nature08942}
  {\bibfield  {journal} {\bibinfo  {journal} {Nature}\ }\textbf {\bibinfo
  {volume} {464}},\ \bibinfo {pages} {847} (\bibinfo {year}
  {2010})}\BibitemShut {NoStop}%
\bibitem [{\citenamefont {Sorella}\ \emph {et~al.}(2012)\citenamefont
  {Sorella}, \citenamefont {Otsuka},\ and\ \citenamefont
  {Yunoki}}]{Sorella2012}%
  \BibitemOpen
  \bibfield  {author} {\bibinfo {author} {\bibfnamefont {S.}~\bibnamefont
  {Sorella}}, \bibinfo {author} {\bibfnamefont {Y.}~\bibnamefont {Otsuka}},\
  and\ \bibinfo {author} {\bibfnamefont {S.}~\bibnamefont {Yunoki}},\
  }\bibfield  {title} {\bibinfo {title} {Absence of a spin liquid phase in the
  {H}ubbard model on the honeycomb lattice},\ }\href
  {https://doi.org/https://doi.org/10.1038/srep00992} {\bibfield  {journal}
  {\bibinfo  {journal} {Sci. Rep.}\ }\textbf {\bibinfo {volume} {2}},\ \bibinfo
  {pages} {992} (\bibinfo {year} {2012})}\BibitemShut {NoStop}%
\bibitem [{\citenamefont {Dopf}\ \emph {et~al.}(1990)\citenamefont {Dopf},
  \citenamefont {Muramatsu},\ and\ \citenamefont {Hanke}}]{DopfPRB1990}%
  \BibitemOpen
  \bibfield  {author} {\bibinfo {author} {\bibfnamefont {G.}~\bibnamefont
  {Dopf}}, \bibinfo {author} {\bibfnamefont {A.}~\bibnamefont {Muramatsu}},\
  and\ \bibinfo {author} {\bibfnamefont {W.}~\bibnamefont {Hanke}},\ }\bibfield
   {title} {\bibinfo {title} {Three-band {Hubbard} model: A {Monte Carlo}
  study},\ }\href {https://doi.org/10.1103/PhysRevB.41.9264} {\bibfield
  {journal} {\bibinfo  {journal} {Phys. Rev. B}\ }\textbf {\bibinfo {volume}
  {41}},\ \bibinfo {pages} {9264} (\bibinfo {year} {1990})}\BibitemShut
  {NoStop}%
\bibitem [{\citenamefont {Kung}\ \emph {et~al.}(2016)\citenamefont {Kung},
  \citenamefont {Chen}, \citenamefont {Wang}, \citenamefont {Huang},
  \citenamefont {Nowadnick}, \citenamefont {Moritz}, \citenamefont {Scalettar},
  \citenamefont {Johnston},\ and\ \citenamefont {Devereaux}}]{KungRB2016}%
  \BibitemOpen
  \bibfield  {author} {\bibinfo {author} {\bibfnamefont {Y.~F.}\ \bibnamefont
  {Kung}}, \bibinfo {author} {\bibfnamefont {C.-C.}\ \bibnamefont {Chen}},
  \bibinfo {author} {\bibfnamefont {Y.}~\bibnamefont {Wang}}, \bibinfo {author}
  {\bibfnamefont {E.~W.}\ \bibnamefont {Huang}}, \bibinfo {author}
  {\bibfnamefont {E.~A.}\ \bibnamefont {Nowadnick}}, \bibinfo {author}
  {\bibfnamefont {B.}~\bibnamefont {Moritz}}, \bibinfo {author} {\bibfnamefont
  {R.~T.}\ \bibnamefont {Scalettar}}, \bibinfo {author} {\bibfnamefont
  {S.}~\bibnamefont {Johnston}},\ and\ \bibinfo {author} {\bibfnamefont
  {T.~P.}\ \bibnamefont {Devereaux}},\ }\bibfield  {title} {\bibinfo {title}
  {Characterizing the three-orbital {Hubbard} model with determinant quantum
  {Monte Carlo}},\ }\href {https://doi.org/10.1103/PhysRevB.93.155166}
  {\bibfield  {journal} {\bibinfo  {journal} {Phys. Rev. B}\ }\textbf {\bibinfo
  {volume} {93}},\ \bibinfo {pages} {155166} (\bibinfo {year}
  {2016})}\BibitemShut {NoStop}%
\bibitem [{\citenamefont {Yi}\ \emph {et~al.}(2021)\citenamefont {Yi},
  \citenamefont {Scalettar},\ and\ \citenamefont {Mondaini}}]{YiPreprint}%
  \BibitemOpen
  \bibfield  {author} {\bibinfo {author} {\bibfnamefont {T.-C.}\ \bibnamefont
  {Yi}}, \bibinfo {author} {\bibfnamefont {R.~T.}\ \bibnamefont {Scalettar}},\
  and\ \bibinfo {author} {\bibfnamefont {R.}~\bibnamefont {Mondaini}},\
  }\bibfield  {title} {\bibinfo {title} {Hamming distance and the onset of
  quantum criticality},\ }\href {https://arxiv.org/abs/2111.12936} {\bibfield
  {journal} {\bibinfo  {journal} {arXiv:2111.12936}\ } (\bibinfo {year}
  {2021})}\BibitemShut {NoStop}%
\bibitem [{\citenamefont {Moreo}\ and\ \citenamefont
  {Scalapino}(1991)}]{moreo1991two}%
  \BibitemOpen
  \bibfield  {author} {\bibinfo {author} {\bibfnamefont {A.}~\bibnamefont
  {Moreo}}\ and\ \bibinfo {author} {\bibfnamefont {D.~J.}\ \bibnamefont
  {Scalapino}},\ }\bibfield  {title} {\bibinfo {title} {Two-dimensional
  negative-{U} {Hubbard} model},\ }\href
  {https://doi.org/10.1103/PhysRevLett.66.946} {\bibfield  {journal} {\bibinfo
  {journal} {Phys. Rev. Lett.}\ }\textbf {\bibinfo {volume} {66}},\ \bibinfo
  {pages} {946} (\bibinfo {year} {1991})}\BibitemShut {NoStop}%
\bibitem [{\citenamefont {Scalapino}\ \emph {et~al.}(1993)\citenamefont
  {Scalapino}, \citenamefont {White},\ and\ \citenamefont
  {Zhang}}]{scalapino1993insulator}%
  \BibitemOpen
  \bibfield  {author} {\bibinfo {author} {\bibfnamefont {D.~J.}\ \bibnamefont
  {Scalapino}}, \bibinfo {author} {\bibfnamefont {S.~R.}\ \bibnamefont
  {White}},\ and\ \bibinfo {author} {\bibfnamefont {S.}~\bibnamefont {Zhang}},\
  }\bibfield  {title} {\bibinfo {title} {Insulator, metal, or superconductor:
  The criteria},\ }\href
  {https://journals.aps.org/prb/abstract/10.1103/PhysRevB.47.7995} {\bibfield
  {journal} {\bibinfo  {journal} {Phys. Rev. B}\ }\textbf {\bibinfo {volume}
  {47}},\ \bibinfo {pages} {7995} (\bibinfo {year} {1993})}\BibitemShut
  {NoStop}%
\bibitem [{\citenamefont {Singer}\ \emph {et~al.}(1996)\citenamefont {Singer},
  \citenamefont {Pedersen}, \citenamefont {Schneider}, \citenamefont {Beck},\
  and\ \citenamefont {Matuttis}}]{singer1996bcs}%
  \BibitemOpen
  \bibfield  {author} {\bibinfo {author} {\bibfnamefont {J.}~\bibnamefont
  {Singer}}, \bibinfo {author} {\bibfnamefont {M.}~\bibnamefont {Pedersen}},
  \bibinfo {author} {\bibfnamefont {T.}~\bibnamefont {Schneider}}, \bibinfo
  {author} {\bibfnamefont {H.}~\bibnamefont {Beck}},\ and\ \bibinfo {author}
  {\bibfnamefont {H.-G.}\ \bibnamefont {Matuttis}},\ }\bibfield  {title}
  {\bibinfo {title} {From {BCS}-like superconductivity to condensation of local
  pairs: A numerical study of the attractive {Hubbard} model},\ }\href
  {https://journals.aps.org/prb/abstract/10.1103/PhysRevB.54.1286} {\bibfield
  {journal} {\bibinfo  {journal} {Phys. Rev. B}\ }\textbf {\bibinfo {volume}
  {54}},\ \bibinfo {pages} {1286} (\bibinfo {year} {1996})}\BibitemShut
  {NoStop}%
\bibitem [{\citenamefont {Kyung}\ \emph {et~al.}(2001)\citenamefont {Kyung},
  \citenamefont {Allen},\ and\ \citenamefont {Tremblay}}]{kyung2001pairing}%
  \BibitemOpen
  \bibfield  {author} {\bibinfo {author} {\bibfnamefont {B.}~\bibnamefont
  {Kyung}}, \bibinfo {author} {\bibfnamefont {S.}~\bibnamefont {Allen}},\ and\
  \bibinfo {author} {\bibfnamefont {A.-M.}\ \bibnamefont {Tremblay}},\
  }\bibfield  {title} {\bibinfo {title} {Pairing fluctuations and pseudogaps in
  the attractive {H}ubbard model},\ }\href
  {https://journals.aps.org/prb/abstract/10.1103/PhysRevB.64.075116} {\bibfield
   {journal} {\bibinfo  {journal} {Phys. Rev. B}\ }\textbf {\bibinfo {volume}
  {64}},\ \bibinfo {pages} {075116} (\bibinfo {year} {2001})}\BibitemShut
  {NoStop}%
\bibitem [{\citenamefont {Paiva}\ \emph {et~al.}(2004)\citenamefont {Paiva},
  \citenamefont {Dos~Santos}, \citenamefont {Scalettar},\ and\ \citenamefont
  {Denteneer}}]{paiva2004critical}%
  \BibitemOpen
  \bibfield  {author} {\bibinfo {author} {\bibfnamefont {T.}~\bibnamefont
  {Paiva}}, \bibinfo {author} {\bibfnamefont {R.~R.}\ \bibnamefont
  {Dos~Santos}}, \bibinfo {author} {\bibfnamefont {R.}~\bibnamefont
  {Scalettar}},\ and\ \bibinfo {author} {\bibfnamefont {P.}~\bibnamefont
  {Denteneer}},\ }\bibfield  {title} {\bibinfo {title} {Critical temperature
  for the two-dimensional attractive {Hubbard} model},\ }\href
  {https://journals.aps.org/prb/abstract/10.1103/PhysRevB.69.184501} {\bibfield
   {journal} {\bibinfo  {journal} {Phys. Rev. B}\ }\textbf {\bibinfo {volume}
  {69}},\ \bibinfo {pages} {184501} (\bibinfo {year} {2004})}\BibitemShut
  {NoStop}%
\bibitem [{\citenamefont {Karakuzu}\ \emph {et~al.}(2018)\citenamefont
  {Karakuzu}, \citenamefont {Seki},\ and\ \citenamefont
  {Sorella}}]{karakuzu2018study}%
  \BibitemOpen
  \bibfield  {author} {\bibinfo {author} {\bibfnamefont {S.}~\bibnamefont
  {Karakuzu}}, \bibinfo {author} {\bibfnamefont {K.}~\bibnamefont {Seki}},\
  and\ \bibinfo {author} {\bibfnamefont {S.}~\bibnamefont {Sorella}},\
  }\bibfield  {title} {\bibinfo {title} {Study of the superconducting order
  parameter in the two-dimensional negative-{U} {H}ubbard model by
  grand-canonical twist-averaged boundary conditions},\ }\href
  {https://journals.aps.org/prb/abstract/10.1103/PhysRevB.98.075156} {\bibfield
   {journal} {\bibinfo  {journal} {Phys. Rev. B}\ }\textbf {\bibinfo {volume}
  {98}},\ \bibinfo {pages} {075156} (\bibinfo {year} {2018})}\BibitemShut
  {NoStop}%
\bibitem [{\citenamefont {Zhang}\ \emph {et~al.}(2019)\citenamefont {Zhang},
  \citenamefont {Mesaros}, \citenamefont {Fujita}, \citenamefont {Edkins},
  \citenamefont {Hamidian}, \citenamefont {Ch'ng}, \citenamefont {Eisaki},
  \citenamefont {Uchida}, \citenamefont {Davis}, \citenamefont {Khatami},\ and\
  \citenamefont {Kim}}]{f_zhang_18}%
  \BibitemOpen
  \bibfield  {author} {\bibinfo {author} {\bibfnamefont {Y.}~\bibnamefont
  {Zhang}}, \bibinfo {author} {\bibfnamefont {A.}~\bibnamefont {Mesaros}},
  \bibinfo {author} {\bibfnamefont {K.}~\bibnamefont {Fujita}}, \bibinfo
  {author} {\bibfnamefont {S.~D.}\ \bibnamefont {Edkins}}, \bibinfo {author}
  {\bibfnamefont {M.~H.}\ \bibnamefont {Hamidian}}, \bibinfo {author}
  {\bibfnamefont {K.}~\bibnamefont {Ch'ng}}, \bibinfo {author} {\bibfnamefont
  {H.}~\bibnamefont {Eisaki}}, \bibinfo {author} {\bibfnamefont
  {S.}~\bibnamefont {Uchida}}, \bibinfo {author} {\bibfnamefont {J.~C.~S.}\
  \bibnamefont {Davis}}, \bibinfo {author} {\bibfnamefont {E.}~\bibnamefont
  {Khatami}},\ and\ \bibinfo {author} {\bibfnamefont {E.-A.}\ \bibnamefont
  {Kim}},\ }\bibfield  {title} {\bibinfo {title} {Machine learning in
  electronic-quantum-matter imaging experiments},\ }\href
  {https://doi.org/10.1038/s41586-019-1319-8} {\bibfield  {journal} {\bibinfo
  {journal} {Nature}\ }\textbf {\bibinfo {volume} {570}},\ \bibinfo {pages}
  {484} (\bibinfo {year} {2019})}\BibitemShut {NoStop}%
\bibitem [{\citenamefont {Khatami}\ \emph {et~al.}(2020)\citenamefont
  {Khatami}, \citenamefont {Guardado-Sanchez}, \citenamefont {Spar},
  \citenamefont {Carrasquilla}, \citenamefont {Bakr},\ and\ \citenamefont
  {Scalettar}}]{e_khatami_20}%
  \BibitemOpen
  \bibfield  {author} {\bibinfo {author} {\bibfnamefont {E.}~\bibnamefont
  {Khatami}}, \bibinfo {author} {\bibfnamefont {E.}~\bibnamefont
  {Guardado-Sanchez}}, \bibinfo {author} {\bibfnamefont {B.~M.}\ \bibnamefont
  {Spar}}, \bibinfo {author} {\bibfnamefont {J.~F.}\ \bibnamefont
  {Carrasquilla}}, \bibinfo {author} {\bibfnamefont {W.~S.}\ \bibnamefont
  {Bakr}},\ and\ \bibinfo {author} {\bibfnamefont {R.~T.}\ \bibnamefont
  {Scalettar}},\ }\bibfield  {title} {\bibinfo {title} {Visualizing strange
  metallic correlations in the two-dimensional {F}ermi-{H}ubbard model with
  artificial intelligence},\ }\href
  {https://doi.org/10.1103/PhysRevA.102.033326} {\bibfield  {journal} {\bibinfo
   {journal} {Phys. Rev. A}\ }\textbf {\bibinfo {volume} {102}},\ \bibinfo
  {pages} {033326} (\bibinfo {year} {2020})}\BibitemShut {NoStop}%
\bibitem [{\citenamefont {Bohrdt}\ \emph {et~al.}(2019)\citenamefont {Bohrdt},
  \citenamefont {Chiu}, \citenamefont {Ji}, \citenamefont {Xu}, \citenamefont
  {Greif}, \citenamefont {Greiner}, \citenamefont {Demler}, \citenamefont
  {Grusdt},\ and\ \citenamefont {Knap}}]{a_bohrdt_18}%
  \BibitemOpen
  \bibfield  {author} {\bibinfo {author} {\bibfnamefont {A.}~\bibnamefont
  {Bohrdt}}, \bibinfo {author} {\bibfnamefont {C.~S.}\ \bibnamefont {Chiu}},
  \bibinfo {author} {\bibfnamefont {G.}~\bibnamefont {Ji}}, \bibinfo {author}
  {\bibfnamefont {M.}~\bibnamefont {Xu}}, \bibinfo {author} {\bibfnamefont
  {D.}~\bibnamefont {Greif}}, \bibinfo {author} {\bibfnamefont
  {M.}~\bibnamefont {Greiner}}, \bibinfo {author} {\bibfnamefont
  {E.}~\bibnamefont {Demler}}, \bibinfo {author} {\bibfnamefont
  {F.}~\bibnamefont {Grusdt}},\ and\ \bibinfo {author} {\bibfnamefont
  {M.}~\bibnamefont {Knap}},\ }\bibfield  {title} {\bibinfo {title}
  {Classifying snapshots of the doped {Hubbard} model with machine learning},\
  }\href {https://doi.org/10.1038/s41567-019-0565-x} {\bibfield  {journal}
  {\bibinfo  {journal} {Nature Physics}\ }\textbf {\bibinfo {volume} {15}},\
  \bibinfo {pages} {921} (\bibinfo {year} {2019})}\BibitemShut {NoStop}%
\bibitem [{\citenamefont {Cheuk}\ \emph {et~al.}(2016)\citenamefont {Cheuk},
  \citenamefont {Nichols}, \citenamefont {Lawrence}, \citenamefont {Okan},
  \citenamefont {Zhang}, \citenamefont {Khatami}, \citenamefont {Trivedi},
  \citenamefont {Paiva}, \citenamefont {Rigol},\ and\ \citenamefont
  {Zwierlein}}]{l_cheuk_16}%
  \BibitemOpen
  \bibfield  {author} {\bibinfo {author} {\bibfnamefont {L.~W.}\ \bibnamefont
  {Cheuk}}, \bibinfo {author} {\bibfnamefont {M.~A.}\ \bibnamefont {Nichols}},
  \bibinfo {author} {\bibfnamefont {K.~R.}\ \bibnamefont {Lawrence}}, \bibinfo
  {author} {\bibfnamefont {M.}~\bibnamefont {Okan}}, \bibinfo {author}
  {\bibfnamefont {H.}~\bibnamefont {Zhang}}, \bibinfo {author} {\bibfnamefont
  {E.}~\bibnamefont {Khatami}}, \bibinfo {author} {\bibfnamefont
  {N.}~\bibnamefont {Trivedi}}, \bibinfo {author} {\bibfnamefont
  {T.}~\bibnamefont {Paiva}}, \bibinfo {author} {\bibfnamefont
  {M.}~\bibnamefont {Rigol}},\ and\ \bibinfo {author} {\bibfnamefont {M.~W.}\
  \bibnamefont {Zwierlein}},\ }\bibfield  {title} {\bibinfo {title}
  {Observation of spatial charge and spin correlations in the {2D}
  {F}ermi-{H}ubbard model},\ }\href {https://doi.org/10.1126/science.aag3349}
  {\bibfield  {journal} {\bibinfo  {journal} {Science}\ }\textbf {\bibinfo
  {volume} {353}},\ \bibinfo {pages} {1260} (\bibinfo {year}
  {2016})}\BibitemShut {NoStop}%
\bibitem [{\citenamefont {Brown}\ \emph {et~al.}(2019)\citenamefont {Brown},
  \citenamefont {Mitra}, \citenamefont {Guardado-Sanchez}, \citenamefont
  {Nourafkan}, \citenamefont {Reymbaut}, \citenamefont {H{\'e}bert},
  \citenamefont {Bergeron}, \citenamefont {Tremblay}, \citenamefont {Kokalj},
  \citenamefont {Huse}, \citenamefont {Schau{\ss}},\ and\ \citenamefont
  {Bakr}}]{p_brown_18}%
  \BibitemOpen
  \bibfield  {author} {\bibinfo {author} {\bibfnamefont {P.~T.}\ \bibnamefont
  {Brown}}, \bibinfo {author} {\bibfnamefont {D.}~\bibnamefont {Mitra}},
  \bibinfo {author} {\bibfnamefont {E.}~\bibnamefont {Guardado-Sanchez}},
  \bibinfo {author} {\bibfnamefont {R.}~\bibnamefont {Nourafkan}}, \bibinfo
  {author} {\bibfnamefont {A.}~\bibnamefont {Reymbaut}}, \bibinfo {author}
  {\bibfnamefont {C.-D.}\ \bibnamefont {H{\'e}bert}}, \bibinfo {author}
  {\bibfnamefont {S.}~\bibnamefont {Bergeron}}, \bibinfo {author}
  {\bibfnamefont {A.-M.~S.}\ \bibnamefont {Tremblay}}, \bibinfo {author}
  {\bibfnamefont {J.}~\bibnamefont {Kokalj}}, \bibinfo {author} {\bibfnamefont
  {D.~A.}\ \bibnamefont {Huse}}, \bibinfo {author} {\bibfnamefont
  {P.}~\bibnamefont {Schau{\ss}}},\ and\ \bibinfo {author} {\bibfnamefont
  {W.~S.}\ \bibnamefont {Bakr}},\ }\bibfield  {title} {\bibinfo {title} {Bad
  metallic transport in a cold atom {F}ermi-{H}ubbard system},\ }\href
  {https://doi.org/10.1126/science.aat4134} {\bibfield  {journal} {\bibinfo
  {journal} {Science}\ }\textbf {\bibinfo {volume} {363}},\ \bibinfo {pages}
  {379} (\bibinfo {year} {2019})}\BibitemShut {NoStop}%
\bibitem [{\citenamefont {Samarakoon}\ and\ \citenamefont
  {Tennant}(2021)}]{SamarakoonJPCM2021}%
  \BibitemOpen
  \bibfield  {author} {\bibinfo {author} {\bibfnamefont {A.~M.}\ \bibnamefont
  {Samarakoon}}\ and\ \bibinfo {author} {\bibfnamefont {D.~A.}\ \bibnamefont
  {Tennant}},\ }\bibfield  {title} {\bibinfo {title} {Machine learning for
  magnetic phase diagrams and inverse scattering problems},\ }\href
  {https://doi.org/10.1088/1361-648x/abe818} {\bibfield  {journal} {\bibinfo
  {journal} {Journal of Physics: Condensed Matter}\ }\textbf {\bibinfo {volume}
  {34}},\ \bibinfo {pages} {044002} (\bibinfo {year} {2021})}\BibitemShut
  {NoStop}%
\bibitem [{\citenamefont {Samarakoon}\ \emph {et~al.}(2020)\citenamefont
  {Samarakoon}, \citenamefont {Barros}, \citenamefont {Li}, \citenamefont
  {Eisenbach}, \citenamefont {Zhang}, \citenamefont {Ye}, \citenamefont
  {Sharma}, \citenamefont {Dun}, \citenamefont {Zhou}, \citenamefont {Grigera},
  \citenamefont {Batista},\ and\ \citenamefont {Tennant}}]{Samarakoon2020}%
  \BibitemOpen
  \bibfield  {author} {\bibinfo {author} {\bibfnamefont {A.~M.}\ \bibnamefont
  {Samarakoon}}, \bibinfo {author} {\bibfnamefont {K.}~\bibnamefont {Barros}},
  \bibinfo {author} {\bibfnamefont {Y.~W.}\ \bibnamefont {Li}}, \bibinfo
  {author} {\bibfnamefont {M.}~\bibnamefont {Eisenbach}}, \bibinfo {author}
  {\bibfnamefont {Q.}~\bibnamefont {Zhang}}, \bibinfo {author} {\bibfnamefont
  {F.}~\bibnamefont {Ye}}, \bibinfo {author} {\bibfnamefont {V.}~\bibnamefont
  {Sharma}}, \bibinfo {author} {\bibfnamefont {Z.~L.}\ \bibnamefont {Dun}},
  \bibinfo {author} {\bibfnamefont {H.}~\bibnamefont {Zhou}}, \bibinfo {author}
  {\bibfnamefont {S.~A.}\ \bibnamefont {Grigera}}, \bibinfo {author}
  {\bibfnamefont {C.~D.}\ \bibnamefont {Batista}},\ and\ \bibinfo {author}
  {\bibfnamefont {D.~A.}\ \bibnamefont {Tennant}},\ }\bibfield  {title}
  {\bibinfo {title} {Machine-learning-assisted insight into spin ice
  {Dy$_2$Ti$_2$O$_7$}},\ }\href {https://doi.org/10.1038/s41467-020-14660-y}
  {\bibfield  {journal} {\bibinfo  {journal} {Nature Communications}\ }\textbf
  {\bibinfo {volume} {11}},\ \bibinfo {pages} {892} (\bibinfo {year}
  {2020})}\BibitemShut {NoStop}%
\bibitem [{\citenamefont {Zhang}\ and\ \citenamefont
  {Batista}(2021)}]{ZhangPRB2021}%
  \BibitemOpen
  \bibfield  {author} {\bibinfo {author} {\bibfnamefont {H.}~\bibnamefont
  {Zhang}}\ and\ \bibinfo {author} {\bibfnamefont {C.~D.}\ \bibnamefont
  {Batista}},\ }\bibfield  {title} {\bibinfo {title} {Classical spin dynamics
  based on {$\mathrm{SU}(N)$} coherent states},\ }\href
  {https://doi.org/10.1103/PhysRevB.104.104409} {\bibfield  {journal} {\bibinfo
   {journal} {Phys. Rev. B}\ }\textbf {\bibinfo {volume} {104}},\ \bibinfo
  {pages} {104409} (\bibinfo {year} {2021})}\BibitemShut {NoStop}%
\bibitem [{\citenamefont {Dahlbom}\ \emph
  {et~al.}(2022{\natexlab{a}})\citenamefont {Dahlbom}, \citenamefont {Zhang},
  \citenamefont {Miles}, \citenamefont {Bai}, \citenamefont {Batista},\ and\
  \citenamefont {Barros}}]{DahlbomPreprint}%
  \BibitemOpen
  \bibfield  {author} {\bibinfo {author} {\bibfnamefont {D.}~\bibnamefont
  {Dahlbom}}, \bibinfo {author} {\bibfnamefont {H.}~\bibnamefont {Zhang}},
  \bibinfo {author} {\bibfnamefont {C.}~\bibnamefont {Miles}}, \bibinfo
  {author} {\bibfnamefont {X.}~\bibnamefont {Bai}}, \bibinfo {author}
  {\bibfnamefont {C.~D.}\ \bibnamefont {Batista}},\ and\ \bibinfo {author}
  {\bibfnamefont {K.}~\bibnamefont {Barros}},\ }\bibfield  {title} {\bibinfo
  {title} {Geometric integration of classical spin dynamics via a mean-field
  schr\"odinger equation},\ }\href
  {https://doi.org/10.1103/PhysRevB.106.054423} {\bibfield  {journal} {\bibinfo
   {journal} {Phys. Rev. B}\ }\textbf {\bibinfo {volume} {106}},\ \bibinfo
  {pages} {054423} (\bibinfo {year} {2022}{\natexlab{a}})}\BibitemShut
  {NoStop}%
\bibitem [{\citenamefont {Dahlbom}\ \emph
  {et~al.}(2022{\natexlab{b}})\citenamefont {Dahlbom}, \citenamefont {Miles},
  \citenamefont {Zhang}, \citenamefont {Batista},\ and\ \citenamefont
  {Barros}}]{DahlbomPreprint2}%
  \BibitemOpen
  \bibfield  {author} {\bibinfo {author} {\bibfnamefont {D.}~\bibnamefont
  {Dahlbom}}, \bibinfo {author} {\bibfnamefont {C.}~\bibnamefont {Miles}},
  \bibinfo {author} {\bibfnamefont {H.}~\bibnamefont {Zhang}}, \bibinfo
  {author} {\bibfnamefont {C.~D.}\ \bibnamefont {Batista}},\ and\ \bibinfo
  {author} {\bibfnamefont {K.}~\bibnamefont {Barros}},\ }\bibfield  {title}
  {\bibinfo {title} {Langevin dynamics of generalized spins as {SU($N$)}
  coherent states},\ }\href {https://arxiv.org/abs/2209.01265} {\bibfield
  {journal} {\bibinfo  {journal} {arXiv:2209.01265}\ } (\bibinfo {year}
  {2022}{\natexlab{b}})}\BibitemShut {NoStop}%
\bibitem [{\citenamefont {Samarakoon}\ \emph {et~al.}(2022)\citenamefont
  {Samarakoon}, \citenamefont {Laurell}, \citenamefont {Balz}, \citenamefont
  {Banerjee}, \citenamefont {Lampen-Kelley}, \citenamefont {Mandrus},
  \citenamefont {Nagler}, \citenamefont {Okamoto},\ and\ \citenamefont
  {Tennant}}]{SamarakoonPRR2022}%
  \BibitemOpen
  \bibfield  {author} {\bibinfo {author} {\bibfnamefont {A.~M.}\ \bibnamefont
  {Samarakoon}}, \bibinfo {author} {\bibfnamefont {P.}~\bibnamefont {Laurell}},
  \bibinfo {author} {\bibfnamefont {C.}~\bibnamefont {Balz}}, \bibinfo {author}
  {\bibfnamefont {A.}~\bibnamefont {Banerjee}}, \bibinfo {author}
  {\bibfnamefont {P.}~\bibnamefont {Lampen-Kelley}}, \bibinfo {author}
  {\bibfnamefont {D.}~\bibnamefont {Mandrus}}, \bibinfo {author} {\bibfnamefont
  {S.~E.}\ \bibnamefont {Nagler}}, \bibinfo {author} {\bibfnamefont
  {S.}~\bibnamefont {Okamoto}},\ and\ \bibinfo {author} {\bibfnamefont {D.~A.}\
  \bibnamefont {Tennant}},\ }\bibfield  {title} {\bibinfo {title} {Extraction
  of interaction parameters for
  {$\ensuremath{\alpha}\text{\ensuremath{-}}{\mathrm{RuCl}}_{3}$} from neutron
  data using machine learning},\ }\href
  {https://doi.org/10.1103/PhysRevResearch.4.L022061} {\bibfield  {journal}
  {\bibinfo  {journal} {Phys. Rev. Research}\ }\textbf {\bibinfo {volume}
  {4}},\ \bibinfo {pages} {L022061} (\bibinfo {year} {2022})}\BibitemShut
  {NoStop}%
\bibitem [{\citenamefont {Marzari}\ \emph {et~al.}(2012)\citenamefont
  {Marzari}, \citenamefont {Mostofi}, \citenamefont {Yates}, \citenamefont
  {Souza},\ and\ \citenamefont {Vanderbilt}}]{MarzariRMP2012}%
  \BibitemOpen
  \bibfield  {author} {\bibinfo {author} {\bibfnamefont {N.}~\bibnamefont
  {Marzari}}, \bibinfo {author} {\bibfnamefont {A.~A.}\ \bibnamefont
  {Mostofi}}, \bibinfo {author} {\bibfnamefont {J.~R.}\ \bibnamefont {Yates}},
  \bibinfo {author} {\bibfnamefont {I.}~\bibnamefont {Souza}},\ and\ \bibinfo
  {author} {\bibfnamefont {D.}~\bibnamefont {Vanderbilt}},\ }\bibfield  {title}
  {\bibinfo {title} {Maximally localized wannier functions: Theory and
  applications},\ }\href {https://doi.org/10.1103/RevModPhys.84.1419}
  {\bibfield  {journal} {\bibinfo  {journal} {Rev. Mod. Phys.}\ }\textbf
  {\bibinfo {volume} {84}},\ \bibinfo {pages} {1419} (\bibinfo {year}
  {2012})}\BibitemShut {NoStop}%
\bibitem [{\citenamefont {Aryasetiawan}\ \emph {et~al.}(2011)\citenamefont
  {Aryasetiawan}, \citenamefont {Miyake},\ and\ \citenamefont
  {Sakuma}}]{Aryasetiawan}%
  \BibitemOpen
  \bibfield  {author} {\bibinfo {author} {\bibfnamefont {F.}~\bibnamefont
  {Aryasetiawan}}, \bibinfo {author} {\bibfnamefont {T.}~\bibnamefont
  {Miyake}},\ and\ \bibinfo {author} {\bibfnamefont {R.}~\bibnamefont
  {Sakuma}},\ }\bibfield  {title} {\bibinfo {title} {The constrained {RPA}
  method for calculating the {H}ubbard $u$ from first-principles},\ }in\ \href
  {https://www.cond-mat.de/events/correl11/manuscripts/aryasetiawan.pdf} {\emph
  {\bibinfo {booktitle} {The {LDA}+{DMFT} approach to strongly correlated
  matter}}},\ \bibinfo {editor} {edited by\ \bibinfo {editor} {\bibfnamefont
  {E.}~\bibnamefont {Pavarini}}, \bibinfo {editor} {\bibfnamefont
  {E.}~\bibnamefont {Koch}}, \bibinfo {editor} {\bibfnamefont {D.}~\bibnamefont
  {Vollhardt}},\ and\ \bibinfo {editor} {\bibfnamefont {A.}~\bibnamefont
  {Lichtenstein}}}\ (\bibinfo  {publisher} {Forschungszentrum J{\"u}lich},\
  \bibinfo {year} {2011})\BibitemShut {NoStop}%
\bibitem [{\citenamefont {Eichstaedt}\ \emph {et~al.}(2019)\citenamefont
  {Eichstaedt}, \citenamefont {Zhang}, \citenamefont {Laurell}, \citenamefont
  {Okamoto}, \citenamefont {Eguiluz},\ and\ \citenamefont
  {Berlijn}}]{EichstaedtPRB2019}%
  \BibitemOpen
  \bibfield  {author} {\bibinfo {author} {\bibfnamefont {C.}~\bibnamefont
  {Eichstaedt}}, \bibinfo {author} {\bibfnamefont {Y.}~\bibnamefont {Zhang}},
  \bibinfo {author} {\bibfnamefont {P.}~\bibnamefont {Laurell}}, \bibinfo
  {author} {\bibfnamefont {S.}~\bibnamefont {Okamoto}}, \bibinfo {author}
  {\bibfnamefont {A.~G.}\ \bibnamefont {Eguiluz}},\ and\ \bibinfo {author}
  {\bibfnamefont {T.}~\bibnamefont {Berlijn}},\ }\bibfield  {title} {\bibinfo
  {title} {Deriving models for the {K}itaev spin-liquid candidate material
  {$\ensuremath{\alpha}\text{\ensuremath{-}}{\mathrm{RuCl}}_{3}$} from first
  principles},\ }\href {https://doi.org/10.1103/PhysRevB.100.075110} {\bibfield
   {journal} {\bibinfo  {journal} {Phys. Rev. B}\ }\textbf {\bibinfo {volume}
  {100}},\ \bibinfo {pages} {075110} (\bibinfo {year} {2019})}\BibitemShut
  {NoStop}%
\bibitem [{\citenamefont {Kulik}\ \emph {et~al.}(2022)\citenamefont {Kulik},
  \citenamefont {Hammerschmidt}, \citenamefont {Schmidt}, \citenamefont
  {Botti}, \citenamefont {Marques}, \citenamefont {Boley}, \citenamefont
  {Scheffler}, \citenamefont {Todorovi{\'{c}}}, \citenamefont {Rinke},
  \citenamefont {Oses}, \citenamefont {Smolyanyuk}, \citenamefont {Curtarolo},
  \citenamefont {Tkatchenko}, \citenamefont {Bart{\'{o}}k}, \citenamefont
  {Manzhos}, \citenamefont {Ihara}, \citenamefont {Carrington}, \citenamefont
  {Behler}, \citenamefont {Isayev}, \citenamefont {Veit}, \citenamefont
  {Grisafi}, \citenamefont {Nigam}, \citenamefont {Ceriotti}, \citenamefont
  {Sch{\"u}tt}, \citenamefont {Westermayr}, \citenamefont {Gastegger},
  \citenamefont {Maurer}, \citenamefont {Kalita}, \citenamefont {Burke},
  \citenamefont {Nagai}, \citenamefont {Akashi}, \citenamefont {Sugino},
  \citenamefont {Hermann}, \citenamefont {No{\'{e}}}, \citenamefont {Pilati},
  \citenamefont {Draxl}, \citenamefont {Kuban}, \citenamefont {Rigamonti},
  \citenamefont {Scheidgen}, \citenamefont {Esters}, \citenamefont {Hicks},
  \citenamefont {Toher}, \citenamefont {Balachandran}, \citenamefont {Tamblyn},
  \citenamefont {Whitelam}, \citenamefont {Bellinger},\ and\ \citenamefont
  {Ghiringhelli}}]{Kulik_2022}%
  \BibitemOpen
  \bibfield  {author} {\bibinfo {author} {\bibfnamefont {H.~J.}\ \bibnamefont
  {Kulik}}, \bibinfo {author} {\bibfnamefont {T.}~\bibnamefont
  {Hammerschmidt}}, \bibinfo {author} {\bibfnamefont {J.}~\bibnamefont
  {Schmidt}}, \bibinfo {author} {\bibfnamefont {S.}~\bibnamefont {Botti}},
  \bibinfo {author} {\bibfnamefont {M.~A.~L.}\ \bibnamefont {Marques}},
  \bibinfo {author} {\bibfnamefont {M.}~\bibnamefont {Boley}}, \bibinfo
  {author} {\bibfnamefont {M.}~\bibnamefont {Scheffler}}, \bibinfo {author}
  {\bibfnamefont {M.}~\bibnamefont {Todorovi{\'{c}}}}, \bibinfo {author}
  {\bibfnamefont {P.}~\bibnamefont {Rinke}}, \bibinfo {author} {\bibfnamefont
  {C.}~\bibnamefont {Oses}}, \bibinfo {author} {\bibfnamefont {A.}~\bibnamefont
  {Smolyanyuk}}, \bibinfo {author} {\bibfnamefont {S.}~\bibnamefont
  {Curtarolo}}, \bibinfo {author} {\bibfnamefont {A.}~\bibnamefont
  {Tkatchenko}}, \bibinfo {author} {\bibfnamefont {A.~P.}\ \bibnamefont
  {Bart{\'{o}}k}}, \bibinfo {author} {\bibfnamefont {S.}~\bibnamefont
  {Manzhos}}, \bibinfo {author} {\bibfnamefont {M.}~\bibnamefont {Ihara}},
  \bibinfo {author} {\bibfnamefont {T.}~\bibnamefont {Carrington}}, \bibinfo
  {author} {\bibfnamefont {J.}~\bibnamefont {Behler}}, \bibinfo {author}
  {\bibfnamefont {O.}~\bibnamefont {Isayev}}, \bibinfo {author} {\bibfnamefont
  {M.}~\bibnamefont {Veit}}, \bibinfo {author} {\bibfnamefont {A.}~\bibnamefont
  {Grisafi}}, \bibinfo {author} {\bibfnamefont {J.}~\bibnamefont {Nigam}},
  \bibinfo {author} {\bibfnamefont {M.}~\bibnamefont {Ceriotti}}, \bibinfo
  {author} {\bibfnamefont {K.~T.}\ \bibnamefont {Sch{\"u}tt}}, \bibinfo
  {author} {\bibfnamefont {J.}~\bibnamefont {Westermayr}}, \bibinfo {author}
  {\bibfnamefont {M.}~\bibnamefont {Gastegger}}, \bibinfo {author}
  {\bibfnamefont {R.~J.}\ \bibnamefont {Maurer}}, \bibinfo {author}
  {\bibfnamefont {B.}~\bibnamefont {Kalita}}, \bibinfo {author} {\bibfnamefont
  {K.}~\bibnamefont {Burke}}, \bibinfo {author} {\bibfnamefont
  {R.}~\bibnamefont {Nagai}}, \bibinfo {author} {\bibfnamefont
  {R.}~\bibnamefont {Akashi}}, \bibinfo {author} {\bibfnamefont
  {O.}~\bibnamefont {Sugino}}, \bibinfo {author} {\bibfnamefont
  {J.}~\bibnamefont {Hermann}}, \bibinfo {author} {\bibfnamefont
  {F.}~\bibnamefont {No{\'{e}}}}, \bibinfo {author} {\bibfnamefont
  {S.}~\bibnamefont {Pilati}}, \bibinfo {author} {\bibfnamefont
  {C.}~\bibnamefont {Draxl}}, \bibinfo {author} {\bibfnamefont
  {M.}~\bibnamefont {Kuban}}, \bibinfo {author} {\bibfnamefont
  {S.}~\bibnamefont {Rigamonti}}, \bibinfo {author} {\bibfnamefont
  {M.}~\bibnamefont {Scheidgen}}, \bibinfo {author} {\bibfnamefont
  {M.}~\bibnamefont {Esters}}, \bibinfo {author} {\bibfnamefont
  {D.}~\bibnamefont {Hicks}}, \bibinfo {author} {\bibfnamefont
  {C.}~\bibnamefont {Toher}}, \bibinfo {author} {\bibfnamefont {P.~V.}\
  \bibnamefont {Balachandran}}, \bibinfo {author} {\bibfnamefont
  {I.}~\bibnamefont {Tamblyn}}, \bibinfo {author} {\bibfnamefont
  {S.}~\bibnamefont {Whitelam}}, \bibinfo {author} {\bibfnamefont
  {C.}~\bibnamefont {Bellinger}},\ and\ \bibinfo {author} {\bibfnamefont
  {L.~M.}\ \bibnamefont {Ghiringhelli}},\ }\bibfield  {title} {\bibinfo {title}
  {Roadmap on machine learning in electronic structure},\ }\href
  {https://doi.org/10.1088/2516-1075/ac572f} {\bibfield  {journal} {\bibinfo
  {journal} {Electronic Structure}\ }\textbf {\bibinfo {volume} {4}},\ \bibinfo
  {pages} {023004} (\bibinfo {year} {2022})}\BibitemShut {NoStop}%
\bibitem [{\citenamefont {Brockherde}\ \emph {et~al.}(2017)\citenamefont
  {Brockherde}, \citenamefont {Vogt}, \citenamefont {Li}, \citenamefont
  {Tuckerman}, \citenamefont {Burke},\ and\ \citenamefont
  {M{\"u}ller}}]{Brockherde2017}%
  \BibitemOpen
  \bibfield  {author} {\bibinfo {author} {\bibfnamefont {F.}~\bibnamefont
  {Brockherde}}, \bibinfo {author} {\bibfnamefont {L.}~\bibnamefont {Vogt}},
  \bibinfo {author} {\bibfnamefont {L.}~\bibnamefont {Li}}, \bibinfo {author}
  {\bibfnamefont {M.~E.}\ \bibnamefont {Tuckerman}}, \bibinfo {author}
  {\bibfnamefont {K.}~\bibnamefont {Burke}},\ and\ \bibinfo {author}
  {\bibfnamefont {K.-R.}\ \bibnamefont {M{\"u}ller}},\ }\bibfield  {title}
  {\bibinfo {title} {Bypassing the kohn-sham equations with machine learning},\
  }\href {https://doi.org/10.1038/s41467-017-00839-3} {\bibfield  {journal}
  {\bibinfo  {journal} {Nature Communications}\ }\textbf {\bibinfo {volume}
  {8}},\ \bibinfo {pages} {872} (\bibinfo {year} {2017})}\BibitemShut {NoStop}%
\bibitem [{\citenamefont {Chandrasekaran}\ \emph {et~al.}(2019)\citenamefont
  {Chandrasekaran}, \citenamefont {Kamal}, \citenamefont {Batra}, \citenamefont
  {Kim}, \citenamefont {Chen},\ and\ \citenamefont
  {Ramprasad}}]{Chandrasekaran2018}%
  \BibitemOpen
  \bibfield  {author} {\bibinfo {author} {\bibfnamefont {A.}~\bibnamefont
  {Chandrasekaran}}, \bibinfo {author} {\bibfnamefont {D.}~\bibnamefont
  {Kamal}}, \bibinfo {author} {\bibfnamefont {R.}~\bibnamefont {Batra}},
  \bibinfo {author} {\bibfnamefont {C.}~\bibnamefont {Kim}}, \bibinfo {author}
  {\bibfnamefont {L.}~\bibnamefont {Chen}},\ and\ \bibinfo {author}
  {\bibfnamefont {R.}~\bibnamefont {Ramprasad}},\ }\bibfield  {title} {\bibinfo
  {title} {Solving the electronic structure problem with machine learning},\
  }\href {https://doi.org/10.1038/s41524-019-0162-7} {\bibfield  {journal}
  {\bibinfo  {journal} {npj Computational Materials}\ }\textbf {\bibinfo
  {volume} {5}},\ \bibinfo {pages} {22} (\bibinfo {year} {2019})}\BibitemShut
  {NoStop}%
\bibitem [{\citenamefont {Wang}\ \emph {et~al.}(2021)\citenamefont {Wang},
  \citenamefont {Ye}, \citenamefont {Wang}, \citenamefont {He}, \citenamefont
  {Huang},\ and\ \citenamefont {Chang}}]{Wang2021}%
  \BibitemOpen
  \bibfield  {author} {\bibinfo {author} {\bibfnamefont {Z.}~\bibnamefont
  {Wang}}, \bibinfo {author} {\bibfnamefont {S.}~\bibnamefont {Ye}}, \bibinfo
  {author} {\bibfnamefont {H.}~\bibnamefont {Wang}}, \bibinfo {author}
  {\bibfnamefont {J.}~\bibnamefont {He}}, \bibinfo {author} {\bibfnamefont
  {Q.}~\bibnamefont {Huang}},\ and\ \bibinfo {author} {\bibfnamefont
  {S.}~\bibnamefont {Chang}},\ }\bibfield  {title} {\bibinfo {title} {Machine
  learning method for tight-binding hamiltonian parameterization from ab-initio
  band structure},\ }\href {https://doi.org/10.1038/s41524-020-00490-5}
  {\bibfield  {journal} {\bibinfo  {journal} {npj Computational Materials}\
  }\textbf {\bibinfo {volume} {7}},\ \bibinfo {pages} {11} (\bibinfo {year}
  {2021})}\BibitemShut {NoStop}%
\bibitem [{\citenamefont {Yu}\ \emph {et~al.}(2020)\citenamefont {Yu},
  \citenamefont {Yang}, \citenamefont {Wu},\ and\ \citenamefont
  {Marom}}]{Yu2020}%
  \BibitemOpen
  \bibfield  {author} {\bibinfo {author} {\bibfnamefont {M.}~\bibnamefont
  {Yu}}, \bibinfo {author} {\bibfnamefont {S.}~\bibnamefont {Yang}}, \bibinfo
  {author} {\bibfnamefont {C.}~\bibnamefont {Wu}},\ and\ \bibinfo {author}
  {\bibfnamefont {N.}~\bibnamefont {Marom}},\ }\bibfield  {title} {\bibinfo
  {title} {Machine learning the hubbard u parameter in dft+u using bayesian
  optimization},\ }\href {https://doi.org/10.1038/s41524-020-00446-9}
  {\bibfield  {journal} {\bibinfo  {journal} {npj Computational Materials}\
  }\textbf {\bibinfo {volume} {6}},\ \bibinfo {pages} {180} (\bibinfo {year}
  {2020})}\BibitemShut {NoStop}%
\bibitem [{\citenamefont {Schattauer}\ \emph {et~al.}(2022)\citenamefont
  {Schattauer}, \citenamefont {Todorovi{\'c}}, \citenamefont {Ghosh},
  \citenamefont {Rinke},\ and\ \citenamefont {Libisch}}]{Schattauer2022}%
  \BibitemOpen
  \bibfield  {author} {\bibinfo {author} {\bibfnamefont {C.}~\bibnamefont
  {Schattauer}}, \bibinfo {author} {\bibfnamefont {M.}~\bibnamefont
  {Todorovi{\'c}}}, \bibinfo {author} {\bibfnamefont {K.}~\bibnamefont
  {Ghosh}}, \bibinfo {author} {\bibfnamefont {P.}~\bibnamefont {Rinke}},\ and\
  \bibinfo {author} {\bibfnamefont {F.}~\bibnamefont {Libisch}},\ }\bibfield
  {title} {\bibinfo {title} {Machine learning sparse tight-binding parameters
  for defects},\ }\href {https://doi.org/10.1038/s41524-022-00791-x} {\bibfield
   {journal} {\bibinfo  {journal} {npj Computational Materials}\ }\textbf
  {\bibinfo {volume} {8}},\ \bibinfo {pages} {116} (\bibinfo {year}
  {2022})}\BibitemShut {NoStop}%
\bibitem [{\citenamefont {Liu}\ \emph {et~al.}(2017{\natexlab{b}})\citenamefont
  {Liu}, \citenamefont {Shen}, \citenamefont {Qi}, \citenamefont {Meng},\ and\
  \citenamefont {Fu}}]{LiuPRB2017}%
  \BibitemOpen
  \bibfield  {author} {\bibinfo {author} {\bibfnamefont {J.}~\bibnamefont
  {Liu}}, \bibinfo {author} {\bibfnamefont {H.}~\bibnamefont {Shen}}, \bibinfo
  {author} {\bibfnamefont {Y.}~\bibnamefont {Qi}}, \bibinfo {author}
  {\bibfnamefont {Z.~Y.}\ \bibnamefont {Meng}},\ and\ \bibinfo {author}
  {\bibfnamefont {L.}~\bibnamefont {Fu}},\ }\bibfield  {title} {\bibinfo
  {title} {Self-learning {Monte Carlo} method and cumulative update in fermion
  systems},\ }\href {https://doi.org/10.1103/PhysRevB.95.241104} {\bibfield
  {journal} {\bibinfo  {journal} {Phys. Rev. B}\ }\textbf {\bibinfo {volume}
  {95}},\ \bibinfo {pages} {241104} (\bibinfo {year}
  {2017}{\natexlab{b}})}\BibitemShut {NoStop}%
\bibitem [{\citenamefont {Huang}\ \emph
  {et~al.}(2017{\natexlab{b}})\citenamefont {Huang}, \citenamefont {Yang},\
  and\ \citenamefont {Wang}}]{HuangPRE2017}%
  \BibitemOpen
  \bibfield  {author} {\bibinfo {author} {\bibfnamefont {L.}~\bibnamefont
  {Huang}}, \bibinfo {author} {\bibfnamefont {Y.-f.}\ \bibnamefont {Yang}},\
  and\ \bibinfo {author} {\bibfnamefont {L.}~\bibnamefont {Wang}},\ }\bibfield
  {title} {\bibinfo {title} {Recommender engine for continuous-time quantum
  {Monte Carlo} methods},\ }\href {https://doi.org/10.1103/PhysRevE.95.031301}
  {\bibfield  {journal} {\bibinfo  {journal} {Phys. Rev. E}\ }\textbf {\bibinfo
  {volume} {95}},\ \bibinfo {pages} {031301} (\bibinfo {year}
  {2017}{\natexlab{b}})}\BibitemShut {NoStop}%
\bibitem [{\citenamefont {Nagai}\ \emph {et~al.}(2017)\citenamefont {Nagai},
  \citenamefont {Shen}, \citenamefont {Qi}, \citenamefont {Liu},\ and\
  \citenamefont {Fu}}]{NagaiPRB2017}%
  \BibitemOpen
  \bibfield  {author} {\bibinfo {author} {\bibfnamefont {Y.}~\bibnamefont
  {Nagai}}, \bibinfo {author} {\bibfnamefont {H.}~\bibnamefont {Shen}},
  \bibinfo {author} {\bibfnamefont {Y.}~\bibnamefont {Qi}}, \bibinfo {author}
  {\bibfnamefont {J.}~\bibnamefont {Liu}},\ and\ \bibinfo {author}
  {\bibfnamefont {L.}~\bibnamefont {Fu}},\ }\bibfield  {title} {\bibinfo
  {title} {Self-learning {Monte Carlo} method: Continuous-time algorithm},\
  }\href {https://doi.org/10.1103/PhysRevB.96.161102} {\bibfield  {journal}
  {\bibinfo  {journal} {Phys. Rev. B}\ }\textbf {\bibinfo {volume} {96}},\
  \bibinfo {pages} {161102} (\bibinfo {year} {2017})}\BibitemShut {NoStop}%
\bibitem [{\citenamefont {Xu}\ \emph {et~al.}(2017)\citenamefont {Xu},
  \citenamefont {Qi}, \citenamefont {Liu}, \citenamefont {Fu},\ and\
  \citenamefont {Meng}}]{XuPRB2017}%
  \BibitemOpen
  \bibfield  {author} {\bibinfo {author} {\bibfnamefont {X.~Y.}\ \bibnamefont
  {Xu}}, \bibinfo {author} {\bibfnamefont {Y.}~\bibnamefont {Qi}}, \bibinfo
  {author} {\bibfnamefont {J.}~\bibnamefont {Liu}}, \bibinfo {author}
  {\bibfnamefont {L.}~\bibnamefont {Fu}},\ and\ \bibinfo {author}
  {\bibfnamefont {Z.~Y.}\ \bibnamefont {Meng}},\ }\bibfield  {title} {\bibinfo
  {title} {Self-learning quantum {Monte Carlo} method in interacting fermion
  systems},\ }\href {https://doi.org/10.1103/PhysRevB.96.041119} {\bibfield
  {journal} {\bibinfo  {journal} {Phys. Rev. B}\ }\textbf {\bibinfo {volume}
  {96}},\ \bibinfo {pages} {041119} (\bibinfo {year} {2017})}\BibitemShut
  {NoStop}%
\bibitem [{\citenamefont {Shen}\ \emph {et~al.}(2018)\citenamefont {Shen},
  \citenamefont {Liu},\ and\ \citenamefont {Fu}}]{ShenPRB2018}%
  \BibitemOpen
  \bibfield  {author} {\bibinfo {author} {\bibfnamefont {H.}~\bibnamefont
  {Shen}}, \bibinfo {author} {\bibfnamefont {J.}~\bibnamefont {Liu}},\ and\
  \bibinfo {author} {\bibfnamefont {L.}~\bibnamefont {Fu}},\ }\bibfield
  {title} {\bibinfo {title} {Self-learning {Monte Carlo} with deep neural
  networks},\ }\href {https://doi.org/10.1103/PhysRevB.97.205140} {\bibfield
  {journal} {\bibinfo  {journal} {Phys. Rev. B}\ }\textbf {\bibinfo {volume}
  {97}},\ \bibinfo {pages} {205140} (\bibinfo {year} {2018})}\BibitemShut
  {NoStop}%
\bibitem [{\citenamefont {Liu}\ \emph {et~al.}(2018)\citenamefont {Liu},
  \citenamefont {Xu}, \citenamefont {Qi}, \citenamefont {Sun},\ and\
  \citenamefont {Meng}}]{LiuPRB2018}%
  \BibitemOpen
  \bibfield  {author} {\bibinfo {author} {\bibfnamefont {Z.~H.}\ \bibnamefont
  {Liu}}, \bibinfo {author} {\bibfnamefont {X.~Y.}\ \bibnamefont {Xu}},
  \bibinfo {author} {\bibfnamefont {Y.}~\bibnamefont {Qi}}, \bibinfo {author}
  {\bibfnamefont {K.}~\bibnamefont {Sun}},\ and\ \bibinfo {author}
  {\bibfnamefont {Z.~Y.}\ \bibnamefont {Meng}},\ }\bibfield  {title} {\bibinfo
  {title} {Itinerant quantum critical point with frustration and a non-{F}ermi
  liquid},\ }\href {https://doi.org/10.1103/PhysRevB.98.045116} {\bibfield
  {journal} {\bibinfo  {journal} {Phys. Rev. B}\ }\textbf {\bibinfo {volume}
  {98}},\ \bibinfo {pages} {045116} (\bibinfo {year} {2018})}\BibitemShut
  {NoStop}%
\bibitem [{\citenamefont {Chen}\ \emph
  {et~al.}(2018{\natexlab{a}})\citenamefont {Chen}, \citenamefont {Xu},
  \citenamefont {Liu}, \citenamefont {Batrouni}, \citenamefont {Scalettar},\
  and\ \citenamefont {Meng}}]{ChenPRB2018}%
  \BibitemOpen
  \bibfield  {author} {\bibinfo {author} {\bibfnamefont {C.}~\bibnamefont
  {Chen}}, \bibinfo {author} {\bibfnamefont {X.~Y.}\ \bibnamefont {Xu}},
  \bibinfo {author} {\bibfnamefont {J.}~\bibnamefont {Liu}}, \bibinfo {author}
  {\bibfnamefont {G.}~\bibnamefont {Batrouni}}, \bibinfo {author}
  {\bibfnamefont {R.}~\bibnamefont {Scalettar}},\ and\ \bibinfo {author}
  {\bibfnamefont {Z.~Y.}\ \bibnamefont {Meng}},\ }\bibfield  {title} {\bibinfo
  {title} {Symmetry-enforced self-learning {Monte Carlo} method applied to the
  {Holstein} model},\ }\href {https://doi.org/10.1103/PhysRevB.98.041102}
  {\bibfield  {journal} {\bibinfo  {journal} {Phys. Rev. B}\ }\textbf {\bibinfo
  {volume} {98}},\ \bibinfo {pages} {041102} (\bibinfo {year}
  {2018}{\natexlab{a}})}\BibitemShut {NoStop}%
\bibitem [{\citenamefont {Li}\ \emph {et~al.}(2019)\citenamefont {Li},
  \citenamefont {Dee}, \citenamefont {Khatami},\ and\ \citenamefont
  {Johnston}}]{LiPRB2019}%
  \BibitemOpen
  \bibfield  {author} {\bibinfo {author} {\bibfnamefont {S.}~\bibnamefont
  {Li}}, \bibinfo {author} {\bibfnamefont {P.~M.}\ \bibnamefont {Dee}},
  \bibinfo {author} {\bibfnamefont {E.}~\bibnamefont {Khatami}},\ and\ \bibinfo
  {author} {\bibfnamefont {S.}~\bibnamefont {Johnston}},\ }\bibfield  {title}
  {\bibinfo {title} {Accelerating lattice quantum {Monte Carlo} simulations
  using artificial neural networks: Application to the {Holstein} model},\
  }\href {https://doi.org/10.1103/PhysRevB.100.020302} {\bibfield  {journal}
  {\bibinfo  {journal} {Phys. Rev. B}\ }\textbf {\bibinfo {volume} {100}},\
  \bibinfo {pages} {020302} (\bibinfo {year} {2019})}\BibitemShut {NoStop}%
\bibitem [{\citenamefont {Tanaka}\ and\ \citenamefont
  {Tomiya}(2017{\natexlab{b}})}]{TanakaPreprint}%
  \BibitemOpen
  \bibfield  {author} {\bibinfo {author} {\bibfnamefont {A.}~\bibnamefont
  {Tanaka}}\ and\ \bibinfo {author} {\bibfnamefont {A.}~\bibnamefont
  {Tomiya}},\ }\bibfield  {title} {\bibinfo {title} {Towards reduction of
  autocorrelation in {HMC} by machine learning},\ }\href
  {https://arxiv.org/abs/1712.03893} {\bibfield  {journal} {\bibinfo  {journal}
  {arXiv:1712.03893}\ } (\bibinfo {year} {2017}{\natexlab{b}})}\BibitemShut
  {NoStop}%
\bibitem [{\citenamefont {Kohshiro}\ and\ \citenamefont
  {Nagai}(2021)}]{Kohshiro2021}%
  \BibitemOpen
  \bibfield  {author} {\bibinfo {author} {\bibfnamefont {H.}~\bibnamefont
  {Kohshiro}}\ and\ \bibinfo {author} {\bibfnamefont {Y.}~\bibnamefont
  {Nagai}},\ }\bibfield  {title} {\bibinfo {title} {Effective
  {Ruderman--Kittel--Kasuya--Yosida}-like interaction in diluted
  double-exchange model: Self-learning {Monte Carlo} approach},\ }\href
  {https://doi.org/10.7566/JPSJ.90.034711} {\bibfield  {journal} {\bibinfo
  {journal} {Journal of the Physical Society of Japan}\ }\textbf {\bibinfo
  {volume} {90}},\ \bibinfo {pages} {034711} (\bibinfo {year}
  {2021})}\BibitemShut {NoStop}%
\bibitem [{\citenamefont {Monroe}\ and\ \citenamefont
  {Shen}(2022)}]{Monroe2022}%
  \BibitemOpen
  \bibfield  {author} {\bibinfo {author} {\bibfnamefont {J.~I.}\ \bibnamefont
  {Monroe}}\ and\ \bibinfo {author} {\bibfnamefont {V.~K.}\ \bibnamefont
  {Shen}},\ }\bibfield  {title} {\bibinfo {title} {Learning efficient,
  collective {Monte Carlo} moves with variational autoencoders},\ }\bibfield
  {booktitle} {\emph {\bibinfo {booktitle} {Journal of Chemical Theory and
  Computation}},\ }\href {https://doi.org/10.1021/acs.jctc.2c00110} {\bibfield
  {journal} {\bibinfo  {journal} {Journal of Chemical Theory and Computation}\
  }\textbf {\bibinfo {volume} {18}},\ \bibinfo {pages} {3622} (\bibinfo {year}
  {2022})}\BibitemShut {NoStop}%
\bibitem [{\citenamefont {Levine}\ \emph {et~al.}(2019)\citenamefont {Levine},
  \citenamefont {Sharir}, \citenamefont {Cohen},\ and\ \citenamefont
  {Shashua}}]{y_levine_19}%
  \BibitemOpen
  \bibfield  {author} {\bibinfo {author} {\bibfnamefont {Y.}~\bibnamefont
  {Levine}}, \bibinfo {author} {\bibfnamefont {O.}~\bibnamefont {Sharir}},
  \bibinfo {author} {\bibfnamefont {N.}~\bibnamefont {Cohen}},\ and\ \bibinfo
  {author} {\bibfnamefont {A.}~\bibnamefont {Shashua}},\ }\bibfield  {title}
  {\bibinfo {title} {Quantum entanglement in deep learning architectures},\
  }\href {https://doi.org/10.1103/PhysRevLett.122.065301} {\bibfield  {journal}
  {\bibinfo  {journal} {Phys. Rev. Lett.}\ }\textbf {\bibinfo {volume} {122}},\
  \bibinfo {pages} {065301} (\bibinfo {year} {2019})}\BibitemShut {NoStop}%
\bibitem [{\citenamefont {Sharir}\ \emph {et~al.}(2021)\citenamefont {Sharir},
  \citenamefont {Shashua},\ and\ \citenamefont {Carleo}}]{o_sharir_22}%
  \BibitemOpen
  \bibfield  {author} {\bibinfo {author} {\bibfnamefont {O.}~\bibnamefont
  {Sharir}}, \bibinfo {author} {\bibfnamefont {A.}~\bibnamefont {Shashua}},\
  and\ \bibinfo {author} {\bibfnamefont {G.}~\bibnamefont {Carleo}},\ }\href
  {https://doi.org/10.48550/ARXIV.2103.10293} {\bibinfo {title} {Neural tensor
  contractions and the expressive power of deep neural quantum states}}
  (\bibinfo {year} {2021})\BibitemShut {NoStop}%
\bibitem [{\citenamefont {Chen}\ \emph
  {et~al.}(2018{\natexlab{b}})\citenamefont {Chen}, \citenamefont {Cheng},
  \citenamefont {Xie}, \citenamefont {Wang},\ and\ \citenamefont
  {Xiang}}]{PhysRevB.97.085104}%
  \BibitemOpen
  \bibfield  {author} {\bibinfo {author} {\bibfnamefont {J.}~\bibnamefont
  {Chen}}, \bibinfo {author} {\bibfnamefont {S.}~\bibnamefont {Cheng}},
  \bibinfo {author} {\bibfnamefont {H.}~\bibnamefont {Xie}}, \bibinfo {author}
  {\bibfnamefont {L.}~\bibnamefont {Wang}},\ and\ \bibinfo {author}
  {\bibfnamefont {T.}~\bibnamefont {Xiang}},\ }\bibfield  {title} {\bibinfo
  {title} {Equivalence of restricted boltzmann machines and tensor network
  states},\ }\href {https://doi.org/10.1103/PhysRevB.97.085104} {\bibfield
  {journal} {\bibinfo  {journal} {Phys. Rev. B}\ }\textbf {\bibinfo {volume}
  {97}},\ \bibinfo {pages} {085104} (\bibinfo {year}
  {2018}{\natexlab{b}})}\BibitemShut {NoStop}%
\bibitem [{\citenamefont {Zheng}\ \emph {et~al.}(2019)\citenamefont {Zheng},
  \citenamefont {He}, \citenamefont {Regnault},\ and\ \citenamefont
  {Bernevig}}]{PhysRevB.99.155129}%
  \BibitemOpen
  \bibfield  {author} {\bibinfo {author} {\bibfnamefont {Y.}~\bibnamefont
  {Zheng}}, \bibinfo {author} {\bibfnamefont {H.}~\bibnamefont {He}}, \bibinfo
  {author} {\bibfnamefont {N.}~\bibnamefont {Regnault}},\ and\ \bibinfo
  {author} {\bibfnamefont {B.~A.}\ \bibnamefont {Bernevig}},\ }\bibfield
  {title} {\bibinfo {title} {Restricted boltzmann machines and matrix product
  states of one-dimensional translationally invariant stabilizer codes},\
  }\href {https://doi.org/10.1103/PhysRevB.99.155129} {\bibfield  {journal}
  {\bibinfo  {journal} {Phys. Rev. B}\ }\textbf {\bibinfo {volume} {99}},\
  \bibinfo {pages} {155129} (\bibinfo {year} {2019})}\BibitemShut {NoStop}%
\end{thebibliography}%
\end{document}